THÈSE DE DOCTORAT
DE SORBONNE UNIVERSITÉ

Spécialité : Physique Théorique

École doctorale n°564 : Physique en Île-de-France

réalisée

au Laboratoire de Physique Théorique et Hautes Énergies

sous la direction de Grégory SCHEHR

présentée par

# Mathis GUÉNEAU

pour obtenir le grade de :

DOCTEUR DE SORBONNE UNIVERSITÉ

Sujet de la thèse :

## Non-Equilibrium Dynamics and First-Passage Properties of Stochastic Processes: From Brownian Motion to Active Particles

soutenue le 16 juin 2025

devant le jury composé de :

| | | |
|---|---|---|
| M. | Olivier BÉNICHOU | Président |
| M. | Yariv KAFRI | Rapporteur |
| M. | Hugo TOUCHETTE | Rapporteur |
| M | Martin R. EVANS | Examinateur |
| M$^{me}$ | Cécile MONTHUS | Examinatrice |
| M. | Grégory SCHEHR | Directeur de thèse |
| M. | Satya N. MAJUMDAR | Membre invité |

# Contents

# Remerciements

Si cette thèse a pu se dérouler de manière aussi épanouissante, tant sur les plans scientifique qu'humain, c'est en grande partie grâce à mon directeur de thèse, Grégory Schehr, que je tiens à remercier tout particulièrement. Merci de m'avoir accordé ta confiance en acceptant de m'encadrer. Ton accompagnement rigoureux, tes qualités humaines et ta bienveillance ont été déterminants dans l'aboutissement de ce travail. Tu as été un guide précieux, aussi bien pour la réalisation de nos projets et articles que pour tes conseils avisés concernant mon orientation professionnelle et la recherche de postdoctorat. J'ai énormément appris de ta vision de la recherche, de ta rigueur et de ton inlassable pédagogie, que j'espère un jour approcher. Je n'aurais pu espérer meilleure thèse, et j'ai déjà hâte de commencer nos futures collaborations. Encore merci Grégory.

Si vous êtes un jour à court d'idées et à la recherche d'un problème intéressant à résoudre, prenez un café avec Satya N. Majumdar : vous en ressortirez avec des questions passionnantes et stimulantes ! Merci Satya pour nos nombreuses collaborations et discussions tout au long de ces trois années de thèse. Ta bonne humeur et ton rire contagieux ont contribué à rendre cette aventure agréable. Je te suis également très reconnaissant pour tes précieux conseils et ton soutien dans ma recherche de postdoctorat.

Merci aux membres du jury – Olivier Bénichou, Martin R. Evans, Yariv Kafri, Cécile Monthus et Hugo Touchette – d'avoir accepté d'évaluer mon travail de thèse. Une mention particulière à Yariv Kafri et Hugo Touchette, que je remercie chaleureusement d'avoir accepté le rôle de rapporteurs.

À peine quelques mois après le début de la thèse, sans vraiment le savoir, on s'est lancé dans une belle aventure, Léo ! Nos efforts ont abouti à deux publications dont je suis très fier. J'ai énormément appris à tes côtés au fil de ces trois années partagées dans le même bureau. Je garde de précieux souvenirs : les pauses café où l'on se plaignait de la recherche de postdoc, les séances de bateau à Beg Rohu, ou encore la splendide randonnée sur le désert de Platé aux Houches avec Julien B. Je te souhaite le meilleur pour ton postdoc à Chicago. J'espère que nos chemins continueront de se croiser régulièrement en conférence – et pourquoi pas, dans quelques années, à nouveau en France.

Merci à Marco Biroli pour nos nombreuses discussions et les souvenirs partagés lors des différentes écoles d'été. Je te souhaite aussi beaucoup de bonheur à Chicago, et j'espère qu'on finira par réussir à creuser ce fameux problème sur le resetting ! Francesco Mori, Benjamin Debruyne, Bertrand Lacroix-A-Chez-Toine, Ricardo Marino et Giuseppe Del Vecchio Del Vecchio : vous faites partie de la grande famille des thésards et postdocs de Grégory et Satya, et vous m'avez tous, à un moment ou un autre, donné des conseils précieux – merci à vous. Merci également à Pierre Le Doussal, pour les nombreux repas partagés à l'Ardoise et la relecture de nos travaux avec Léo.

Merci à Diyar, Greivin, Jules et Tom, avec qui j'ai partagé ces trois années de thèse au LPTHE – et de très nombreux repas au CROUS, d'abord à 11h30, puis, malheureusement pour mon estomac, à 13h. On a aussi partagé la préparation (et le rangement!) du thé du lundi pendant 2 ans. Je vous souhaite à tous le meilleur pour la suite, en espérant qu'on continuera à se recroiser. Un merci tout particulier à Yann pour ses conseils et sa bienveillance pendant nos presque trois années dans le même bureau. Je tiens également à saluer l'ensemble des thésards du LPTHE que j'ai eu le plaisir de côtoyer durant cette thèse : Andrei, Andriani, Anthony, Arno, Ayla, Daniele, les Francesco, Grégoire, Jordan, Matteo, Pierre, Simon, Vincent, Wenqi et Yehudi.

Je remercie également l'équipe administrative du LPTHE – Carole, Françoise, Laurent, Marwa, Sofiane – ainsi que Michela Petrini et Marco Picco pour leur aide précieuse tout au long de ces années. Merci aussi à l'ensemble des membres permanents et postdocs du laboratoire, en particulier à Mark Goodsell, qui a accepté de faire partie de mon comité de suivi de thèse, et à Laeticia Cugliandolo pour les discussions que nous avons eues.

Merci à nos voisins du LPTMC pour les nombreux séminaires. En particulier, je remercie Aurélien Grabsch pour sa présence dans mon comité de thèse, son aide précieuse lors de mes candidatures aux postdocs, ainsi que pour sa bienveillance et nos nombreux échanges. Merci également à Olivier Bénichou, à qui nous avons eu le plaisir de présenter nos travaux avec Léo. Enfin, un grand merci à Jérémie Klinger pour nos échanges enrichissants et tes conseils éclairants sur les postdocs.

Un peu plus loin, à Orsay, j'ai toujours été très bien accueilli au LPTMS. Merci à tous ses membres, et plus particulièrement à Claudine Le Vaou et Alberto Rosso. Merci aussi à Maximilien Bernard et Alessandro Pacco, avec qui j'ai eu plaisir à échanger. Je tiens également à remercier Christophe Texier, qui m'a fait découvrir la physique statistique au magistère de physique d'Orsay, ainsi que Bertrand Lacroix-A-Chez-Toine, qui m'avait donné les TDs à l'époque – sans que je me doute un instant que nos chemins se recroiseraient plus tard.

Merci à Giulio Biroli ainsi qu'à tous les organisateurs et enseignants de l'école d'été de Beg Rohu 2023. Merci également à celles et ceux de la session d'été aux Houches, "Theory of Large Deviations and Applications", en juillet 2024. En particulier, je remercie Hugo Touchette pour les discussions que nous avons eues durant ce mois à la montagne. Les moments passés lors de ces deux écoles d'été comptent parmi mes meilleurs souvenirs de thèse. J'adresse aussi mes remerciements à Martin R. Evans et aux organisateurs de la conférence "New Vistas in Stochastic Resetting" à Édimbourg.

Je ne peux pas faire l'impasse sur les belles années passées à la fac. Merci à Bruno Espagnon, Patrice Hello, Pierre-Yves Le Gall, Claire Marrache, Sylvia Matzen, Nicolas Pavloff, Patrick Puzo, David Ros, ainsi qu'à toute l'équipe du magistère de physique d'Orsay et de la PMCP, pour la qualité de leurs enseignements et leur aide précieuse dans la préparation du concours de l'X. Merci à Asmaa Abada et Alexandre Zabi de m'avoir encadré lors de mes premiers stages de recherche : vous avez contribué à éveiller mon envie de me lancer dans cette thèse.

Enfin, pour l'après-thèse, je remercie chaleureusement Frank Jülicher et Christina Kurzthaler pour leur confiance et pour m'offrir cette belle opportunité de postdoc au MPI-PKS à Dresde.

Je souhaite maintenant remercier mes amis, rencontrés à Orsay il y a déjà une dizaine d'années (!). Merci à Chrysostome et Thomas, pour ces premières années à la fac, les concerts, et toutes les jam sessions improvisées pendant les pauses du midi. Merci à Kuzco, Jeff, Guillaume et Thibault qui me font encore et toujours rire depuis les $9m^2$ du CROUS. Merci à Charley pour ton amitié et pour m'avoir fait découvrir le plaisir des randonnées en montagne et des nuits passées en refuge. Merci à Émile pour tous ces moments partagés – des révisions d'oraux de l'X aux randonnées, en passant par le ski avec Charley et Victor, et aussi pour le stage en gendarmerie avec Victor.

Enfin, un immense merci à Victor, partenaire infatigable de tant d'aventures : des marches-cafés aux longues courses à pied, toujours suivies d'un bon repas ; des vacances sportives à la montagne à la découverte de Stockholm et de ses fikas, où nous avons mis à l'épreuve nos

# List of Publications

# Index of Notations and Abbreviations

| | |
|---|---|
| ABP | Active Brownian Particle |
| AOUP | Active Ornstein-Uhlenbeck Particle |
| BLT | Bilateral Laplace Transform |
| CTRW | Continuous Time Random Walk |
| CLT | Central Limit Theorem |
| DBM | Dyson Brownian Motion |
| EVS | Extreme Value Statistics |
| FDT | Fluctuation-Dissipation Theorem |
| fBM | fractional Brownian Motion |
| i.i.d. | independent and identically distributed |
| LDF | Large Deviation Function |
| LDT | Large Deviation Theory |
| MFPT | Mean First-Passage Time |
| MSD | Mean Square Displacement |
| NESS | Non-Equilibrium Steady State |
| OU | Ornstein-Uhlenbeck |
| PDF | Probability Distribution Function |
| rBM | resetting Brownian Motion |
| RTP | Run-and-Tumble Particle |
| SCGF | Scaled Cumulant Generating Function |
| SDE | Stochastic Differential Equation |
| | |
| $\Gamma(z)$ | Gamma function, $\Gamma(z) = \int_0^\infty t^{z-1} e^{-t}\, dt$ |
| $(a)_n$ | Pochhammer symbol (rising factorial), $(a)_n = \frac{\Gamma(a+n)}{\Gamma(a)}$ |
| $_pF_q$ | Generalized hypergeometric function, $_pF_q(a_1,\ldots,a_p; b_1,\ldots,b_q; z) = \sum_{k=0}^\infty \frac{(a_1)_k \cdots (a_p)_k}{(b_1)_k \cdots (b_q)_k} \frac{z^k}{k!}$ |
| $\boxed{\text{result}}$ | Boxed results indicate important findings that represent original contributions of this thesis |

# Résumé en Français

Considérez un simple lancer de pièce. Naturellement, pour prédire son issue, nous utilisons une description probabiliste, en attribuant une probabilité égale 1/2 à chacun des deux résultats possibles – pile ou face. Pourtant, si nous connaissions précisément les conditions initiales du lancer – la vitesse initiale, sa direction, l'orientation de la pièce, ainsi que les facteurs environnementaux tels que la température de l'air ou la vitesse du vent – alors, en principe, nous pourrions intégrer exactement les équations du mouvement afin de prédire le résultat avec certitude. Cependant, résoudre directement ces équations devient rapidement une tâche extrêmement complexe. C'est pourquoi nous adoptons une approche probabiliste. Bien que simple, cet exemple illustre de manière frappante l'élégance et la puissance du raisonnement probabiliste : face à des systèmes complexes et sensibles aux conditions initiales, une approche probabiliste devient non seulement pratique, mais indispensable.

En physique, un tel raisonnement statistique est à l'origine de concepts fondamentaux – par exemple, la définition de la température d'un gaz comme mesure de l'énergie cinétique moyenne de ses constituants microscopiques via le théorème d'équipartition. Mais la portée des descriptions probabilistes dépasse largement la physique : elles ont fourni des éclairages puissants dans des domaines aussi divers que la biologie moléculaire, pour décrire la diffusion de moteurs moléculaires le long de brins d'ADN [7], les marchés financiers, pour modéliser la dynamique des prix d'actifs [8, 9], et même l'apprentissage automatique, pour caractériser la dynamique d'apprentissage dans des réseaux de neurones [10].

La théorie des probabilités et la physique statistique constituent donc des pierres angulaires de la science moderne, permettant des descriptions efficaces et fidèles des phénomènes macroscopiques. Ces descriptions ne peuvent toutefois pas reposer uniquement sur les propriétés microscopiques des constituants individuels, car des comportements fondamentalement nouveaux et inattendus émergent aux échelles plus grandes – une idée capturée par la célèbre formule d'Anderson, "More is Different" [11]. Parmi les nombreux phénomènes qui mettent en évidence la puissance de la formulation statistique, le mouvement brownien est un exemple particulièrement marquant et fondamental.

Le mouvement brownien est l'un des processus stochastiques les plus fondamentaux, doté d'applications d'une grande portée en physique, mathématiques, chimie, finance et biologie [12]. Observé pour la première fois par le botaniste Robert Brown en 1827 alors qu'il étudiait des grains de pollen suspendus dans l'eau, le mouvement brownien a reçu sa fondation théorique grâce au travail précurseur d'Albert Einstein en 1905 [13], suivi un an plus tard par Smoluchowski [14], qui montrèrent tous deux que ce mouvement erratique pouvait être compris comme le résultat d'une *marche aléatoire*. Le travail d'Einstein fournit une explication simple mais profonde du déplacement erratique de particules microscopiques en suspension dans un fluide, le reliant aux collisions moléculaires et offrant l'une des premières preuves directes de l'existence des atomes. Cette percée fut ensuite confirmée expérimentalement par Jean Perrin, dont les travaux consolidèrent la théorie atomique de la matière [15]. Au-delà de la physique, le mouvement brownien est devenu un concept essentiel en théorie des probabilités et dans l'étude des processus stochastiques. Dans sa thèse de 1900, Louis Bachelier fut le premier à modéliser le processus stochastique aujourd'hui connu sous le nom de mouvement brownien dans un contexte mathématique, l'appliquant à la spéculation financière [8]. Norbert Wiener en formalisa la structure mathématique dans les années 1920, menant à la théorie moderne des processus stochastiques à temps continu, avec des applications allant de la science du climat [16] aux modèles de diffusion en intelligence artificielle [17].

Plus précisément, le mouvement brownien désigne le déplacement erratique d'une particule passive suspendue dans un fluide maintenu à une température constante $T$ – c'est-à-dire un bain à l'équilibre thermodynamique. On qualifie la particule de passive car, en l'absence du fluide, elle resterait au repos. Dans ce cadre, son mouvement résulte purement d'innombrables collisions microscopiques avec les molécules environnantes. Chaque déplacement individuel provoqué par ces collisions est aléatoire, symétrique et de variance finie. D'un point de vue mathématique, la position de la particule à un instant donné peut donc être vue comme la somme de nombreux petits déplacements indépendants. Cette observation relie naturellement le mouvement brownien à l'un des piliers de la théorie des probabilités : le théorème central limite (TCL), qui stipule que la somme d'un grand nombre de variables aléatoires indépendantes et identiquement distribuées converge en distribution vers une loi gaussienne – la fameuse "courbe en cloche" [18]. La position d'une particule subissant un mouvement brownien est donc décrite par une distribution gaussienne. Une caractéristique clé de ce processus est que la distribution de probabilité *s'élargit* au cours du temps, reflétant la nature diffusive du mouvement. Cet élargissement est quantitativement décrit par le déplacement quadratique moyen (DQM), qui croît linéairement avec le temps – une signature de la diffusion normale.

Si cette description de la diffusion passive est élégante et puissante, elle repose cependant sur une hypothèse cruciale et souvent limitante : le fait que le bain environnant soit à l'équilibre thermique. Dans de nombreux systèmes réels, cette hypothèse ne peut être envisagée que localement et temporairement, et elle échoue à rendre compte de la dynamique à des échelles spatiales ou temporelles plus grandes. En réalité, de nombreux systèmes expérimentaux présentent un DQM linéaire – propriété appelée diffusion fickienne – alors que leurs distributions de positions s'écartent nettement de la gaussienne caractéristique du mouvement brownien [19]. Un exemple frappant est la diffusion de billes microscopiques à travers des réseaux enchevêtrés de filaments d'actine, où l'environnement – un maillage de filaments protéiques – introduit des hétérogénéités spatiales conduisant à des statistiques non gaussiennes [20]. Ces écarts proviennent de la présence de désordre, d'effets de mémoire et d'inhomogénéités spatiales dans le milieu. Dans les systèmes biologiques, comme l'intérieur d'une cellule vivante, la complexité est encore plus prononcée : la cellule fonctionne comme un système actif, loin de l'équilibre, animé par la consommation d'énergie environnante et des interactions biochimiques complexes. Dans de tels environnements, l'approximation d'un simple bain thermique à l'équilibre n'est plus valide.

Pour aller au-delà du mouvement brownien tout en conservant un DQM linéaire et en tenant compte du désordre environnemental, on peut recourir à des modèles de diffusion aléatoire dans lesquels la diffusivité elle-même fluctue dans le temps ou l'espace. Fait intéressant, dans de tels systèmes, les fluctuations typiques obéissent encore au théorème central limite et paraissent gaussiennes, mais les queues des distributions, parfois accessibles expérimentalement, dévient fortement – présentant souvent une décroissance exponentielle plutôt que gaussienne. Le cadre mathématique approprié pour caractériser ces fluctuations rares et atypiques, au-delà du domaine de validité du TCL, est celui de la *théorie des grandes déviations* (TGD) [22, 24], qui fournit des outils permettant de dépasser l'approximation gaussienne et de quantifier précisément le comportement statistique complet de ces systèmes hors équilibre.

La théorie des grandes déviations s'est imposée comme un cadre puissant et unificateur en physique statistique pour l'analyse des fluctuations rares dans les systèmes complexes, tant à l'équilibre qu'hors équilibre. La TGD est désormais couramment appliquée à une grande variété de systèmes physiques. Par exemple, elle a été utilisée pour étudier la diffusion "single-file" [25], où les particules se déplacent dans des canaux étroits sans pouvoir se dépasser ; la théorie des matrices aléatoires [26–31], où les grandes déviations décrivent les fluctuations des valeurs propres extrêmes ; ainsi que les systèmes désordonnés [32–34], et plus récemment, la réinitialisation stochastique [35–38]. Dans tous ces cas, la TGD fournit un moyen rigoureux de calculer les probabilités exponentiellement petites d'événements rares, encapsulées dans un objet unique : la fonction de taux (ou fonction de Cramér), qui quantifie la probabilité d'une fluctuation donnée dans la limite de grande taille ou de long temps.

De plus, la fonction de taux contient plus qu'une simple information quantitative : elle peut également refléter des changements qualitatifs dans le comportement du système. En particulier, les singularités de la fonction de taux signalent souvent des transitions de phase, lorsque le système passe d'un régime à un autre. Cette connexion rappelle celle de la thermodynamique classique, où l'énergie libre (une fonction génératrice de cumulants, reliée à la fonction de taux par transformée de Legendre) devient non analytique aux points critiques. De telles singularités ont été identifiées dans des fonctions de grandes déviations décrivant des transitions dynamiques [22, 23, 38]. Ainsi, la TGD fournit à la fois une théorie quantitative des fluctuations et un cadre pour détecter et caractériser les transitions de phase dans les systèmes loin de l'équilibre — faisant d'elle un outil essentiel dans l'arsenal du physicien statisticien moderne.

Un autre thème central de cette thèse est l'étude des propriétés de premier passage : en substance, la question de savoir quand un processus stochastique atteindra pour la première fois une certaine frontière. Les événements de premier passage sont cruciaux dans de nombreux contextes, déterminant, par exemple, les vitesses de réaction en chimie [39], les temps d'extinction en dynamique des populations, et les probabilités de franchissement en finance. Ils offrent également une prise quantitative sur les processus de recherche (combien de temps un chercheur aléatoire met-il pour trouver une cible) [40, 41] et de survie (combien de temps jusqu'à ce qu'un état absorbant soit atteint). Malgré leur importance, les calculs de premier passage sont notoirement difficiles pour les processus hors équilibre et non markoviens. Dans les modèles markoviens classiques comme la diffusion normale, de nombreux résultats analytiques existe grâce à l'utilisation de la propriété d'absence de mémoire et à des outils comme l'équation de Fokker-Planck [42] ou la théorie du renouvellement [43]. Cependant, de nombreux systèmes d'intérêt actuel échappent à ce paradigme classique.

La matière active en est un exemple emblématique : elle désigne les systèmes où les unités individuelles (particules) consomment de l'énergie pour générer un mouvement persistant. Les exemples vont des micro-organismes biologiques tels que les bactéries et les cellules jusqu'aux colloïdes synthétiques et aux robots. Au cours des deux dernières décennies, les particules actives ont été intensément étudiées en raison de leur capacité à s'auto-organiser et à manifester des comportements collectifs loin de l'équilibre [44–52]. Dans cette thèse, toutefois, nous nous concentrons sur la dynamique stochastique d'une particule active unique – en particulier, la particule "run-and-tumble" (RTP), qui est l'un des modèles les plus simples de mouvement actif [53]. Une RTP se déplace en ligne droite à vitesse constante ("run") pendant un temps distribué exponentiellement, puis se réoriente aléatoirement (un "tumble") et repart en ligne droite dans une nouvelle direction. Contrairement au mouvement brownien (qui ne conserve pas la mémoire de sa vitesse), la vitesse d'une RTP possède un temps de persistance caractéristique (la durée du run), rendant le processus non markovien en termes de position. Pour les processus à mémoire et ceux maintenus hors équilibre, les techniques analytiques standard échouent souvent, et seuls quelques résultats exacts sont connus. Cette thèse est motivée par la volonté d'approfondir notre compréhension de ces processus, en comblant le fossé entre la théorie bien établie de la diffusion et celle de la matière active et d'autres processus stochastiques complexes.

Une autre déviation intrigante de la diffusion standard est introduite par la réinitialisation stochastique [35, 54]. Ici, un mécanisme externe réinitialise de façon intermittente le processus vers un état donné (par exemple, ramener une particule diffusante à l'origine à des temps aléatoires). Cette procédure maintient le système hors équilibre et celui-ci atteint une distribution stationnaire fondamentalement non boltzmannienne. La réinitialisation a suscité un intérêt considérable non seulement pour les nouveaux états stationnaires qu'elle produit, mais aussi pour son impact frappant sur les propriétés de premier passage. Notamment, un processus de recherche diffusif soumis à la réinitialisation peut atteindre un temps de recherche optimal – il existe un taux de réinitialisation qui minimise le temps moyen pour trouver une cible, caractéristique ayant des applications potentielles dans les stratégies de recherche et l'optimisation. Par exemple, l'idée de redémarrer un processus a trouvé des applications naturelles en informatique et en physique numérique, où elle est utilisée pour améliorer la vitesse de convergence des algorithmes [55], renforcer l'entraînement des réseaux neuronaux [56, 57], et accélérer les simulations de dynamique moléculaire [58].

Enfin, l'une des questions fondamentales dans les problèmes de premier passage concerne le temps qu'un processus stochastique mettra pour quitter un intervalle donné pour la première fois – cela définit un observable connu sous le nom de *probabilité de sortie* (ou *de franchissement*). Fait remarquable, il s'avère que cette question est profondément liée à un problème apparemment distinct : déterminer la distribution spatiale d'une particule confinée dans le même intervalle. Autrement dit, la probabilité qu'une particule sorte d'un intervalle par une frontière particulière peut être exactement mise en correspondance avec la probabilité de trouver une particule située dans une région spécifique du même intervalle. Cette connexion profonde et quelque peu surprenante est encapsulée par un cadre mathématique connu sous le nom de *dualité de Siegmund* [59]. Essentiellement, la dualité de Siegmund fournit une correspondance entre les propriétés de premier passage d'un processus stochastique et les caractéristiques spatiales d'un processus "dual" soigneusement construit. Dans cette thèse, nous étendons ce concept puissant de dualité de Siegmund à une large classe de processus stochastiques pertinents pour la physique.

Après avoir présenté les motivations principales de cette thèse, nous proposons à présent un aperçu structuré de son organisation et de ses principales contributions.

## Aperçu de la thèse

**Partie I – Revue.** La première partie de cette thèse (chapitres 1 à 4) passe en revue les notions fondamentales nécessaires à la compréhension des contributions originales présentées plus loin dans la thèse. Elle débute par l'étude classique du mouvement brownien, en explorant ses propriétés et son formalisme mathématique, puis aborde la diffusion non gaussienne, en soulignant la façon dont les écarts à la gaussienne apparaissent dans les milieux hétérogènes. Nous introduisons la réinitialisation stochastique comme mécanisme essentiel maintenant les systèmes hors équilibre et présentons la "run-and-tumble particle" comme modèle paradigmatique de matière active. Cette synthèse complète apporte également les techniques analytiques de base, et les chapitres 2, 3 et 4 se concluent en reliant explicitement ces idées aux nouveaux résultats développés dans les parties suivantes.

**Partie II – Distributions stationnaires avec bruit réinitialisé.** Dans la partie II (chapitres 5 à 7), nous présentons de nouveaux résultats sur l'influence de la réinitialisation stochastique sur les distributions stationnaires, en nous concentrant sur le cas où la réinitialisation agit directement sur le bruit plutôt que sur la position de la particule. Nous considérons des particules soumises à la fois à un confinement harmonique et à des protocoles de réinitialisation appliqués à leur bruit. En utilisant une *approche à la Kesten* [60], nous dérivons une équation intégrale pour la distribution stationnaire de la position. Fait remarquable, cette équation peut être résolue exactement dans certains cas spécifiques. Même lorsqu'une solution exacte n'est pas disponible, il est néanmoins possible d'extraire analytiquement les propriétés clés de l'état stationnaire hors équilibre. Le cadre développé ici s'applique notamment aux RTP généralisées – dans lesquelles la vitesse devient une variable aléatoire, restant constante pendant chaque run – et aux modèles de "switching diffusion" introduits dans la partie III, pour lesquels nous résolvons exactement l'équation intégrale correspondante.

**Partie III – Switching Diffusion et grandes déviations.** Dans la partie III (chapitres 8 à 10), nous étudions la *switching diffusion*, un processus où une particule alterne aléatoirement entre différents coefficients de diffusion, modélisant le transport dans des milieux hétérogènes et désordonnés. La diffusivité $D(t)$ reste constante pendant des intervalles aléatoires et est tirée d'une distribution $W(D)$ à chaque renouvellement. Ce modèle appartient à la classe des modèles de diffusivité aléatoire, qui reproduisent la diffusion fickienne $\langle x^2(t) \rangle \sim 2Dt$ tout en exhibant des distributions de déplacements non gaussiennes. Nous dérivons des expressions exactes pour les moments dépendant du temps et montrons que, à long terme, les cumulants sont proportionnels aux cumulants libres d'une variable aléatoire tirée de $W(D)$ – une connexion pour le moins surprenante. À l'aide de techniques de grandes déviations, nous calculons la fonction de taux et la fonction génératrice de cumulants pour une large classe de $W(D)$, révélant des transitions dynamiques entre régimes gaussien et non gaussien. Nous étendons également le modèle au confinement harmonique et obtenons l'état stationnaire hors équilibre exact pour un $W(D)$ arbitraire (en espace de Fourier), où le lien avec les cumulants libres subsiste.

**Partie IV – Temps de premier passage pour les particules run-and-tumble dans un potentiel.** Dans la partie IV (chapitres 11 à 13), nous nous concentrons sur les propriétés de premier passage des particules run-and-tumble se déplaçant dans un potentiel arbitraire. Nous dérivons des expressions exactes pour le temps moyen de premier passage (MFPT) vers l'origine d'une RTP soumise à une force externe arbitraire – un problème difficile en raison du caractère persistant du mouvement et de la présence de points fixes où la vitesse effective s'annule. Pour une classe de potentiels de la forme $V(x) = \alpha|x|^p$ avec $p > 1$, nous montrons que le MFPT présente un minimum en un taux de tumble optimal. Nous caractérisons en détail cet optimum, fournissant un éclairage sur la manière d'optimiser le processus de recherche. D'autres applications sont également discutées, notamment le calcul du taux de Kramers pour les RTP.

**Partie V – Dualité de Siegmund.** La dernière partie de la thèse (chapitres 14 à 17) présente et développe la dualité de Siegmund comme concept mathématique reliant distributions spatiales et propriétés de premier passage des processus stochastiques. Nous étendons cette dualité – précédemment connue pour la diffusion – à une large classe de modèles étudiés tout au long de cette thèse, comprenant la réinitialisation stochastique, les modèles de diffusion aléatoire et les particules actives. En construisant des processus duaux explicites pour chacun de ces scénarios, nous montrons comment cette dualité peut être utilisée par les physiciens, tant analytiquement que numériquement.

# Introduction and Overview

Consider the simple act of tossing a coin. Naturally, to predict the outcome, we use a probabilistic description, assigning equal probability 1/2 to each possible result – heads or tails. However, if we knew precisely the initial conditions of the toss – the coin's initial velocity, direction, orientation, as well as environmental factors such as air temperature or wind speed – then, in principle, we could exactly integrate the equations of motion to predict the outcome with certainty. Yet, directly solving the equations that govern all relevant variables quickly becomes an overwhelmingly complex task. Thus, we adopt a probabilistic approach. Although simple, this example strikingly illustrates the elegance and power of probabilistic reasoning: when dealing with complex systems sensitive to initial conditions, a probabilistic approach becomes not just convenient, but indispensable.

In physics, such statistical reasoning underpins foundational concepts, for instance, defining the temperature of a gas as the measure of the average kinetic energy of its microscopic constituents through the equipartition theorem. But the reach of probabilistic descriptions extends far beyond physics. It has provided powerful insights into fields as diverse as molecular biology, to describe the diffusion of molecular motors along DNA strands [7], financial markets, to model asset price dynamics [8, 9], and even machine learning, to characterize learning dynamics in complex neural architectures [10].

The mathematical frameworks of probability theory and statistical physics thus serve as cornerstones of modern science, enabling effective and faithful descriptions of macroscopic phenomena. Crucially, these descriptions cannot rely solely on the microscopic properties of individual constituents, because fundamentally new and unexpected behaviors emerge at larger scales – an idea captured by Anderson's celebrated phrase, "More is Different" [11]. Among the many phenomena that highlight the power of statistical thinking, Brownian motion is a particularly vivid and foundational example.

Brownian motion is one of the most fundamental stochastic processes, with far-reaching applications across physics, mathematics, chemistry, finance, and biology [12]. First observed by the botanist Robert Brown in 1827 while studying pollen grains suspended in water, Brownian motion later received its theoretical foundation through the seminal work of Albert Einstein in 1905 [13], followed a year later by Smoluchowski [14], both of whom demonstrated that the erratic motion could be understood as the result of an underlying *random walk*. Einstein's work provided a simple yet profound explanation for the erratic motion of microscopic particles suspended in a fluid, linking it to molecular collisions and offering one of the first direct pieces of evidence for the existence of atoms. This breakthrough was later confirmed experimentally by Jean Perrin, whose work solidified the atomic theory of matter [15]. Beyond physics, Brownian motion has become an essential concept in probability theory and stochastic processes. In his 1900 thesis, Louis Bachelier was the first to model the stochastic process now known as Brownian motion in a mathematical context, applying it to financial markets [8]. Norbert Wiener formalized its mathematical structure in the 1920s, leading to the modern theory of continuous-time stochastic processes, including applications in climate science [16], and even diffusion models in artificial intelligence [17].

To be more specific, Brownian motion refers to the erratic movement of a passive particle suspended in a fluid maintained at a constant temperature $T$ – that is, in a thermal bath at equilibrium. The particle is said to be passive because, in the absence of the fluid, it would remain at rest. In this context, its motion arises purely from countless microscopic collisions with the surrounding fluid molecules. Each individual displacement caused by these collisions is random, symmetric, and of finite variance. From a mathematical perspective, the position of the particle at a given time can thus be seen as the sum of many such small, independent displacements. This observation naturally connects Brownian motion to one of the cornerstones of probability theory: the Central Limit Theorem (CLT), which states that the sum of a large number of independent, identically distributed random variables converges in distribution to a Gaussian – the well-known "bell shaped curve" [18]. The position of a particle undergoing Brownian motion is therefore described by a Gaussian distribution. A key feature of this process is that the probability distribution *spreads* over time, reflecting the diffusive nature of the motion. This spreading is quantitatively captured by the mean-squared displacement (MSD), which grows linearly with time – a hallmark of normal diffusion.

While this description of passive diffusion is elegant and powerful, it relies on a crucial and often limiting assumption: that the surrounding bath is at thermal equilibrium. In many real-world systems, this assumption may hold only locally and temporarily, but it fails to capture the dynamics at larger spatial or temporal scales. In fact, numerous experimental systems exhibit a linear MSD – a feature known as Fickian diffusion – yet their position distributions deviate markedly from the Gaussian characteristic of Brownian motion [19]. A striking example is the diffusion of microscopic beads through entangled F-actin networks, where the underlying environment – a mesh of protein filaments – introduces spatial heterogeneities that lead to non-Gaussian statistics [20]. These deviations stem from the presence of disorder, memory effects, and spatial inhomogeneities in the medium. In biological systems, such as within a living cell, the complexity is even more pronounced: the cell operates as an active, far-from-equilibrium system driven by energy consumption and intricate biochemical interactions. In such settings, approximating the environment as a simple equilibrium thermal bath is no longer valid.

To go beyond Brownian motion while retaining a linear MSD and accounting for environmental disorder, one can use random diffusion models in which the diffusivity itself fluctuates in time or space. Interestingly, in such systems, typical fluctuations still tend to obey the Central Limit Theorem and appear Gaussian, but the tails of the distributions, which are sometimes accessible in experiments, deviate significantly – displaying in many cases exponential rather than Gaussian decay. The appropriate mathematical framework to characterize such rare, atypical fluctuations beyond the range of validity of the CLT is that of *large deviation theory* (LDT) [21–24], which provides tools to go beyond the Gaussian approximation and quantify precisely the full statistical behavior of these non-equilibrium systems.

Large deviation theory has emerged as a powerful and unifying framework in statistical physics for analyzing rare fluctuations in complex systems, both in and out of equilibrium. LDT is now routinely applied to a wide variety of physical systems. For example, it has been used to study single-file diffusion [25], where particles move in narrow channels and cannot overtake each other; random matrix theory [26–31], where large deviations describe fluctuations of extreme eigenvalues; as well as disordered systems [32–34], and more recently, stochastic resetting [35–38]. In all these cases, LDT provides a principled way to compute the exponentially small probabilities of rare events, encapsulated in a single object: the rate function (or Cramér function), which quantifies how unlikely a given fluctuation is in the large-size or long-time limit.

Moreover, the rate function carries more than just quantitative information – it can also reflect qualitative changes in the system's behavior. In particular, singularities in the rate function often signal phase transitions, where the system switches between different regimes. This connection mirrors that of classical thermodynamics, where the free energy (a scaled cumulant generating function related to the rate function via Legendre transform) becomes non-analytic at critical points. Such singularities have been identified in large deviation functions describing dynamical transitions [22, 23, 38]. In this way, LDT provides both a quantitative theory of fluctuations and a framework for detecting and characterizing phase transitions in systems far from equilibrium – making it an essential tool in the modern statistical physicist's repertoire.

Another central theme in this thesis is the study of first-passage properties: essentially, the question of when a stochastic process will reach a certain boundary for the first time. First-passage events are crucial in many contexts, determining, for example, reaction rates in chemistry [39], extinction times in population dynamics, and hitting probabilities in finance. They also provide a quantitative handle on search problems (how long until a random searcher finds a target) [40, 41] and survival problems (how long until an absorbing state is reached). Despite their importance, first-passage calculations are notoriously difficult for non-equilibrium and non-Markovian processes. In classical Markovian models like normal diffusion, a wealth of analytical results exists thanks to simplifications like the memoryless property and tools like the Fokker-Planck equation [42] or renewal theory [43]. However, many systems of current interest fall outside this classical paradigm.

Active matter is a prime example: it refers to systems where individual units (particles) consume energy to generate sustained motion. Examples range from biological microswimmers like bacteria and cells to synthetic colloidal particles and robots. In the past two decades, active particles have been intensely studied due to their ability to self-organize and exhibit collective behaviors far from equilibrium [44–52]. In this thesis, however, we focus on the stochastic dynamics of a single active particle – in particular, the run-and-tumble particle (RTP), which is one of the simplest active motion models [53]. An RTP moves in a straight line with a constant speed ("run") for an exponentially distributed time, then randomly re-orients (a "tumble") and runs again in a new direction. Unlike Brownian motion (which has no memory of its velocity), the RTP's velocity has a characteristic persistence time (the run duration), making the process non-Markovian in terms of position. For processes with memory and those maintained out of equilibrium, standard analytical techniques often fail, and only few exact results are known. This thesis is motivated by the aim to deepen our understanding of these processes, bridging the gap between well-established diffusion theory and the frontier of active matter and other complex stochastic systems.

Another intriguing deviation from standard diffusion is introduced by stochastic resetting [35, 54]. Here, an external mechanism intermittently resets the process to a given state (for example, returning a diffusing particle to the origin at random times). This procedure keeps driving the system out of equilibrium, and the system approaches a stationary distribution that is fundamentally non-Boltzmann. Resetting has attracted considerable interest not only for the novel steady states it produces but also for its striking impact on first-passage properties. Notably, a diffusive search process under resetting can achieve an optimal search time – there exists a resetting rate that minimizes the mean time to find a target, a feature with potential applications in search strategies and optimization. For instance, the idea of restarting a process has found natural applications in computer science and computational physics, where it is used to improve the convergence rate of algorithms [55], enhance the training of neural networks [56, 57], and accelerate molecular dynamics simulations [58].

Finally, one of the fundamental questions in first-passage problems concerns the time it takes for a stochastic process to leave a given interval for the first time – this defines an observable known as the *exit* (or *hitting*) *probability*. Remarkably, it turns out that this question is deeply related to an apparently distinct problem: determining the spatial distribution of a particle confined within the same interval. In other words, the probability for a particle to exit an interval through a particular boundary can be mapped exactly onto the probability of finding a particle located within a specific region of the same interval. This profound and somewhat surprising connection is encapsulated by a beautiful mathematical framework known as *Siegmund duality* [59]. In essence, Siegmund duality provides a mapping between the first-passage properties of one stochastic process and the spatial characteristics of a carefully constructed "dual" process. In this thesis, we extend this powerful concept of Siegmund duality to a broad class of stochastic processes relevant to physics.

Having introduced the primary motivations of this thesis, we now provide a structured overview of its organization and main contributions.

# Overview of the Thesis

**Part I – Review.** The first part of this thesis (chapters 1–4) reviews foundational topics central to understanding the original contributions presented later. It begins with classical Brownian motion, exploring its properties and mathematical formalism, and then moves to discuss non-Gaussian diffusion, emphasizing how deviations from Gaussian statistics appear in heterogeneous environments. We introduce stochastic resetting as an essential mechanism driving systems out of equilibrium and discuss the run-and-tumble particle as a paradigmatic model for active matter. This comprehensive survey also introduces fundamental analytical techniques, and chapters 2, 3, and 4 conclude by explicitly connecting these foundational ideas to the new results presented in the subsequent parts of this thesis.

**Part II – Steady-State Distributions with Resetting Noise.** In Part II (chapters 5–7), we present new results on the influence of stochastic resetting on steady-state distributions, specifically focusing on the case where resetting acts directly on the noise rather than on the position of the particle. We consider particles evolving under the combined influence of confinement (a harmonic potential) and resetting protocols applied to their noise. By using a *Kesten approach* [60], we derive an integral equation for the stationary distribution of the position of the particle. Remarkably, these equations can be solved exactly in some specific case. Even when an exact solution is not available, we are still able to analytically extract key properties of the non-equilibrium steady state. The framework developed in this part applies in particular to generalized RTPs – where the velocity becomes a random variable, remaining constant during each run – and to switching diffusion models introduced in Part III, for which we are able to solve the corresponding integral equation exactly.

**Part III – Switching Diffusion and Large Deviations.** In Part III (Chapters 8–10), we study *switching diffusion*, a process in which a particle randomly alternates between different diffusion coefficients, modeling transport in disordered and heterogeneous media. The diffusivity $D(t)$ remains constant for random time intervals and is drawn from a distribution $W(D)$ at each renewal. This model belongs to the class of random diffusivity models, which reproduce Fickian diffusion $\langle x^2(t) \rangle \sim 2Dt$ while exhibiting non-Gaussian displacement distributions. We derive exact expressions for the time-dependent moments and show that, in the long-time limit, the cumulants are proportional to the free cumulants of a random variable drawn from $W(D)$ – which turns out to be a quite surprising connection. Using large deviation techniques, we compute the rate function and the scaled cumulant-generating function for a large class of distributions $W(D)$, uncovering dynamical transitions between Gaussian and non-Gaussian regimes. We also extend the model to include harmonic confinement and obtain the exact non-equilibrium steady state for arbitrary $W(D)$ (in Fourier space), where the link to free cumulants remains.

**Part IV – First-Passage Times for Run-and-Tumble Particles within a Potential.** In Part IV (chapters 11–13), we focus on the first-passage properties of run-and-tumble particles moving inside an arbitrary potential. We derive exact expressions for the mean first-passage time (MFPT) to the origin for an RTP subjected to an arbitrary external force – a challenging problem due to the persistent motion of the particle and the presence of turning points in the dynamics, where the effective velocity vanishes. For a class of confining potentials of the form $V(x) = \alpha |x|^p$ with $p > 1$, we show that the MFPT exhibits a minimum at an optimal tumbling rate. We characterize this optimum in detail, providing insight into how the search process can be optimized. Additional applications are also discussed, including the computation of Kramers' rate for RTPs.

**Part V – Siegmund Duality.** The final part of the thesis (chapters 14–17) introduces and develops Siegmund duality as a mathematical concept linking spatial distributions and first-passage properties of stochastic processes. We extend the concept of Siegmund duality – previously known for diffusion – to encompass a large class of stochastic models studied throughout this thesis, including stochastic resetting, random diffusion models, and active particles. By constructing explicit dual processes for each of these scenarios, we show how it can be used by physicists, both analytically and numerically.

# Part I

# From Passive to Active Dynamics

**Abstract**

Part I lays the theoretical foundations of continuous-time stochastic processes, beginning with an introduction to Brownian motion and a presentation of its fundamental properties, including first-passage times. Building on this classical framework, we then explore scenarios in which diffusion deviates from Gaussian behavior – introducing models of random diffusivity, such as diffusing diffusivity models. These models generate non-Gaussian fluctuations in the displacement distributions while preserving a linear in time mean squared displacement, thereby illustrating how deviations from the classical Brownian paradigm can naturally emerge in disordered or complex environments. We then turn to stochastic resetting, a mechanism in which a diffusing particle intermittently returns to a given location. Resetting drives the system into a non-equilibrium steady state and significantly modifies first-passage properties – leading for instance to a finite optimal search time, in contrast to classical free diffusion where the mean first-passage time diverges. Next, we introduce the run-and-tumble particle (RTP) as a toy model of an active particle. Unlike memoryless Brownian motion, the RTP exhibits persistent motion with a finite correlation time. These features give rise to rich dynamical behavior and non-Boltzmann steady states. Throughout Part I, each chapter not only surveys established results on these stochastic processes but also highlights their connections to the original contributions developed in this thesis.

# Chapter 1

# Introduction to Brownian Motion

In this chapter, we introduce Brownian motion and the mathematical tools used to study it. We begin with Einstein's derivation of the diffusion equation and then present the Langevin equation, which models Brownian motion as a stochastic differential equation (SDE). From there, we derive the Fokker-Planck equation, which describes how probability distributions evolve over time [42]. We then explore key properties such as mean-square displacement (MSD) and propagators, and discuss how confinement leads to equilibrium distributions. Finally, we examine first-passage properties, which describe the probability of a particle reaching a threshold for the first time – essential for understanding chemical reactions, search processes, and extreme value statistics [40, 43, 61].

While this chapter focuses on equilibrium properties, many real-world systems operate far from equilibrium. This motivates the study of more complex stochastic processes, which we will explore in later chapters.

## 1.1 Historical Derivation from Albert Einstein

Let us consider $N$ independent particles, all starting their motion at the origin. We will work in a one-dimensional setting, though extending to multiple dimensions is straightforward. Each particle follows a time-dependent trajectory over the interval $[0, t]$ and is subjected to collisions with the fluid particles which are in thermal equilibrium (i.e., the fluid temperature is constant). Einstein made the following assumptions to describe their motion:

1. We introduce a small time interval $\tau \ll t$, such that the sub-trajectories within $[0, \tau]$, $[\tau, 2\tau]$, ..., $[t - \tau, t]$ are considered independent.

2. Over a time duration $\tau$, the displacement $x(t + \tau) - x(t) = \Delta$ is assumed stationary and to be drawn from a probability distribution function (PDF) $\varphi(\Delta)$.

3. The PDF $\varphi(\Delta)$ is assumed to be symmetric due to spatial isotropy. Furthermore, it is nonzero only for small values of $\Delta$, as we assume that the displacement in a single time step $\tau$ is microscopic.

Now, we introduce the PDF $f(x, t)$ such that $f(x, t)dx$ represents the number of particles within the interval $[x, x + dx]$ at time $t$. Our goal is to derive an evolution equation for the density $f(x, t)$. At time $t + \tau$, suppose there are $f(x, t + \tau)dx$ particles between $x$ and $x + dx$. This means that at time $t$, there are $\int_{-\infty}^{+\infty} f(x - \Delta, t)dx\, \varphi(\Delta)d\Delta$ particles inside $[x, x + dx]$. Therefore, we have

$$f(x, t + \tau)dx = \int_{-\infty}^{+\infty} f(x - \Delta, t)dx\, \varphi(\Delta)d\Delta \,. \tag{1.1.1}$$

Since we are in a regime where both $\tau$ and $\Delta$ are small, we expand $f(x, t+\tau)$ and $f(x-\Delta, t)$ to obtain

$$f(x,t)+\tau\frac{\partial f(x,t)}{\partial t} = f(x,t)\int_{-\infty}^{+\infty}\varphi(\Delta)d\Delta - \frac{\partial f(x,t)}{\partial x}\int_{-\infty}^{+\infty}\Delta\varphi(\Delta)d\Delta + \frac{\partial^2 f(x,t)}{\partial x^2}\int_{-\infty}^{+\infty}\frac{\Delta^2}{2}\varphi(\Delta)d\Delta\,. \tag{1.1.2}$$

The first integral on the right-hand side is simply 1 due to the normalization of $\varphi(\Delta)$. The second integral vanishes due to the symmetry of $\varphi(\Delta)$. Thus, we are left with

$$\frac{\partial f(x,t)}{\partial t} = \left(\frac{1}{\tau}\int_{-\infty}^{+\infty}\frac{\Delta^2}{2}\varphi(\Delta)d\Delta\right)\frac{\partial^2 f(x,t)}{\partial x^2} = D\frac{\partial^2 f(x,t)}{\partial x^2}\,, \tag{1.1.3}$$

where we recognize the diffusion equation with diffusion coefficient

$$D = \frac{1}{\tau}\int_{-\infty}^{+\infty}\frac{\Delta^2}{2}\varphi(\Delta)d\Delta. \tag{1.1.4}$$

Since for $x \neq 0$, we have $f(x, t=0) = 0$ and the total number of particles is conserved, i.e.,

$$\int_{-\infty}^{+\infty} f(x,t)dx = N, \tag{1.1.5}$$

the solution to the diffusion equation is given by (this can be easily seen by going in Fourier space)

$$f(x,t) = \frac{N}{\sqrt{4\pi Dt}}e^{-\frac{x^2}{4Dt}}\,. \tag{1.1.6}$$

When $N = 1$, this result corresponds to the well-known Gaussian probability distribution of Brownian motion, with a mean squared displacement (MSD) given by:

$$\text{MSD}[x(t)] = \langle(x(t) - x(t=0))^2\rangle = 2Dt\,, \tag{1.1.7}$$

which can be directly measured in experiments. Furthermore, Einstein also derived a theoretical expression for the diffusion coefficient of a diffusing particle immersed in a fluid:

$$D = \frac{RT}{6\mathcal{N}_A \pi \eta r}\,, \tag{1.1.8}$$

where $R$ is the universal gas constant, $T$ is the temperature of the fluid, $\mathcal{N}_A$ is Avogadro's number, $\eta$ is the dynamic viscosity of the medium, and $r$ is the radius of the particle. This result, called the Stokes-Einstein relation, implies that by measuring the MSD of a Brownian particle, one can determine the value of $\mathcal{N}_A$. Between 1907 and 1909, in a series of experiments [15], Jean Perrin obtained several key results:

- $\langle x^2(t)\rangle \propto Dt$, confirming the diffusing scaling of Brownian motion.
- $D \propto \frac{1}{r}$, validating Einstein's theoretical prediction.
- An experimental determination of Avogadro's number $\mathcal{N}_A$.

These findings definitively settled the long-standing debate on whether matter was continuous or discrete, providing strong evidence for the existence of atoms. This work earned Jean Perrin the Nobel Prize in Physics in 1926.

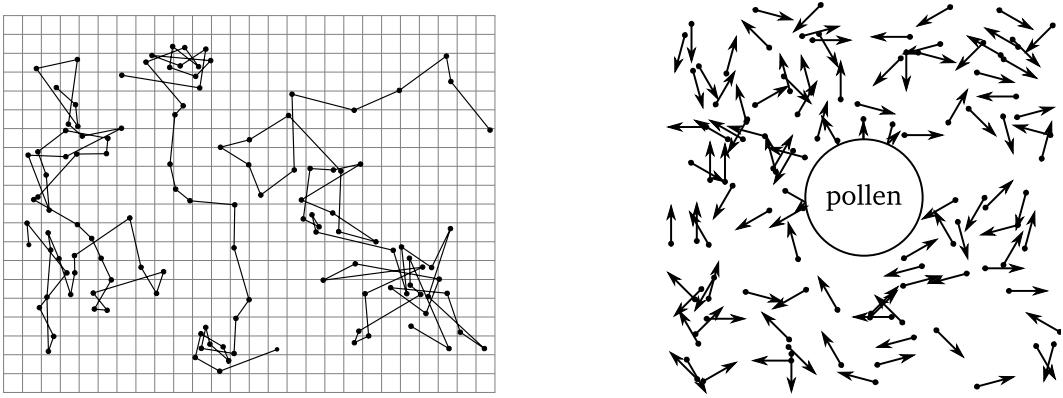

**Figure 1.1: Left.** Three trajectories of colloidal particles are displayed, and successive positions are joined by segments – reproduced from [15]. **Right.** A grain of pollen in a fluid that undergoes collisions with the particles of the fluid.

## 1.2 The Langevin Equation

In this section, we follow the approach to Brownian motion proposed by P. Langevin [62] and aim to derive a differential equation describing the motion of the particle. This equation, known as the Langevin equation, is a *stochastic differential equation* (SDE) as it incorporates the random forces acting on the particle. It is a fundamental tool for describing stochastic processes in physics.

We consider the motion of a particle, such as a pollen grain, in a fluid – see Fig. 1.1. We assume that the particles composing the fluid are much smaller than the pollen grain. From Newton's second law of motion, the equation of motion is given by

$$m\,\ddot{\mathbf{x}}(t) + \gamma\,\dot{\mathbf{x}}(t) = \mathbf{F}(x,t) \quad , \quad \mathbf{x}(t) \equiv (x(t), y(t), z(t)) \quad , \quad \dot{\mathbf{x}}(t) \equiv \frac{d\mathbf{x}(t)}{dt}, \quad (1.2.1)$$

where boldface letters indicate vectors in three-dimensional space, and $m$ is the mass of the pollen grain. The force $\mathbf{F}(x,t)$ represents any external force acting on the particle, such as gravity, and $\gamma$ is the viscous drag coefficient.

This naive description of the particle's motion leads to a paradox. When no external force is applied, i.e., $\mathbf{F}(x,t) = 0$, the solution is given by

$$\dot{\mathbf{x}}(t) = \dot{\mathbf{x}}(0)\exp\left(-\frac{\gamma}{m}t\right) \underset{t\to+\infty}{\longrightarrow} 0. \quad (1.2.2)$$

However, this result contradicts the equipartition theorem of thermodynamics, which states that the kinetic energy of a particle suspended in a fluid is given by

$$\frac{1}{2}m\langle\dot{\mathbf{x}}^2(t)\rangle = \frac{3}{2}k_B T \implies \langle\dot{\mathbf{x}}^2(t)\rangle = \frac{3k_B T}{m} > 0. \quad (1.2.3)$$

Thus, the deterministic equation of motion fails to account for the persistent kinetic energy expected from thermal fluctuations, highlighting the need for a stochastic description of Brownian motion. To address this issue, P. Langevin introduced a random force into the equation of motion to account for the erratic collisions between the fluid particles and the pollen grain. The modified equation reads

$$m\,\ddot{\mathbf{x}}(t) + \gamma\,\dot{\mathbf{x}}(t) = \mathbf{F}(x,t) + \boldsymbol{\eta}(t), \quad (1.2.4)$$

where $\boldsymbol{\eta}(t)$ represents a stochastic force known as (thermal) *white noise*. On average, the random thermal forces cancel out over time and do not produce a net force, which is consistent with the fact that the fluid is in thermal equilibrium. The forces are also considered to be uncorrelated at different times, meaning the noise has no memory and is "white". Hence, the stochastic force $\eta(t)$ satisfies the following properties:

$$\langle \eta^i(t) \rangle = 0 \quad , \quad \langle \eta^i(t_1) \eta^j(t_2) \rangle = A \delta_{i,j} \delta(t_1 - t_2) \,, \tag{1.2.5}$$

where the upper indices denote the spatial components, $\delta(x)$ is the Dirac delta function, and $A$ is a constant to be determined below.

To reconcile the equipartition theorem with Newton's description of the dynamics of the pollen grain, we again consider $\mathbf{F}(x, t) = 0$. In this case, the solution of Eq. (1.2.4) is given by

$$\dot{\mathbf{x}}(t) = \dot{\mathbf{x}}(0) \exp\left(-\frac{\gamma}{m} t\right) + \int_0^t \exp\left(-\frac{\gamma}{m}(t - t')\right) \frac{\boldsymbol{\eta}(t')}{m} dt' \,. \tag{1.2.6}$$

From this explicit solution, we can compute the variance of the velocity averaged over the white noise (which is the only source of randomness). It is given by

$$\langle \dot{\mathbf{x}}^2(t) \rangle = \dot{\mathbf{x}}(0)^2 \exp\left(-2\frac{\gamma}{m} t\right) \tag{1.2.7}$$

$$+ \int_0^t \int_0^t \exp\left(-\frac{\gamma}{m}(t - t_1)\right) \exp\left(-\frac{\gamma}{m}(t - t_2)\right) \frac{\langle \boldsymbol{\eta}(t_1) \boldsymbol{\eta}(t_2) \rangle}{m^2} dt_1 dt_2 \,. \tag{1.2.8}$$

We are interested in the large-time behavior, as the equipartition theorem holds in a thermodynamic equilibrium setting. One can show that the first term is subleading, and that we have

$$\langle \dot{\mathbf{x}}^2(t) \rangle \underset{t \to \infty}{\approx} \frac{3A}{m^2} \int_0^t \exp\left(-2\frac{\gamma}{m}(t - t_1)\right) dt_1 \approx \frac{3A}{2m\gamma} \,. \tag{1.2.9}$$

Here, we have used the isotropy property of the correlations:

$$\langle \boldsymbol{\eta}(t_1) \boldsymbol{\eta}(t_2) \rangle = 3 \langle \eta^x(t_1) \eta^x(t_2) \rangle = 3A \delta(t_1 - t_2) \,. \tag{1.2.10}$$

For this to be consistent with the equipartition theorem given in Eq. (1.2.3), we require

$$A = 2\gamma k_B T \,. \tag{1.2.11}$$

Therefore, the two-time correlations of a one-dimensional thermal white noise have an intensity proportional to the temperature and are given by

$$\langle \eta(t_1) \eta(t_2) \rangle = 2\gamma k_B T \delta(t_1 - t_2) \,. \tag{1.2.12}$$

From now on, we set $k_B = 1$ and $\gamma = 1$, leading to $D = T$. In all the systems studied throughout this thesis, we will neglect the inertial term in Eq. (1.2.4) (see Appendix A for more details) resulting in the so-called *Langevin equation*, which we write in one dimension as

$$\dot{x}(t) = F(x) + \eta(t) \,, \tag{1.2.13}$$

where $F(x) = -V'(x)$ represents an external force derived from a potential $V(x)$, and $\eta(t)$ is a white noise term with zero mean $\langle \eta(t) \rangle = 0$ and two-time correlations

$$\langle \eta(t_1) \eta(t_2) \rangle = 2D \delta(t_1 - t_2) \,. \tag{1.2.14}$$

Equation (1.2.13) is our first example of a continuous-time *stochastic differential equation*.

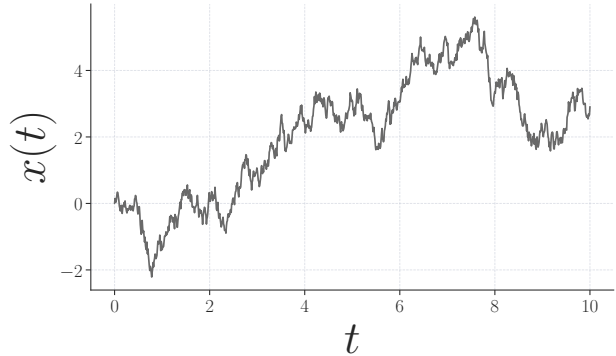

**Figure 1.2:** We show a single Brownian trajectory obtained by simulating the discretized version of the Langevin equation (1.3.1), as detailed in Appendix C.5. The simulation parameters are set to $dt = 10^{-2}$ and $D = 1$.

## 1.3 Basic Properties of the Free Brownian Motion

The standard definition of Brownian motion typically describes the motion of a *free* particle, meaning that it is not subject to any external force. In this case, the Langevin equation for free Brownian motion is

$$\dot{x}(t) = \eta(t)\,. \tag{1.3.1}$$

We show a Brownian trajectory in Fig. 1.2. To simplify notation, we will use $x(t)$ to denote both the stochastic process of Brownian motion and the value of a specific trajectory at time $t$. Here and in the following sections, we will consider a one-dimensional Brownian motion. A $d$-dimensional Brownian motion describes the motion of a particle in which each coordinate evolves independently as a one-dimensional Brownian motion. While the spatial properties of a $d$-dimensional Brownian motion can be generalized straightforwardly, the first-passage properties are more complex [63]. In this discussion, we will focus on the one-dimensional case.

A *Brownian motion* $x(t)$ is modeled as a stochastic process with the following properties

- It starts at the origin: $x(0) = 0$.

- For any time $t$ and any $\tau > 0$, the increment $x(t + \tau) - x(t)$ is statistically independent of the past trajectory $x(t')$ for all $0 < t' < t$.

- The increment $x(t + \tau) - x(t)$ is a Gaussian random variable with mean 0 and variance $2D\tau$, i.e., it follows $\mathcal{N}(0, 2D\tau)$.

- $x(t)$ is a continuous function of $t$ but is nowhere differentiable.

A direct consequence of this definition is that Brownian motion is a *Markov process*, meaning that for any $t > 0$ and $\Delta t > 0$ and $x \in \mathbb{R}$,

$$\mathbb{P}\left(x(t + \Delta t) = x | x(\tau) = y \text{ for } 0 \leq \tau \leq t\right) = \mathbb{P}\left(x(t + \Delta t) = x | x(t) = y\right). \tag{1.3.2}$$

In particular, since all increments are completely independent of the rest of the trajectory, we even have

$$\mathbb{P}\left(x(t + \Delta t) = x | x(t) = y\right) = \mathbb{P}\left(x(t + \Delta t) = x\right). \tag{1.3.3}$$

Since Brownian motion has independent and symmetric increments, reversing time does not change the statistical properties of the process. This means Brownian motion is also time-reversible.

Another important point is that $x(t)$ is not differentiable at any point. This is why, in mathematics, the stochastic differential equation (SDE) of Brownian motion is written in the form $dX_t = dW_t$ where $W_t$ is a Wiener process. This equation is not written as $\frac{dX_t}{dt} = \frac{dW_t}{dt}$ since the derivative of a Wiener process does not exist in the usual sense. However, in physics, we often use the Langevin equation, such as Eq. (1.3.1), where we explicitly write the derivative of $x(t)$. To justify this, we provide in Appendix B a derivation of white noise as the velocity of a Brownian motion. In Appendix C, we discuss the necessity of a convention when discretizing Brownian motion and introduce the Itô and Stratonovich conventions. Additionally, we explain how to simulate a Brownian motion. Below, we give some important properties of the free Brownian motion.

### 1.3.1 Two-time correlation function of a Brownian motion

The two-time correlation function of a stochastic process $\langle x(t)x(t')\rangle$ is an important quantity that quantifies how the value at one time $t$ is correlated with its value at another time $t'$. To derive the two-time correlation function of a Brownian motion $x(t)$, let us consider two times $t' > t$. First, we can write

$$x(t') = x(t) + (x(t') - x(t)). \tag{1.3.4}$$

Taking the expectation of the product $x(t)x(t')$, we obtain

$$\langle x(t)x(t')\rangle = \langle x(t)[x(t) + (x(t') - x(t))]\rangle \tag{1.3.5}$$
$$= \langle x^2(t)\rangle + \langle x(t)(x(t') - x(t))\rangle. \tag{1.3.6}$$

From the definition of Brownian motion, we have $\langle x^2(t)\rangle = 2Dt$. Additionally, since $x(t)$ is independent of the increment $(x(t') - x(t))$, their expectation satisfies $\langle x(t)(x(t') - x(t))\rangle = 0$. Thus, we obtain

$$\langle x(t)x(t')\rangle = 2Dt. \tag{1.3.7}$$

By symmetry, for $t > t'$, one can show that $\langle x(t)x(t')\rangle = 2Dt'$. Therefore, the two-time correlation function of Brownian motion is given by

$$\langle x(t)x(t')\rangle = 2D\min(t, t'). \tag{1.3.8}$$

### 1.3.2 Free propagator, Moments and Cumulants

The Langevin equation for a free Brownian motion, $\dot{x}(t) = \eta(t)$, is solved by

$$x(t) = \int_{t_0}^{t} \eta(t')dt', \tag{1.3.9}$$

which is a linear combination of Gaussian random variables. Hence, $x(t)$ itself follows a Gaussian distribution and is fully determined by its mean, $\langle x(t)\rangle = 0$, and its variance, $\langle x^2(t)\rangle = 2D(t-t_0)$. Therefore, if we assume $x(0) = x_0$ at initial time $t_0$, the free Gaussian propagator is given by

$$G_0(x, t|x_0, t_0) = \frac{e^{-\frac{(x-x_0)^2}{4D(t-t_0)}}}{\sqrt{4\pi D(t-t_0)}}, \tag{1.3.10}$$

where $G_0(x, t|x_0, t_0)dx$ represents the probability of finding the particle inside the interval $[x, x+dx]$ at time $t$. For simplicity, let us consider the case where $x_0 = 0$ and $t_0 = 0$. In particular, we have the interesting property that

$$\frac{x(t)}{\sqrt{t}} \stackrel{d}{\sim} \mathcal{N}(0, 2D), \tag{1.3.11}$$

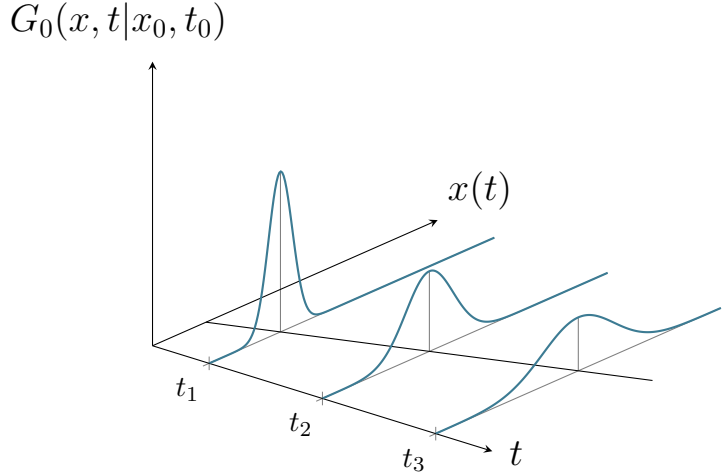

**Figure 1.3:** A visualization of the probability distribution $G_0(x,t|x_0,t_0)$ evolving over different times $t_1 < t_2 < t_3$. This shows the spread over time of the PDF with a MSD growing as $MSD[x(t)] \sim \sqrt{t}$. This figure is adapted from [64].

which means that Brownian motion scales as a diffusive process. In the Eq. (1.3.11), $\stackrel{d}{\sim}$ means that $\frac{x(t)}{\sqrt{t}}$ has for probability distribution $\mathcal{N}(0, 2D)$. Additionally, the distribution is symmetric with respect to the change $x \to -x$. We show the evolution of $G_0(x,t|x_0,t_0)$ over time in Fig. 1.3, along with some Brownian trajectories in the left panel of Fig. 1.4.

The moments of the Gaussian distribution $G_0(x,t|0,0)$ are given by

$$\langle x^{2n}(t) \rangle = (2Dt)^n \, (2n-1)!! \,. \tag{1.3.12}$$

Meanwhile, its cumulants – denoted by $\langle x^n(t) \rangle_c$ – are given by

$$\langle x(t) \rangle_c = \langle x(t) \rangle = 0 \,, \tag{1.3.13}$$
$$\langle x^2(t) \rangle_c = \langle x^2(t) \rangle - \langle x(t) \rangle^2 = 2Dt \,, \tag{1.3.14}$$
$$\langle x^n(t) \rangle_c = 0 \quad \text{for } n \geq 3 \,. \tag{1.3.15}$$

Since free Brownian motion is a non-stationary process it does not correspond to an equilibrium dynamics. However, in the presence of a sufficiently confining potential, we will show in the next section that its steady-state distribution is the Boltzmann distribution meaning that it represents an equilibrium steady state.

### 1.3.3 Self-Similarity

Consider a free Brownian motion $x(t) \stackrel{d}{\sim} \mathcal{N}(0, 2Dt)$. For $a > 0$, we have

$$x(at) \stackrel{d}{\sim} \sqrt{a} \, x(t) \stackrel{d}{\sim} \mathcal{N}(0, 2Dat) \,, \tag{1.3.16}$$

which implies that Brownian motion is a self-similar process. More precisely, $X(t)$ is self-similar if for any $a > 0$, there exists a $b > 0$ such that

$$X(at) \stackrel{d}{\sim} bX(t) \,. \tag{1.3.17}$$

Physically, the self-similarity property refers to the fact that the process looks statistically similar at different time scales – it is scale invariant. In other words, if you zoom in or out on a Brownian motion trajectory, the statistical properties remain the same. This invariance implies that Brownian motion lacks a characteristic time or length scale. This self-similarity property also applies to other processes, such as fractional Brownian motion, which we introduce in Section 1.5.6.

## 1.4 Brownian Motion Moving Within a Potential: an Equilibrium Dynamics

We have seen that free Brownian motion is characterized by a time-dependent propagator that does not reach a stationary state, thus describing a non-equilibrium dynamics. However, in this section, we consider the dynamics of a Brownian particle moving within a potential $V(x)$. We will see that, in this case, the motion of the particle corresponds to an equilibrium dynamics, which we will define. The Langevin equation governing this motion is given by

$$\dot{x}(t) = F(x) + \eta(t), \tag{1.4.1}$$

where $\eta(t)$ is a Gaussian white noise with zero mean and two-time correlations given by Eq. (1.2.14), and the particle is subjected to the force $F(x) = -V'(x)$. We will focus exclusively on potentials $V(x)$ satisfying

$$\int_{-\infty}^{+\infty} dx \, \exp\left(\frac{-V(x)}{D}\right) < +\infty. \tag{1.4.2}$$

In this case, the probability distribution of the position of the particle, $x(t)$, will reach a stationary state which is one of the main interests of this section. We will derive the *forward Fokker-Planck equation* associated with Eq. (1.4.1). The Fokker-Planck equation is a differential equation that describes the evolution of the PDF of the position of a particle over time. This will allow us to analyze the spatial properties of $x(t)$, including the propagator $p(x,t)$ and its equilibrium properties. We will always consider Itō's convention – see Appendix C and [65] for a discussion.

### 1.4.1 Forward Fokker-Planck Equation

Let us consider $x(t)$ evolving according to Eq. (1.4.1), and the probability density function $p(x,t)$ such that $p(x,t)dx$ represents the probability of finding the particle in the interval $[x, x+dx]$ at time $t$. For a function $f(x(t))$, we have

$$\frac{d}{dt}\langle f(x(t))\rangle = \frac{d}{dt}\int_{-\infty}^{+\infty} p(x,t)f(x)dx = \int_{-\infty}^{+\infty} \frac{\partial p(x,t)}{\partial t} f(x)dx. \tag{1.4.3}$$

We can also compute the left-hand side of Eq. (1.4.3) using Itō's chain rule (see Eq. (C.7)) and then compare both results. This gives

$$\frac{d}{dt}\langle f(x(t))\rangle = \left\langle f'(x(t))\frac{dx}{dt}\right\rangle + D\langle f''(x(t))\rangle \tag{1.4.4}$$

$$= \left\langle f'(x(t))\left(F(x) + \eta(t)\right)\right\rangle + D\langle f''(x(t))\rangle \tag{1.4.5}$$

$$= \langle f'(x(t))F(x)\rangle + D\langle f''(x(t))\rangle, \tag{1.4.6}$$

where we have used the Langevin equation (1.4.1) in the second line, and the fact that the noise and the position are uncorrelated in the third equality. Now, combining Eq. (1.4.3) and (1.4.6), we obtain

$$\int_{-\infty}^{+\infty} \frac{\partial p(x,t)}{\partial t} f(x)dx = \int_{-\infty}^{+\infty} p(x,t)\left(F(x)\frac{\partial f(x)}{\partial x} + D\frac{\partial^2 f(x)}{\partial x^2}\right) dx. \tag{1.4.7}$$

We can perform two integrations by parts on the right-hand side, using the fact that $p(x,t)$ vanishes at $\pm\infty$ since it is a probability density function. This yields

$$\int_{-\infty}^{+\infty} \frac{\partial p(x,t)}{\partial t} f(x)dx = \int_{-\infty}^{+\infty} \left[-\frac{\partial}{\partial x}\left(F(x)p(x,t)\right) + D\frac{\partial^2 p(x,t)}{\partial x^2}\right] f(x)dx. \tag{1.4.8}$$

Since this equation must hold for any function $f(x)$, we obtain the *forward Fokker-Planck equation*:
$$\frac{\partial p(x,t)}{\partial t} = -\frac{\partial}{\partial x}\left(F(x)p(x,t)\right) + D\frac{\partial^2 p(x,t)}{\partial x^2}. \qquad (1.4.9)$$

This equation is called *forward* because it describes the time evolution of the probability density function $p(x,t)$. It allows us to predict how the distribution of positions evolves forward in time, starting from an initial condition. Here, we have specifically
$$p(x,t=0) = \delta(x) \quad , \quad \int_{-\infty}^{+\infty} p(x,t)dx = 1, \qquad (1.4.10)$$

where the first condition comes from the fact that $x(t=0) = 0$, and the second is simply the normalization.

In Appendix D, we derive the forward Fokker-Planck equation using an alternative approach that is more intuitive. This method involves averaging over all possible paths a particle can take from position $x'$ at time $t$ to position $x$ at time $t + dt$.

### 1.4.2 Steady State

Let us look for a stationary solution of the forward Fokker-Planck equation given in Eq. (1.4.9). In order to do this, we suppose that $p(x,t) \to p(x)$ when $t \to +\infty$. We therefore need to solve the equation
$$D\frac{\partial^2 p(x)}{\partial x^2} = \frac{\partial}{\partial x}\left(F(x)p(x)\right). \qquad (1.4.11)$$

It is straightforward to show that the zero current solution is given by
$$p(x) = \frac{1}{Z}e^{-\frac{V(x)}{D}} \quad , \quad Z = \int_{-\infty}^{+\infty} dy\, e^{-\frac{V(y)}{D}}, \qquad (1.4.12)$$

which is the Boltzmann-Gibbs distribution.

### 1.4.3 Detailed Balance, Time Reversibility, and Equilibrium

We introduce $J(x,t)$, the *probability flux* at position $x$ and time $t$. Intuitively, $J(x,t)$ describes how probability flows in space due to both diffusion and drift (represented by $F(x)$). The Fokker-Planck equation relates the probability flux to the time evolution of the probability distribution:
$$\frac{\partial p(x,t)}{\partial t} = -\frac{\partial J(x,t)}{\partial x}, \qquad (1.4.13)$$

where, for Brownian motion, the probability flux is given by
$$J(x,t) = D\frac{\partial p(x,t)}{\partial x} - F(x)p(x,t). \qquad (1.4.14)$$

For a **stationary state**, the probability distribution no longer changes over time, meaning $\frac{\partial p(x,t)}{\partial t} = 0$. This implies that the probability flux is *constant*. However, stationarity alone does not necessarily mean equilibrium, as a system can be in a steady state with a nonzero constant flux.

A system is **time-reversible** if the probability of observing a trajectory forward in time is the same as observing the reverse trajectory. Mathematically, this requires no net probability flow, i.e., $J(x) = 0$. This condition implies that probability is not preferentially moving in any direction, ensuring that the system is in *detailed balance*.

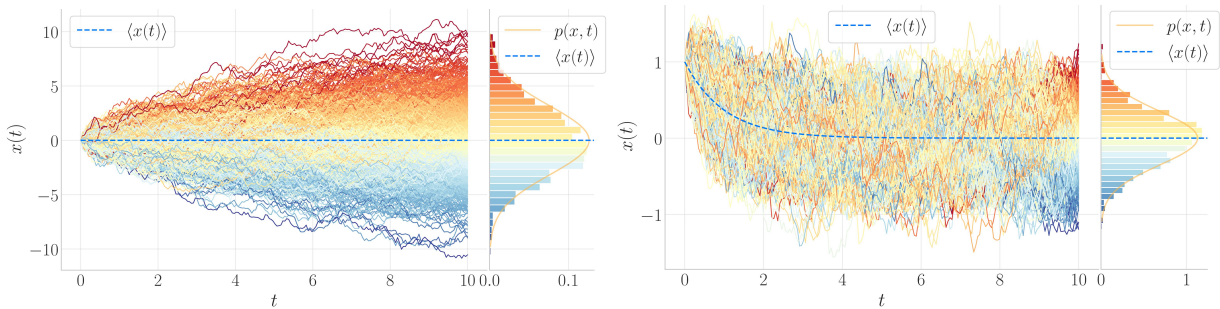

**Figure 1.4: Left.** We show $N = 1000$ trajectories of Brownian motion with a diffusion coefficient $D = 1$. The corresponding histogram is displayed, illustrating convergence toward the Gaussian distribution. **Right.** The same plot is shown for Ornstein-Uhlenbeck processes with $\mu = 1$ and $D = 0.5$. Here, the distribution $p(x,t)$ converges toward a steady-state equilibrium distribution $p(x)$, which can be observed as the mean and variance stabilize at constant values. For visualization, we have used `aleatory` package in python.

The condition of **detailed balance** states that, at equilibrium, transitions between any two states $x$ and $x'$ must be balanced:

$$p(x)W(x \to x') = p(x')W(x' \to x), \tag{1.4.15}$$

where $W(x \to x')$ is the transition rate (probability per unit time) for moving from state $x$ to state $x'$. This ensures that there is no net probability flow, which directly leads to $J(x) = 0$. Thus, detailed balance guarantees both stationarity and time reversibility, leading to the key result

$$\text{Equilibrium} = \text{Detailed Balance} = \text{Stationarity} + \text{Time Reversibility}. \tag{1.4.16}$$

For a system obeying the *Boltzmann-Gibbs distribution*, we have

$$J(x) = D\frac{\partial p(x)}{\partial x} - F(x)p(x) = \frac{D}{Z}\frac{\partial}{\partial x}e^{-\frac{V(x)}{D}} + \frac{V'(x)}{Z}e^{-\frac{V(x)}{D}} = 0\,, \tag{1.4.17}$$

where we used the fact that $F(x) = -V'(x)$. This confirms that Brownian motion in a potential, when reaching a stationary state, follows equilibrium dynamics.

For an arbitrary time $t$, when the system has not yet reached equilibrium, we typically have a nonzero probability flux

$$J(x,t) = D\frac{\partial p(x,t)}{\partial x} - F(x)p(x,t) \neq 0. \tag{1.4.18}$$

In this case, the probability distribution $p(x,t)$ is still evolving, meaning probability is actively redistributing in state space. This breaks time reversibility, as there is a preferred direction for probability flow.

### 1.4.4 Ornstein-Uhlenbeck Process

In the presence of a harmonic potential $V(x) = \frac{\mu x^2}{2}$, with $\mu > 0$, the particle is subjected to a force $F(x) = -V'(x)$, and the resulting stochastic process is known as the *Ornstein-Uhlenbeck (OU) process* [66]. It models the dynamics of a particle undergoing Brownian motion with friction due to collisions with other particles, which is more accurate than simple Brownian motion. The frictional force pulls the velocity back toward the mean, ultimately leading to a stationary state at large times. The equation of motion reads

$$\dot{x}(t) = -\mu x + \eta(t)\,, \tag{1.4.19}$$

and we suppose that the particle starts its motion at $x_0$ at initial time $t_0$. The explicit solution can be derived by solving this differential equation with initial condition $x(0) = x_0$. This leads to

$$x(t) = e^{-\mu(t-t_0)}x_0 + \int_{t_0}^{t} dt'\, e^{-\mu(t-t')}\eta(t')\,. \tag{1.4.20}$$

Since the integral represents a linear combination of Gaussian random variables, the quantity $x(t) - e^{-\mu(t-t_0)}x_0$ follows a normal distribution $\mathcal{N}(0, \sigma^2)$, where

$$\sigma^2 = \int_{t_0}^{t} dt_1 \int_{t_0}^{t} dt_2\, e^{-\mu(t-t_1)} e^{-\mu(t-t_2)} \langle \eta(t_1)\eta(t_2) \rangle \tag{1.4.21}$$

$$= 2D \int_{t_0}^{t} dt_1 e^{-2\mu(t-t_1)} = \frac{D}{\mu}\left(1 - e^{-2\mu(t-t_0)}\right). \tag{1.4.22}$$

Hence, $x(t)$ follows the Gaussian distribution

$$x(t) \stackrel{d}{\sim} \mathcal{N}\left(e^{-\mu(t-t_0)}x_0, \frac{D}{\mu}\left(1 - e^{-2\mu(t-t_0)}\right)\right), \tag{1.4.23}$$

with the associated probability density function

$$p(x,t|x_0,t_0) = \sqrt{\frac{\mu}{2\pi D\left(1 - e^{-2\mu(t-t_0)}\right)}} \exp\left[-\frac{\mu\left(x - e^{-\mu(t-t_0)}x_0\right)^2}{2D\left(1 - e^{-2\mu(t-t_0)}\right)}\right]. \tag{1.4.24}$$

In the long-time limit $t \to +\infty$, the system reaches a stationary distribution, which is given by the Boltzmann-Gibbs distribution

$$p(x) = \lim_{t \to +\infty} p(x,t) = \sqrt{\frac{\mu}{2\pi D}} e^{-\frac{\mu x^2}{2D}}\,. \tag{1.4.25}$$

This stationary distribution reflects the equilibrium state of a Brownian particle in a harmonic potential at temperature $T \sim D$. The Ornstein-Uhlenbeck process is thus one of the simplest examples of a stochastic process that exhibits relaxation to equilibrium. We show trajectories of OU processes in Fig. 1.4.

### 1.4.5 Fluctuation-Dissipation Theorem

The *fluctuation-dissipation theorem* (FDT), fundamentally states that the way in which a system returns to equilibrium after a small perturbation is directly related to the magnitude of its intrinsic fluctuations at equilibrium due to thermal noise [67]. It provides deep insights into both equilibrium and non-equilibrium dynamics.

Let us consider a Brownian dynamics $x(t)$, with $x(t=t_0) = x_0$, subject to an additional small perturbation $h(t)$. The corresponding Langevin equation is given by

$$\dot{x}(t) = F(x) + h(t) + \eta(t)\,. \tag{1.4.26}$$

We introduce the two-time correlation function in the absence of the perturbation,

$$C(t,t') = \left\langle x(t)x(t') \right\rangle \bigg|_{h=0}, \tag{1.4.27}$$

and the response function, which quantifies how the system reacts to an infinitesimally small external perturbation $h(t')$ applied at time $t'$

$$R(t,t') = \left\langle \frac{\delta x(t)}{\delta h(t')} \right\rangle \bigg|_{h=0}. \tag{1.4.28}$$

The fluctuation-dissipation theorem states that when $t_0 \to -\infty$, both $C(t,t')$ and $R(t,t')$ become independent of $x_0$ and stationary, i.e., $C(t,t') \to C_{st}(t-t')$ and $R(t,t') \to R_{st}(t-t')$, and satisfy the fundamental relation

$$R_{st}(\tau) = -\frac{1}{D}\frac{dC_{st}(\tau)}{d\tau}. \tag{1.4.29}$$

When a system is at equilibrium, the fluctuation-dissipation relation holds. However, for out-of-equilibrium dynamics, deviations from FDT can serve as a quantitative measure of how far the system is from equilibrium.

Let us illustrate this result for the Ornstein-Uhlenbeck process where $F(x) = -\mu x$ with $\mu > 0$. The explicit solution of the Langevin equation is given by

$$x(t) = e^{-\mu(t-t_0)}x_0 + \int_{t_0}^{t} dt'\, e^{-\mu(t-t')} \left[\eta(t') + h(t')\right]. \tag{1.4.30}$$

From this dynamics, the response function is found to be

$$R(t,t') = \left\langle \frac{\delta x(t)}{\delta h(t')} \right\rangle \bigg|_{h=0} = e^{-\mu(t-t')}\Theta(t-t')\Theta(t'-t_0), \tag{1.4.31}$$

where $\Theta(x)$ is the Heaviside function. It can also be shown that for any $t > 0$ and $t' > 0$, the correlation function is given by

$$C(t,t') = x_0^2\, e^{-\mu(t+t'-t_0)} + \frac{D}{\mu}\left[e^{-\mu|t-t'|} - e^{\mu(t+t'-2t_0)}\right]. \tag{1.4.32}$$

Taking the limit $t_0 \to -\infty$, we obtain the stationary correlation and response functions

$$R(t,t') \underset{t_0 \to -\infty}{\longrightarrow} e^{-\mu|t-t'|}\Theta(t-t') = R_{st}(t-t') \tag{1.4.33}$$

$$C(t,t') \underset{t_0 \to -\infty}{\longrightarrow} \frac{D}{\mu}e^{-\mu|t-t'|} = C_{st}(t-t'). \tag{1.4.34}$$

Thus, for $\tau > 0$, we verify that Eq. (1.4.29) holds, which is consistent with the fact that Ornstein-Uhlenbeck process is an equilibrium process. However, this result does not hold in the free Brownian limit $\mu \to 0$. This implies that free Brownian motion does not correspond to an equilibrium dynamics.

For out-of-equilibrium dynamics, it has been shown that the fluctuation-dissipation theorem (1.4.29) holds in a modified form, where a time-scale-dependent effective temperature plays the same role as the thermodynamic temperature [68]. In my first paper [6], we have derived the FDT for an out-of-equilibrium process driven by a resetting noise. We will introduce stochastic resetting in the Chapter 3.

## 1.5 First-Passage Properties of Brownian Motion

In this section, we will derive key observables related to the study of first-passage properties and compute them in the specific case of Brownian motion. The tools presented here are, in fact, both general and powerful, making them essential for studying the first-passage properties of stochastic processes. They play a central role in my Ph.D. research, and introducing them in the context of Brownian motion is natural and pedagogically valuable. For comprehensive surveys on the topic, see [43, 63, 69, 70].

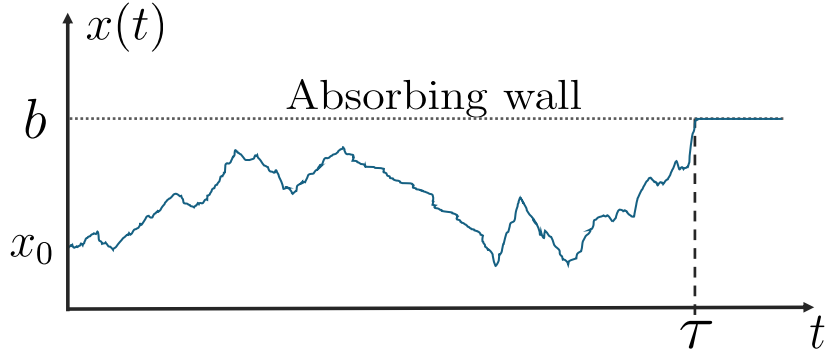

**Figure 1.5:** An important aspect of the study of stochastic processes is understanding their first-passage properties. Typically, this involves computing the survival probability $Q_b(x_0, t)$ of a particle within a given region of space, such as $]-\infty, b]$ (as shown in the figure), i.e. the probability that the particles stayed below $b$ for a duration $t$. The particle starts its motion at $x_0 < b$, and we model the threshold $b$ as an absorbing wall. Another crucial observable is the mean first-passage time (MFPT) to the barrier at $b$. In the figure, the first-passage time of the trajectory is denoted by $\tau$. Since $\tau$ is a random variable, one typically computes its average, $\langle \tau(x_0) \rangle$.

### 1.5.1 Survival Probability for Free Brownian motion

A key quantity in the study of first-passage properties of a stochastic process is the *survival probability*. It represents the probability that a stochastic process $x(t)$ remains below a threshold $b$, where $x(0) < b$, up to time $t$ (see Fig. 1.5). Typically, this threshold is modeled as an absorbing barrier at position $b$, meaning that if the particle reaches $b$, it is "killed", hence the term survival. We denote this probability as

$$Q_b(x_0, t) = \mathbb{P}\left(x(t) < b\right). \tag{1.5.1}$$

It is quite straightforward to relate the survival probability to the first-passage properties of $x(t)$. To proceed, we note that $Q(x_0, t)$ is also the probability that the first-passage time to $b$, denoted by $T$, is larger than $t$, i.e.,

$$Q_b(x_0, t) = \text{Prob.}(T > t) = 1 - \text{Prob.}(T < t). \tag{1.5.2}$$

The survival probability is therefore directly related to the cumulative distribution function of the random variable $T$, leading to the important relation

$$F_b(x_0, t) = -\partial_t Q_b(x_0, t), \tag{1.5.3}$$

where $F_b(x_0, t)$ is the probability density function (PDF) of the first-passage time $T$. This means that $F_b(x_0, t)\, dt$ represents the probability that the process $x(t)$, initially located at $x_0$, reaches the threshold $b$ for the first time in the interval $[t, t + dt]$.

**Remark.** The survival probability is also an important observable in the study of extreme value statistics (EVS). Suppose one is interested in the statistics of the maximum of the process $x(t)$. The probability that the maximum of $x(t)$ over the interval $[0, t]$ remains below a threshold $b$ is simply given by

$$\mathbb{P}\left(\max_{0 < t' < t} x(t') < b\right) = Q_b(x_0, t). \tag{1.5.4}$$

Thus, the probability density function of the maximum of $x(t)$ over $[0, t]$ is given by $\partial_b Q_b(x_0, t)$. For details on extreme value statistics, we refer readers to the following references [61, 71–73].

For a free Brownian motion $x(t)$, the survival probability reads

$$Q_b(x_0, t) = \int_{-\infty}^{b} dx\, p_b(x, t | x_0), \tag{1.5.5}$$

with $p_b(x,t|x_0)$ the Brownian propagator with an absorbing wall at $x = b$. It can be calculated using the "image method" [43]. Indeed, the idea is to say there is a Brownian propagator at initial position $x_0$, and we subtract a second Brownian propagator - the image with respect to the wall - at initial position $2b - x_0$ such that $p_b(b,t|x_0) = 0$. Over time, this image propagator subtracts the mass of the particles absorbed by the wall. That way, we have

$$p_b(x,t|x_0) = \frac{1}{\sqrt{4\pi D t}} \left( e^{-\frac{(x-x_0)^2}{4Dt}} - e^{-\frac{(x-(2b-x_0))^2}{4Dt}} \right), \quad (1.5.6)$$

and we directly integrate it form $-\infty$ to $b$ in order to obtain the survival probability

$$Q_b(x_0,t) = \mathrm{erf}\left(\frac{b-x_0}{\sqrt{4Dt}}\right). \quad (1.5.7)$$

The first-passage time density is given by

$$F_b(x_0,t) = -\partial_t Q_b(x_0,t) = \frac{(b-x_0)}{\sqrt{4\pi Dt^3}} e^{-\frac{(b-x_0)^2}{4Dt}}. \quad (1.5.8)$$

The distribution of first-passage times has a heavy tail $F_b(x_0,t) \underset{t\to+\infty}{\approx} t^{-3/2}$, meaning that very long times have a non-negligible probability. As a consequence, the first moment that we name the Mean First-Passage Time (MFPT) – see Fig. 1.5 – to the barrier $b$ is infinite

$$\langle \tau(x_0) \rangle = \int_0^{+\infty} dt\, t\, F_b(x_0,t) = \int_0^{+\infty} dt \frac{(b-x_0)}{\sqrt{4\pi Dt}} e^{-\frac{(b-x_0)^2}{4Dt}} = +\infty. \quad (1.5.9)$$

This is also the case of higher order moments. This phenomenon arises because, although short first-passage times are possible, there is a significant probability that the particle takes an exceedingly long time to reach $b$. These long times contribute heavily to the average, causing it to diverge.

However, it is interesting to note that as $t \to +\infty$, $Q_b(x_0,t)$ approaches 0, indicating that the probability of the particle not having reached $b$ decreases to zero. This implies that the particle will almost surely reach $b$ in finite time. This result highlights the counterintuitive nature of certain stochastic processes, where an event is almost certain to happen, yet the expected time for its occurrence is infinite.

### 1.5.2 Backward Fokker-Planck Equation

We aim to derive a differential equation for the survival probability that a Brownian particle, starting its motion at $x_0$ with $x_0 < b$, remains below $b$ up to time $t + dt$. We denote this probability as $Q_b(x_0, t+dt)$. One approach is to recognize that this probability can be expressed as the average over all possible displacements from $x_0$ during the infinitesimal time interval $dt$, as determined by the discretized Langevin equation:

$$x(dt) = x_0 + F(x_0)dt + \sqrt{2Ddt}\xi, \quad (1.5.10)$$

where $\xi \overset{d}{\sim} \mathcal{N}(0,1)$ is a standard normal random variable. Since we are considering displacements from $t = 0$ to $t = dt$ to determine the probability at time $t + dt$, we refer to this equation as the *backward Fokker-Planck* equation. By averaging over $\xi$, we obtain

$$Q_b(x_0, t+dt) = \left\langle Q_b\left(x_0 + F(x_0)dt + \sqrt{2Ddt}\xi, t\right) \right\rangle_\xi \quad (1.5.11)$$

$$= Q_b(x_0,t) + dt F(x_0)\frac{\partial Q_b(x_0,t)}{\partial x_0} + dt D \frac{\partial^2 Q_b(x_0,t)}{\partial x_0^2}. \quad (1.5.12)$$

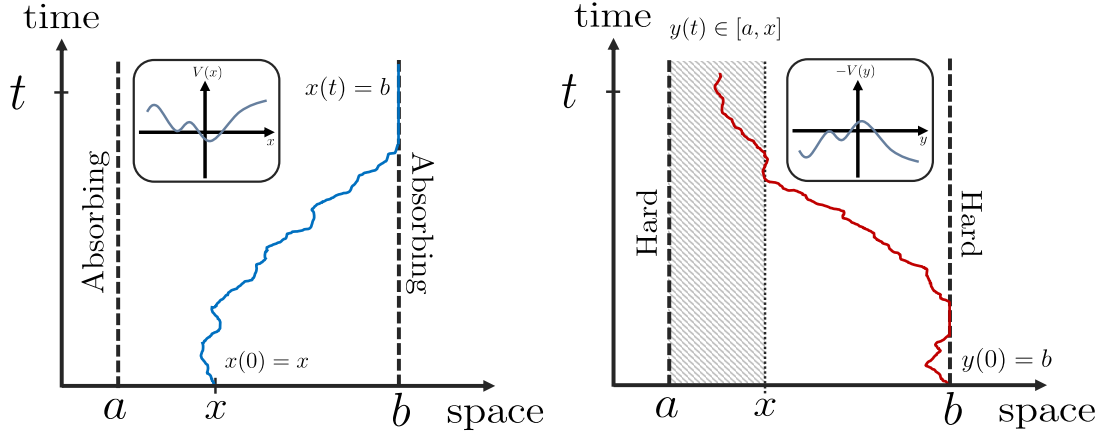

**Figure 1.6: Left.** We show a trajectory of a stochastic process that starts its motion at $x$ and reaches the absorbing boundary $b$ before time $t$. It moves within a potential $V(x)$. This trajectory contributes to the exit probability $E_b(x,t)$, which is the probability that the particle reaches $b$ before $a$, and before time $t$. **Right.** Here, we show the trajectory of the Siegmund dual with hard walls at $a$ and $b$, moving within the reversed potential $-V(x)$. It starts at $b$, and the trajectory contributes to $\Phi(x,t)$, which is the probability of finding the particle in $[a,x]$ at time $t$. Siegmund duality states that $E_b(x,t) = \Phi(x,t)$.

Taking the limit $dt \to 0$, we obtain the backward Fokker-Planck equation

$$\frac{\partial Q_b(x_0,t)}{\partial t} = F(x_0)\frac{\partial Q_b(x_0,t)}{\partial x_0} + D\frac{\partial^2 Q_b(x_0,t)}{\partial x_0^2}, \tag{1.5.13}$$

which is supplemented by the initial conditions

$$Q_b(x_0, t=0) = 1 \quad , \quad Q_b(b, t=0) = 0. \tag{1.5.14}$$

The first condition follows from the fact that, since the particle starts at $x_0 < b$, it has certainly survived at $t = 0$. The second condition indicates that if the particle is initially at $b$, it is immediately absorbed and thus "killed." For this reason, $b$ is referred to as an absorbing boundary (or absorbing wall), meaning that once the particle reaches $b$ for the first time, it remains there indefinitely. One can check that Eq. (1.5.7) is indeed solution of the differential equation (1.5.13) when $F(x) = 0$.

It is crucial to note that forward Fokker-Planck equations describe the evolution of an observable as a function of the position $x$ at time $t$, while backward Fokker-Planck equations describe the evolution as a function of the starting position $x_0$ and the duration $t$. Both formulations play a fundamental role in the study of continuous-time stochastic processes.

### 1.5.3 Exit Probability from an Interval / Splitting Probability

Another important observable for studying first-passage properties of stochastic processes is the *exit probability* from an interval $[a,b]$. Specifically, what is the probability $E_b(x_0,t)$ that a particle, starting at $x_0$, exits the interval $[a,b]$ through $b$ before time $t$? See the left panel of Fig. 1.6 for an illustration. Naturally, this is the complementary probability to the survival probability within the interval $[a,b]$, which satisfies

$$Q_{[a,b]}(x_0,t) = 1 - E_b(x_0,t) - E_a(x_0,t), \tag{1.5.15}$$

where $E_a(x_0,t)$ is the probability that the particle exits through $a$ instead of $b$.

For a Brownian particle with absorbing boundaries at $a$ and $b$, it is straightforward to show, using the same approach as above, that the backward Fokker-Planck equation for $E_b(x_0,t)$ reads

$$\frac{\partial E_b(x_0,t)}{\partial t} = F(x_0)\frac{\partial E_b(x_0,t)}{\partial x_0} + D\frac{\partial^2 E_b(x_0,t)}{\partial x_0^2}, \tag{1.5.16}$$

with boundary conditions (for all $t$)

$$E_b(a,t) = 1 \quad , \quad E_b(b,t) = 0 \,. \tag{1.5.17}$$

In the long-time limit, the particle will almost surely reach one of the two absorbing walls. The limit

$$E_b(x_0) = \lim_{t \to \infty} E_b(x_0, t) \tag{1.5.18}$$

is referred to as the *splitting probability* or *hitting probability*, which represents the probability that the particle eventually exits the interval $[a, b]$ through $b$. From Eq. (1.5.16), the splitting probability satisfies the equation

$$\frac{\partial^2 E_b(x_0)}{\partial x_0^2} = -\frac{F(x_0)}{D} \frac{\partial E_b(x_0)}{\partial x_0} \,, \tag{1.5.19}$$

with boundary conditions

$$E_b(a) = 0 \quad , \quad E_b(b) = 1 \,. \tag{1.5.20}$$

The solution is given by

$$E_b(x_0) = \frac{\int_a^{x_0} dz\, e^{\frac{V(z)}{D}}}{\int_a^b dz\, e^{\frac{V(z)}{D}}} \,. \tag{1.5.21}$$

When $V(x) = 0$, the exit probability of a free Brownian motion is linear and given by

$$E_b(x_0) = \frac{x_0 - a}{b - a} \,. \tag{1.5.22}$$

### 1.5.4 A First Glimpse at Siegmund Duality

Interestingly, the splitting probability in Eq. (1.5.21) has the same form as the cumulative distribution function of a Brownian motion with reflective (or hard) walls at $a$ and $b$, but in the presence of an inverted potential $-V(x)$. This connection was noted, for instance, in [74].

It can actually be shown that this result also holds at finite time. To see this, we consider a Brownian motion $y(t)$ evolving in the inverted potential $-V(x)$ with reflective walls at $a$ and $b$. We also impose the initial condition $y(0) = b$, which is necessary to establish the connection. The motion obeys the following Langevin equation

$$\dot{y}(t) = -F(y) + \eta(t) \,. \tag{1.5.23}$$

The probability density function of its position satisfies the forward Fokker-Planck equation

$$\frac{\partial p(y,t)}{\partial t} = \frac{\partial}{\partial y} \left( F(y) p(y,t) \right) + D \frac{\partial^2 p(y,t)}{\partial y^2} \,. \tag{1.5.24}$$

We define its cumulative distribution function as

$$\Phi(x,t) = \int_a^x dy\, p(y,t) \,, \tag{1.5.25}$$

which corresponds to the probability of finding the particle inside the interval $[a, x]$ at time $t$ – see the right panel of Fig. 1.6. Using the relation $p(x,t) = \partial_x \Phi(x,t)$ in Eq. (1.5.24), we obtain

$$\frac{\partial}{\partial x} \left[ -\frac{\partial \Phi(x,t)}{\partial t} + \left( F(x) \frac{\partial \Phi(x,t)}{\partial x} \right) + D \frac{\partial^2 \Phi(x,t)}{\partial x^2} \right] = 0 \,. \tag{1.5.26}$$

For a reflecting boundary, the imposed condition is that no probability flux crosses the boundary. This means that the probability current must vanish at $x = a$ and $x = b$

$$J(a,t) = 0 = D\frac{\partial p(x,t)}{\partial x}\bigg|x = a - F(a)p(a,t) = D\frac{\partial^2 \Phi(x,t)}{\partial x^2}\bigg|x = a - F(a)\frac{\partial \Phi(x,t)}{\partial x}\bigg|x = a\,, \quad (1.5.27)$$

with the same condition holding for $J(b,t)$. Using Eq. (1.4.13), one can show that $\partial_t \Phi(a,t) = \partial_t \Phi(b,t) = 0$. Thus, integrating Eq. (1.5.26) leads to:

$$\frac{\partial \Phi(x,t)}{\partial t} = F(x)\frac{\partial \Phi(x,t)}{\partial x} + D\frac{\partial^2 \Phi(x,t)}{\partial x^2}\,. \quad (1.5.28)$$

This is precisely the backward Fokker-Planck equation satisfied by the exit probability $E_b(x,t)$ (see Eq. (1.5.16)). Moreover, $\Phi(x,t)$ and $E_b(x,t)$ satisfy the same boundary conditions, namely $E_b(a,t) = \Phi(a,t) = 0$ and $E_b(b,t) = \Phi(b,t) = 1$, as well as the same initial condition $E_b(x,0) = \Phi(x,0) = \mathbb{1}_{x \geq b}$ (which justifies the choice $y(0) = b$). Since these conditions uniquely determine the functions, we conclude that:

$$E_b(x,t) = \Phi(x,t)\,. \quad (1.5.29)$$

This result is a consequence of the so-called *Siegmund duality* [59, 75, 76]. More generally, two stochastic processes $x(t)$ and $y(t)$ (where $t$ is either discrete or continuous), initialized at $x(0) = x$ and $y(0) = y$, are said to be Siegmund duals if, at any time $t$,

$$\mathbb{P}(x(t) \geq y|x(0) = x) = \mathbb{P}(y(t) \leq x|y(0) = y)\,. \quad (1.5.30)$$

If $x(t)$ evolves between two absorbing boundaries at $a$ and $b$, then its Siegmund dual $y(t)$ evolves between two reflective boundaries at the same locations. Applying Eq. (1.5.30) to the case $y = b$ leads to Eq. (1.5.29). This result is particularly useful, as it establishes a direct connection between the first-passage properties of a stochastic process and the spatial properties of its Siegmund dual. Consequently, knowing one of the two functions in Eq. (1.5.29) allows for the determination of the other. For ergodic processes, this relationship can also be leveraged to numerically compute the exit probability more efficiently by evaluating the cumulative distribution function of the dual process instead [3, 4].

A key aspect of my PhD research was extending Siegmund duality to out-of-equilibrium stochastic processes, such as stochastic resetting, random diffusion models, and active particles. These three topics will be introduced in the following chapters. Along with PhD student Léo Touzo, we worked independently on this problem and published two papers on the subject [3, 4]. Our results will be discussed in Part V.

### 1.5.5 Kramer's Law

Although all the moments of the first-passage density of a free Brownian motion are infinite, the mean first-passage time is well-defined for a Brownian motion moving within a specific class of potentials $V(x)$ and can be explicitly computed. Below, we derive this result and highlight an important application known as Kramers' law. In many physical, chemical, and biological systems, transitions between stable states occur due to thermal fluctuations overcoming an energy barrier. A classic example is a Brownian particle in a double-well potential, where the MFPT to escape a potential well plays a crucial role in reaction rate theory. In this section, we derive the well-known Arrhenius-Kramers law, which characterizes the exponential dependence of the MFPT on the barrier height in the weak noise limit [39, 70, 77].

Let us consider a Brownian particle with diffusion coefficient $D$, starting at $x(0) = x_0 > 0$, moving in a potential with a local minimum at $x = 1$ and a local maximum at $x = 0$, typically a double-well potential – see Fig. 1.7 for an illustration. In the weak noise limit ($D \ll 1$), an important problem is to determine the average time required for the particle to transition from one side of the double well (starting at $x_0 > 0$) to the other. This problem can be reformulated as follows: what is the Mean First-Passage Time (MFPT) to the origin starting from $x_0 > 0$?

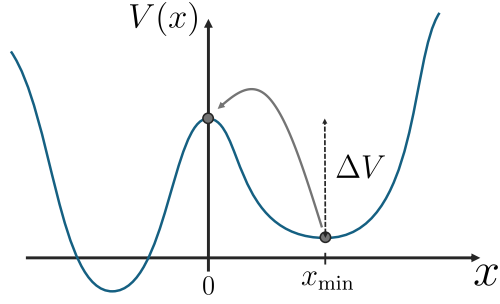

**Figure 1.7:** Consider a potential with a local minimum located at $x_\mathrm{min} > 0$ and a local maximum at the origin $x = 0$. We want to estimate the average time needed for a particle to escape from this local minimum. This is given by the mean first-passage time to the origin. For a Brownian particle, it is simply proportional to $\exp(\Delta V/D)$ where $\Delta V$ is the barrier height and $D$ the diffusion coefficient.

In the absence of noise, i.e., when the dynamics follows $\dot{x}(t) = f(x)$, regardless of the initial position $x_0$, the particle will always fall into the potential minimum at $x = 1$ and remain trapped there indefinitely, never crossing the barrier. When weak noise is introduced, the force will still drive the particle towards the minimum in a finite time. However, once at the minimum, the particle can escape through thermally activated random fluctuations. The MFPT is thus dominated by the mean time needed for the particle to escape from $x = 1$ towards $x = 0$.

To derive the MFPT for a Brownian particle subjected to a force $F(x)$, we start with the backward Fokker-Planck equation for the survival probability

$$\frac{\partial Q(x_0, t)}{\partial t} = F(x_0) \frac{\partial Q(x_0, t)}{\partial x_0} + D \frac{\partial^2 Q(x_0, t)}{\partial x_0^2}, \tag{1.5.31}$$

with the initial conditions

$$Q(x_0, t=0) = 1 \quad , \quad Q(0, t=0) = 0. \tag{1.5.32}$$

Deriving with respect to time, and multiplying Eq. (1.5.31) by $-t$, we obtain

$$-t \frac{\partial^2 Q(x_0, t)}{\partial t^2} = F(x_0) \frac{\partial}{\partial x_0} \left( -t \frac{\partial Q(x_0, t)}{\partial t} \right) + D \frac{\partial^2}{\partial x_0^2} \left( -t \frac{\partial Q(x_0, t)}{\partial t} \right). \tag{1.5.33}$$

Now we use the following identities

$$\int_0^{+\infty} dt' \, t' \left( -\frac{\partial Q(x_0, t')}{\partial t'} \right) = \int_0^{+\infty} t' F_0(x_0, t') = \langle \tau(x) \rangle. \tag{1.5.34}$$

where we recall that $F_0(x_0, t)$ is the first-passage time density at $x = 0$, and $\langle \tau(x) \rangle$ its first moment (the MFPT). Therefore, integrating with respect to $t$ Eq. (1.5.33) yields

$$-\int_0^{+\infty} dt' \, t' \frac{\partial^2 Q(x_0, t')}{\partial t'^2} = F(x_0) \frac{\partial}{\partial x_0} \langle \tau(x_0) \rangle + D \frac{\partial^2}{\partial x_0^2} \langle \tau(x_0) \rangle, \tag{1.5.35}$$

where the left-hand side simplifies to

$$-\int_0^{+\infty} dt' \, t' \frac{\partial^2 Q(x_0, t')}{\partial t^2} = [t F(x_0, t)]_0^{+\infty} - \int_0^{+\infty} dt' \frac{\partial Q(x_0, t')}{\partial t'} \tag{1.5.36}$$

$$= [Q(x_0, t)]_0^{+\infty} = -1. \tag{1.5.37}$$

In the end, we obtain a differential equation for the MFPT

$$F(x_0)\frac{\partial}{\partial x_0}\langle\tau(x_0)\rangle + D\frac{\partial^2}{\partial x_0^2}\langle\tau(x_0)\rangle = -1. \tag{1.5.38}$$

Using the same procedure, it is possible to show that the $n^{\text{th}}$ moment of $F(x_0,t)$ denoted $\langle\tau^n\rangle$ obeys the following recursive equation known as "Pontryagin equation" [77]

$$F(x_0)\frac{\partial}{\partial x_0}\langle\tau^n(x_0)\rangle + D\frac{\partial^2}{\partial x_0^2}\langle\tau^n(x_0)\rangle = -n\langle\tau^{n-1}\rangle. \tag{1.5.39}$$

To solve Eq. (1.5.38), we can use the fact that $F(x) = -V'(x)$, and multiply it by $e^{-\frac{V(x)}{D}}$. It leads to

$$\left(\frac{\partial}{\partial x_0}e^{-\frac{V(x_0)}{D}}\right)\frac{\partial}{\partial x_0}\langle\tau(x_0)\rangle + e^{-\frac{V(x_0)}{D}}\frac{\partial^2}{\partial x_0^2}\langle\tau(x_0)\rangle = -\frac{1}{D}e^{-\frac{V(x_0)}{D}}. \tag{1.5.40}$$

which can be re-written as

$$\frac{\partial}{\partial x_0}\left(e^{-\frac{V(x_0)}{D}}\frac{\partial}{\partial x_0}\langle\tau(x_0)\rangle\right) = -\frac{1}{D}e^{-\frac{V(x_0)}{D}}. \tag{1.5.41}$$

We therefore have

$$\frac{\partial}{\partial x_0}\langle\tau(x_0)\rangle = -\frac{1}{D}e^{\frac{V(x_0)}{D}}\int_a^{x_0}dy\,e^{-\frac{V(y)}{D}} + C, \tag{1.5.42}$$

where $C$ is a real constant, and we consider a reflective boundary at $x = a > x_0$. This reflective boundary is here to fix the constant $C$. Indeed, since at a reflective boundary, we have

$$\left.\frac{\partial}{\partial x_0}\langle\tau(x_0)\rangle\right|_{x=a} = 0, \tag{1.5.43}$$

the constant $C$ is simply zero. One can take the limit $a \to +\infty$, and a last integration of Eq. (1.5.42) gives

$$\langle\tau(x_0)\rangle = \frac{1}{D}\int_{-\infty}^{x_0}dz\,e^{\frac{V(z)}{D}}\int_z^{+\infty}dy\,e^{-\frac{V(y)}{D}} + D, \tag{1.5.44}$$

with $D \in \mathbb{R}$ another integration constant that we determine using the fact that $\langle\tau(0)\rangle = 0$. This is because the particle is instantaneously absorbed when $x_0 = 0$. In the end, we obtain

$$\langle\tau(x_0)\rangle = \frac{1}{D}\int_0^{x_0}dz\int_z^{\infty}dy\,\exp\left(\frac{V(z) - V(y)}{D}\right). \tag{1.5.45}$$

We are interested in the weak noise limit, i.e., the limit $D \to 0$ of Eq. (1.5.45). Performing the change of variable $y = z + u$, we have

$$\langle\tau(x_0)\rangle = \frac{1}{D}\int_0^{x_0}dz\int_0^{\infty}du\,\exp\left(\frac{V(z) - V(z+u)}{D}\right). \tag{1.5.46}$$

In the limit $D \to 0$, this double integral can be evaluated by the saddle point method, leading to

$$\log(\langle\tau(x_0)\rangle) \underset{D\to 0}{\sim} \frac{\max_{(z,u)}(V(z) - V(z+u))}{D}. \tag{1.5.47}$$

As $V(x)$ is decreasing on $[0,1]$, the maximum is reached for $(z^* = 0, u^* = 1)$ which represents the MFPT to reach the origin starting from the minimum of the well as announced. Therefore, we arrive at the Kramers' law

$$\langle \tau(x_0) \rangle \underset{D \to 0}{\sim} \exp\left(\frac{\Delta E}{D}\right) \quad , \quad \Delta E = V(0) - V(1) \,, \tag{1.5.48}$$

where $\Delta E$ is thus the height of the barrier.

A natural question that arises is how Kramers' law extends to non-equilibrium dynamics [78]. One class of stochastic processes that has attracted significant interest in this context involves active particles [79–81]. In Chapter 13, we derive Kramers' law for a simple model of active systems known as the run-and-tumble particle [2]. In particular, we show that while the MFPT still exhibits an exponential divergence, it does so with respect to an effective barrier height, which depends on the specific parameters of the active dynamics.

### 1.5.6 Beyond Brownian Motion: Persistence Exponent

Obtaining the full expression of $Q_b(x_0, t)$ is usually a complex task. In general, in the absence of confinement, the survival probability exhibits an algebraic decay at large times, i.e., $Q_b(x_0, t) \underset{t \to \infty}{\approx} t^{-\theta}$ where $\theta$ is called the persistence exponent. However, $\theta$ is known to be difficult to compute, and only a few exact results are available. For simple Brownian motion, using Eq. (1.5.7), we have $Q_b(x_0, t) \underset{t \to \infty}{\approx} t^{-1/2}$, and so $\theta = 1/2$. The persistence exponent can also be measured numerically and experimentally in certain systems. For more details, we refer to two review articles and references therein [63, 82]. Below, we present two widely studied non-Markovian processes in both mathematics and physics, for which the persistence exponent is known. Additionally, in a recent work, the persistence exponent has been determined for models of self-interacting random walks [83].

**Random Acceleration Process**

The random acceleration process is an integrated Brownian motion, defined as follows

$$\ddot{x}(t) = \eta(t) \,. \tag{1.5.49}$$

Discretizing the dynamics gives

$$x(t) \underset{dt \to 0}{\approx} 2x(t - dt) - x(t - 2dt) + 2D dt \xi \,, \tag{1.5.50}$$

where $\xi \sim N(0, 1)$. Here, the position at time $t$ depends only on its values at the two previous times, $t - dt$ and $t - 2dt$. This makes the process one of the simplest examples of a non-Markovian continuous-time stochastic process. However, when considering the vector $(x(t), v(t) = \dot{x}(t))$, the dynamics becomes Markovian, allowing the derivation of Fokker-Planck equations. The persistence exponent of the random acceleration process is known to be $\theta = 1/4$ [63, 84, 85].

**Fractional Brownian Motion**

Fractional Brownian motion (fBM) $x(t)$ is a continuous-time centered Gaussian process with two-time correlations function [86]

$$\langle x(t_1) x(t_2) \rangle = t_1^{2H} + t_2^{2H} - |t_1 - t_2|^{2H} \,, \tag{1.5.51}$$

where $0 < H < 1$ the Hurst parameter. The fBM increments are generated by a fractional Gaussian noise $\xi_f(t)$ which is a stationary non-Markovian process with correlations

$$\langle \xi_f(t_1) \xi_f(t_2) \rangle = 2H(2H - 1)|t_1 - t_2|^{2H-2} \,. \tag{1.5.52}$$

When $H = 1/2$, the dynamics becomes a simple Brownian motion. For the fBM, the persistence exponent is known to be $\theta = 1 - H$ [87, 88].

## 1.6 Conclusion

In this chapter, we introduced Brownian motion and derived various results related to its spatial properties and first-passage behavior. We also demonstrated that, in the presence of external confinement, Brownian motion exhibits equilibrium dynamics, with its stationary state described by the Boltzmann-Gibbs measure. Brownian motion is a well-understood stochastic process from both mathematical and physical perspectives, and numerous analytical tools have been developed to study its properties. While this chapter has focused primarily on equilibrium Brownian motion, many real-world systems operate far from equilibrium. These include active matter, driven stochastic processes, and transport in disordered systems, where detailed balance is broken, leading to rich non-equilibrium steady states (NESS). In the next chapters, we will extend our analysis to such out-of-equilibrium stochastic processes, exploring how concepts like large deviations [22, 23] and anomalous diffusion [89] emerge in complex systems.

# Chapter 2

# When Brownian Diffusion is Not Gaussian

## 2.1 Introduction

The simplest model for molecular transport is certainly Brownian motion (introduced in Chapter 1), where a passive particle is set into random motion by collisions with surrounding fluid particles in thermal equilibrium. Mathematically, this process results in Fickian diffusion, characterized by a linear relationship between the mean-square displacement (MSD) and time, and a Gaussian distribution for the position of the particle. However, many real-world systems exhibit deviations from this simple behavior, known as anomalous diffusion, where the MSD is not linear but follows a different power-law dependence on time

$$\text{MSD}[x(t)] = \langle (x(t) - x(t=0))^2 \rangle \propto t^\alpha. \tag{2.1.1}$$

Depending on the value of $\alpha$ we usually distinguish different types of motion:

- $\alpha = 1$ corresponds to normal (Fickian) diffusion.
- $\alpha < 1$ indicates subdiffusion (slower than normal diffusion).
- $\alpha > 1$ describes superdiffusion (faster than normal diffusion).

There also exist systems exhibiting ultra-slow diffusion, characterized by a MSD growing slower than a power-law. An example is the Sinai model [90], which describes the diffusion of a particle in a quenched random potential $U(x)$, where $U(x)$ is given by a realization of a random walk/Brownian motion [91–96]. At large times, the mean squared displacement behaves as $\overline{\langle x^2(t) \rangle} \propto \log(t)^4$, where the overline denotes averaging over different realizations of the disordered potential.

Anomalous diffusion processes have attracted significant interest across diverse scientific fields, including complex and disordered systems [89, 97], soft materials such as colloids [98], movement ecology [99], or financial markets [100]. However, recent studies have revealed numerous cases displaying standard Brownian scaling ($\alpha = 1$) accompanied by distinctly non-Gaussian fluctuations [19], contradicting the standard kinetic theory of normal diffusion. For instance, experiments on colloids [20, 101] have demonstrated a crossover in the position distribution from Gaussian behavior at short distances to an exponential tail at larger distances – see Fig. 2.1. These observations indicate that even in situations where the overall diffusion appears normal, the underlying processes—such as heterogeneity, intermittency, or complex interactions—can induce significant deviations from Gaussian statistics in the displacement distributions.

To account for environmental disorder, various random-walk models have been introduced [102, 103]. For instance, one can incorporate random trapping times at lattice sites, as considered in Continuous-Time Random Walk (CTRW) models [89, 104, 105]. Alternatively, besides the Sinai diffusion mentioned above, diffusion in a disordered medium can be modeled by assigning random jump rates at each lattice site, preserving a mean squared displacement that scales linearly with time [102, 103]. Theoretical studies of diffusion in static (quenched) disordered environments have also been conducted [106, 107], where the disorder originates from spatial variations of the diffusion coefficient.

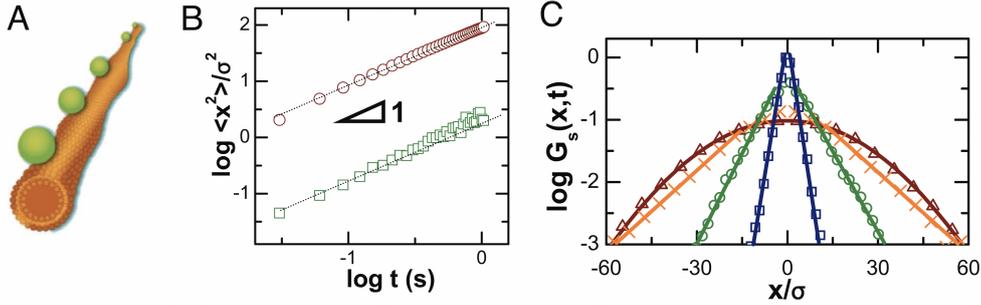

**Figure 2.1:** Colloidal beads of diameter $4\sigma = 100\mu m$ moving along linear lipid tubes, schematically shown in (A). MSD for two different lipid compositions, the lines have unit slope (B). The displacement distribution (C) has exponential tails at earlier times and crosses over to a Gaussian shape at longer times: 60ms (squares), 0.6s (circles), 3s (crosses), and 5.8s (triangles). Reproduced from [20, 110].

A simple way to model this spatial disorder is the following [19] (see also [108–110]): the fluid in which the particle diffuses can be divided into smaller cells, where each cell is locally in equilibrium and has a constant diffusion coefficient $D_i$. If the diffusion coefficients of these cells evolve on a sufficiently long timescale, the overall distribution can be described as a weighted average of Gaussian distributions, leading to

$$p(x,t) = \int_0^{+\infty} dD\, W(D) \frac{e^{-\frac{x^2}{4Dt}}}{\sqrt{4\pi Dt}}, \qquad (2.1.2)$$

where $W(D)$ is the stationary-state probability density for the diffusion coefficients. It is evident that the expression of $p(x,t)$ given in Eq. (2.1.2) is not necessarily Gaussian. In particular, if $W(D)$ is exponential, then $p(x,t)$ as given in Eq. (2.1.2), is itself exponential [111]. Moreover, it is straightforward to see that the MSD is linear in time

$$\langle x^2(t) \rangle = \int_0^{+\infty} dD\, W(D) 2Dt = 2\langle D \rangle t. \qquad (2.1.3)$$

To theoretically capture and describe these "diffusive yet non-Brownian" behaviors, a broad spectrum of models has been proposed, including continuous time random walks and variants of it [112–116] as well as random diffusivity models [117–119]. In the latter models, a key feature is the incorporation of stochasticity or randomness into the time evolution of the diffusion constant $D(t)$. In the context of disordered systems, this random diffusion constant effectively accounts for the spatial heterogeneities present in the system [89]. For such models in the simple one-dimensional setting, the MSD, which is the second cumulant (or variance) of the particle position, typically behaves as $\text{MSD}[x(t)] = \text{Var}[x(t)] \approx 2 D_{\text{eff}} t$, where $D_{\text{eff}}$ is an effective diffusion constant that has been computed for various models.

The implications of environmental heterogeneities on non-Gaussian fluctuations have been investigated experimentally. In [120], the authors studied the diffusion of colloidal particles in an environment where they controlled the degree of disorder/spatial heterogeneities. Although the diffusion remained Fickian, even a small amount of disorder led to the emergence of non-Gaussian behavior, resulting in an exponential displacement distribution in a highly disordered environment. This study thus confirms the hypothesis that non-Gaussianity can arise from spatial heterogeneities.

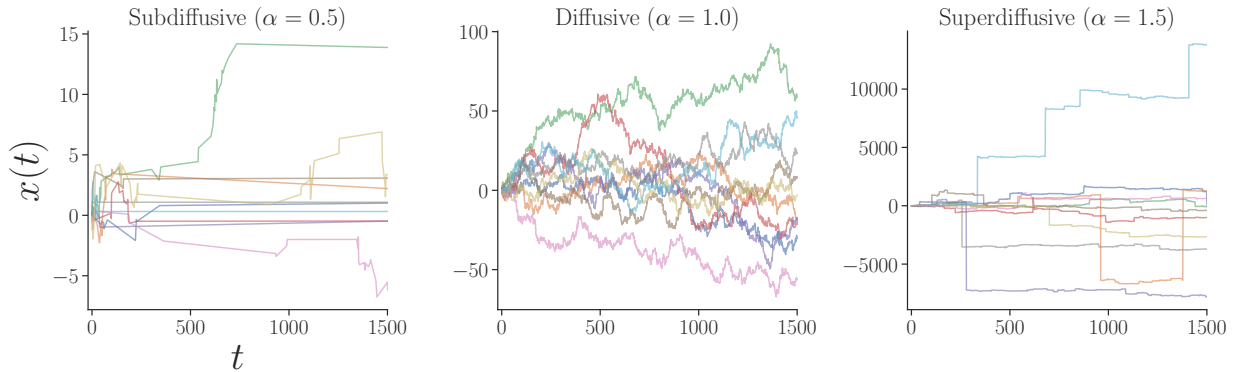

**Figure 2.2:** Trajectories of continuous-time random walks with varying jump and waiting-time distributions, illustrating distinct diffusion behaviors: subdiffusive (left), normal diffusive (middle), and superdiffusive (right).

Finally, in finance, the diffusion coefficient is referred to as volatility [121]. Numerous models with stochastic volatility have been introduced, such as the Heston model [122]. The concept of stochastic volatility was first proposed by H. Johnson and D. Shanno in [123], where they introduced a model in which volatility itself undergoes diffusion. Stochastic volatility models describe how the volatility of an asset price evolves randomly over time, offering a better representation of real market behavior [124].

Below, we review the most studied models recently employed to describe this Fickian yet non-Brownian diffusion.

## 2.2 Continuous Time Random Walk

The first model we introduce here is not a stochastic process with random diffusivity but rather a model of random walks, extensively used to study anomalous diffusion: the continuous-time random walk (CTRW). Initially proposed by Montroll, Weiss, and Scher [104, 105], this framework has found broad application across fields such as physics, biology, and finance [89, 97, 104, 105, 125, 126]. As we will see, CTRW can also describe motions that exhibit linear mean squared displacement alongside non-Gaussian fluctuations.

In a CTRW, a particle performs random jumps whose lengths follow a probability distribution function $f(x)$, occurring at random intervals determined by another PDF $g(t)$. In discrete space, this problem corresponds to an "annealed" random diffusion, where the particle remains trapped for a random waiting time drawn from the distribution $g(t)$ [89][1]. The PDFs $f(x)$ and $g(t)$ are obtained from the joint PDF of the jumps lengths and time intervals $\psi(x,t)$ through

$$f(x) = \int_0^{+\infty} dt\, \psi(x,t), \qquad g(t) = \int_{-\infty}^{+\infty} dx\, \psi(x,t). \qquad (2.2.1)$$

Here, we focus on the case where jump lengths and waiting times are independent, allowing the factorization $\psi(x,t) = f(x)g(t)$.

The CTRW model is commonly used to study both standard diffusion and anomalous diffusion behaviors. Subdiffusion occurs when the waiting-time distribution has a heavy-tailed form, specifically $g(t) \sim t^{-1-\alpha}$ with $0 < \alpha < 1$. This results in the MSD growing sublinearly ($0 < \alpha < 1$ in Eq. (2.1.1)). Conversely, if the jump-length distribution is heavy-tailed and has a diverging variance, the motion becomes superdiffusive ($\alpha > 1$ in Eq. (2.1.1)). Figure 2.2 illustrates simulated trajectories for three distinct values of $\alpha$, demonstrating subdiffusive, diffusive, and superdiffusive behaviors.

---

[1] In the "quenched" version, the particle moves within a fixed realization of the disorder, and the waiting time at each site is constant [89].

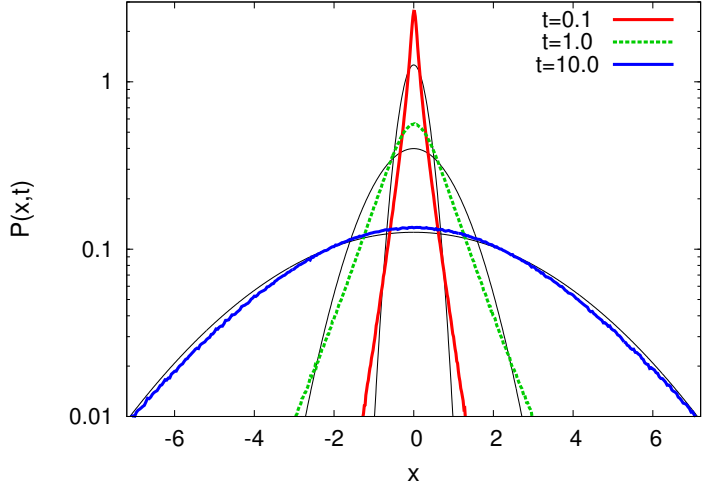

**Figure 2.3:** Probability density function $p(x,t)$ of a diffusing diffusivity model from simulations of the Langevin equations (2.3.1) for three different times. Comparison with Gaussian distribution (2.3.3) (solid lines) demonstrates the strongly non-Gaussian behavior at short times and the almost fully Gaussian shape at longer times. This figure is taken from [111].

Here, we are particularly interested in scenarios characterized by normal (linear) MSD growth. This occurs when both the mean waiting time,

$$\langle \tau \rangle = \int_0^{+\infty} dt\, t g(t), \tag{2.2.2}$$

and the jump-length variance,

$$\langle x^2 \rangle = \int_{-\infty}^{+\infty} dx\, x^2 f(x), \tag{2.2.3}$$

are finite. Under these conditions, the CTRW obeys the central limit theorem, and the typical fluctuations of the positions approaches a Gaussian behavior at large times.

Recent studies on large deviations of CTRWs have uncovered significant non-Gaussian fluctuations, even in the presence of a linear MSD [112–116]. To be more specific, when $f(x)$ decays faster than an exponential for large $x$, and when $g(t) \underset{t\to 0}{\approx} A\, t^a + B\, t^{a+1}$ where $A$ and $B$ are some constant, and $a > 0$, then we have

$$p(x,t) \underset{|x|\to\infty}{\approx} \exp\left[-\alpha |x|\, Z\left(\frac{|x|}{t}\right) + \beta\, t\right], \tag{2.2.4}$$

where $\alpha$ and $\beta$ are some constants, and $Z(y)$ is a slowly varying function which introduce logarithmic corrections at large $|x|/t$ [115]. We will see that a similar large deviation form arises in the switching diffusion model introduced below in Section 2.4, albeit with additional and interesting dynamical transitions (see Chapter 9 for a detailed discussion) [1].

Furthermore, a CTRW model described in [127], with a Lévy-stable waiting-time distribution and an exponential cutoff, exhibits Fickian diffusion at all times but demonstrates non-Gaussian fluctuations with exponential tails in the short-time regime, consistent with experimental observations. Another variant presented in [128] considers random jump rates, effectively capturing scenarios of Fickian yet non-Brownian diffusion.

## 2.3 Diffusing Diffusivity

As suggested in [19], non-Gaussian fluctuations observed in systems with a linear mean squared displacement may originate from spatial heterogeneities in the environment. These heterogeneities imply a position and time-dependent diffusion coefficient, denoted by $D(x,t)$, which has its own complex dynamics. A useful approach to capturing this complexity in a simple way is to represent the diffusion coefficient as a stochastic process that depends solely on time. For instance, recent studies have introduced *diffusing diffusivity models* [111, 117, 129, 130], where the diffusion coefficient itself undergoes diffusion. Here, we focus on the minimal model described in [111]. In this model, the diffusion coefficient is defined as the squared norm of an Ornstein-Uhlenbeck process, ensuring that it remains positive at all times. It is also stationary and it has a given correlation time. This minimal model is governed by the following coupled Langevin equations

$$\frac{dx}{dt} = \sqrt{2D(t)}\,\eta(t), \quad D(t) = \theta^2(t), \quad \frac{d\theta}{dt} = -\theta(t) + \xi(t), \tag{2.3.1}$$

where $D$ represents the diffusion coefficient, and $\xi(t)$ and $\eta(t)$ are Gaussian white noise terms with delta correlations: $\langle \eta(t)\eta(t') \rangle = \delta(t-t')$ and $\langle \xi(t)\xi(t') \rangle = \delta(t-t')$. An alternative approach, introduced in [117], is to consider $D(t)$ as a diffusive process constrained by reflective boundaries at $D = 0$ and $D = D_{\max}$. This method ensures $D$ remains within physically realistic limits. However, the minimal model described earlier is analytically more tractable and thus preferred for theoretical analysis [111]. It is worth noting that this minimal diffusing diffusivity model shares similarities with the Heston model frequently used in mathematical finance [122].

Since $\theta(t)$ is an Ornstein–Uhlenbeck process, it eventually reaches a steady state described by the stationary distribution $f_{st}(\theta) = \frac{1}{\sqrt{\pi}}e^{-\theta^2}$ (see Eq. (1.4.25)). Consequently, the diffusion coefficient $D(t)$ also reaches a steady state with a distribution given by

$$p_{st}(D) = \int_{-\infty}^{+\infty} d\theta\, f_{st}(\theta)\, \delta(\theta^2 - D) = \frac{1}{\sqrt{\pi D}}e^{-D}. \tag{2.3.2}$$

This stationary distribution decays (almost) exponentially, leading to a Laplace-like behavior in the displacement distribution at short times [111]. At large times, however, the displacement distribution approaches a Gaussian shape with a variance determined by the mean stationary diffusion coefficient $\langle D(t) \rangle = \langle D_{\text{st}} \rangle$ [111]:

$$p(x,t) \approx \frac{1}{\sqrt{4\pi\langle D_{\text{st}}\rangle t}} \exp\left(-\frac{x^2}{4\langle D_{\text{st}}\rangle t}\right). \tag{2.3.3}$$

It is also possible to show that the mean squared displacement grows linearly with time and is explicitly given by $\text{MSD}[x(t)] = 2\langle D_{\text{st}}\rangle t$ [111]. In Fig. 2.3, we illustrate the crossover from the short-time Laplace-like distribution to the long-time Gaussian behavior.

## 2.4 Switching Diffusion

While modeling the randomness of $D(t)$ as a diffusive process is a natural first step, recent experiments investigating protein diffusion in various media have revealed the occurrence of sudden changes, or *switches*, in the dynamics of $D(t)$. For example, in [131], G-proteins were shown to switch between four distinct diffusion states. In [132], the dynamics of proteins on the surface of neurons were observed to alternate between "confined" states, where the diffusion coefficient vanishes ($D(t) \sim 0$), and "free" states, where $D(t) > 0$. Similarly, in [133], it has been showed that the motion of dendritic cell-specific intercellular adhesion molecule 3-grabbing nonintegrin (DC-SIGN) experiences changes in diffusivity. These changes are detected using the change-point

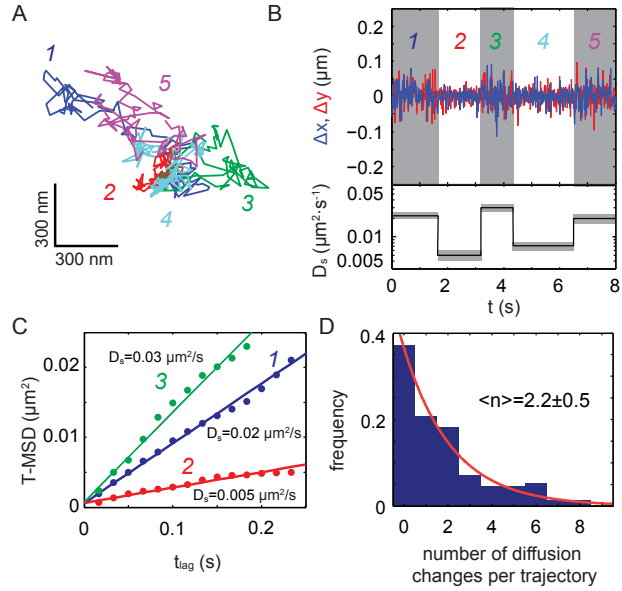

**Figure 2.4:** DC-SIGN motion experiences changes in diffusivity. (a) Representative (wild-type) DC-SIGN trajectory displaying changes of diffusivity. Change-point analysis evidences five different regions represented with different colors. (b) Plot of the $x$ (blue) and $x$ (red) displacements for the trajectory in (a) as a function of time. The shaded areas indicate the regions of different diffusivity. The lower panel displays the corresponding short-time diffusion coefficient as obtained from a linear fit of the time-averaged MSD for the five different regions. Gray areas correspond to the 95% confidence level. (c) Plot of time-averaged MSD versus time lag for the first three regions of the trajectory in (a). (d) Histogram of the number of changes of diffusion per trajectory. Most of the trajectories (63%) display at least one dynamical change, with an average of 2.2 changes per trajectory. This Figure and its caption are taken from [133]

algorithm [134] and the DC-SIGN diffusion on living cells was found to be accurately described by a model with a switching diffusion coefficient – see Fig. 2.4.

Switching diffusion is a model in which a particle, starting from the origin, performs a standard one-dimensional Brownian motion with a diffusion constant $\mathsf{D}_1$ over a time $\tau_1$. Both $\mathsf{D}_1$ and $\tau_1$ are random variables drawn from a joint distribution $P_{\text{joint}}(D, \tau)$. After this time $\tau_1$, the particle resumes its motion from its current position, now performing a new Brownian motion with diffusion constant $\mathsf{D}_2$ for a duration $\tau_2$, which are drawn independently from the same distribution $P_{\text{joint}}(D, \tau)$ as $\mathsf{D}_1$ and $\tau_1$. This process continues iteratively for a fixed period of time $t$ (see Fig. 2.5 for an illustration of this process). Such models have been used to model recent experiments on cytoplasmic membranes (which control the movement of substances in and out of a cell) showing patches of strongly varying diffusivity [132, 135–137].

A simpler version of this model is one in which $\mathsf{D}_i$'s and $\tau_i$'s are independent, that is, $P_{\text{joint}}(D, \tau) = W(D) \, p(\tau)$. More specifically, we are interested in the case where the $\tau_i$'s are exponential random variables with a rate $r$, i.e., $p(\tau) = r \, e^{-r\tau}$, while $W(D)$ is an arbitrary probability distribution function (PDF). A well-known example is the case where $W(D)$ is a superposition of Dirac delta peaks, i.e. $W(D) = \sum_{i=1}^{N} p_i \, \delta(D - D_i)$, with $D_1 > D_2 > \cdots > D_N$ and $\sum_{i=1}^{N} p_i = 1$. This model, sometimes called "composite Markov process" [65], has been studied in various contexts ranging from disordered systems [138, 139], biophysics [131, 140–142], nuclear magnetic resonance [143], finance [144] or movement ecology [145, 146]. In the latter, mixtures of random walks with switching dynamics between them are widely used to model intermittent searches where an animal/a particle can employ different motion modes [40, 145]. In the case $N = 2$ (referred to as the two-state model), the mode with $D = D_2 < D_1$ would then model local search, while the one with $D = D_1$ corresponds to an exploratory motion with larger displacements. Incidentally, this model with $N = 2$ recently appeared in the context of stochastic
3030

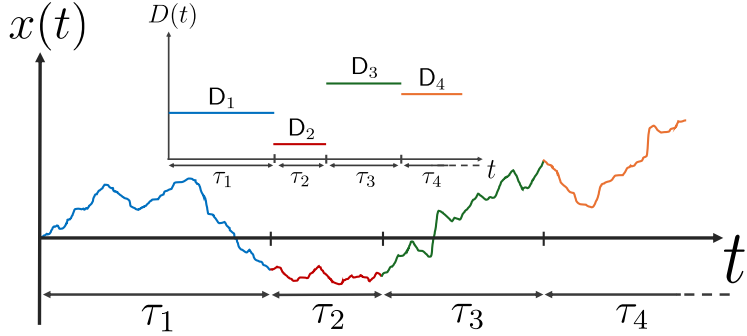

**Figure 2.5:** Trajectory of a switching diffusion process in one-dimension. During each time interval $\tau_i$, the particle performs an independent Brownian motion with a diffusion constant $D_i$. In the simplest case, the $\tau_i$'s are independent exponential random variables, while the $D_i$'s, which are also independent, are drawn from an arbitrary distribution $W(D)$.

resetting with two resetting points [147]. Besides the case of discrete diffusion modes, various studies, both theoretical [118,142,148,149] and experimental [117,120,150], have considered a continuous distribution for $W(D)$ including exponential and gamma distribution [118,120,148,149] but also distribution with a finite support [117,142].

## 2.5 Connection With My Work

The non-Gaussian fluctuations of $x(t)$ are usually captured by the higher-order cumulants of $x(t)$, like the skewness and kurtosis (respectively the third and fourth cumulants). Understanding these higher-order cumulants is thus crucial for characterizing non-Gaussianities of $x(t)$. Cumulants are also interesting because they carry information on the large deviations of $x(t)$ that characterizes its atypical large fluctuations. However, calculating higher-order cumulants is often quite challenging, as it requires evaluating higher-order correlation functions of $x(t)$. Consequently, there are very few results in the literature concerning these cumulants or the large deviations of the position distribution in random diffusivity models.

To understand the relationship between large deviations and cumulants, let us look at a standard illustrative calculation. We consider a diffusive motion characterized by a large deviation function $I(y = x/t)$ for the position distribution $p(x,t)$. We also consider that this distribution is symmetric under the transformation $x \to -x$. Consequently, for large times and $x > 0$, the distribution reads

$$p(x,t) \underset{t\to+\infty}{\asymp} \exp\left[-t\, I\left(\frac{x}{t}\right)\right], \qquad (2.5.1)$$

where $\underset{t\to+\infty}{\asymp}$ is a logarithmic equivalent. Let us define the moment generating function of $p(x,t)$ which is given by

$$\hat{p}(q,t) = \langle e^{qx}\rangle = \int_{-\infty}^{+\infty} dx\, e^{qx}\, p(x,t)\,. \qquad (2.5.2)$$

It satisfies

$$\hat{p}(q,t) \underset{t\to+\infty}{\asymp} \int_{-\infty}^{+\infty} dx\, \exp\left[qx - t\, I\left(\frac{x}{t}\right)\right] = \int_{-\infty}^{+\infty} dy\, \exp\left[t\,(qy - I(y))\right], \qquad (2.5.3)$$

where we have made the change of variables $y = x/t$. For large times, the integral is dominated by the term for which $(qy - I(y))$ reaches its maximum value (assuming it has one). Thus, by performing a saddle-point analysis, we obtain

$$\hat{p}(q,t) \underset{t\to+\infty}{\asymp} \exp\left[t \max_y \left[qy - I(y)\right]\right] \equiv \exp\left[t\,\Psi(q)\right], \qquad (2.5.4)$$

where we have introduced $\Psi(q) = \max_y \left[qy - I(y)\right]$, the Lengendre transform of $I(y)$. In particular, note that one can compute $I(y)$ through [22]

$$I(y) = q^* y - \Psi(q^*) \quad , \quad y = \Psi'(q^*). \qquad (2.5.5)$$

Hence, we find that the cumulants generating function satisfies

$$\log\left[\hat{p}(q,t)\right] = \sum_{n=1}^{+\infty} \frac{q^{2n}}{(2n)!} \langle x^{2n}(t) \rangle_c \underset{t\to+\infty}{\approx} t\,\Psi(q), \qquad (2.5.6)$$

where $\langle x^{2n}(t) \rangle_c$ are the cumulants of the distribution $p(x,t)$, and the equivalence on the right-hand side follows from Eq. (2.5.4). This analysis indicates that the cumulant generating function is related to the Legendre transform of the large deviation function $I(y)$. It also reveals that for a diffusive process, the leading-order behavior of the cumulants $\langle x^{2n}(t) \rangle_c$ is proportional to $t$ at large times.

The aim of our article [1] is to present a detailed analytical study of both the cumulants and the large deviation function for the switching diffusion model with exponential switching, and with an arbitrary distribution $W(D)$ for the diffusion coefficient. It will be discussed in Chapter 9. Notably, we have showed that at short times, the model behaves according to the distribution described in Eq. (2.1.2). For any time $t$, we have found the exact expression of the moments $\langle x^{2n}(t) \rangle$ for any $W(D)$, as well as the leading order in time of the cumulants. Surprisingly, we found that this model has a connection to free probability theory [151]. Indeed, the cumulants scale as $\langle x^{2n}(t) \rangle_c \propto \kappa_n(D)\,t$, where $\kappa_n(D)$ are the free cumulants of $W(D)$ – see Chapter 9. We have performed a detailed analysis of the large deviations in the position distribution at long times. For displacements $x(t) \sim \mathcal{O}(\sqrt{t})$, we showed that the distribution is Gaussian. For a certain class of diffusion coefficient distributions $W(D)$, the large deviation function $I(y)$ exhibits dynamical transitions, including an exponential regime. In Chapter 10, we will also derive exactly the NESS of the same switching diffusion model in the presence of a harmonic potential.

Let us finally point out that we may not require an explicit Gaussian white noise term $\sqrt{2D(t)}\eta(t)$ to obtain Gaussian behavior for typical distances of order $O(\sqrt{t})$, with non-Gaussian tails. Similar behavior can arise from dynamics driven by colored noise. For instance in [36], the area under a resetting Brownian motion (rBM) is studied [35, 54]. The main properties of resetting Brownian motion will be introduced and analyzed in detail in the next chapter. Its dynamics is a standard Brownian motion, however, at exponentially distributed intervals (with resetting rate $r$), the position of the particle is reset instantaneously to a fixed location $X_r$. The model is described by the following equations:

$$\dot{x}(t) = y_r(t) \quad , \quad y_r(t+dt) = \begin{cases} y_r(t) + \sqrt{2D dt}\,\xi & \text{with probability } (1 - r\,dt), \\ X_r & \text{with probability } r\,dt. \end{cases} \qquad (2.5.7)$$

In particular, without loss of generality setting $D = 1/2$, and the initial condition is $x(0) = 0$, the mean squared displacement (MSD) becomes linear at large times [36]

$$\langle x^2(t) \rangle = \frac{2}{r^3}\left(rt - 2 + e^{-rt}(2 + rt)\right) \underset{t\to\infty}{\approx} \frac{2}{r^2} t. \qquad (2.5.8)$$

Interestingly, although this model does not describe diffusion in the strict sense (as there is no explicit Gaussian white noise term $\sqrt{2D}\eta(t)$), the fluctuations of $x(t)$ at large times are Gaussian for typical distances of order $O(\sqrt{t})$, with large-deviation tails [36]. This result suggests that dynamics driven by colored noise, rather than strictly white noise, may also account for experimental observations.

In my first publication [6], we studied the NESS of the model (2.5.7) in the presence of an additional harmonic potential. We have shown that the distribution of the position crosses-over from exponential to Gaussian behavior. In particular, it exhibits Gaussian fluctuations at short distances, while its tails are exponential. We present a detailed analysis of this model in Chapter 6.

# Chapter 3

# Stochastic Resetting

Stochastic processes under resetting have attracted significant interest in recent years [35,54,152, 153]. The resetting mechanism involves a stochastic process $x(t)$ that evolves according to its intrinsic dynamics, but at random times, the process is reset to a specific position $X_r$, where it restarts its evolution. A well-known example is the resetting Brownian motion (rBM), where a Brownian particle is reset at exponentially distributed times [54]. Other variations include the resetting of the position of random walks [154], Lévy flights [155,156], or active particles, such as run-and-tumble particles (RTPs) [157,158]. It is also possible to consider resetting parameters other than the position of a particle, such as the velocity orientation of active particles [159], or a randomly evolving diffusion coefficient in diffusing diffusivity models [160]. Additionally, stochastic resetting has been used in various other models, including the Ising model [161] and models of fluctuating interfaces [162]. These systems can be analyzed analytically using tools such as renewal theory and large deviations [37,38,163,164], and in many cases, knowing the dynamics without resetting is sufficient to formulate equations for the observables under resetting [35].

In the case of resetting Brownian motion, two important properties emerge due to the presence of resetting. First, resetting induces out-of-equilibrium dynamics by breaking detailed balance. As a result, rBM reaches a non-equilibrium stationary state (NESS) characterized by a double-exponential stationary distribution [35, 54]. Second, from a first-passage perspective, as we demonstrate below, there exists an optimal resetting rate $r$ that minimizes search processes [54,157,165–168]. The rBM has been investigated experimentally [165, 166]. In experiments, as it is not possible to teleport the particle, resets are non-instantaneous [165, 166, 169] and it is often easier to reset the particle at fixed periods [165, 166].

Beyond physics, the concept of resetting has found applications in machine learning. For instance, resetting the learning rate in stochastic gradient descent (SGD) has been shown to improve the convergence rate of algorithms [55]. Additionally, resetting strategies have been employed in neural network training [56, 57]. It has also been applied to accelerate molecular dynamics simulations [58], in optimal control problems [170], and in many-body systems [171]. Furthermore, recent connections between stochastic resetting and quantum measurement suggest exciting directions for future research [172–175].

To illustrate the basic framework, let us consider the position of a particle $x(t)$ which is described by a stochastic process. We model the evolution of $x(t)$ as follows. For simplicity, it starts from $x(0) = 0$. It evolves by its own stochastic dynamics, e.g., just a Brownian or a ballistic motion (with random velocity) or even a (random) constant and then gets reset to 0 at random epochs $\{t_1, t_2, t_3, \ldots\}$ – see the left panel of Fig. 3.1. The successive intervals between the resetting events, $\tau_n = t_n - t_{n-1}$ (for $n = 1, 2, \ldots$ with $t_0 = 0$) are statistically independent and each is drawn from a PDF $p(\tau)$ normalized to unity. Clearly, in the case where the "free" evolution between two successive resettings is a Brownian motion, then $x(t)$ is the well-known resetting Brownian motion (rBM), to which the major part of this chapter is devoted. On the other hand, in the case where the "free" evolution is a random constant, say $\pm v_0$ with equal probability, then $x(t)$ corresponds to the telegraphic noise. In this work, we focus on *Poissonian resetting*, where $p(\tau) = r\,e^{-r\,\tau}$ [54], which serves as a simple model for understanding stochastic processes under resetting and their implications.

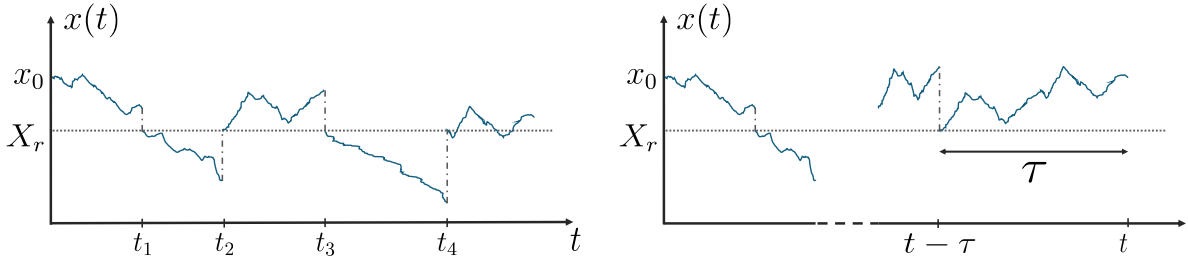

**Figure 3.1: Left.** We present an illustration of a resetting process, such as a resetting Brownian motion. The process starts at position $x_0$, follows its owns specific dynamics, and at random times $t_1, t_2, \cdots$, it is reset to $X_r$ where it restarts its motion until the next resetting time. **Right.** We illustrate the renewal argument used in Eq. (3.1.4), which states that the position of an rBM at time $t$ depends only on the diffusion since the last resetting event at time $t - \tau$.

## 3.1 Resetting Brownian Motion: a Non-Equilibrium Dynamics

In this section, we derive the fundamental properties of rBM and demonstrate that they can be obtained using only the observables of the same process without resetting. This is achieved through a simple renewal argument, which remains valid beyond rBM. We establish two key results. First, rBM reaches a nonequilibrium steady state (NESS), which we characterize completely. Second, its mean first-passage time (MFPT) is finite and, as a function of the resetting rate $r$, exhibits a minimum, allowing us to optimize the mean time required for the rBM to reach a target.

### 3.1.1 Non-Equilibrium Steady State of the resetting Brownian Motion

The dynamics of the resetting Brownian motion is given by

$$x(t+dt) = \begin{cases} x(t) + \sqrt{2Ddt}\,\xi & \text{with probability } (1 - r\,dt)\,, \\ X_r & \text{with probability } r\,dt\,. \end{cases} \qquad (3.1.1)$$

Here, $\xi \stackrel{d}{\sim} \mathcal{N}(0,1)$, and $D$ is the diffusion coefficient. The initial position of the particle is $x(t=0) = x_0$. In the most general case, one could consider the reset position $X_r$ to be drawn from a random distribution [54, 176], but for simplicity we will focus on the case where it is fixed. A schematic representation of a trajectory is shown in the left panel of Fig. 3.1.

Here, we are interested in studying the propagator $p_r(x, t | x_0; X_r)$ which we will simply write $p_r(x, t)$ when $x_0$ is treated as a constant – $X_r$ being always constant. A natural approach is to write the forward Fokker-Planck equation. For this purpose, we write $p_r(x, t + dt)$ with respect to $p_r(x, t)$. In a small time displacement $dt$, the particle can either reset to $X_r$ with probability $rdt$, or follow a Brownian dynamics with probability $1 - rdt$, thus

$$p_r(x, t+dt) = rdt\,\delta(x - X_r) + (1 - rdt)\int_{-\infty}^{+\infty} D\xi\, p_r(x - \sqrt{2Ddt}\xi, t)\,, \qquad (3.1.2)$$

where $\int_{-\infty}^{+\infty} D\xi$ is an integral over the random variable $\xi \stackrel{d}{\sim} \mathcal{N}(0,1)$. A Taylor expansion of the integrand on the right-hand side, and taking the limit $dt \to 0$ leads straightforwardly to the forward Fokker-Planck equation

$$\frac{\partial p_r(x,t)}{\partial t} = D\frac{\partial^2 p_r(x,t)}{\partial x^2} - rp_r(x,t) + r\delta(x - X_r), \qquad (3.1.3)$$

with the initial condition $p_r(x, t = 0) = \delta(x - x_0)$. The first term on the right-hand side corresponds to free diffusion, while $-rp_r(x,t)$ accounts for the loss of probability from $x$ due to resetting to $X_r$. Meanwhile, $r\delta(x - X_r)$ represents the gain of probability at the resetting position.

Although one could directly analyze Eq. (3.1.3), it is often more convenient and intuitive to adopt a *renewal approach* instead. Indeed, let us show that the simple knowledge of the Brownian propagator $G_0(x,t|x_0)$ (the subscript 0 means that it is the propagator without resetting) between resets is sufficient to study $p_r(x,t|x_0)$. We can write a *renewal equation* as follows [35]

$$p_r(x,t|x_0) = e^{-rt}G_0(x,t|x_0) + r\int_0^t d\tau\, e^{-r\tau}G_0(x,\tau|X_r) \quad , \quad G_0(x,t|x_0) = \frac{1}{\sqrt{4\pi Dt}}e^{-\frac{(x-x_0)^2}{4Dt}} \qquad (3.1.4)$$

The first term on the right-hand side accounts for trajectories that experienced no reset up to time $t$. This occurs with probability $e^{-rt}$, and the particle undergoes free diffusion for a duration $t$, leading to the propagator $G_0(x,t|x_0)$. The second term corresponds to trajectories that experienced at least one reset, where the last reset occurred at time $t - \tau$ and no further reset took place during the interval $\tau$. This occurs with probability $re^{-r\tau}$, while the probability of a reset occurring within the small time interval $[t-\tau, t-\tau+d\tau]$ is $rd\tau$. To account for all possible reset times, we integrate over $\tau$. From $t - \tau$ to $t$, the particle diffuses freely, contributing the propagator $G_0(x,\tau|x_0)$ inside the integral. Such a trajectory is illustrated in the right panel of Fig. 3.1. Notice that for different dynamics between resets, Eq. (3.1.4) remains valid with the appropriate replacement of the relevant propagator. This makes it a very general and robust approach.

In the large time limit, the first term in Eq. (3.1.4) is exponentially suppressed, and the steady state is given by

$$\lim_{t\to\infty} p_r(x,t|x_0) = p_r(x) = r\int_0^{+\infty} d\tau\, e^{-r\tau}G_0(x,\tau|X_r). \qquad (3.1.5)$$

Using the expression of $G_0$ in Eq. (3.1.4), it is possible to explicitly evaluate the integral in Eq. (3.1.5), yielding a Laplace distribution

$$p_r(x) = \frac{1}{2}\sqrt{\frac{r}{D}}e^{-\sqrt{\frac{r}{D}}|x-X_r|}. \qquad (3.1.6)$$

Unlike simple Brownian motion, resetting Brownian motion reaches a stationary state even in the absence of a confining potential. However, this stationary state is out of equilibrium, as it does not follow the Boltzmann-Gibbs measure. Detailed balance is violated because the probability flux from a position $x$ toward $X_r$ is greater than the flux from $X_r$ toward $x$. Remarkably, despite being a non-equilibrium stochastic process, resetting Brownian motion can be fully characterized analytically, with surprising simplicity, thanks to its renewal structure.

### 3.1.2 Relaxation to the steady state

It is also insightful to examine how $p_r(x,t|x_0)$, given in Eq. (3.1.4), relaxes toward the stationary state. For simplicity, we set $X_r = x_0 = 0$. To analyze this relaxation, we consider Eq. (3.1.4) in the large-time limit $t \to +\infty$ while keeping the ratio $y = x/t$ fixed. When rescaling the time $\tau = wt$, we obtain

$$p_r(x,t) = \frac{e^{-rt+\frac{x^2}{4Dt}}}{\sqrt{4\pi Dt}} + \frac{r\sqrt{t}}{\sqrt{4\pi D}}\int_0^1 \frac{dw}{\sqrt{w}}\, e^{-t\left[rw + \frac{x^2}{4Dwt^2}\right]}. \qquad (3.1.7)$$

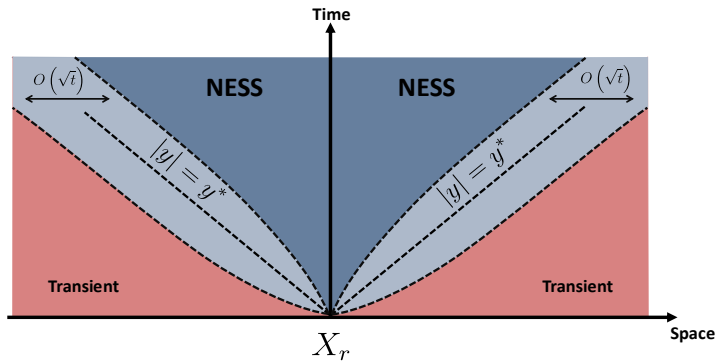

**Figure 3.2:** In the large-time limit, the rate function of the position distribution of an rBM (see Eq. (3.1.9)) exhibits a second-order dynamical transition. This transition separates trajectories that have already reached the non-equilibrium steady state (NESS) from those that remain transient due to insufficient resetting. The two different regimes are separated by the line $x = y^*t$, where $y^* = \sqrt{4Dr}$. A smooth crossover in the region $x \pm y^*t = O(\sqrt{t})$ connects these two regimes, as described by Eq. (3.1.14).

By a saddle-point analysis of the integral in the right-hand side of Eq. (3.1.7), it can be shown that $p_r(x, t)$ follows a *large deviation principle* [38], meaning that it asymptotically takes the form

$$p_r(y = x/t, t) \underset{t \to +\infty}{\approx} A(y, t)\, e^{-tI(y)}, \tag{3.1.8}$$

where $I(y)$ is the *rate function*, given by

$$I(y) = \begin{cases} \sqrt{\frac{r}{D}}|y| & , \quad \text{for } |y| < y^*, \\ r + \frac{y^2}{4D} & , \quad \text{for } |y| > y^*, \end{cases} \tag{3.1.9}$$

with $y^* = \sqrt{4Dr}$. For a detailed derivation, we refer to [35, 38], as well as my first publication [6], where we performed a similar computation. One can observe that both $I(y)$ and its first derivative are continuous at $y^*$, while $I''(y)$ is discontinuous. This discontinuity signals a second-order dynamical transition. The physical interpretation of this result is as follows: the first line in Eq. (3.1.9) describes typical trajectories that have already reached the steady state. In contrast, the second line has a clear physical interpretation: for $|y| > y^*$, we find

$$p_r(x, t) \underset{t \to +\infty}{\approx} e^{-rt} e^{-\frac{x^2}{4Dt}}, \tag{3.1.10}$$

which corresponds to atypical trajectories that have experienced no reset and contribute to the tails at large values of $x$. These trajectories occur with probability $e^{-rt}$ and propagate freely. Thus, the scaling $x \sim O(t)$ acts as a separatrix between typical and atypical trajectories – see Fig. 3.2. This relaxation mechanism toward the stationary state is not exclusive to the rBM but is also observed in generic stochastic resetting processes [38].

The amplitude $A(y, t)$ in (3.1.8) can also be computed and is given by

$$A(y, t) = \begin{cases} \frac{1}{2}\sqrt{\frac{r}{D}} & , \quad \text{for } |y| < y^*, \\ \frac{y^2}{y^2 - 4Dr} \frac{1}{\sqrt{4\pi Dt}} & , \quad \text{for } |y| > y^*. \end{cases} \tag{3.1.11}$$

We see from (3.1.11) that $A(y, t)$ is diverging as $y \to y^*$ from above. Of course, at finite time $t$ these singularities are smoothened out and it is interesting to describe the rounding of these singularities for finite time $t$.

One can also show that there is indeed an interesting crossover around $x = y^*t$ at finite time $t$. The integral on the right-hand side of Eq. (3.1.7) can be evaluated explicitly, yielding

$$p_r(x,t) = \frac{e^{-rt+\frac{x^2}{4Dt}}}{\sqrt{4\pi Dt}} + \frac{1}{4}\sqrt{\frac{r}{D}}e^{-\sqrt{\frac{r}{D}}x}\left[\text{erfc}\left[\frac{x-y^*t}{\sqrt{4Dt}}\right] - e^{2\sqrt{\frac{r}{D}}x}\text{erfc}\left[\frac{x+y^*t}{\sqrt{4Dt}},\right]\right] \quad (3.1.12)$$

where

$$\text{erfc}(z) = \frac{2}{\sqrt{\pi}}\int_z^\infty du\, e^{-u^2}\,. \quad (3.1.13)$$

From the exact formula (3.1.12), we can study the behavior of $p_r(x,t)$ near the critical line $y^*t$ – note that a similar analysis can be performed for the line $-y^*t$. By setting $x = y^*t + z\sqrt{4Dt}$ with $z$ fixed, and examining the large time limit, we obtain

$$p_r(x,t) \approx \frac{1}{4}\sqrt{\frac{r}{D}}e^{-\sqrt{\frac{r}{D}}x}\mathcal{F}\left(\frac{x-y^*t}{\sqrt{4Dt}}\right) \quad,\quad \mathcal{F}(z) = \text{erfc}(z) = \frac{2}{\sqrt{\pi}}\int_z^\infty du\, e^{-u^2}\,. \quad (3.1.14)$$

Using the asymptotic behaviors $\mathcal{F}(z) \to 2$ as $z \to -\infty$ and $\mathcal{F}(z) \sim e^{-z^2}/(\sqrt{\pi}z)$ as $z \to \infty$, one can easily show that this behavior (3.1.14) smoothly matches the behaviors (3.1.8) for $y < y^*$ on the one hand (corresponding to $z \to -\infty$) and $y > y^*$ on the other hand (corresponding to $z \to +\infty$)[2].

Recently, we have observed similar transitions as in Eq. (3.1.9) in the large deviation function of switching diffusion. We will discuss switching diffusion in details in Part III.

### 3.1.3 Survival Probability and Mean First-Passage Time

One of the key properties of stochastic resetting is its ability to optimize first-passage observables, such as the mean first-passage time. For instance, consider a simple Brownian motion starting at $x_0 > 0$ in the presence of an absorbing boundary at the origin. In the absence of resetting, there is a nonzero probability that the particle will diffuse towards $x \to +\infty$, causing the MFPT to diverge. However, introducing stochastic resetting prevents trajectories from diffusing indefinitely by bringing the particle back to the resetting position. For this reason, not only does resetting ensure a finite MFPT, but it also allows for optimization with respect to the resetting rate $r$, making it highly relevant in the context of search processes [40, 176]. To illustrate this, consider the MFPT as a function of $r$. When $r \to 0$, the dynamics reduce to free Brownian motion, and the MFPT diverges. Conversely, when $r \to +\infty$, the particle becomes trapped at the resetting position, leading to another divergence of the MFPT. We will show that for a finite $r$, the MFPT remains finite, implying the existence of an optimal resetting rate $r^*$ that minimizes it.

In the following, we will consider a rBM with initial position $x_0 > 0$ in the presence of an absorbing boundary at the origin.

**Renewal Approach**

As shown for the propagator of a rBM in Eq. (3.1.4), an integral equation for the survival probability can be derived using a renewal argument

$$Q_r(x_0, t) = e^{-rt}Q_0(x_0, t) + r\int_0^t d\tau\, e^{-r\tau}Q_0(X_r, \tau)Q_r(x_0, t-\tau)\,, \quad (3.1.15)$$

where $Q_0(x,t)$ is the survival probability *without resetting*, and $Q_r(x_0,t)$ is the survival probability *with resetting*.

In a time duration $t$, two scenarios are possible:

---
[2] To accurately match the amplitude for $y > y^*$ in Eq. (3.1.11), one must compute the amplitude near the critical line, i.e., $A(y = y^* + z\sqrt{4D/t}, t)$, and take the limit $z \to +\infty$.

- No resetting occurs, meaning the particle diffuses freely for the entire duration $t$, which accounts for the first term in Eq. (3.1.15).

- At least one reset occurs, with the last reset taking place at time $t - \tau$. The probability that the particle resets at $t - \tau$ and does not reset again until $t$ is $r d\tau e^{-r\tau}$. In this case, the particle must survive without resetting for a time $\tau$, contributing the factor $Q_0(X_r, \tau)$ inside the integral, and must have survived for $t - \tau$ under resetting, contributing $Q_r(x_0, t - \tau)$.

To solve Eq. (3.1.15), we observe that the integral contains a convolution in time, which suggests the use of the *Laplace transform*. The Laplace transform of the survival probability is defined as

$$\tilde{Q}_r(x_0, s) = \int_0^{+\infty} dt\, e^{-st} Q_r(x_0, t)\,, \tag{3.1.16}$$

while its inverse is given by the Bromwich integral

$$Q_r(x_0, t) = \frac{1}{2\pi i} \int_{\gamma - i\infty}^{\gamma + i\infty} ds\, e^{st} \tilde{Q}_r(x_0, s)\,, \tag{3.1.17}$$

where $\gamma$ is a real number chosen such that all singularities of $\tilde{Q}_r(x_0, s)$ in the complex $s$-plane lie to the left of the Bromwich contour $(\gamma - i\infty, \gamma + i\infty)$. Applying the Laplace transform to Eq. (3.1.15) leads to

$$\tilde{Q}_r(x_0, s) = \tilde{Q}_0(x_0, r + s) + r \tilde{Q}_0(X_r, r + s) \tilde{Q}_r(x_0, s)\,. \tag{3.1.18}$$

Rearranging, we obtain the final expression

$$\tilde{Q}_r(x_0, s) = \frac{\tilde{Q}_0(x_0, r + s)}{1 - r \tilde{Q}_0(X_r, r + s)}\,. \tag{3.1.19}$$

This result is general and holds for any stochastic process subjected to resetting. For an rBM, using the result

$$\tilde{Q}_0(x_0, s) = \frac{1 - e^{-\sqrt{\frac{s}{D}} x_0}}{s}\,, \tag{3.1.20}$$

we obtain

$$\tilde{Q}_r(x_0, s) = \frac{1 - e^{-\sqrt{\frac{r+s}{D}} x_0}}{s + r e^{-\sqrt{\frac{r+s}{D}} X_r}}\,. \tag{3.1.21}$$

By definition, the MFPT is given by

$$\langle \tau_r(x_0) \rangle = -\int_0^{+\infty} dt\, t \frac{\partial Q_r(x_0, t)}{\partial t} = [-t Q_r(x_0, t)]_0^{+\infty} + \int_0^{+\infty} dt\, Q_r(x_0, t) = \tilde{Q}_r(x_0, s = 0)\,, \tag{3.1.22}$$

where we assume that $Q_r(x_0, t)$ decays faster than $t$ as $t \to +\infty$[3]. Thus, the MFPT reads (for $X_r = 0$)

$$\langle \tau_r(x_0) \rangle = \frac{1}{r} \left( e^{\sqrt{\frac{r}{D}} x_0} - 1 \right)\,. \tag{3.1.23}$$

We show a plot of $\langle \tau_r(x_0) \rangle$ as a function of $r$ in Fig. 3.3. As the resetting rate increases ($r \to \infty$), the exponential term in the MFPT grows without bound, causing the MFPT to diverge. Conversely, when the resetting rate approaches zero, the MFPT behaves as $\langle \tau_r(x_0) \rangle \sim r^{-1/2}$, which also leads to divergence. Therefore, an optimal resetting rate that minimizes the MFPT must exist. To determine this optimal value, we set the derivative of $\langle \tau_r(x_0) \rangle$ with respect to $r$ to zero and solve for $r$

$$\frac{\partial \langle \tau_r(x_0) \rangle}{\partial r} = 0 \,, \tag{3.1.24}$$

which leads to solving the transcendental equation

$$\frac{\gamma}{2} = 1 - e^{-\gamma} \,, \tag{3.1.25}$$

where $\gamma = \sqrt{r/D}\, x_0$. The unique nonzero solution is given by

$$\gamma^* = 1.5936 \cdots \,. \tag{3.1.26}$$

---

[3] It is possible to show that $Q_r(x_0, t)$ decays exponentially fast in the large time limit [177].

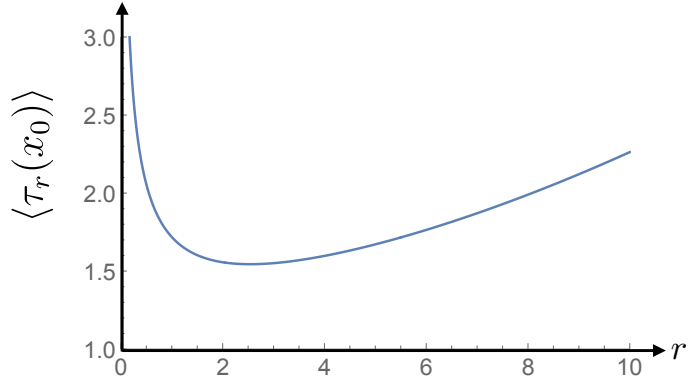

**Figure 3.3:** We show here a plot of the MFPT $\langle \tau_r(x_0) \rangle$ given in Eq. (3.1.23) for $D = 1$ and $x_0 = X_r = 1$. It exhibits a minimum for an optimal resetting rate $r^*$ determined by Eq. (3.1.26). The figure is adapted from [35].

Thus, the minimum MFPT is achieved when the ratio between the initial distance to the target $x_0$ and the mean displacement between resets equals $\gamma^*$. Hence, the optimal MFPT is given by

$$\langle \tau_r^{\text{opt}}(x_0) \rangle = \left[ \frac{e^{\gamma^*} - 1}{\gamma^{*2}} \right] \frac{x_0^2}{D} \approx 1.54414 \cdots \frac{x_0^2}{D} \,. \tag{3.1.27}$$

In Chapter 4, we will introduce the run-and-tumble particle (RTP), a stochastic process driven by telegraphic noise. The particle moves at a constant speed and randomly tumbles, changing direction at times drawn from an exponential distribution with a fixed rate. In the presence of an arbitrary potential, we will derive the exact expression for the MFPT and demonstrate that, similar to rBM, it can be optimized with respect to the tumbling rate for a broad class of potentials – see Chapter 12 and [2, 5].

**Fokker-Planck Approach**

If we consider the evolution of the survival probability from $t = 0$ to $t = dt$, a similar decomposition as in Eq. (3.1.2) leads to the backward Fokker-Planck equation where the initial position $x_0 > 0$ is considered as a variable. It is given by

$$\frac{\partial Q(x_0, t)}{\partial t} = D \frac{\partial^2 Q(x_0, t)}{\partial x_0^2} - r Q(x_0, t) + r Q(X_r, t) \,, \tag{3.1.28}$$

and it can be solved in Laplace space with boundary and initial conditions $Q(0, t) = 0$ and $Q(x_0, 0) = 1$. Deriving Eq. (3.1.28) with respect to time, and multiplying it by $-t$ gives a differential equation for the MFPT

$$-1 = D\frac{\partial^2 \langle \tau_r(x_0)\rangle}{\partial x_0^2} - r\langle \tau_r(x_0)\rangle + r\langle \tau_r(X_r)\rangle\,, \tag{3.1.29}$$

where the boundary conditions are $\langle \tau_r(0)\rangle = 0$, and $\langle \tau_r(x \to +\infty)\rangle$ is finite. Note that in a presence of a force $F(x)$ acting on the particle, we have an additional term

$$-1 = D\frac{\partial^2 \langle \tau_r(x_0)\rangle}{\partial x_0^2} + F(x_0)\frac{\partial \langle \tau_r(x_0)\rangle}{\partial x_0} - r\langle \tau_r(x_0)\rangle + r\langle \tau_r(X_r)\rangle\,. \tag{3.1.30}$$

It would be interesting to solve this equation, for instance, to examine how the incorporation of stochastic resetting affects Kramers' law. If we consider a Brownian particle trapped in a local minimum between two barriers, resetting the particle – say, to the minimum of the well – is unlikely to be beneficial. The reason is that, in the presence of confinement, the Brownian particle is already confined and will not escape to infinity as a free Brownian motion would. However, more sophisticated resetting scheme could be beneficial [178].

### 3.1.4 Optimal Diffusive Search: Equilibrium vs Non-Equilibrium

Interestingly, in the large time limit, the distribution of the position of the particle $p_r(x, t)$ at time $t$ converges to a stationary distribution given by [35, 54]

$$\lim_{t \to \infty} p_r(x, t) = p_{\text{st}}(x) = \frac{\alpha_0}{2}\exp\left[-\alpha_0|x - x_0|\right]\,, \quad \text{where} \quad \alpha_0 = \sqrt{r/D}\,. \tag{3.1.31}$$

Therefore, although the rBM is a non-equilibrium process, the stationary distribution (3.1.31) can nevertheless be expressed as an effective Boltzmann weight $p_{\text{st}}(x) \propto \exp\left[-U_{\text{eff}}(x)\right]$ with the effective potential

$$U_{\text{eff}}(x) = \alpha_0 |x - x_0|\,. \tag{3.1.32}$$

Hence, if we consider the following equilibrium Langevin dynamics

$$\dot{x}(t) = -D\partial_x U_{\text{eff}}(x) + \eta(t)\,, \tag{3.1.33}$$

where $\eta(t)$ is a Gaussian white noise with zero mean and $\langle \eta(t)\eta(t')\rangle = \delta(t-t')$, then the stationary state of Eq. (3.1.33) is characterized by the same stationary distribution $p_{\text{st}}(x) \propto \exp[-U_{\text{eff}}(x)]$ – although of course the two dynamics (3.1.1) and (3.1.33) are quite different.

In Ref. [179], the authors asked the following question: since the non-equilibrium resetting dynamics (3.1.1) and the equilibrium Langevin process (3.1.33) lead to the same stationary state, can one compare the MFPT to the origin of these two processes?

It is possible to compute the MFPT to the origin of Eq. (3.1.33) and it is given by [179]

$$\langle T_r(x_0)\rangle = \left[\frac{2(e^\gamma - 1) - \gamma)}{\gamma^2}\right]\frac{x_0^2}{D}\,, \tag{3.1.34}$$

where we recall that $\gamma = \alpha_0 x_0 = \sqrt{r/D}\,x_0$. We can now compare both Eq. (3.1.27) and Eq. (3.1.34) using that $e^\gamma - 1 \geq \gamma$. One sees that we have $\langle T_r(x_0)\rangle \geq \langle \tau_r(x_0)\rangle$ for fixed values of $x_0$ and $D$. In particular, it is possible to compute the optimal value of $\langle T_r(x_0)\rangle$ and show that

$$\frac{\langle \tau_r^{\text{opt}}(x_0)\rangle}{\langle T_r^{\text{opt}}(x_0)\rangle} < 1\,. \tag{3.1.35}$$

Therefore, the optimal "nonequilibrium" MFPT (i.e., for the rBM) is always smaller than the "equilibrium" MFPT evaluated at its optimal value. Loosely speaking, "nonequilibrium offers a better search strategy than equilibrium" [179].

In [2], we address the efficiency of the search strategy offered by a free RTP subjected to resetting compared to a confined RTP, following the same approach as in Ref. [179]. In this study, we demonstrate that the resetting search procedure is again more efficient, as indicated by its smaller MFPT – see Chapter 13.

## 3.2 Gas of Resetting Particles

Although my research has focused on a single-particle system, it is also interesting to study many-particle systems, or "gases of particles". A system composed of $N$ non-interacting particles can be interpreted in two ways: as a gas of $N$ particles or as the $N$-dimensional counterpart of a single-particle system, where each particle's position corresponds to a coordinate in the many-dimensional representation of the system. For such systems, several observables are of interest, including the gas density, the gap distribution between particles, and the rightmost (or leftmost) particle in the context of extreme value statistics [61]. It is also useful to compute the full counting statistics (i.e., the number of particles within a given interval) of the gas for quantifying the level of order within the gas, as exemplified by the concept of hyperuniformity [180].

In [181, 182], the authors recently studied a Brownian gas that is correlated by stochastic resetting. The system consists of $N$ independent Brownian particles, all of which are reset simultaneously to the origin – see the left panel of Fig. 3.4. This simultaneous resetting introduces strong correlations within the system, which persist even in the non-equilibrium steady state (NESS). Another advantage of studying a gas of particles subjected to resetting lies in the analytical framework it offers. Specifically, the gas exhibits a solvable structure, which stems from the fact that the positions of the particles are conditionally independent and identically distributed (c.i.i.d.) variables [182, 183]. To be more specific, one can write the joint probability distribution function (JPDF) of the positions of the $N$ particles using a renewal argument. All the particles start their motion at the origin, and we want to compute the JPDF of the positions of the particles $\{x_i\}$ at time $t$. With a probability $e^{-rt}$ there will be no resetting in $[0, t]$ – in that case, the joint distribution at time $t$ will be simply $P_0[\{x_i\}, t] e^{-rt}$ where $P_0[\{x_i\}, t]$ is the JPDF of the positions of the particles when there is no reset, i.e. $r = 0$. For $r = 0$, the particles evolve as $N$ independent Brownian motions and their joint distribution just becomes a product of $N$ independent Gaussians, given by

$$P_0[\{x_i\}, t] = \prod_{i=1}^{N} \frac{1}{\sqrt{4\pi D t}} e^{-\frac{x_i^2}{4Dt}} . \qquad (3.2.1)$$

When there is at least one resetting event in $[0, t]$, we remark that the state of the system at time $t$ depends only on the time elapsed since the last resetting before $t$. This is because every resetting event brings back all the particles to the origin, and hence we only need to keep track of the time since the last resetting. This idea is illustrated in the left panel of Fig. 3.4 where $t$ is the observation time and $t - \tau$ is the time at which the last resetting occurs. The argument is then the same than for the simple case of rBM given in Eq. (3.1.4). Since the evolution between $t - \tau$ and $t$ is free (i.e., without resetting), clearly the joint distribution of the positions at time $t$ is simply $P_0[\{x_i\}, \tau]$. However, $\tau$ itself is a random variable, with a probability density $r\, e^{-r\tau}$ and $\tau$ can vary from 0 to $t$. Hence we need to multiply $P_0[\{x_i\}, \tau]$ by $r\, e^{-r\tau}\, d\tau$ and integrate $\tau$ from 0 to $t$. Adding these two contributions, i.e., no-resetting event and the multiple resettings, we get the joint distribution at time $t$ as

$$P_r[\{x_i\}, t] = e^{-rt} P_0[\{x_i\}, t] + r \int_0^t d\tau e^{-r\tau} P_0[\{x_i\}, \tau] . \qquad (3.2.2)$$

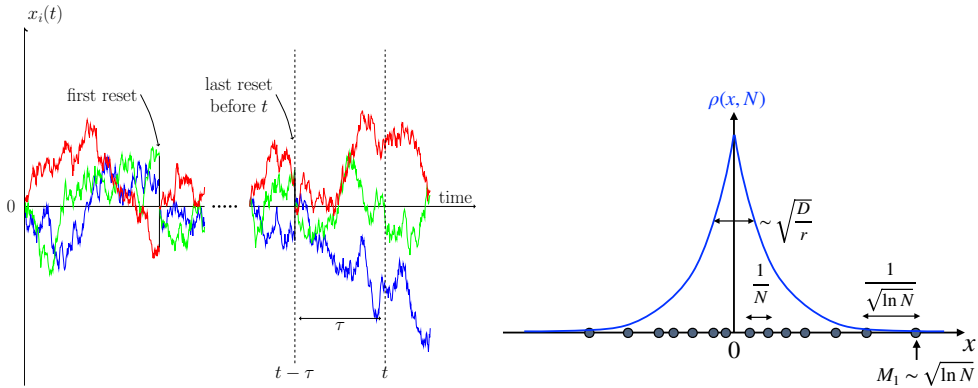

**Figure 3.4: Left panel.** Schematic trajectories of $N = 3$ Brownian motions undergoing simultaneous resetting to the origin at random times. The observation time is marked by $t$ and the time of the last reset before $t$ is marked by $t - \tau$. During the last period $\tau$, the particles evolve independently as free Brownian motions. **Right panel.** The solid blue line shows the average density $\rho(x, N) = \sqrt{\frac{r}{4D}} e^{-\sqrt{r/D}|x|}$. The positions of the particles in a typical sample are shown schematically on the line with most particles living over a distance $\sqrt{D/r}$ around the origin. The typical spacing in the bulk $\sim 1/N$, while it is of order $\sim 1/\sqrt{\ln N}$ near the extreme edges of the sample. The typical position of the rightmost particle $M_1 \sim \sqrt{\ln N}$ for large $N$. These figures were taken from [182].

In the stationary state, only the second term remains, and the joint law of the positions of the $N$ particles is then given by

$$P_{\text{stat}}[\{x_i\}] = \int_0^\infty d\tau\, p(\tau) \prod_{i=1}^N \frac{1}{\sqrt{4\pi D\tau}} e^{-\frac{x_i^2}{4D\tau}}, \qquad (3.2.3)$$

where $p(\tau) = r\, e^{-r\tau}$ is the distribution of the resetting times. In Eq. (3.2.3), the integrand inside the integral has a factorized form, however, since they are conditioned on sharing the same resetting time, they are c.i.i.d., as indicated by the average over $p(\tau)$. It is also possible to derive the average density of the gas in the stationary state, i.e. the average fraction of particles in the interval $[x, x + dx]$ denoted by $\rho(x, N) = 1/N \langle \sum_{i=1}^N \delta(x - x_i) \rangle$, which is also the one-point function. A straightforward computation allows showing that the average density corresponds to the position distribution of a single resetting Brownian motion – a double-exponential distribution given in Eq. (3.1.6) (which does not mean that the particles are uncorrelated as seen in Eq. (3.2.3)). We show a schematic description of the average density in the right panel of Fig. 3.4.

Such a gas composed of $N$ particles is particularly interesting, even in one dimension, because its properties can be computed explicitly. In fact, there are very few strongly correlated systems that we can study analytically. One notable example is the Dyson's log-gas [184], which has deep connections with random matrix theory. In this model, particles interact through a pairwise logarithmic potential while experiencing independent white noise. When subjected to a quadratic external potential, the system corresponds to the well-known Gaussian $\beta$-ensemble [185, 186]. Indeed, in this case, in the stationary state, the particle positions are in one-to-one correspondence the eigenvalues of a Gaussian random matrix. It is worth noting that, unlike the Dyson log-gas, the average density of the resetting gas, $\rho(x, N)$, is independent of $N$ and extends over the entire real line. In contrast, the Dyson log-gas density is confined to a finite interval, following the famous Wigner semi-circle law [185, 186]. It is worth noting that a very recent study introduced the "resetting Dyson Brownian motion" [187], in which the gas evolves according to the standard Dyson log-gas dynamics, but all particles are simultaneously reset to their initial position at a constant rate $r$.

Such analysis is also relevant for switching processes, as explored in [183], where the authors investigated a one-dimensional gas of $N$ noninteracting diffusing particles in a harmonic trap. In their study, the trap's stiffness alternates between two values, $\mu_1$ and $\mu_2$, with constant rates $r_1$ and $r_2$, respectively. The positions of the particles remain c.i.i.d., enabling a thorough analytical study. Furthermore, it can be shown that the position of a particle at the moment of a switch is a Kesten variable [60], meaning it satisfies a specific recursive relation of the form $x_n = U_n x_{n-1} + V_n$, where $U_n$ and $V_n$ are (possibly correlated) random variables. This Kesten approach, introduced in my first paper [6], enables the derivation of an integral equation for the steady-state distribution of the particle positions [183].

## 3.3 Connection With My Work

The tools developed for the study of stochastic resetting can be extended to a broader class of stochastic processes known as *switching processes*. Consider a process that evolves deterministically from a starting position but undergoes random resets to a new position drawn from a given probability distribution. This framework includes, for instance, the telegraphic noise $\sigma(t)$, where $\sigma(t) = \pm 1$ and switches at exponentially distributed times. In [6], we analyzed the steady state of a particle evolving in a harmonic potential under the influence of such switching processes. This idea also motivated our study of switching diffusion [1], in which the diffusion coefficient of a Brownian motion randomly switches values at exponentially distributed times (see Section 2.4 and Part III). Regarding the study of optimal search strategies, we were also interested in exploring them in the context of run-and-tumble particles (i.e., particles subjected to a telegraphic noise). We have addressed these questions in [2, 5] where we showed that the MFPT of a confined RTP can indeed be optimized. Furthermore, in [3], we have extended Siegmund duality (discused before in Section 1.5.4) to processes with stochastic resetting. This extension shows that Eq. (1.5.29), which links spatial properties to first-passage properties in Brownian motion, also applies to rBM (under some technical assumptions) [3]. We will discuss it in Chapter 16.

# Chapter 4

# The Run-and-Tumble Particle: a toy model of active particles

Active particle systems are an exciting and rapidly evolving field of research that offers a unique opportunity to study the emergent collective behavior of out-of-equilibrium systems [46–52]. They encompass a wide variety of physical situations, ranging from micro-organisms like bacteria [188, 189] or synthetic Janus particles [190, 191] all the way to flocks of birds [192, 193] and fish schools [194, 195]. These systems are composed of self-driven particles that can convert energy into directed motion, leading to a wide range of phenomena, including clustering and jamming [196–200], motility induced phase separation [201–204], and absence of well-defined pressure [205]. However, these many-body phenomena are still difficult to describe analytically starting from microscopic models, for which very few exact results have been obtained (see however [53, 198, 206–208]). A comprehensive classification of active systems can be found in Fig. 4.1.

Hence, several recent works focused on the study of the dynamics of a single or of a few active particles. Indeed, it was realized that active systems exhibit many intriguing features even at the single particle level. In particular, it was shown that, in the presence of external confining potentials, active particles behave very differently from their passive counterparts, exhibiting non-Boltzmann stationary states, clustering near the boundaries of the confining region [198, 209–220] and unusual dynamical and first-passage properties [221–226]. Paradigmatic models in this context include for instance the active Brownian particle (ABP), see e.g. [215], the active Ornstein-Uhlenbeck particle (AOUP), see e.g. [227, 228], or the run-and-tumble particle (RTP), see e.g. [53]. In all these cases, the stochastic dynamics of the active particle is *non-Markovian* and driven by a correlated noise [229] – at variance with a passive particle which is driven by a white noise. For such non-Markovian dynamics, characterizing the interplay between a confining potential and a colored/correlated noise yields challenging questions, such as: What is the nature of the stationary state? How does the system reach the stationary state?

Another important aspect of studying active particles is the characterization of their first passage properties [43, 63, 69]. For instance, understanding how small organisms navigate toward targets like food, or how sperm seek an egg cell presents a substantial challenge [40, 41, 230]. Extracting these properties for active particles is often complicated, primarily due to the persistence of their motion stemming from the colored nature of their stochastic noise – being correlated in time, unlike white noise [229].

This thesis focuses on the study of RTPs. While the spatial properties of confined RTPs have been well characterized recently, exact analytical results for their first-passage properties remain relatively sparse. Important observables for characterizing these first-passage properties include the survival probabilityand the exit probability, which encodes all the information of the first-passage time distribution [43, 63, 69]. For a free RTP, these quantities have been computed in one [222, 226, 230–233] and higher [225, 234–236] dimensions. Recent studies have focused on the addition of partially absorbing boundaries [237–240]. Further works have also considered the case of RTPs in the presence of external potentials [4, 5, 216, 241–243].

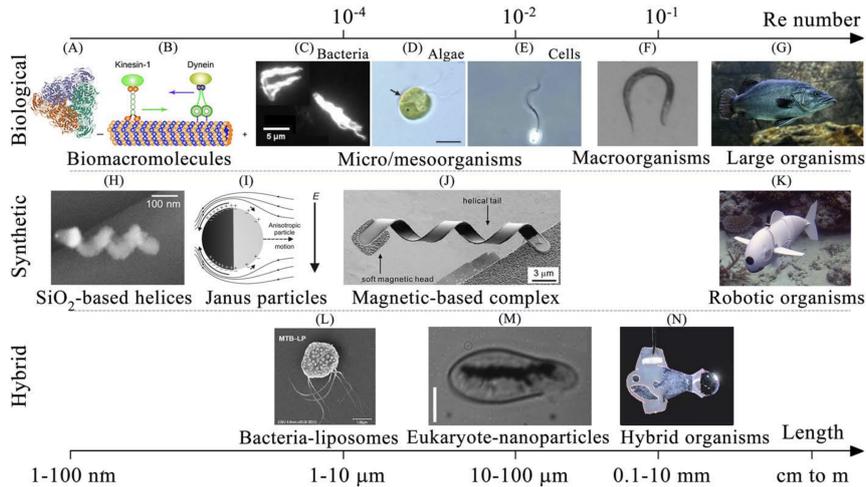

**Figure 4.1:** Classification of active systems across biological, synthetic, and hybrid domains, organized by their characteristic length scales and Reynolds numbers (Re). **Top row (Biological)**: Examples range from biomacromolecules (A–B), to microorganisms such as bacteria and algae (C–D), cells and macroorganisms (E–F), up to large swimming organisms (G). **Middle row (Synthetic):** Includes engineered active systems such as nanopropellers (H), Janus particles (I), artificial bacterial flagella (J), and robotic swimmers (K). **Bottom row (Hybrid):** Systems combining biological and synthetic components, such as bacteria-liposomes (L), eukaryote-nanoparticle (M), and bio-hybrid fish (N). This figure is reproduced from [52], with permission from the Royal Society of Chemistry (Copyright Clearance Center Order License ID: 1593649-1). For further details, the reader is referred to the references therein [52].

In this chapter, we introduce the main models of active particles, with a particular focus on the RTP. We present its key properties, both in free motion and when subject to an external potential. In Section 4.5, we explain how the work in this thesis relates to active particles and outline its main contributions.

## 4.1 Langevin equation for active particles: some active noises

In the most general context, active particles are systems moving within a thermal bath, thus subjected to white noise $\eta(t)$, and potentially influenced by an external force $F(x,t)$. The key difference from simple Brownian motion is that active particles, much like living organisms, have the capacity to extract energy from their environment to self-propel. To characterize this self-propulsion, we introduce an "active noise" $v(t)$, which stochastically represents the orientation and/or magnitude of this active force. For simplicity, we consider a one-dimensional system, whose dynamics can be described by the Langevin equation

$$m\ddot{x}(t) + \nu\,\dot{x}(t) = F(x,t) + v(t) + \sqrt{2D}\,\eta(t)\,, \qquad (4.1.1)$$

where $m$ is the particle mass, $\nu$ is the viscous drag coefficient, and the noise $\eta(t)$ satisfies $\langle \eta(t)\eta(t')\rangle = \delta(t-t')$. Generally, the active noise $v(t)$ is colored, with its two-time correlation function typically decaying exponentially.

Although studying the inertial motion of active systems is natural and interesting [244–247], this thesis focuses on the overdamped limit, which is more relevant for small active systems (low Reynolds number in Fig. 4.1). We will demonstrate that overdamped motion requires subtle and intricate analysis, leading to rich and complex behaviors. In the overdamped limit, taking $\nu = 1$, the Langevin equation simplifies to

$$\dot{x}(t) = F(x,t) + v(t) + \sqrt{2D}\,\eta(t)\,. \tag{4.1.2}$$

In the following sections, we introduce the three most studied models of active particles.

### 4.1.1 Active Ornstein-Ulhenbeck Particles

The self-propulsion velocity $v(t)$ of Active Ornstein-Uhlenbeck Particles (AOUP) [227,228] evolves according to an Ornstein-Uhlenbeck process (discussed in detail in Section 1.4.4). It is governed by the Langevin equation

$$\nu \frac{dv(t)}{dt} = -v(t) + \sqrt{2D_a}\,\xi(t)\,, \tag{4.1.3}$$

where $\nu$ is the persistence time, and $D_a$ is the diffusion coefficient associated with the active velocity fluctuations (the subscript $a$ stands for "active"). The velocity distribution reaches a steady state at long times, described by the Boltzmann-Gibbs measure

$$p_{\rm eq}(v) = \sqrt{\frac{\nu}{2\pi D_a}}\,e^{-\frac{\nu v^2}{2D_a}}\,. \tag{4.1.4}$$

The two-time correlation function of the active noise can be computed by writing the explicit solution of Eq. (4.1.3) and averaging over the noise as done in Section 1.4.4. It is given explicitly by

$$\langle v(t_1)v(t_2)\rangle = v_0^2 e^{-\frac{(t_1+t_2)}{\nu}} + \frac{D_a}{\nu}\left[e^{-\frac{|t_1-t_2|}{\nu}} - e^{-\frac{(t_1+t_2)}{\nu}}\right]\,. \tag{4.1.5}$$

This model has been used to characterize the dynamics of passive colloidal particles immersed in active bacterial suspensions [248,249], and to describe collective cell movements observed experimentally [250,251]. In Fig. 4.2, we show a schematic trajectory of an AOUP in two dimensions.

When an AOUP moves under the influence of an external potential $U(x)$, with the equation of motion given by $\dot{x}(t) = -U'(x) + v(t)$, obtaining exact analytical solutions for its dynamics becomes challenging. Nevertheless, perturbative methods have been developed to study the distribution of the position of AOUPs [228]. Furthermore, using large-deviation theory, it is possible to study the mean first-passage time (MFPT) across a barrier, extending Kramers' law (see Section 1.5.5). Specifically, for an AOUP trapped near a local minimum at $x_{\min}$, the mean time required to cross a barrier, $\langle \tau \rangle$, where the maximum is located at the origin, follows the large-deviation form in the small-noise limit $D_a \to 0$ [79]

$$\langle \tau \rangle \underset{D_a \to 0}{\asymp} e^{\frac{\Phi_\nu(0)}{D_a}}\,, \tag{4.1.6}$$

where $\Phi_\nu(x)$ represents the quasi-potential. Such large deviation form, captured by the quasi-potential, is a characteristic feature of active particle dynamics [2,79–81,252–255].

In [2], we derived exact analytical expressions for the MFPT of one-dimensional run-and-tumble particles within arbitrary potentials. Using these exact expressions, we were able to find the expression of the quasi-potential and to extend Kramer's law for RTPs. In Chapter 13, we present our explicit results.

### 4.1.2 Active Brownian Particle

Active Brownian particles (ABPs) move at a constant speed $v_0$, with their orientation angle continuously fluctuating [46,199]. More precisely, the orientation angle of their velocity vector follows Brownian motion itself. In two dimensions, the dynamics is described by the following equations

$$\frac{d}{dt}\begin{bmatrix}x(t)\\y(t)\end{bmatrix} = v_0\begin{bmatrix}\cos(\theta(t))\\\sin(\theta(t))\end{bmatrix},\quad \frac{d\theta(t)}{dt} = \sqrt{2D_r}\,\eta(t)\,, \tag{4.1.7}$$

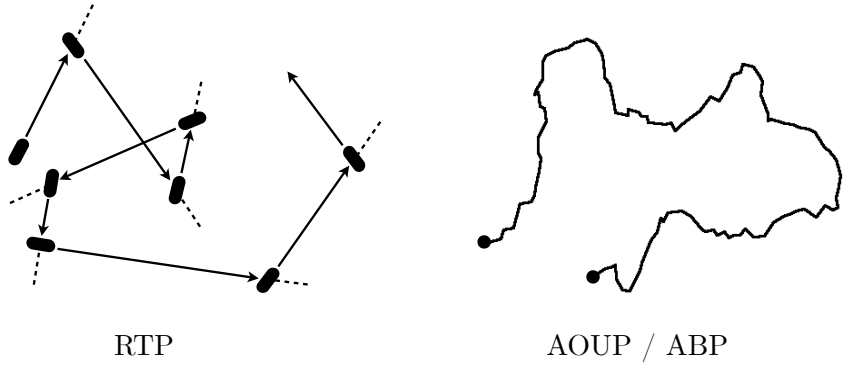

**Figure 4.2: Left.** Schematic trajectory of an RTP in two dimensions. The particle moves ballistically between successive reorientations. **Right.** Schematic trajectory of a two-dimensional AOUP or ABP, showing qualitatively similar behavior. This figure is adapted from Ref. [50].

where $\eta(t)$ represents Gaussian white noise and $D_r$ is the rotational diffusion coefficient. Although deriving analytical results for ABPs remains challenging, several recent studies have successfully computed observables such as the intermediate scattering function [159, 256] – an experimentally measurable quantity – or characterized the distributions of the position [223, 257]. Additionally, a perturbative approach has been developed to calculate the first-passage density of a two-dimensional ABP reaching an absorbing boundary in the presence of Gaussian white noise [258].

In one dimension, the ABP reduces to a simple run-and-tumble particle [259], which we introduce in the next section. However, studying one of the component of a multidimensional ABP system is also of considerable interest, for instance, when a first-passage time problem reduces to a one-dimensional case [223, 258, 260].

### 4.1.3 The Run-and-Tumble Particle and the Telegraphic Noise

In one dimension, the RTP is driven by an active force $v(t) = v_0\,\sigma(t)$, where $v_0$ is a constant velocity and $\sigma(t)$ is a telegraphic noise [261]. This noise alternates between the values $\pm 1$, mimicking the particle's tumbling, i.e., random changes in direction. This model is used to describe the motion of bacteria such as *E. Coli* [262, 263]. Typically, Poissonian tumblings are considered[4], meaning the intervals $\tau$ between successive tumbling events are exponentially distributed with rate $\gamma > 0$, such that $p(\tau) = \gamma\,e^{-\gamma\tau}$. The mean time between two tumbles is called the *persistence time* and is given by $1/\gamma$. The dynamics of the telegraphic noise $\sigma(t)$ can be explicitly described as follows: during the infinitesimal time interval $[t, t+dt]$, the noise evolves according to

$$\sigma(t+dt) = \begin{cases} \sigma(t) & \text{with probability } (1-\gamma\,dt) \\ -\sigma(t) & \text{with probability } \gamma\,dt \end{cases}. \qquad (4.1.8)$$

Initially, the state is chosen as $\sigma(0) = +1$ or $\sigma(0) = -1$, each with equal probability $1/2$. We show in Fig. 4.3 a trajectory of a telegraphic noise. The two-time correlation function of the

---

[4] Traditionally, bacterial run times, such as those observed in *Escherichia coli*, were assumed to follow exponential distributions (e.g., [262, 263]). However, recent experimental measurements suggest alternative distributions may more accurately describe their behavior [264–267]. Typically, these distributions initially increase from zero to a maximum and then decay, often exponentially [268], similar to log-normal distributions [266] or gamma distributions [265, 267].

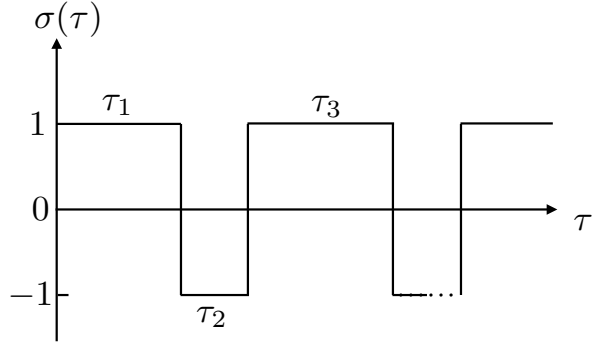

**Figure 4.3:** The plot depicts a trajectory of a telegraphic noise, characterized by changes between the values $\pm 1$ at time intervals $\tau_i$ that follow an exponential distribution $p(\tau_i) = \gamma e^{-\gamma \tau_i}$. This figure is sourced from [216].

active noise is given by

$$\langle v(t_1)v(t_2)\rangle = v_0^2 e^{-2\gamma|t_1-t_2|}, \qquad (4.1.9)$$

where we write $v_0 = v(t=0)$. The RTP in $d$-dimensions is also characterized by alternating phases of ballistic motion and random changes in direction (see Fig. 4.2 for a schematic trajectory for $d = 2$). These directional changes can occur in either a continuous or discrete manner – see [225, 269]. Note also that the RTP model can be further generalized, for instance, by introducing randomness in the velocity of each run, drawing velocities from an arbitrary distribution, or by assigning different transition rates between states.

In my thesis, I investigated the first-passage properties of one-dimensional RTPs moving within an arbitrary potential [2, 4, 5]. A key advantage of RTPs is that their motion is piecewise deterministic. Specifically, between two consecutive tumbles occurring at times $t_1$ and $t_2$, the evolution of the particle is governed by one of the two states $\sigma(t) = \pm 1$ for $t_1 < t < t_2$, following equations of motion

$$\dot{x}(t) = F(x) + v_0, \quad \text{or} \quad \dot{x}(t) = F(x) - v_0. \qquad (4.1.10)$$

This property makes RTPs analytically more tractable compared to AOUPs or ABPs, although deriving exact results for relevant observables remains challenging. As the vector $(x(t), \sigma(t))$ is a Markovian process, it is possible to use tools from renewal theory and Fokker-Planck equations.

We present the main results regarding the dynamics of free RTPs in Section 4.2, and discuss the case of RTPs moving within a potential in Section 4.3. The latter corresponds to the model we studied in [2, 5], where we derived the mean first-passage time for RTPs subjected to an arbitrary force. Recent studies have also investigated interacting RTPs, a topic we discuss in Section 4.4.

## 4.2 Properties of a Free RTP – the Persistent Random Walk

In this section, we introduce the main properties of the free run-and-tumble particle, whose dynamics is described by the equation

$$\dot{x}(t) = v_0 \sigma(t), \qquad (4.2.1)$$

where $\sigma(t)$ is a telegraphic noise defined in Eq. (4.1.8). This model represents the continuous-time analog of the persistent random walk [270–272] on a one-dimensional lattice. The motion of the particle on the lattice follows the simple rule

$$x_n = x_{n-1} + \sigma_n, \qquad (4.2.2)$$

where the steps $\sigma_n = \pm 1$ evolve according to the Markovian dynamics

$$\sigma_n = \begin{cases} \sigma_{n-1} & \text{with probability } p \\ -\sigma_{n-1} & \text{with probability } 1-p \end{cases}. \qquad (4.2.3)$$

The parameter $p \in [0,1]$ controls the "persistence" of the random walk. When $p = 1/2$, we retrieve the well-known symmetric random walk with uncorrelated steps. For $p > \frac{1}{2}$, the steps are correlated positively, leading to persistence in the motion of the walker, while for $p < \frac{1}{2}$ the steps are negatively correlated.

In the following, we consider the continuous-time RTP governed by the Langevin equation (4.2.1). Let us denote by $P(x,t,+)$ and $P(x,t,-)$ the probability densities of an RTP in the $+$ or $-$ state at time $t$, respectively. These densities are normalized such that

$$\int_{-\infty}^{+\infty} dx \left[ P(x,t,+) + P(x,t,-) \right] = 1. \qquad (4.2.4)$$

To derive the corresponding forward Fokker-Planck equations, we first write the evolution equation for $P(x,t,+)$ as

$$P(x, t+dt, +) = (1 - \gamma dt) P(x - v_0 dt, t, +) + \gamma dt\, P(x + v_0 dt, t, -). \qquad (4.2.5)$$

The first term on the right-hand side represents the scenario where the particle does not tumble between times $t$ and $t+dt$, which occurs with probability $(1-\gamma dt)$. In this case, to be at position $x$ at time $t+dt$, the particle must have been at position $x - v_0 dt$ at time $t$. The second term corresponds to the particle undergoing a tumble with probability $\gamma dt$. A similar equation can be written for $P(x,t,-)$. Expanding these equations to first order in $dt$, we obtain the telegraphic equations [222, 270, 273]

$$\partial_t P(x,t,+) = -v_0\, \partial_x P(x,t,+) - \gamma\, P(x,t,+) + \gamma\, P(x,t,-), \qquad (4.2.6)$$
$$\partial_t P(x,t,-) = v_0\, \partial_x P(x,t,-) + \gamma\, P(x,t,+) - \gamma\, P(x,t,-). \qquad (4.2.7)$$

Finally, going in Laplace space with respect to the variable $t$, it is possible to derive the averaged density $P(x,t) = P(x,t,+) + P(x,t,-)$ [222, 270, 273], which gives the probability of finding an RTP (irrespective of its initial state) within the interval $[x, x+dx]$ at time $t$

$$P(x,t) = \frac{e^{-\gamma t}}{2} \Bigg[ \delta(x - v_0 t) + \delta(x + v_0 t)$$
$$+ \left( \frac{\gamma}{2v_0} I_0\left( \sqrt{\gamma^2 t^2 - \frac{\gamma^2 x^2}{v_0^2}} \right) + \frac{\gamma t}{2\sqrt{v_0^2 t^2 - x^2}} I_1\left( \sqrt{\gamma^2 t^2 - \frac{\gamma^2 x^2}{v_0^2}} \right) \right) \Theta(v_0 t - |x|) \Bigg], \qquad (4.2.8)$$

where $I_n$ denotes the modified Bessel function of the first kind of order $n$, and $\Theta$ is the Heaviside step function. The presence of the two delta functions at $\pm v_0$ is due to particles that have not experienced any tumble during the entire time interval $t$. Consequently, there is a finite probability mass proportional to $e^{-\gamma t}$ (the probability of experiencing no tumble for a duration $t$) located exactly at positions $\pm v_0 t$. Figure 4.4 shows a plot of the propagator given by Eq. (4.2.8) at different times. In the limit $t \to +\infty$ and $x \to +\infty$, with the ratio $x/\sqrt{t}$ held fixed, the distribution approaches a Gaussian form characteristic of diffusive behavior [222]

$$P(x,t) \underset{t \to +\infty}{\approx} \frac{e^{-\frac{x^2}{4 D_{\text{eff}} t}}}{\sqrt{4\pi D_{\text{eff}} t}}, \qquad (4.2.9)$$

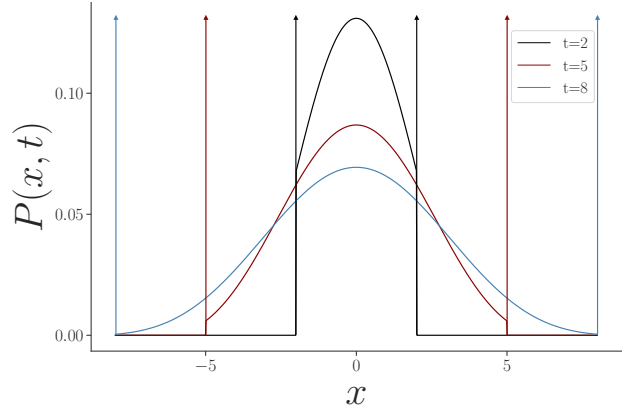

**Figure 4.4:** We present here a plot of the propagator $P(x,t)$ for the free RTP, as given by Eq. (4.2.8). The arrows indicate the delta-function masses located at positions $\pm v_0 t$, corresponding to particles that have experienced no tumbles up to time $t$. At long times, the distribution progressively approaches a Gaussian shape.

where we introduce the effective diffusion coefficient $D_{\text{eff}} = \frac{v_0^2}{2\gamma}$. In contrast, when the ratio $z = x/t$ is held fixed, the distribution satisfies a large deviation form given by (when $\gamma = 1$ and $v_0 = 1$) [274, 275]

$$P(x,t) \underset{t\to +\infty}{\approx} \exp\left[-t I\left(z = \frac{x}{t}\right)\right] \quad , \quad I(z) = \frac{1 - \sqrt{1-z^2}}{2}. \qquad (4.2.10)$$

The coexistence of a Gaussian profile on the scale $x \sim \sqrt{t}$ and a large deviation principle on the scale $x \sim t$ is also a feature of switching diffusion, which will be discussed in Part III.

At fixed time $t$, the RTP has a diffusive limit (meaning that $x(t)$ represents the position of a Brownian particle) when $D_{\text{eff}}$ is kept fixed while $\gamma \to +\infty$ and $v_0 \to +\infty$. In this case, it is possible to show that the correlations of the telegraphic noise become delta correlated [216]

$$\langle v(t_1) v(t_2) \rangle = v_0^2 e^{-2\gamma |t_1 - t_2|} \approx 2 D_{\text{eff}} \delta(t_1 - t_2). \qquad (4.2.11)$$

where $v(t) = v_0 \sigma(t)$.

## 4.3 Properties of a Confined RTP

In this section, we will focus on the case where the run-and-tumble particle is moving within a confining potential $V(x)$. When the persistence time $1/\gamma$ is finite, we will see that the RTP reaches a NESS which we can well characterize. The Langevin equation now reads

$$\dot{x}(t) = F(x) + v_0 \sigma(t), \qquad (4.3.1)$$

where $F(x) = -V'(x)$. Using a similar argument to the one in Eq. (4.2.5), one can derive the forward Fokker-Planck equations for the distributions $P(x, t, \pm)$ in the presence of the force and they are given by

$$\frac{\partial P(x,t,+)}{\partial t} = -\frac{\partial}{\partial x}\left[(F(x) + v_0) P(x,t,+)\right] - \gamma P(x,t,+) + \gamma P(x,t,-), \qquad (4.3.2)$$

$$\frac{\partial P(x,t,-)}{\partial t} = -\frac{\partial}{\partial x}\left[(F(x) - v_0) P(x,t,-)\right] + \gamma P(x,t,+) - \gamma P(x,t,-). \qquad (4.3.3)$$

In the large time limit, the term on the left-hand side vanishes and we obtain

$$\frac{d}{dx}\left[(F(x) + v_0)P(x,+)\right] + \gamma P(x,+) - \gamma P(x,-) = 0\,, \tag{4.3.4}$$

$$\frac{d}{dx}\left[(F(x) - v_0)P(x,-)\right] - \gamma P(x,+) + \gamma P(x,-) = 0. \tag{4.3.5}$$

Now, consider a confining potential of the form $V(x) = \alpha |x|^p$, with parameters $\alpha > 0$ and $p > 0$. In this case, the dynamics given by Eq. (4.3.1) exhibit two turning points, $x_\pm$, defined by $F(x_-) = v_0$ and $F(x_+) = -v_0$. For positions greater than $x_+$ (resp. less than $x_-$), the particle's velocity is always directed toward the origin. Therefore, at large times, the particle becomes confined within the region $[x_-, x_+]$, as illustrated in the left panel of Fig. 4.5. Within the interval $x_- < x < x_+$, the non-equilibrium steady state can be explicitly derived (see Ref. [216] and references therein), and it is given by

$$P(x) = \frac{A}{v_0^2 - F(x)^2} \exp\left(2\gamma \int_0^x dy \, \frac{F(y)}{v_0^2 - F(y)^2}\right), \tag{4.3.6}$$

where $A$ is a normalization constant. In the diffusive limit, i.e., when $D_{\text{eff}}$ is fixed, $\gamma \to +\infty$ and $v_0 \to +\infty$, it is easy to show (see Eq. (4.3.6)) that the stationary state $P(x) \propto e^{-V(x)/D_{\text{eff}}}$ is the Boltzmann-Gibbs measure.

In the right panel of Fig.(4.5), we plot Eq.(4.3.6) for different values of $\gamma$ in the case of the quadratic potential $V(x) = \mu x^2/2$. The expression of the steady state reads

$$P(x) = \frac{2}{4^{\gamma/\mu} B(\gamma/\mu, \gamma/\mu)} \frac{\mu}{v_0} \left[1 - \left(\frac{\mu x}{v_0}\right)^2\right]^{\frac{\gamma}{\mu}-1}, \tag{4.3.7}$$

where $B(\alpha, \beta)$ is the beta function. Interestingly, there is a shape transition of the PDF at the value $\gamma_c = \mu$. For $\gamma > \gamma_c$, the distribution exhibits a U-shaped profile, characteristic of a qualitatively "passive" phase. However, when $\gamma < \gamma_c$, the persistence time is large, causing RTPs to remain longer in their current state. This increased persistence results in higher probabilities at the turning points $x_\pm$, leading to divergences at these points. Such a strongly non-Boltzmann distribution characterizes an "active" phase. When $\gamma = \gamma_c$ the distribution is uniform.

### 4.3.1 Derivation of the backward Fokker-Planck equations for the survival probabilities $Q^\pm(x,t)$

We denote by $Q^+(x,t)$ and $Q^-(x,t)$ the survival probabilities of RTPs with respectively $\sigma(0) = +1$, or $\sigma(0) = -1$. It is the probability that the particle stayed in the positive region up to time $t$ having started its motion at position $x(0)$. Let us derive the backward Fokker-Planck equation for $Q^+$, and $Q^-$. For this purpose, one can first write the discretized version of Eq. (4.3.1). If at time $t$, the particle is in state $\sigma(t) = +1$, then one has

$$x(t + dt) = \begin{cases} x(t) + [F(x) + v_0]\, dt & , \text{ with } \text{ probability } 1 - \gamma\, dt, \text{ and } \sigma(t + dt) = +1 \\ x(t) & , \text{ with } \text{ probability } \gamma\, dt, \text{ and } \sigma(t + dt) = -1\,. \end{cases} \tag{4.3.8}$$

On the opposite case, if at time $t$, the particle is in state $\sigma(t) = -1$, one has

$$x(t + dt) = \begin{cases} x(t) + [F(x) - v_0]\, dt & , \text{ with } \text{ probability } 1 - \gamma\, dt, \text{ and } \sigma(t + dt) = -1 \\ x(t) & , \text{ with } \text{ probability } \gamma\, dt, \text{ and } \sigma(t + dt) = +1\,. \end{cases} \tag{4.3.9}$$

Now suppose that the RTP starts its motion at time $t = 0$ at position $x(0) = x_0$. We want to write the probability $Q^\pm(x, t + dt)$ that it survives for a duration $t + dt$. If the initial state is $\sigma(0) = +1$, we have two possibilities as explained in Fig. 4.6. First, with probability $(1 - \gamma \, dt)$, there is no change of state such that $\sigma(dt) = +1$. From time $t = 0$ to time $t = dt$, the particle has speed $\dot{x} = F(x) + v_0$ so that at time $dt$, its position is $x_0 + [F(x) + v_0] \, dt$. On the other hand, with probability $\gamma \, dt$, the particle tumbles and $\sigma(dt) = -1$ while its new position is $x_0 + [F(x) - v_0] \, dt$. In both cases, after a duration $dt$, the particle still needs to survive for a time $t$. Hence, we have

$$Q^+(x_0, t + dt) = (1 - \gamma \, dt) \, Q^+(x_0 + [F(x_0) + v_0] \, dt, t) + \gamma \, dt \, Q^-(x_0 + [F(x) - v_0] \, dt, t). \quad (4.3.10)$$

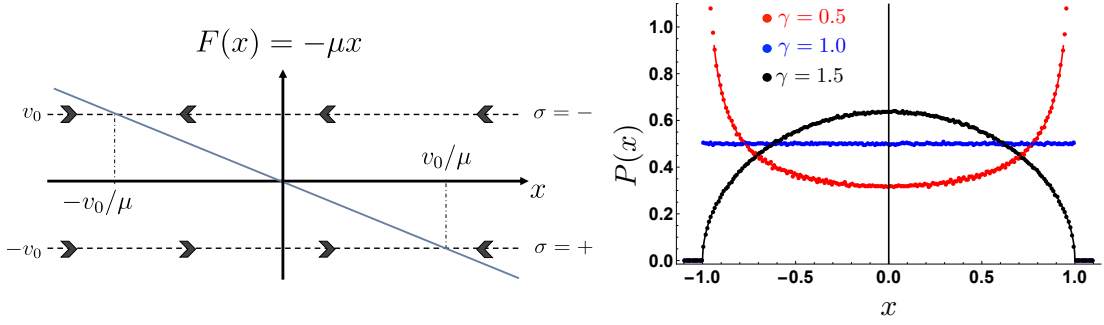

**Figure 4.5: Left.** Illustration of the confinement of the particle within the finite box $[x_-, x_+]$, where $F(x_\pm) = \mp v_0$ for the harmonic potential $V(x) = \mu x^2/2$. In this case, the two turning points are $x_\pm = \pm v_0/\mu$. The arrows represent the direction of the drift felt by the particle in state $\sigma(t) = \pm 1$ – see Eq. (4.3.1). **Right.** Numerical verification of the theoretical result of the steady state distribution in (4.3.7) for the harmonic potential $V(x) = \alpha x^2 = \mu x^2/2$, setting $v_0 = 1$ and $\mu = 1$, i.e. $\alpha = 1/2$. In this case, the shape-transition can be accessed by varying the parameter $\gamma$ (as discussed in the text), for fixed $\alpha = 1/2$. The critical value of $\gamma$ is $\gamma_c = 2\alpha = 1$. For $\gamma < \gamma_c = 1$, the theory predicts an active-like phase where $P(x)$ is U-shaped, while for $\gamma > \gamma_c = 1$, a passive-like phase with a bell-shaped $P(x)$ (more appropriately an inverted U-shaped $P(x)$). Exactly at the critical point $\gamma = \gamma_c = 1$, one would obtain a flat distribution $P(x)$ as predicted in (4.3.7) with $\phi = 0$. We performed simulations for three different values of $\gamma = 1/2$ (red), 1 (blue) and 3/2 (black) to probe these three different cases. The figure on the right panel is taken from [216]

One can Taylor expand the first term at first order in $dt$, which gives

$$Q^+(x_0 + [F(x_0) + v_0] \, dt, t) = Q^+(x_0, t) + [F(x_0) + v_0] \, dt \, \partial_{x_0} Q^+(x_0, t). \quad (4.3.11)$$

Injecting it back in (4.3.10), and keeping only terms of order $dt$, we obtain

$$Q^+(x_0, t+dt) = Q^+(x_0, t) + [F(x_0) + v_0] \, dt \, \partial_{x_0} Q^+(x_0, t) - \gamma \, dt \, Q^+(x_0, t) + \gamma \, dt \, Q^-(x_0, t). \quad (4.3.12)$$

Finally, taking the limit $dt \to 0$ leads to

$$\partial_t Q^+(x_0, t) = [F(x_0) + v_0] \, \partial_{x_0} Q^+(x_0, t) - \gamma \, Q^+(x_0, t) + \gamma \, Q^-(x_0, t). \quad (4.3.13)$$

An analogous reasoning gives us the equation for $Q^-(x_0, t)$, namely

$$\partial_t Q^-(x_0, t) = [F(x_0) - v_0] \, \partial_{x_0} Q^-(x_0, t) - \gamma \, Q^-(x_0, t) + \gamma \, Q^+(x_0, t). \quad (4.3.14)$$

These coupled backward Fokker-Planck equations are difficult to solve, and exact solutions are limited. The probabilities $Q^\pm(x_0, t)$ have been derived explicitly in the Laplace domain (with respect to time) only for special cases, such as a harmonic potential in Ref. [216] and a linear potential in Ref. [226]. However, even extracting the mean first-passage time from these expressions remains a challenging task. This difficulty motivates the development of analytical methods to compute first-passage properties of RTPs subjected to external forces which we have developed in [2, 5] for the MFPT.

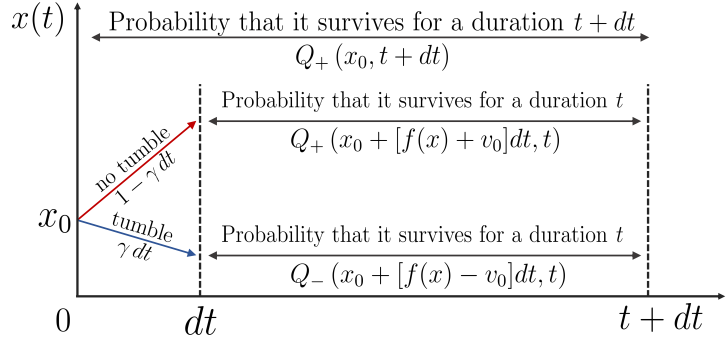

**Figure 4.6:** Illustration for the derivation of the backward Fokker-Plank equation for an RTP starting in state + at position $x_0$. The probability for the RTP to survives for a duration $t + dt$ can be written as follows: between time 0 and time $dt$, the RTP tumbles with probability $\gamma\, dt$ while it stays in the same state with probability $1 - \gamma\, dt$. In both cases, it must still survive for a duration $t$.

## 4.4 Interacting RTP: Some Exactly Solvable Models in Continuous Space

Even though precisely characterizing the dynamics of a single RTP analytically is already challenging, recent developments have extended this analysis to systems of interacting RTPs. For example, in [198], the authors examined a lattice-based model of run-and-tumble particles with hard-core repulsion. In continuous space, various models of interacting RTP gases have been proposed, such as the active Dyson Brownian motion (DBM) introduced in [206, 276–278]. The active DBM describes a gas of $N$ interacting RTPs through a logarithmic interaction potential. It is defined by the evolution equation for the positions $x_i(t)$ of $N$ particles[5]

$$\dot{x}_i(t) \;=\; -\lambda x_i(t) + \frac{2}{N} \sum_{j \neq i} \frac{g\, \delta_{\sigma,\sigma'}}{x_i(t) - x_j(t)} + v_0 \sigma_i(t) + \sqrt{\frac{2T}{N}} \xi_i(t) \;. \tag{4.4.1}$$

Each particle can be in two internal states $\sigma_i(t) = \pm 1$ of velocities respectively $\pm v_0$, and flips its sign with a constant rate $\gamma$. In addition, each particle is submitted to an external potential $V(x) = \frac{\lambda}{2} x^2$ and to a thermal noise at temperature $T/N$, where the $\xi_i(t)$'s are independent standard white noises. The particles in the same state interact via a repulsive pairwise logarithmic potential (i.e. a $1/x$ force) of strength $g > 0$. The simplest observables to study are the densities of each species $\sigma = \pm 1$

$$\rho_\sigma(x,t) = \frac{1}{N} \sum_i \delta_{\sigma_i(t),\sigma} \delta(x - x_i(t)) \;, \tag{4.4.2}$$

as well as the total density $\rho_s(x,t) = \rho_+(x,t) + \rho_-(x,t)$, normalized to unity, and $\rho_d(x,t) = \rho_+(x,t) - \rho_-(x,t)$. A qualitative phase diagram of the densities $\rho_\pm$ is shown in Fig. 4.7.

This model can be studied analytically using the hydrodynamics description provided by the Dean-Kawasaki (DK) equation in the large $N$ limit. This DK approach [279, 280] has been extended to derive evolution equations for the densities $\rho_\pm(x,t)$ defined in (4.4.2) in presence of the active noise. As shown in [206], they take the form at large $N$

$$\partial_t \rho_\sigma(x,t) \;=\; \partial_x \left[ \rho_\sigma(x,t) \left( x - v_0 \sigma - 2g \fint dy \frac{\rho_\sigma(y,t)}{x - y} \right) \right] + \gamma \left( \rho_{-\sigma}(x,t) - \rho_\sigma(x,t) \right) + O(1/\sqrt{N}) \tag{4.4.3}$$

---

[5] The choice to allow only particles in the same state to interact was made to enable particles to cross each other.

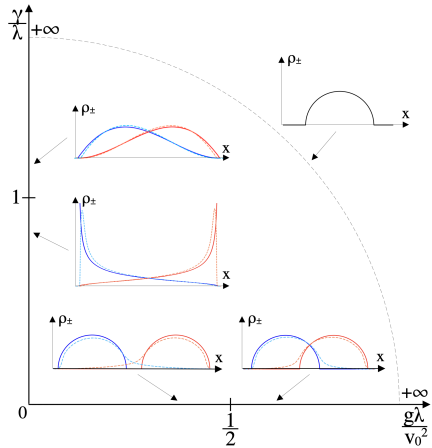

**Figure 4.7:** Shape of the particle density in the plane $(g\lambda/v_0^2, \gamma/\lambda)$ in different limits. The density $\rho_+$ is plotted in red and $\rho_-$ in blue. When the two coincide they are plotted in black. The light dashed curves represent the density slightly away from the limit considered. The dashed circular line in the diagram symbolizes infinity. The diffusive limit, which requires a specific scaling between $v_0$ and $\gamma$, is not shown here. This figure is taken from [206].

for $\sigma = \pm 1$ and where $\fint$ denotes the Cauchy principal value. It is possible to show that this model interpolates naturally between the independent RTPs limit and the usual DBM [206].

A generalization of the active DBM model was explored in [276], where $N$ identical RTPs interact via a repulsive power-law potential of the form $1/|x_i - x_j|^s$, with $s > -1$, on a circle. In the limit $s \to 0$, the active Riesz gas reduces to the active Dyson Brownian motion. Additionally, one strength of these one-dimensional interacting particle models is that their properties can be studied through numerical simulations.

## 4.5 Connection With My Work

In this chapter, we discussed different theoretical models of active particles, focusing mainly on the run-and-tumble particle, which is the central model of active system studied in my thesis. My research aims to better understand RTPs dynamics when influenced by external forces.

In particular, in our works [2, 5], we derived exact analytical expressions for the mean first-passage time of a run-and-tumble particle subjected to an arbitrary force $F(x)$. A key aspect of these studies is the crucial role played by the turning points of the dynamics – positions where the velocity of the RTP vanishes in one of the internal states $\sigma(t) = \pm 1$ (see Fig. 4.5 for an example in the case of a harmonic potential). Depending on the nature of the turning point – whether it occurs at $F(x) = +v_0$ or $F(x) = -v_0$ – and on its stability, determined by the sign of the force gradient $F'(x)$ (i.e., stable for $F'(x) < 0$, unstable for $F'(x) > 0$), we identify distinct phases, each characterized by a different expression for the MFPT. Using these exact expressions, we were able to show that for a confining potential $V(x) = \alpha|x|^p$ ($\alpha > 0$), and for generic $p > 1$, there exists an optimal rate $\gamma_{\text{opt}}$ that minimizes the MFPT, which we characterize precisely [5]. We have also extended Kramer's law for a 1D RTP, i.e., the mean time required for an RTP to cross a potential barrier [2]. These results are presented in detail in Part IV.

Additionally, we explored the steady-state behavior of RTPs confined by harmonic potentials using the approach of Kesten variables [6]. This method allowed us to exactly calculate the stationary distribution of RTPs moving within a harmonic potential for periodic tumbling[6]. We also extended our analysis to a generalized RTP, where each time the particle changes direction, its velocity is chosen randomly from a distribution $W(v)$. This result which has not yet been published, will be described in Chapter 7.2.

---

[6] It also reproduces the already known NESS in the case of Poissonian tumbling – see Eq. (4.3.7).

Finally, another important contribution of this thesis is the duality relations found in active systems. In [4], we established a relation between two apparently different problems: the probability that a particle escapes from an interval through an absorbing boundary before a certain time $t$ (the exit probability), and the cumulative probability distribution of its position at the same time $t$ when instead trapped between two hard walls. This relation, known as the Siegmund duality, was introduced for simple Brownian motion (see Section 1.5.4) and we have extended it to RTPs in [4]. In [3], we showed that this duality is quite general and holds true not only for RTPs but also for other models such as AOUPs and coordinates of ABPs. A detailed presentation of these results is given in Part V.

# Part II

# Resetting Noise in a Harmonic Potential: A Kesten Approach to Steady State Distributions

**Abstract**

We consider a single particle trapped in a one-dimensional harmonic potential, and subjected to a driving noise with memory, that is represented by a resetting stochastic process. The finite memory of this driving noise makes the dynamics of this particle "active". At some chosen times (deterministic or random), the noise is reset to an arbitrary position and restarts its motion. We focus on two resetting protocols: periodic resetting, where the period is deterministic, and Poissonian resetting, where times between resets are exponentially distributed with a rate $r$. Between the different resetting epochs, we can express recursively the position of the particle. The random relation obtained takes a simple Kesten form that can be used to derive an integral equation for the stationary distribution of the position. We begin by illustrating periodic resetting in the cases of Brownian motion, ballistic motion, and telegraphic noise and then analyze in detail the distribution for resetting Brownian motion (with Poissonian resetting), as published in [6]. Finally, we compute the NESS of a run-and-tumble particle with velocities drawn from an arbitrary distribution at each tumble – an unpublished result. Note that in the next chapter, this Kesten approach is also applied to compute explicitly the NESS in the case of switching diffusion.

# Chapter 5

# General Approach

This part of the thesis focuses on the active dynamics of a single particle on a line, whose position is denoted by $x(t)$, in the presence of a harmonic potential $V(x) = \mu\, x^2/2$ and subjected to an active noise $y(t)$ – which is independent of $x(t)$. The overdamped equation of motion thus reads

$$\frac{dx(t)}{dt} = -\mu\, x(t) + y(t)\,, \qquad (5.0.1)$$

starting from $x(0) = 0$ for simplicity. For a passive particle, $y(t)$ is just a white noise $y(t) = \eta(t)$, of zero mean $\langle \eta(t) \rangle = 0$ and with delta correlation $\langle \eta(t)\eta(t') \rangle \propto \delta(t-t')$: in this case $x(t)$ is just the standard Ornstein-Uhlenbeck process. On the other hand, if $y(t)$ is itself a OU process, i.e., it evolves via $\dot{y} = -\gamma\, y + \zeta(t)$ where $\zeta(t)$ is a Gaussian white noise of zero mean, then Eq. (5.0.1) represents the so called active Ornstein-Uhlenbeck process (see e.g. [228]). Another example is the RTP dynamics where $y(t) = v_0\,\sigma(t)$ where $v_0$ is a constant and $\sigma(t)$ is a telegraphic noise that flips between the two values $\sigma(t) = \pm 1$ at a constant rate $\gamma/2$ [216, 261, 270]. In the two latter cases, AOUP and RTP, the correlation of the noise decays exponentially, i.e., $\langle y(t)y(t') \rangle \sim e^{-\gamma|t-t'|}$ for large times $t, t' \gg 1/\gamma$. As $\gamma$ decreases, the dynamics of the particle thus crosses over from a passive behavior, as $\gamma \to \infty$, to a strongly active one as $\gamma \to 0$. In the case $\gamma \to \infty$, the noise $y(t)$ behaves essentially as a white noise with some effective temperature $T^*$ (and consequently $x(t)$ is essentially a passive OU process), while when $\gamma \to 0$, $x(t)$ is "slaved" to the noise, i.e., $x(t) \approx y(t)/\mu$. A central observable is the probability density function $p(x,t)$ of the position at time $t$. In the presence of a harmonic potential (5.0.1), it is natural to expect that $p(x,t)$ converges to a stationary distribution $p(x, t \to \infty) = p(x)$. From the remark above, in the limit $\gamma \to \infty$, one thus expects that $p(x)$ is given by the Gibbs-Boltzmann weight associated to the quadratic potential $V(x) = \mu\, x^2/2$, i.e., a Gaussian form $p(x) \propto e^{-\mu x^2/(2T^*)}$, where $T^*$ is a model-dependent constant that represents an effective temperature. On the other hand, in the limit $\gamma \to 0$, assuming that the noise $y(t)$ admits a stationary distribution $p_{\text{noise}}(y)$, one anticipates that $p(x) \approx (1/\mu)p_{\text{noise}}(x/\mu)$. In the case of the AOUP the stationary distribution of the noise is yet another Gaussian, while for the RTP, $p_{\text{noise}}(y)$ is just a sum of two Dirac delta functions $p_{\text{noise}}(y) = (1/2)\delta(y-v_0) + (1/2)\delta(y+v_0)$.

Computing the full crossover between the passive regime $p(x) \propto e^{-\mu x^2/(2T^*)}$ as $\gamma \to \infty$ and the strongly active one $p(x) \approx (1/\mu)p_{\text{noise}}(x/\mu)$ as $\gamma \to 0$ for arbitrary active noise is of course quite challenging. In this chapter, we describe a rather general method, based on an approach via Kesten variables, that allows us to describe analytically this crossover of the stationary PDF $p(x)$ for a rather wide class of active noises. We call them "resetting noises", since they bear strong similarities with resetting stochastic processes [35, 54, 152, 153, 165, 166, 281]. We model the evolution of $y(t)$ as follows. For simplicity it starts from $y(0) = 0$. It evolves by its own stochastic dynamics, e.g., just a Brownian or a ballistic motion (with random velocity) or even a (random) constant and then gets reset to 0 at random epochs $\{t_1, t_2, t_3, \ldots\}$ – see Fig. 5.1. The successive intervals between the resetting events, $\tau_n = t_n - t_{n-1}$ (for $n = 1, 2, \ldots$ with $t_0 = 0$) are statistically independent and each is drawn from a PDF $p_{\text{int}}(\tau)$ normalized to unity. Clearly, in the case where the "free" evolution between two successive resettings is a Brownian motion, then $y(t)$ is the well known resetting Brownian motion (rBM), which we study in Section 7.1. On the other hand, in the case where the "free" evolution is a random constant, say $\pm v_0$ with equal probability, then $y(t)$ corresponds to the telegraphic noise and, consequently, $x(t)$ in (5.0.1) corresponds to the dynamics of the RTP in the presence of a harmonic potential, which was studied e.g. in [216]. Here we will consider two resetting protocols, namely (i) *Poissonian resetting* where $p_{\text{int}}(\tau) = r\, e^{-r\tau}$ [54] and (ii) *periodic resetting* where $p_{\text{int}}(\tau) = \delta(\tau - T)$ with $T$ being the period [282]. In analogy with the AOUP and the RTP discussed above, one thus has $r \sim \gamma$ in the Poissonian case, while $T \sim 1/\gamma$ for periodic resetting. We will see that in the Poissonian case, $x(t)$ approaches a stationary distribution at long times, while in the periodic case $x(t)$ approaches a 'time-periodic' stationary state with period $T$.

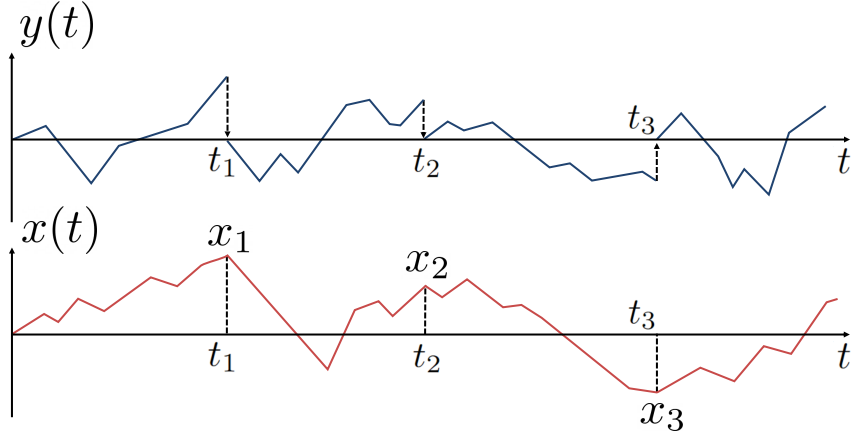

**Figure 5.1:** A schematic realization of the process $x(t)$ and the noise $y(t)$ in Eq. (5.0.1). The noise $y(t)$ is reset at epochs $\{t_1, t_2, t_3, \ldots\}$ at which the process $x(t)$ takes values $\{x_1, x_2, x_3, \ldots\}$. Here, both the process $x(t)$ and the noise $y(t)$ start at the origin.

To study the dynamics in Eq. (5.0.1) driven by an active resetting noise $y(t)$ described above and depicted schematically in Fig. 5.1, it is convenient to introduce $y_n(t)$ which denotes the $y$-process between two successive resetting epochs $t_{n-1}$ and $t_n$. Clearly, $y_n(t)$'s for different $n$'s are statistically independent. Here we present an approach, based on Kesten variables, to study the stationary distribution of the $x$-process. For a given realization of the resetting epochs $\{t_1, t_2, \ldots\}$, let $x_n$ denote the position of the $x$-process evolving via Eq. (5.0.1) at the epoch $t_n$, starting from $x_0 = 0$, see Fig. 5.1. We want to find out the fixed point limiting distribution of $x_n$ as $n \to \infty$. In the Poissonian resetting case, this will coincide with the stationary distribution of the time series $x(t)$. In the periodic case, this will give the limiting distribution of the $x$-process at the end point of a period. Integrating Eq. (5.0.1) from $t_{n-1}$ to $t_n$ we get a random recursion relation

$$x_n = x_{n-1}\, e^{-\mu \tau_n} + e^{-\mu \tau_n} \int_0^{\tau_n} d\tau\, y_n(\tau)\, e^{\mu \tau}. \tag{5.0.2}$$

Note that there are two sources of randomness in this equation. One comes from the realization of the process $y_n(\tau)$ between the two epochs and the second from the randomness of the time interval $\tau_n$ itself. Interestingly, this recursion relation is of the generalised Kesten form [60, 283–289]

$$x_n = U_n\, x_{n-1} + V_n \tag{5.0.3}$$

where $U_n$ and $V_n$ are random variables that may be correlated for a given $n$, but are uncorrelated for different values of $n$. In our case,

$$U_n = e^{-\mu \tau_n}\quad, \quad \text{and}\quad V_n = e^{-\mu \tau_n} \int_0^{\tau_n} d\tau\, y_n(\tau)\, e^{\mu \tau} \tag{5.0.4}$$

are correlated for a given $n$, since the same random variable $\tau_n$ appears in both $U_n$ and $V_n$, but they are uncorrelated for different $n$'s. In general, finding the stationary distribution of the generalised Kesten recursion (5.0.3) is known to be very hard. However, one can make progress in the case where $U_n$ and $V_n$ are jointly distributed according to the joint distribution $P(U,V)$ which is independent of $n$. In this case, using Eq. (5.0.3), one can write down a recursion relation for the position distribution $p(x,n)$ after $n$ "steps", namely

$$p(x,n) = \int dU \int dV \int_{-\infty}^{\infty} dx'\, P(U,V)\, p(x', n-1)\, \delta(x - Ux' - V), \tag{5.0.5}$$

where the integration bounds over $U$ and $V$ depend on the joint distribution $P(U,V)$. Assuming then that $p(x,n)$ approaches a fixed-point, i.e., $p(x,n) \to p(x)$, as $n \to \infty$, it follows from Eq.(5.0.5) that the stationary position distribution satisfies the following integral equation [6]

$$\boxed{p(x) = \int dU \int dV \int_{-\infty}^{\infty} dx'\, P(U,V)\, p(x')\, \delta(x - U\, x' - V)}. \tag{5.0.6}$$

The solution of Eq. (5.0.6) is not known for general $P(U,V)$. For instance, if we use $y_n(t) = B_n(t)$, i.e., a pure Brownian motion (starting from 0) between resettings for the noise, and Poissonian resetting $p_{\text{int}}(\tau) = r\, e^{-r\tau}$, we can easily compute explicitly the joint distribution $P(U,V)$ of $U_n$ and $V_n$. Although it is difficult to solve explicitly the integral equation (5.0.6) in this case, we will see that it is possible to understand the model in details via general methods that can be extended to other resetting noises. Remarkably, we also show that this approach via Kesten variables allows to recover the results for the stationary distribution of the RTP in a harmonic potential (obtained, e.g. in Ref. [216] via a completely different method), and also generalize it to more general velocity distribution (see Section 7.2 for details).

The rest of this part is organized as follows. In Section 6, we first study the case of periodic resetting and provide explicit solutions of Eq. (5.0.6) for various resetting noises: Brownian, ballistic, and telegraphic. In Section 7.1, we provide a detailed study of the stationary state in the case where the noise $y(t)$ is the rBM with Poissonian resetting. In this case, the relaxation to the stationary state and time-dependent properties can be analyzed, however, these aspects are not discussed in this thesis, and we instead refer the reader to [6]. Finally, in Section 7.2, using the Kesten approach, we derive the steady-state distribution of an RTP whose velocity is drawn from an arbitrary distribution at each tumble.

Another type of noise to which the Kesten approach applies is switching diffusion, introduced in Section 2.4. In this model, the diffusion coefficient switches to a new value drawn from a distribution $W(D)$ at each resetting event. When the particle evolves in a harmonic potential, we were able to derive the exact non-equilibrium steady state in Fourier space for any choice of $W(D)$, and further show that this distribution is connected to free cumulants. We present it in detail in Chapter 10 of the next part, which is devoted to switching diffusion.

# Chapter 6

# Steady State Distribution with Periodic Resetting of the Noise

## 6.1 Brownian Noise with Periodic Resetting

Here we consider the case where $y_n(\tau) = B_n(\tau)$ is a standard Brownian motion with diffusion coefficient $D$, starting from 0. In the case of periodic resetting protocol, where $p_{\text{int}}(\tau) = \delta(\tau - T)$ with period $T$ fixed, Eq. (5.0.2) reads

$$x_n = x_{n-1}\, e^{-\mu T} + e^{-\mu T} \int_0^T d\tau\, B_n(\tau)\, e^{\mu \tau}. \tag{6.1.1}$$

Here $T$ is just a constant. Hence, by iterating this equation, one sees that $x_n$ is just a linear combination of terms each involving a Brownian motion and consequently $x_n$ is a Gaussian variable for any $n$. Its average is clearly zero, $\langle x_n \rangle = 0$ and we only need to compute its variance $\langle x_n^2 \rangle$. Here, $\langle \cdots \rangle$ denotes an average over the Brownian noise, which is the only source of randomness here. Squaring Eq. (6.1.1) and taking average gives

$$\langle x_n^2 \rangle = \langle x_{n-1}^2 \rangle\, e^{-2\mu T} + 2D\, e^{-2\mu T} \int_0^T d\tau_1 \int_0^T d\tau_2 \min(\tau_1, \tau_2)\, e^{\mu(\tau_1 + \tau_2)}, \tag{6.1.2}$$

where we have used $\langle B(\tau_1) B(\tau_2) \rangle = 2D \min(\tau_1, \tau_2)$. Note that we have used the fact that $x_{n-1}$ and $B_n(\tau)$ are independent, together with $\langle x_{n-1} \rangle = \langle B_n(\tau) \rangle = 0$. Performing the integral in (6.1.2) explicitly, we get

$$\langle x_n^2 \rangle = \langle x_{n-1}^2 \rangle\, e^{-2\mu T} + \frac{D}{\mu^3} \left[ 2\mu T - 3 + 4\, e^{-\mu T} - e^{-2\mu T} \right]. \tag{6.1.3}$$

As $n \to \infty$, the sequence $\{x_n\}$ reaches a stationary Gaussian distribution

$$p(x) = \lim_{n \to \infty} p(x, n) = \frac{1}{\sqrt{2\pi\sigma^2}} e^{-\frac{x^2}{2\sigma^2}}, \tag{6.1.4}$$

where the variance $\sigma$ is obtained by taking the limit $n \to \infty$ in Eq. (6.1.3). The fixed point variance $\sigma^2 = \lim_{n \to \infty} \langle x_n^2 \rangle$ is then given explicitly by

$$\sigma^2 = \langle x_\infty^2 \rangle = \frac{D}{\mu^3} \frac{\left[ 2\mu T - 3 + 4\, e^{-\mu T} - e^{-2\mu T} \right]}{\left[ 1 - e^{-2\mu T} \right]}. \tag{6.1.5}$$

In Fig. 6.1, we show a plot of $\sigma^2$ vs $T$ for $\mu = 1$ and $D = 1$. It has the asymptotic behaviors for small and large $T$

$$\sigma^2 \approx \begin{cases} \frac{DT^2}{3\mu} & \text{as}\quad T \to 0 \\[1em] \frac{2DT}{\mu^2} & \text{as}\quad T \to \infty \end{cases} \tag{6.1.6}$$

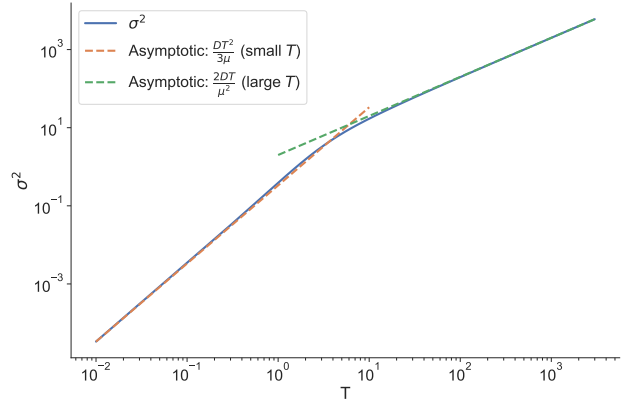

**Figure 6.1:** The solid blue line shows a plot, on a log-log scale, of $\sigma^2$ given in Eq. (6.1.5) vs $T$ for $\mu = 1$ and $D = 1$. The dashed orange and green lines show the asymptotic behaviors of $\sigma^2$ for small and large $T$ respectively, as given in Eq. (6.1.6).

Note that the limit $T \to 0$ (rapid resetting) corresponds to the strongly 'passive' limit, while $T \to \infty$ (rare resetting) corresponds to the strongly 'active' limit. Roughly speaking, these two limits in the periodic resetting correspond respectively to the limits $r \to \infty$ and $r \to 0$ of the Poissonian resetting, since the two protocols are qualitatively similar with the identification $T \sim 1/r$. Thus, as the activity $T$ (period $T$ can be taken as an activity strength) increases, the stationary limit distribution of $x_\infty$, while staying Gaussian, has an increasing variance as a function of $T$. Thus the probability mass spreads from the center of the trap outwards as the activity $T$ increases. Therefore activity enhances fluctuations.

Another case with periodic resetting that can be easily solved along the same line is when the particle is driven by an Ornstein-Uhlenbeck noise between resets. As in the case of Brownian motion, the stationary state is a centered Gaussian distribution, with a variance that can also be computed explicitly (see the appendices of [6]).

## 6.2 Ballistic & Telegraphic Noises with Periodic Resetting

### 6.2.1 Ballistic noise

Another solvable case corresponds to the ballistic model for the reset process $y(t)$. In this case, after each reset a random velocity $v$ is chosen independently from a symmetric distribution $w(v)$ and the process $y(t) = v\,t$ evolves ballistically with this velocity till it gets reset at the next epoch. Thus in this model the evolution of the noise between the $(n-1)$-th reset and the $n$-th reset is described by $y_n(t) = v_n\,t$ where $v_n$'s are independent and identically distributed (IID) random variables each drawn from $w(v)$. In this case, the recursion relation (5.0.2) reads

$$x_n = x_{n-1}\,e^{-\mu\,\tau_n} + v_n\,e^{-\mu\,\tau_n}\int_0^{\tau_n} d\tau\,\tau\,e^{\mu\,\tau} = x_{n-1}\,e^{-\mu\,\tau_n} + \frac{v_n}{\mu^2}\left[\mu\,\tau_n - 1 + e^{-\mu\,\tau_n}\right]. \qquad (6.2.1)$$

Now, this is again of the generalised Kesten form in Eq. (5.0.3) with $U_n = e^{-\mu\tau_n}$ and $V_n = \frac{v_n}{\mu^2}\left[\mu\,\tau_n - 1 + e^{-\mu\tau_n}\right]$. For periodic protocol, $p_{\text{int}}(\tau_n) = \delta(\tau_n - T)$, Eq. (6.2.1) becomes

$$x_n = x_{n-1}\,e^{-\mu T} + \frac{v_n}{\mu^2}\left[\mu T - 1 + e^{-\mu T}\right], \qquad (6.2.2)$$

where $v_n$ is the only random variable left. This recursion relation is of the form

$$x_n = a\,x_{n-1} + b\,v_n, \qquad (6.2.3)$$

where $a = e^{-\mu T} \leq 1$ and $b = \left(\mu T - 1 + e^{-\mu T}\right)/\mu^2 \geq 0$ are constants, while $v_n$'s are IID random variables, each drawn from a symmetric distribution $w(v)$. Interestingly, Eq. (6.2.3) can be thought of as a discrete-time version of an OU process [290–292], also known as the AR(1) process (autoregressive process of order 1) in finance [293]. In this case, as we will see now, the limiting distribution of $x_\infty$ can be computed explicitly, at least formally, for arbitrary distribution $w(v)$, leading to nontrivial distributions for the steady state $p(x, n \to \infty)$ [290–292].

To proceed, it is convenient to define $\tilde{v}_n = b\, v_n$, whose PDF is simply $\phi(\tilde{v}_n) = b^{-1}w(\tilde{v}_n/b)$, such that the recursion relation becomes

$$x_n = a\, x_{n-1} + \tilde{v}_n\,. \tag{6.2.4}$$

The integral equation (5.0.5) then reads

$$p(x, n+1) = \int_{-\infty}^{+\infty} dy\, \phi(x - ay)\, p(y, n)\,, \tag{6.2.5}$$

where $p(x, n)$ is the distribution of the position after $n$ resets. Thanks to the convolution structure of Eq. (6.2.5), we have, in Fourier space,

$$\hat{p}(k, n+1) = \hat{\phi}(k)\hat{p}(ak, n)\,, \tag{6.2.6}$$

where, for any function $f(x)$, we define its Fourier transform $\hat{f}(k)$ as

$$\hat{f}(k) = \int_{-\infty}^{+\infty} dx\, e^{i\,kx} f(x)\,. \tag{6.2.7}$$

In addition, since $x(0) = 0$ one has $p(x, 0) = \delta(x)$ which implies that the initial condition of the recursion relation (6.2.6) is simply $\hat{p}(k, 0) = 1$. This recursion relation (6.2.6) has the following solution (see e.g. [290])

$$\hat{p}(k, n) = \prod_{m=0}^{n} \hat{\phi}(a^m k) = \prod_{m=0}^{n} \hat{w}(a^m\, b\, k)\,, \tag{6.2.8}$$

where we recall that $a = e^{-\mu T} \in (0, 1)$. Taking the $n \to +\infty$ limit gives

$$\boxed{\hat{p}(k) = \lim_{n \to \infty} \hat{p}(k, n) = \prod_{m=0}^{+\infty} \hat{w}(a^m b\, k)}\,. \tag{6.2.9}$$

Of course, it is not possible to invert explicitly this Fourier transform for arbitrary distribution $w(v)$. However, there is one interesting case for which this inversion can be performed. This is the case of Lévy distributions of index $\alpha$ and parameter $\lambda$, for which $\hat{w}(k) = e^{-|\lambda k|^\alpha}$ with $0 < \alpha \leq 2$. In particular, $\alpha = 1$ (respectively $\alpha = 2$) corresponds to the Cauchy distribution (respectively the Gaussian distribution) in real space. For these cases, Eq. (6.2.9) reads

$$\hat{p}(k) = \prod_{m=0}^{+\infty} \hat{w}(a^m b\, k) = \prod_{m=0}^{+\infty} \exp\left[-|\lambda\, b\, k|^\alpha (a^\alpha)^m\right] = \exp\left[-|\lambda\, b\, k|^\alpha \sum_{m=0}^{+\infty} (a^\alpha)^m\right]\,. \tag{6.2.10}$$

Hence, performing the sum of the geometric series (recalling that $a^\alpha < 1$), we have

$$\hat{p}(k) = \exp\left[-\frac{|\lambda\, b\, k|^\alpha}{1 - a^\alpha}\right]\,. \tag{6.2.11}$$

It is again a Lévy distribution with the same index $\alpha$ as the PDF of the velocity $w(v)$, but with a different parameter $\lambda/(1 - a^\alpha)^{1/\alpha}$. Inverting the Fourier transform in (6.2.11) one finds

$$p(x) = \frac{1}{\ell}\mathcal{L}_\alpha\left(\frac{x}{\ell}\right)\,,\quad \ell = \frac{\lambda b}{(1 - a^\alpha)^{1/\alpha}} \tag{6.2.12}$$

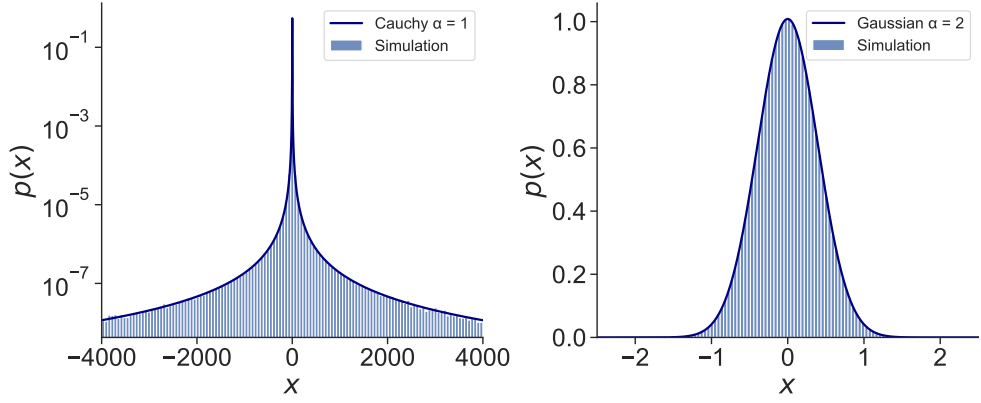

**Figure 6.2:** Plot of the stationary distribution for a particle evolving via (5.0.1) subjected to a resetting ballistic noise $y(t) = v\,t$, where $v$ is distributed according to a Cauchy distribution (left panel) and a Gaussian distribution (right panel). The histograms have been obtained by solving numerically the recursion relations (6.2.2) after a large number $n$ of iterations, while the solid line corresponds to our exact analytical results in (6.2.12).

where
$$\mathcal{L}_\alpha(z) = \int_{-\infty}^{+\infty} \frac{dq}{2\pi} e^{-iqz - |q|^\alpha} \tag{6.2.13}$$

is a Lévy distribution. In Fig. 6.2, we have compared our analytical prediction (6.2.12) with numerical simulation for the cases $\alpha = 1$ and $\alpha = 2$, showing a very good agreement.

Actually, these results can be generalized to a wider class of resetting noises of the form $y(t) = v\,h(t)$, with $h(t)$ an arbitrary function – see [6]. In this case, the above analysis, starting from the mapping to the AR(1) process in (6.2.3) remains the same with the substitution $b = e^{-\mu T} \int_0^T h(\tau)\,e^{\mu \tau}\,d\tau$, while $a$ remains unchanged.

### 6.2.2 Telegraphic noise

An interesting generalization corresponds to $h(t) = 1$ and while $w(v) = (1/2)\delta(v - v_0) + (1/2)\delta(v + v_0)$ which corresponds to a periodic telegraphic noise, which we now analyse in detail, since it is very similar to a standard RTP model. In the standard RTP model, the time $\tau$ between two flips of the noise is a continuous random variable with an exponential distribution. In the present model, this time takes discrete values $\tau = kT$ with $k = 1, 2, \ldots$ with probability Prob.$(\tau = kT) = 2^{-k} = e^{-\gamma_\text{eff}\,\tau}$, with an effective "tumbling rate"
$$\gamma_\text{eff} = \frac{\ln 2}{T}\,. \tag{6.2.14}$$

From the analysis performed above, using $\hat{w}(k) = \cos(v_0\,k)$ together with $b = e^{-\mu T} \int_0^T h(\tau)\,e^{\mu \tau}\,d\tau = 1/\mu\left(1 - e^{-\mu T}\right)$, the Fourier transform of the stationary PDF of the position $\hat{p}(k)$ is given by

$$\boxed{\hat{p}(k) = \prod_{m=0}^{+\infty} \cos\left\{\frac{v_0}{\mu}\left[\left(e^{-\mu T}\right)^m \left(1 - e^{-\mu T}\right)\right] k\right\}}\,. \tag{6.2.15}$$

As $\left(e^{-\mu T}\right)^m < 1$ the product converges and is well defined. Interestingly, such distributions (6.2.15) have appeared in various contexts in the mathematics [294, 295] and physics [296, 297] literature and they are known to exhibit a rich and intriguing behavior. To understand it better, it is useful to come back to the Kesten recursion relation in Eq. (6.2.3) which reads, in this case
$$x_n = e^{-\mu T}\,x_{n-1} \pm \frac{v_0}{\mu}\left(1 - e^{-\mu T}\right)\,. \tag{6.2.16}$$

If we rescale the position such that $\tilde{x}_n = x_n \left[\frac{v_0}{\mu}\left(1 - e^{-\mu T}\right)\right]^{-1}$, the dynamics is as follows

$$\tilde{x}_n = \lambda\, \tilde{x}_{n-1} + \epsilon_n \quad , \quad \epsilon_n = \begin{cases} +1 & , \quad \text{with proba. } 1/2 \\ \\ -1 & , \quad \text{with proba. } 1/2 \end{cases}, \quad (6.2.17)$$

with $\lambda = e^{-\mu T}$. Hence Eq. (6.2.17) is an AR(1) process – i.e., a discrete version of the OU process – with Bernoulli $\pm 1$ jumps $\epsilon_n$. This recursion relation (6.2.17) can be solved explicitly

$$\tilde{x}_0 = 0 \quad , \quad \tilde{x}_n = \sum_{m=0}^{n-1} \lambda^m\, \epsilon_{n-m}\,. \quad (6.2.18)$$

Note that if one denotes $\tilde{\epsilon}_m = \epsilon_{n-m}$, which are also independent Bernoulli random variables, one can then interpret $\tilde{x}_n$ as a random walk with shrinking steps, since the size of the $m$-th step is $\lambda^m$ (and we recall that $\lambda < 1$). This is precisely the problem that was studied in Ref. [296]. From Eq. (6.2.18), since $\epsilon_{n-m} = \pm 1$, it is clear that the PDF of $\tilde{x}_n$ has a finite support $[-(1-\lambda^n)/(1-\lambda), +(1-\lambda^n)/(1-\lambda)]$. Indeed, the maximum value of $\tilde{x}_n$ in (6.2.18) is attained for $\epsilon_{n-m} = +1$ for $m = 0, 1, \cdots, n-1$, leading to $\tilde{x}_n = \sum_{m=0}^{n-1} \lambda^m = +(1-\lambda^n)/(1-\lambda)$. Similarly, the minimum corresponds to $\epsilon_{n-m} = -1$ for $m = 0, 1, \cdots, n-1$, leading to the minimal value $\tilde{x}_n = -\sum_{m=0}^{n-1} \lambda^m = +(1-\lambda^n)/(1-\lambda)$. In the limit $n \to \infty$, the support of the stationary PDF of $\tilde{x}_n$ is thus $[-(1-\lambda)^{-1}, (1-\lambda)^{-1}]$. Recalling that $x_n = (v_0/\mu)(1-\lambda)\,\tilde{x}_n$, we thus obtain that the stationary PDF $p(x)$ has support over $[-v_0/\mu, +v_0/\mu]$. Interestingly, in the case of the standard RTP, the support of the stationary distribution is exactly the same. However, the PDF in the present case turns out to be more exotic.

To state the main results about $p(x)$, it is useful to rewrite (6.2.15) as

$$\hat{p}(k) = \prod_{m=0}^{+\infty} \cos\left(\lambda^m \tilde{k}\right), \quad \text{with } \lambda = e^{-\mu T} \quad \text{and} \quad \tilde{k} = k\frac{v_0}{\mu}\left(1 - e^{-\mu T}\right). \quad (6.2.19)$$

In the math literature, such distribution (6.2.19) are called "infinite Bernoulli convolutions" [294, 295], since they correspond to the convolution of an infinite number of Bernoulli random variables $\pm \lambda^n$ – see Eq. (6.2.18). For generic values of $\lambda$, this infinite product cannot be expressed in a closed form. There are however special values of $\lambda$ for which this infinite product simplifies and consequently $\hat{p}(k)$ can be inverted. For the simplest case $\lambda = 1/2$, thanks to the so called Viete's formula, Eq. (6.2.19) reads

$$\hat{p}(k) = \prod_{m=0}^{+\infty} \cos\left(\frac{\tilde{k}}{2^m}\right) = \frac{\sin 2\tilde{k}}{2\tilde{k}}, \quad (6.2.20)$$

which means that $p(x)$ is simply the uniform distribution over $[-v_0/\mu, +v_0/\mu]$ for $\lambda = 1/2$. In fact, it turns out that $p(x)$ exhibits quite different behaviors as $\lambda$ crosses this special value $\lambda_c = 1/2$. For $\lambda > 1/2$, the distribution is regular for almost all $\lambda$ and it has typically a bell-shape, see Fig. 6.3 (in this region, explicit expressions for $p(x)$ can be obtained for $\lambda = 2^{-1/m}$, with $m = 1, 2, \ldots$ [296]): this is a regime that we thus call "passive". In this regime, the density near the boundaries at $\pm v_0/\mu$ vanishes as $p(x) \sim |x \pm v_0/\mu|^\nu$ with $\nu = \ln 2/|\ln \lambda| - 1$ [294]. In terms of the effective tumbling rate (6.2.14) it thus reads $\nu = \gamma_{\text{eff}}/\mu - 1$, exactly as in the standard RTP with the same tumbling rate (see e.g. [216]). On the other hand, for $\lambda < 1/2$ is quite singular. In particular, the support of the distribution is a Cantor set and is a fractal [296]. In this case, the stationary distribution $p(x)$ exhibits peaks (see Fig. 6.3) and as $\lambda$ approaches 0, the support gets restricted to the two points $-v_0/\mu$ and $v_0/\mu$ resulting in particle concentration at the edge of the support. This is a regime that we call "active". Recalling that $\lambda = e^{-\mu T}$, we see from Eq. (6.2.14) that this transition between the active and the passive regime occurs when the effective tumbling rate $\gamma_{\text{eff}}$ crosses the "critical" value $\gamma_{\text{eff},c} = \mu$, exactly as in the standard RTP model with the same effective tumblings rate (see e.g. [216]). It is thus interesting to see that this transition at $\lambda = 1/2$, which is rather well known in the mathematics literature, has a physical interpretation as a transition between a passive phase for $\lambda > 1/2$ and an active one for $\lambda < 1/2$.

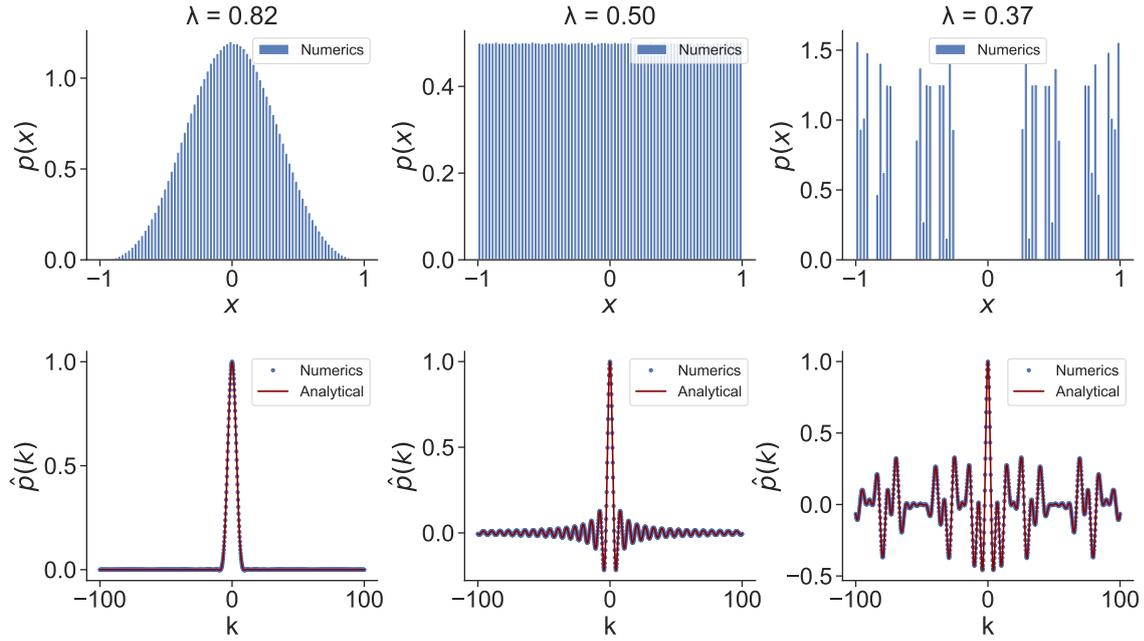

**Figure 6.3: Top panel**: Plots of the stationary PDF of the position for periodic run-and-tumble particles in the stationary state governed by the recursion relation (6.2.16) for different values of the parameter $\lambda = e^{-\mu T}$. The speed $v_0$ and the strength of the potential $\mu$ are taken to be unity. The support of the distribution is $[-v_0/\mu, v_0/\mu] = [-1, 1]$. When $\lambda > 1/2$ (left panel), the steady state has a bell shape (passive regime), while if $\lambda < 1/2$ (right panel), its support is a Cantor set, and as $\lambda$ decreases, particles are more and more concentrated at the edge of the support, i.e. $x = \pm 1$ (active regime). The transition between the two regimes occurs at $\lambda = 1/2$, when the distribution becomes uniform (middle panel). **Bottom panel:** Plots of the Fourier transform of the stationary PDF of the position for the same values of $\lambda$. The red curves correspond to the analytical product of cosines (6.2.15) – truncated at a large number of terms. The blue points correspond to a numerical evaluation of the Fourier transform of the density $\langle e^{ikx} \rangle$. This has been obtained by averaging over $10^7$ trajectories.

# Chapter 7

# Steady State Distribution with Poissonian Resetting of the Noise

## 7.1 Brownian Noise with Poissonian Resetting

In this section, we focus on the dynamics of a one-dimensional particle in a harmonic potential $V(x) = \mu x^2/2$ that starts its motion at $x(0) = 0$ and is subjected to a Poissonian resetting Brownian noise. The position of the particle $x(t)$ thus evolves through the Langevin equation

$$\frac{dx(t)}{dt} = -\mu\, x(t) + \tilde{y}_r(t) \quad , \quad \tilde{y}_r(t) = r\, y_r(t)\,, \tag{7.1.1}$$

with $y_r(t)$ being a resetting Brownian motion (rBM) with resetting rate $r$ [35, 54]. Note that in Eq. (7.1.1) the factor $r$ in the noise term has been added such that the noise has the dimension of a velocity. More precisely we consider the case where the rBM starts at the origin, i.e., $y_r(0) = 0$, and it is reset at exponential random times also at the origin (as, e.g., in the top panel of Fig. 5.1). During the infinitesimal time interval $[t, t+dt]$, the rBM thus evolves via [35, 54]

$$y_r(t+dt) = \begin{cases} 0 & \text{with probability } r\, dt\,, \\ y_r(t) + \xi(t)\, dt & \text{with probability } (1-r\, dt)\,, \end{cases} \tag{7.1.2}$$

where $\xi(t)$ is a Gaussian white noise of zero mean $\langle \xi(t)\rangle = 0$ and delta-correlations, i.e., $\langle \xi(t)\xi(t')\rangle = 2D\,\delta(t-t')$. We denote by $\{t_1, t_2, t_3, \ldots\}$ the random times (or epochs) at which the rBM gets reset to 0. For such a dynamics (7.1.2), the time intervals $\tau_n = t_n - t_{n-1}$ (for $n = 1, 2, \ldots$ with $t_0 = 0$) are statistically independent and distributed according an exponential distribution $p_{\text{int}}(\tau_n) = r\, e^{-r\,\tau_n}$: this is called *Poissonian resetting*. Therefore we see that the dynamics described by Eqs. (7.1.1) and (7.1.2) can be described in the framework relying on Kesten variables as described in Section 5. However, at variance with the periodic resetting studied in the previous section, we will see that the integral equation (5.0.6) is quite complicated to solve explicitly for Poissonian resetting. Nonetheless, we will see how it can be used to obtain useful detailed information on the stationary state of the Langevin equation (7.1.1).

Let us start with a qualitative description of the late time dynamics described by (7.1.1) and (7.1.2). At large time, the resetting Brownian noise $\tilde{y}_r(t)$ converges towards a stationary state, whose limiting PDF is a symmetric exponential distribution, namely [54]

$$P(\tilde{y}_r) = \frac{1}{2\sqrt{r\,D}}\, e^{-\frac{|\tilde{y}_r|}{\sqrt{r\,D}}}\,. \tag{7.1.3}$$

Besides, its two-time correlation function is given by [298]

$$\tilde{C}_r(t_1, t_2) = \langle \tilde{y}_r(t_1)\tilde{y}_r(t_2)\rangle = 2\,D\,r\, e^{-r\,(t_2-t_1)}\,(1 - e^{-r\,t_1}) \quad , \quad t_2 > t_1\,. \tag{7.1.4}$$

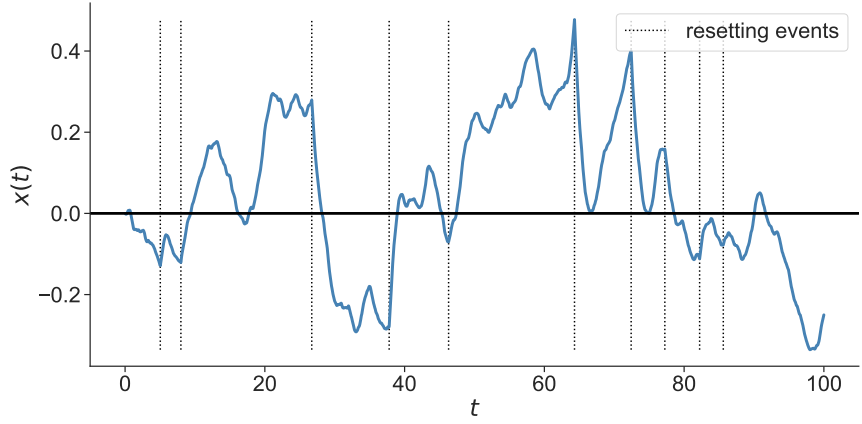

**Figure 7.1:** A typical realisation of a trajectory for the equation of motion Eq. (7.1.1). Right after a resetting of the noise, one observes a relaxation of the trajectory towards the origin, due to the confining potential. On the other hand it is harder to see the relaxation of the noise $\tilde{y}_r(t)$ itself. While a pure Brownian motion would diverge at large time, one can observe that the resets have a tendency to confine the position of the particle $x(t)$ near the origin. Indeed, when a reset happens, the noise is reset at value 0 and the speed of the particle is thus $\dot{x} = -\mu x$ such that it is directed toward the origin. If the position of the particle is positive (respectively negative) and increasing (respectively decreasing) before the reset, it gets reoriented towards the origin just after the reset. This explains the cusps that we can clearly observe on the trajectory around a resetting event. Here the parameters are $D = 1$, $\mu = 1$, $r = 0.1$, $x_0 = 1$.

In particular, in the stationary state where $t_1 \to \infty$, keeping $t_2 - t_1 = \tau$ fixed, the correlation function $\tilde{C}_r(t_1, t_2)$ decays as a pure exponential, i.e.,

$$\lim_{t_1 \to \infty} \tilde{C}_r(t_1, t_1 + \tau) = 2\,D\,r\,e^{-r\tau}\;. \tag{7.1.5}$$

Hence in the stationary regime, the two-time correlations (7.1.5) are very similar to the AOUP or the RTP model, but the one time distribution (7.1.3) is different from both models – since it is Gaussian for the AOUP while it is the sum of two delta-peaks at $\pm v_0$ for the RTP. Because of the similarities with these two models, it is rather natural to expect that PDF of the position $x(t)$ will converge to a stationary distribution $p(x)$ which is the main focus of this section.

Before we proceed to the detailed analysis of $p(x)$, it is useful to note that there are two time scales in the problem: (i) $\tau_\mu = 1/\mu$, which characterises the time scale of the relaxation within the confining harmonic potential and (ii) $\tau_r = 1/r$ which measures the correlation time of the resetting Brownian noise [see Eq. (7.1.5)]. Hence it is convenient to use the dimensionless ratio

$$\beta = \frac{\tau_\mu}{\tau_r} = \frac{r}{\mu}\;. \tag{7.1.6}$$

In the limit $\tau_r \ll \tau_\mu$, i.e. $\beta \to \infty$, the system is said to be "passive" while for $\tau_r \gg \tau_\mu$, i.e., $\beta \to 0$, the system is strongly active. We will see that the parameter $\beta$ in (7.1.6) indeed controls the crossover between the passive to active regimes (see Fig. 7.2).

### 7.1.1  Integral equation of the stationary distribution via Kesten variables

Here, we apply the method of Chapter 5 to derive an integral equation for the stationary distribution of the position. We decompose the motion as a sum of sub motions between resetting events that occurred at times $\{t_1, t_2, \ldots\}$. We define $\tau_n = t_n - t_{n-1}$, and integrate Eq. (7.1.1) from $t_{n-1}$ to $t_n$ such that we get a random recursion relation

$$x_n = x_{n-1}\,e^{-\mu \tau_n} + r\,e^{-\mu \tau_n} \int_0^{\tau_n} d\tau\,B(\tau)\,e^{\mu \tau}\;, \tag{7.1.7}$$

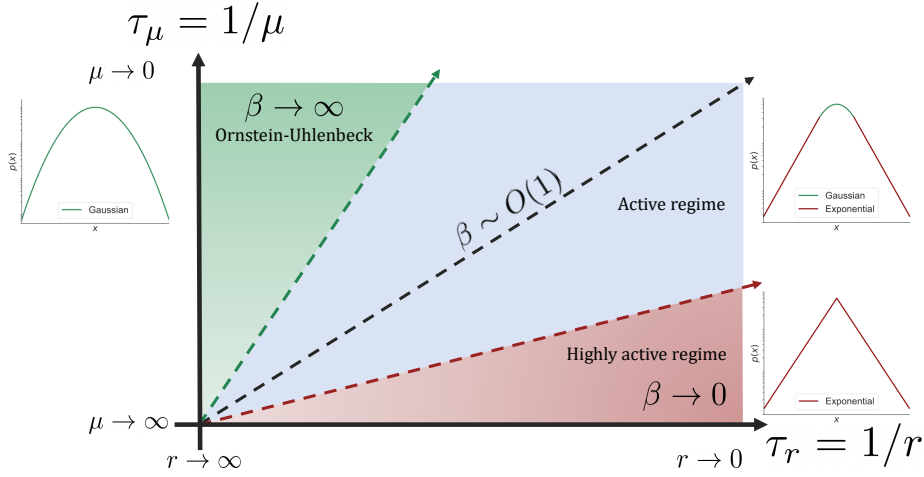

**Figure 7.2:** Schematic description of the different regimes of the model described by (7.1.1) in the plane ($\tau_r = 1/r, \tau_\mu = 1/\mu$). When $\tau_\mu \gg \tau_r$, i.e., in the strongly passive limit $\beta \to \infty$, $x(t)$ is an effective Ornstein-Uhlenbeck process with a diffusion constant $D_{\rm OU} = 2D$. In this case the stationary distribution is a Gaussian [see the second line of Eq. (7.1.36)]. On the other hand, when $\tau_\mu \ll \tau_r$, i.e., in the strongly active limit $\beta \to 0$ the motion is "slaved" to the noise $\tilde{y}_r(t)$ and the distribution is thus a double-exponential [see the first line of Eq. (7.1.36)]. Finally, for intermediate values of $\beta = \tau_\mu/\tau_r = r/\mu$ of order $O(1)$, the tails of distribution are exponential (7.1.53), while it has a Gaussian shape around the center. For a more comprehensive analysis of these regimes, please refer to Figs. 7.3, and 7.5.

where $B(\tau)$ is a Brownian motion starting from 0, since the noise is reset to zero at each resetting. As noticed in Chapter 5, Eq. (7.1.7) is of the generalised Kesten form

$$x_n = U_n\, x_{n-1} + V_n\,, \tag{7.1.8}$$

with $U_n = e^{-\mu \tau_n}$ and $V_n = r\, e^{-\mu \tau_n} \int_0^{\tau_n} B(\tau)\, e^{\mu \tau}\, d\tau$. The stationary state of the position of the particle is given by the following integral equation as in Eq. (5.0.6)

$$p(x) = \int_0^1 dU \int_{-\infty}^\infty dV \int_{-\infty}^\infty dx'\, P(U,V)\, p(x') \delta(x - U\, x' - V)\,, \tag{7.1.9}$$

and $P(U,V)$ is the joint distribution of $U_n$ and $V_n$. It can be computed using the Bayes's rule

$$P(U,V) = P(U) P(V|U)\,, \tag{7.1.10}$$

where $P(V|U)$ is the conditional PDF of $V$ given $U$. Using the fact that $p_{\rm int}(\tau) = r\, e^{-r\tau}$, we deduce that $U_n = e^{-\mu \tau_n}$ is distributed over the interval $[0,1]$ according to the PDF

$$P(U) = \beta\, U^{\beta-1}\,, \quad 0 \leq U \leq 1 \quad \text{with} \quad \beta = \frac{r}{\mu}\,. \tag{7.1.11}$$

Because $V_n$ at fixed $\tau_n$ (equivalently at fixed $U_n$) is a linear functional of Brownian motion, it is clear that the distribution of $V_n$, given $U_n$, is a Gaussian random variable. Its mean is clearly zero, while its variance is given by [see Eq. (6.1.3)]

$$\left\langle V_n^2 \right\rangle_{U_n} = e^{-2\mu \tau_n} \frac{D}{r} \beta^3 \left[ e^{2\mu \tau_n}(2\mu \tau_n - 3) + 4 e^{\mu \tau_n} - 1 \right]\,, \tag{7.1.12}$$

where $\langle \cdots \rangle_{U_n}$ denotes an average at fixed $U_n$ (equivalently at fixed $\tau_n$). We can then rewrite the variance $\left\langle V_n^2 \right\rangle_{U_n}$ in (7.1.12) as a function of $U_n = e^{-\mu \tau_n}$, leading to

$$\left\langle V_n^2 \right\rangle_{U_n} = \alpha(U_n) = \frac{D}{r} \beta^3 \left[ 4 U_n - U_n^2 - 2 \ln(U_n) - 3 \right]\,. \tag{7.1.13}$$

This shows that
$$P(V|U) = \frac{1}{\sqrt{2\pi\alpha(U)}} e^{-\frac{V^2}{2\alpha(U)}} . \quad (7.1.14)$$

Finally, Eq. (7.1.10) becomes
$$P(U,V) = P(U)P(V|U) = \beta\, U^{\beta-1} \frac{1}{\sqrt{2\pi\alpha(U)}} e^{-\frac{V^2}{2\alpha(U)}} , \quad (7.1.15)$$

and the integral equation (7.1.9) reads explicitly
$$p(x) = \int_0^1 dU \int_{-\infty}^\infty dV \int_{-\infty}^\infty dx'\, \beta\, U^{\beta-1} \frac{1}{\sqrt{2\pi\alpha(U)}} e^{-\frac{V^2}{2\alpha(U)}} p(x')\, \delta(x - Ux' - V) , \quad (7.1.16)$$

where $\alpha(U)$ is given in (7.1.13). Note that we can check that it is normalized by integrating (7.1.16) over $x$. To proceed, it is useful to go to Fourier space and introduce the Fourier transform of $p(x)$
$$\hat{p}(k) = \int_{-\infty}^{+\infty} dx\, p(x)\, e^{ikx} . \quad (7.1.17)$$

By taking the Fourier transform of Eq. (7.1.16) with respect to $x$, one gets
$$\hat{p}(k) = \int_0^1 dU \int_{-\infty}^\infty dV \int_{-\infty}^\infty dx'\, \beta\, U^{\beta-1} \frac{1}{\sqrt{2\pi\alpha(U)}} e^{-\frac{V^2}{2\alpha(U)}} p(x') e^{ik(Ux'+V)} . \quad (7.1.18)$$

We can perform the integration over $x'$ to simplify further the expression [using Eq. (7.1.17)]
$$\hat{p}(k) = \int_0^1 dU \int_{-\infty}^\infty dV\, \beta\, U^{\beta-1} \frac{1}{\sqrt{2\pi\alpha(U)}} e^{-\frac{V^2}{2\alpha(U)}} \hat{p}(kU) e^{ikV} . \quad (7.1.19)$$

Finally, performing the integral over $V$, one obtains
$$\boxed{\hat{p}(k) = \int_0^1 dU\, \beta\, U^{\beta-1} e^{-\frac{k^2\alpha(U)}{2}} \hat{p}(kU)} , \quad (7.1.20)$$

with
$$\alpha(U) = \frac{D}{r}\beta^3 \left[4U - U^2 - 2\ln(U) - 3\right] . \quad (7.1.21)$$

Its asymptotic behaviors are given by
$$\alpha(U) = \begin{cases} -\frac{D}{r}\beta^3 \left(2\ln(U) + 3 + \mathcal{O}(U)\right) , & U \to 0 , \\ \frac{2D}{3r}\beta^3 (1-U)^3 + \mathcal{O}((1-U)^4) , & U \to 1 . \end{cases} \quad (7.1.22)$$

Although it seems very hard to solve this integral equation (7.1.20), we will see below that many useful information about the stationary distribution $p(x)$ can however be extracted from it.

We end this section by noting that a similar integral equation (7.1.20) can be derived in the case where the noise $\tilde{y}_r(t)$ in Eq. (7.1.1) is a Gaussian stochastic process (not necessarily Brownian motion) subjected to Poissonian resetting (see e.g. [299]). In this case the conditional PDF $P(V|U)$ will still be a Gaussian, as in Eq. (7.1.14), but with a different function $\alpha(U)$. In the appendices of [6], we use this property to treat the case where the noise is a resetting Ornstein-Uhlenbeck process. An interesting case where we can solve Eq. (7.1.20) exactly is when the noise follows a switching diffusion process. This result is presented in detail in Chapter 10.

### 7.1.2 Moments of the stationary distribution

We start by deriving a recursion relation that allows to compute the moment of the stationary distribution. This can be done by expanding the left and right hand sides of Eq. (7.1.20) in powers of $k$ and identify the corresponding coefficients of $k^m$ on both sides. At this stage, it is useful to recall that both $x(t)$ and $\tilde{y}_r(t)$ in Eq. (7.1.1) start from the origin, i.e., $x(0) = 0$ and $\tilde{y}_r(0) = 0$. Therefore, since the time evolution of both processes are symmetric under $x \to -x$ and $\tilde{y}_r \to -\tilde{y}_r$, one expects that the PDF $p(x,t)$ is symmetric, i.e., $p(x,t) = p(-x,t)$ at all times. Therefore, in particular, the stationary PDF is also symmetric, i.e., $p(x) = p(-x)$ which implies that only the even moments $\langle x^{2n} \rangle$ are nonzero. The power expansion in $k$ of the left hand side (LHS) of Eq. (7.1.20) thus reads

$$\hat{p}(k) = \sum_{n=0}^{+\infty} \frac{(ik)^{2n}}{(2n)!} \left\langle x^{2n} \right\rangle . \tag{7.1.23}$$

Similarly, by expanding the right hand side of Eq. (7.1.20) one gets

$$\sum_{n=0}^{+\infty} \frac{(ik)^{2n}}{(2n)!} \left\langle x^{2n} \right\rangle = \int_0^1 dU\, \beta U^{\beta-1} \sum_{m=0}^{+\infty} \frac{(ik)^{2m} \left[\alpha(U)\right]^m}{2^m m!} \sum_{p=0}^{+\infty} \frac{(ik)^{2p} U^{2p}}{(2p)!} \left\langle x^{2p} \right\rangle . \tag{7.1.24}$$

One can group the two sums on the r.h.s and lighten the expression,

$$\sum_{n=0}^{+\infty} \frac{(ik)^{2n}}{(2n)!} \left\langle x^{2n} \right\rangle = \sum_{m=0}^{+\infty} \sum_{p=0}^{+\infty} \int_0^1 dU\, \beta U^{\beta-1+2p} \frac{\left[\alpha(U)\right]^m}{2^m m! (2p)!} (ik)^{2(m+p)} \left\langle x^{2p} \right\rangle . \tag{7.1.25}$$

By identifying the coefficient of the term $k^{2n}$ on both sides of (7.1.25) one obtains, after straightforward manipulations

$$\left\langle x^{2n} \right\rangle = (2n)! \sum_{p=0}^{n} \int_0^1 dU\, \beta U^{\beta-1+2p} \frac{\left[\alpha(U)\right]^{n-p}}{2^{n-p}(n-p)!(2p)!} \left\langle x^{2p} \right\rangle . \tag{7.1.26}$$

By isolating the terms $\propto \langle x^{2n} \rangle$ to the LHS, we finally obtain the recursion relation

$$\boxed{\left\langle x^{2n} \right\rangle = \beta\, (\beta + 2n)\, (2n-1)! \sum_{p=0}^{n-1} \frac{\left\langle x^{2p} \right\rangle}{2^{n-p}\,(n-p)!\,(2p)!} \int_0^1 dU\, U^{\beta-1+2p} \left[\alpha(U)\right]^{n-p}}, \tag{7.1.27}$$

where we recall $\alpha(U) = \frac{D}{r}\beta^3 \left[4U - U^2 - 2\ln(U) - 3\right]$. It seems difficult to obtain an explicit expression of $\langle x^{2n} \rangle$ for any arbitrary value of $n$ but the recursion relation in (7.1.27) allows to compute the first few moments:

$$\langle x^2 \rangle = \frac{2D\beta^2}{r(1+\beta)}, \tag{7.1.28}$$

$$\langle x^4 \rangle = \frac{12 D^2 \beta^4 (12 + 19\beta + \beta^2)}{r^2 (1+\beta)^2 (2+\beta)(3+\beta)}, \tag{7.1.29}$$

$$\langle x^6 \rangle = \frac{120 D^3 \beta^6 (4320 + 16344\beta + 22614\beta^2 + 12443\beta^3 + 2393\beta^4 + 61\beta^5 + \beta^6)}{r^3 (1+\beta)^3 (2+\beta)^2 (3+\beta)^2 (4+\beta)(5+\beta)} . \tag{7.1.30}$$

In fact, we see that the moments take the form

$$\left\langle x^{2n} \right\rangle = L^{2n} R_n(\beta) \quad \text{with} \quad L = \sqrt{\frac{D}{r}} \frac{\beta}{\sqrt{1+\beta}}, \tag{7.1.31}$$

and $R_n$ being a rational function of $\beta$. This form (7.1.31) thus suggests to interpret $L$ as the characteristic length scale of the stationary PDF $p(x)$.

### 7.1.3 Stationary distribution of the position

This result (7.1.31) suggests that the stationary distribution takes the following scaling form

$$p(x) = \frac{1}{L} \mathcal{F}_s \left( z = \frac{x}{L}; \beta = \frac{r}{\mu} \right) , \qquad (7.1.32)$$

where the subscript '$s$' refers to 'stationary'. We thus consider $\mathcal{F}_s(z; \beta)$ as a function of the variable $z$, depending on the parameter $\beta$. This scaling form can be easily shown directly from Eq. (7.1.20). Indeed this scaling form implies that $\hat{p}(k)$ reads

$$\hat{p}(k) = \hat{\mathcal{F}}_s \left( q = k\, L, \beta = \frac{r}{\mu} \right) \quad , \quad \text{where} \quad \hat{\mathcal{F}}_s(q, \beta) = \int_{-\infty}^{\infty} dz\, e^{iqz} \mathcal{F}_s(z; \beta) . \qquad (7.1.33)$$

Inserting this form in (7.1.20) one finds that $\hat{\mathcal{F}}_s(q; \beta)$ satisfies

$$\hat{\mathcal{F}}_s(q; \beta) = \beta \int_0^1 dU\, U^{\beta-1}\, e^{-\frac{1}{2}\beta(1+\beta)q^2 \tilde{\alpha}(U)}\, \hat{\mathcal{F}}_s(q\,U; \beta) \quad , \quad \tilde{\alpha}(U) = 4U - U^2 - 2\ln(U) - 3 . \qquad (7.1.34)$$

Note that the asymptotic behaviours of $\tilde{\alpha}(U)$ are simply given by

$$\tilde{\alpha}(U) = \begin{cases} -2\ln(U) - 3 + \mathcal{O}(U) \quad , & U \to 0 , \\ \frac{2}{3}(1-U)^3 + \mathcal{O}((1-U)^4) \quad , & U \to 1 . \end{cases} \qquad (7.1.35)$$

This form (7.1.34) will be useful below to discuss the asymptotic behaviors as $\beta \to \infty$ or $\beta \to 0$.

Of course, the choice of the couple $L$ and $\mathcal{F}_s(z; \beta)$ such that $p(x)$ can be written as in Eq. (7.1.32) is not unique but, as we will show below, it ensures that the scaling function $\mathcal{F}_s(z; \beta)$ has a well defined limit both when $\beta \to 0$ and $\beta \to \infty$. Namely we show below that in these two limit $\mathcal{F}_s(z; \beta)$ behaves as

$$\boxed{\mathcal{F}_s\left(z = \frac{x}{L}; \beta\right) \approx \begin{cases} \frac{1}{2} e^{-|z|} , & \text{as} \quad \beta \to 0 \\ \frac{1}{2\sqrt{\pi}} e^{-\frac{z^2}{4}} , & \text{as} \quad \beta \to +\infty \end{cases}} . \qquad (7.1.36)$$

For intermediate values of $\beta$, this suggests that the distribution crosses over continuously from an exponential to a Gaussian distribution – see Figs. 7.2 and 7.3.

**The passive limit $\beta \to \infty$.**

In this limit, we anticipate, and check it below, that $\mathcal{F}_s(z; \beta)$ admits the following expansion

$$\mathcal{F}_s(z; \beta) = \mathcal{F}_{s,0}(z) + \frac{1}{\beta} \mathcal{F}_{s,1}(z) + \frac{1}{\beta^2} \mathcal{F}_{s,2}(z) + O(1/\beta^3) , \qquad (7.1.37)$$

and similarly for $\hat{\mathcal{F}}_s(z; \beta)$, i.e.

$$\hat{\mathcal{F}}_s(q; \beta) = \hat{\mathcal{F}}_{s,0}(q) + \frac{1}{\beta} \hat{\mathcal{F}}_{s,1}(q) + \frac{1}{\beta^2} \hat{\mathcal{F}}_{s,2}(q) + O(1/\beta^3) . \qquad (7.1.38)$$

We next insert this expansion (7.1.38) in Eq. (7.1.34). It turns out that for $\beta \to \infty$, the integral over $U$ in (7.1.34) is dominated by $U \approx 1$. By performing the change of variable $U = 1 - u/\beta$, and injecting the expansion in Eq. (7.1.34), it is then rather straightforward to obtain a set of

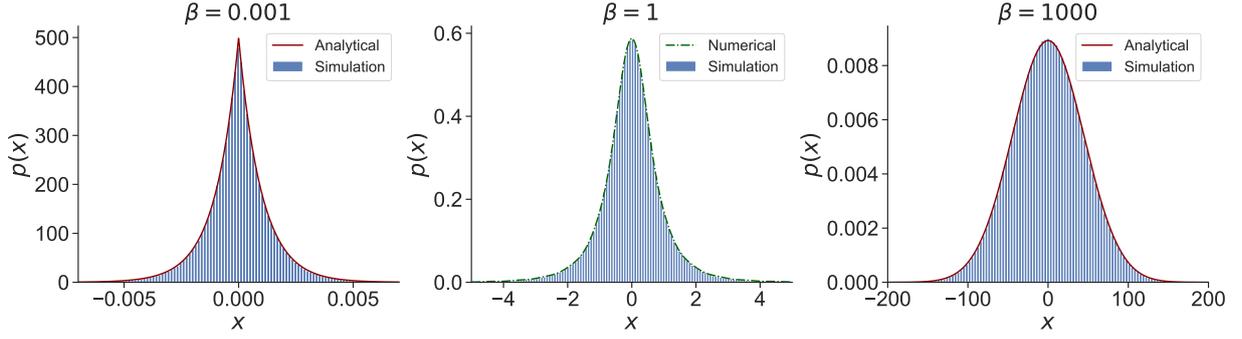

**Figure 7.3:** These plots show the rich behaviour of the stationary distribution of the position for a particle trapped in a harmonic potential subjected to a colored noise, namely a rBM as in Eq. (7.1.1). The plot on the left panel corresponds to the strongly active limit $\tau_r \gg \tau_\mu$ ($\beta \to 0$) studied in section 7.1.3. The distribution is a double-exponential - see Eq. (7.1.48). It is highly out of equilibrium as in that case, we showed that the effective temperature of the system diverges - see Eq. (7.1.58). The strongly passive limit $\tau_r \ll \tau_\mu$ ($\beta \to \infty$) is shown on the right panel. In this limit, the rBM behaves as a white noise, and the underlying distribution is therefore a Gaussian - see Eq. (7.1.45). The two red lines on the left and right panels correspond to the theoretical predictions, respectively given in Eq. (7.1.48) and (7.1.45). In the middle panel, for a particular value of $\beta$, we compare our simulations to a numerical solution of the integral equation (7.1.16). The agreement is perfect. For details on the numerical procedure, see [6].

differential equations that can be solved recursively. The first two of this hierarchy of equations read

$$\hat{\mathcal{F}}'_{s,0}(q) + 2q\hat{\mathcal{F}}_{s,0}(q) = 0 \tag{7.1.39}$$

$$\hat{\mathcal{F}}'_{s,1}(q) + 2q\hat{\mathcal{F}}_{s,1}(q) = 2q\left(20q^2 + 2\right)\hat{\mathcal{F}}_{s,0}(q) + \left(8q^2 + 1\right)\hat{\mathcal{F}}'_{s,0}(q) + q\hat{\mathcal{F}}''_{s,0}(q), \tag{7.1.40}$$

supplemented by the boundary conditions, obtained from $\hat{\mathcal{F}}_s(q = 0; \beta) = 1$ (since the PDF $\mathcal{F}_s(z; \beta)$ is normalized to unity)

$$\hat{\mathcal{F}}_{s,0}(0) = 1 \quad , \quad \hat{\mathcal{F}}_{s,1}(0) = 0. \tag{7.1.41}$$

The solution to these equations (7.1.39)-(7.1.41) reads

$$\hat{\mathcal{F}}_{s,0}(q) = e^{-q^2} \quad , \quad \hat{\mathcal{F}}_{s,1}(q) = 7q^4 e^{-q^2}. \tag{7.1.42}$$

Taking the inverse Fourier transform, one finds

$$\mathcal{F}_{s,0}(z) = \frac{1}{2\sqrt{\pi}} e^{-\frac{z^2}{4}} \quad , \quad \mathcal{F}_{s,1}(z) = \frac{7}{32\sqrt{\pi}} \left(z^4 - 12z^2 + 12\right) e^{-\frac{z^2}{4}}. \tag{7.1.43}$$

The leading term $\mathcal{F}_{s,0}(z)$ thus gives the result announced in the second line of Eq. (7.1.36), while $\mathcal{F}_{s,1}(z)$ provides the first correction to this limiting behavior.

This Gaussian limiting behavior of $\mathcal{F}_s(z; \beta)$ in the limit $\beta = r/\mu \to \infty$ can be easily understood by considering the limit $r \to \infty$ at fixed $\mu$. Indeed, in this limit, the noise $\tilde{y}_r(t)$ in Eq. (7.1.1) converges to a white noise in the sense that

$$\langle \tilde{y}_r(t)\tilde{y}_r(t + |t_2 - t_1|)\rangle \underset{t \to \infty}{\sim} 4D \frac{r}{2} e^{-r|t_2 - t_1|} \xrightarrow[r \to \infty]{} 4D\,\delta(t_2 - t_1). \tag{7.1.44}$$

In this limit, one thus expects that, at large time, Eq. (7.1.1) behaves similarly to a (passive) Ornstein-Uhlenbeck process with a diffusion constant $D_{\rm OU} = 2D$. It is then natural to expect that the system will eventually converge to a Boltzmann equilibrium described by the stationary PDF

$$p(x) = \sqrt{\frac{\mu}{2\pi D_{\rm OU}}} e^{-\frac{\mu}{2D_{\rm OU}} x^2} = \sqrt{\frac{\mu}{4\pi D}} e^{-\frac{\mu}{4D} x^2}. \tag{7.1.45}$$

In the large $\beta$ limit, one has $L \sim \sqrt{D/\mu}$ (see Eq. (7.1.31)), hence $p(x)$ in Eq. (7.1.45) can be re-written as $p(x) \sim e^{-z^2}/(L\sqrt{4\pi})$ with $z = x/L = \sqrt{\mu/D}\,x$, which yields the second line of Eq. (7.1.36). Note however that even in this limit $r \to \infty$, the noise term $\tilde{y}_r(t)$ remains non-Gaussian (see Eq. (7.1.3)), but, as shown by our explicit computation [see Eq. (7.1.43)], this does not modify the nature of the stationary state.

**The active limit $\beta \to 0$.**

To analyse this limit, it is convenient to perform the change of variable $z = U^\beta$ in Eq. (7.1.34), which yields

$$\hat{\mathcal{F}}_s(q; \beta) = \int_0^1 dz\, z^{q^2 + \beta q^2} e^{-\frac{1}{2}\beta(1+\beta)q^2(4z^{1/\beta} - z^{2/\beta} - 3)} \hat{\mathcal{F}}_s(q\, z^{1/\beta}; \beta)\,. \tag{7.1.46}$$

Since $z \in [0,1]$, $z^{1/\beta} \ll 1$ in the limit $\beta \to 0$ and one thus expects from this equation (7.1.46) that $\hat{\mathcal{F}}_s(q, \beta)$ admits an expansion in powers of $\beta$ as $\beta \to 0$. This expansion is however a bit cumbersome and we restrict our analysis here to the leading term $\hat{\mathcal{F}}_s(q; \beta = 0)$. Indeed using $\hat{\mathcal{F}}_s(q\, z^{1/\beta}; \beta) \to \hat{\mathcal{F}}(0,0) = 1$ as $\beta \to 0$, one easily obtains from Eq. (7.1.46) that $\hat{\mathcal{F}}(q, 0)$ is simply given by

$$\hat{\mathcal{F}}_s(q; 0) = \int_0^1 dz\, z^{q^2} = \frac{1}{1+q^2}\,. \tag{7.1.47}$$

By inverting this Fourier transform, one immediately obtains the result announced in the first line of Eq. (7.1.36).

This limiting behavior in the limit $\beta = r/\mu \to 0$ can be rationalised by considering the limit $r \to 0$ for fixed $\mu$. In this limit, the typical time between two successive resettings of the noise $\tilde{y}_r(t)$ is $\tau_r \approx 1/r \gg 1$. Therefore, after such a long time, the amplitude of the resetting Brownian noise is $\tilde{y}_r(\tau_r) \sim r\sqrt{\tau_r} = \sqrt{r}$. The equation of motion (7.1.1) thus suggests that the process $x(t)$ is "slaved" to the noise, i.e., $x(\tau_r) \sim \tilde{y}_r(\tau_r)/\mu \sim \sqrt{r}$, such that the term in the left hand side of Eq. (7.1.1), i.e., $\dot{x}(\tau_r) \sim r/\sqrt{\tau_r} \sim r^{3/2}$, is indeed subleading in the limit $r \to 0$. Using the fact that the stationary PDF is a symmetric exponential given in Eq. (7.1.3) it follows that $x(t) \sim \tilde{y}_r(t)/\mu$ is given by

$$p(x) \sim \frac{1}{2}\frac{\mu}{\sqrt{D\,r}} e^{-\frac{\mu}{\sqrt{D\,r}}|x|}\,. \tag{7.1.48}$$

In the limit $\beta \to 0$, the length is $L \sim \sqrt{D\,r}/\mu$ (see Eq. (7.1.31)) and therefore the PDF in (7.1.48) can be re-written as $p(x) = 1/(2L)\,e^{-|z|}$, as given in first line of Eq. (7.1.36).

### 7.1.4 Asymptotic behaviours of the stationary distribution $p(x)$

To extract the large $x$ behavior of $p(x)$, it turns out to be more convenient to start from the integral equation in real space, i.e., Eq. (7.1.16). Performing the integral over $V'$ in (7.1.16), one obtains

$$p(x) = \beta \int_0^1 \frac{dU}{\sqrt{2\pi\alpha(U)}} U^{\beta-1} \int_{-\infty}^\infty dx'\, e^{-\frac{(x-Ux')^2}{2\alpha(U)}} p(x')\,. \tag{7.1.49}$$

The function $p(x)$ is symmetric in $x$, which can be checked self-consistently from Eq. (7.1.49). Hence below we will focus on $x \geq 0$ only. In the limit of large $x$, one can simply replace $e^{-\frac{(x-Ux')^2}{2\alpha(U)}}$ by $e^{-\frac{x^2}{2\alpha(U)}}$, to leading order for large $x$. This is because, in that limit, the integral in Eq. (7.1.49) is dominated by $U = O(e^{-x}) \ll 1$. Therefore, one can thus expand, for small $U$,

$$\int_{-\infty}^\infty dx'\, e^{-\frac{(x-Ux')^2}{2\alpha(U)}} p(x') = \int_{-\infty}^\infty dx'\, p(x') e^{-\frac{x^2}{2\alpha(U)}} \left(1 + \frac{U}{\alpha(U)} xx' + \cdots\right)\,. \tag{7.1.50}$$

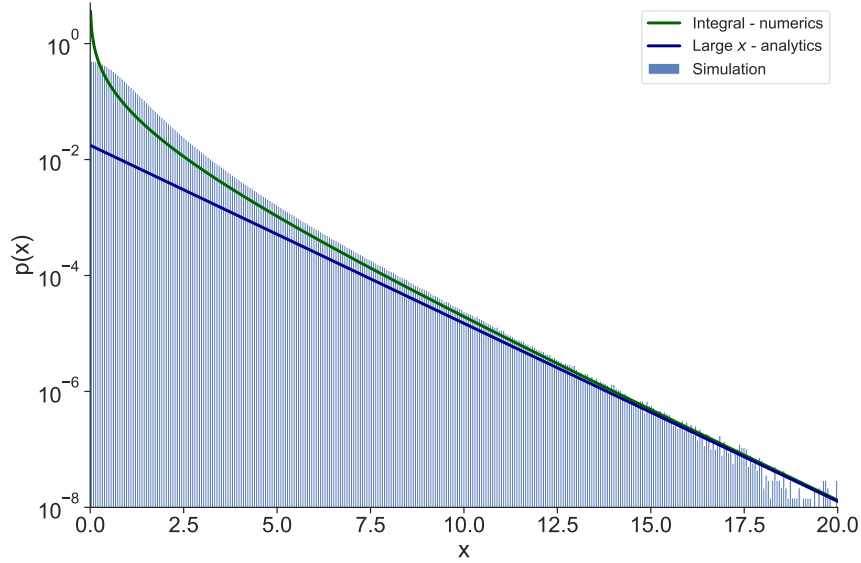

**Figure 7.4:** Plot of the stationary distribution $p(x)$ vs $x$ for $\beta = 2$. The histogram shows the result of numerical simulations of Eq. (7.1.1). The green solid line corresponds to the function $h(x)$ in Eq. (7.1.51), which is supposed to be a reasonably good approximation of $p(x)$ for $x$ large enough, while the blue straight line corresponds to our analytical prediction of the exact large $x$ behavior (7.1.52). There are no fitting parameters. As expected, both solid lines agree very well with the numerical results for large $x$ (here for $x \geq 12.5$), while the green one gives a very good estimate of $p(x)$ already for $x \simeq 7.5$. The parameter used for this plot are $\mu = 1$, $r = 2$, $D = 1$, and $x_0 = 0$.

Keeping only the first term in this expansion and performing then the integral over $x'$, using $\int_{-\infty}^{\infty} dx' p(x') = 1$, one obtains

$$p(x) \approx h(x) \quad \text{with} \quad h(x) = \beta \int_0^1 \frac{dU}{\sqrt{2\pi\alpha(U)}} U^{\beta-1} e^{-\frac{x^2}{2\alpha(U)}} . \tag{7.1.51}$$

The large $x$ behavior can finally be obtained via a saddle-point analysis [see [6] for details], leading to

$$p(x) \approx \frac{e^{-3\beta/2}}{2\beta} \sqrt{r/D}\, e^{-\frac{x}{\beta}\sqrt{r/D}} \quad , \quad x \to \infty . \tag{7.1.52}$$

From this, we can read off the large $z$ behavior of the scaling function $\mathcal{F}_s(z;\beta)$ defined in Eq. (7.1.32), namely

$$\mathcal{F}_s(z;\beta) \approx \frac{e^{-3\frac{\beta}{2}}}{2\sqrt{1+\beta}} e^{-\frac{z}{\sqrt{1+\beta}}} \quad , \quad z \to \infty . \tag{7.1.53}$$

One can also ask about the small $x$ behavior of $p(x)$. In fact one can easily see, by expanding the right hand side of Eq. (7.1.16) around $x = 0$ that $p(x)$ behaves for small $x$ as $p(x) \approx p(0) - a_2 x^2$, where $a_2$ si a constant we do not specify. Hence, one can summarize the asymptotic behavior of $p(x)$ in the limits of small and large $x$ as

$$\boxed{p(x) \approx \begin{cases} p(0) - a_2 x^2 \quad , & x \to 0 , \\ \\ \frac{e^{-3\beta/2}}{2\beta}\sqrt{r/D}\, e^{-\frac{x}{\beta}\sqrt{r/D}} \quad , & x \to \infty . \end{cases}} \tag{7.1.54}$$

We recall that $p(x)$ is symmetric around $x = 0$.

In Fig. 7.4, we show a comparison between our numerical simulations and our predictions of the large $x$ behaviour of $p(x)$ in Eqs. (7.1.51) and (7.1.52), showing a very nice agreement. In the middle panel of Fig. 7.5, we show a plot of $p(x)$ as a function of $x$ for a moderate value of $\beta = 10$ in log linear scale. These figures, in agreement with Eq. (7.1.54) show that the stationary PDF crosses over from a Gaussian behavior in the center (i.e., for moderate value of $x$) to an exponential tail behavior (7.1.52) for large $x$. In the large $\beta$ limit, by matching the Gaussian behavior in (7.1.36) with the exponential tail in (7.1.53), one finds that this crossover occurs for $z \sim \sqrt{\beta}$, i.e., for $x \sim \beta$.

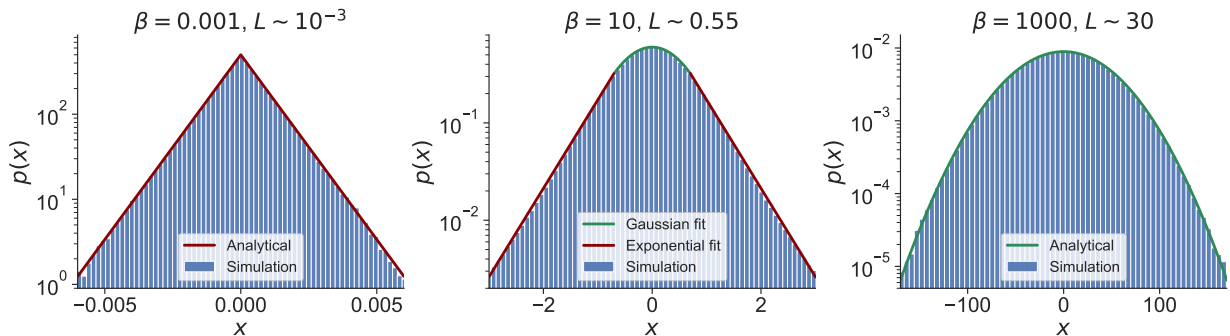

**Figure 7.5:** When $\beta$ approaches the value 0, we showed that the distribution becomes a symmetric double exponential, while when $\beta \to +\infty$, $p(x)$ converges to a Gaussian - see the plots on the left and right panel (the plots shown on these two panels are the same as the one shown in Fig. 7.3 but there are shown here in a log-linear scale). In blue, we show the histograms of simulations of many trajectories – the solid red and green lines correspond respectively to the double exponential form (7.1.48) and the Gaussian form (7.1.45). The middle panel shows a plot of $p(x)$ for an intermediate value of $\beta = 10$. The solid green line in the center corresponds to a Gaussian fit, while the red solid line corresponds to an exponential fit. It shows a nice crossover from a Gaussian to an exponential behavior.

Let us comment on the fact that the two limits $z \to \infty$ and $\beta \to \infty$ do not commute. This can be easily seen by comparing Eqs. (7.1.53) and the second line of Eq. (7.1.36). This is certainly an interesting point that may deserve further investigations and that probably requires a finer investigation of the integral equation (7.1.20).

Let us conclude this subsection by remarking that the crossover from a Gaussian form at small $x$ to an exponential decay at large $x$ has been seen in a variety of theoretical models, for example in the time-dependent position distribution in models of random diffusion such as diffusing diffusivity, [111, 115, 300] and also in certain experimental systems [20, 301]. Random diffusion models were introduced in Chapter 2.

### 7.1.5 Violation of the fluctuation-dissipation theorem in the steady-state

We end our study of this non-equilibrium steady state (NESS) by studying the violation of the fluctuation-dissipation theorem (FDT) – note that it has been recently studied for a pure Brownian motion under resetting in [302]. For that purpose, it is useful to compute the stationary correlation $C_{\rm st}(\tau)$ and response $R_{\rm st}(\tau)$ functions defined respectively as

$$C_{\rm st}(\tau) = \lim_{t \to \infty} \langle x(t) x(t+\tau) \rangle \quad , \quad R_{\rm st}(\tau) = \lim_{t \to \infty} \left\langle \frac{\delta \tilde{x}(t+\tau)}{\delta h(t)} \right\rangle \bigg|_{h=0} , \qquad (7.1.55)$$

where $\tilde{x}(t)$ evolves as in (7.1.1) in the presence of an additional external force field $h(t)$, i.e., $\dot{\tilde{x}}(t) = -\mu \tilde{x}(t) + r \, y_r(t) + h(t)$.

For the present model (7.1.1), $C_{\rm st}(\tau)$ and $R_{\rm st}(\tau)$ can be computed straightforwardly since the equation of motion is linear, leading to (see Appendix [6] for details)

$$C_{\rm st}(\tau) = \frac{2 D r \, (\mu \, e^{-r\tau} - r e^{-\mu \tau})}{\mu(\mu^2 - r^2)} \quad , \quad R_{\rm st}(\tau) = e^{-\mu \tau} . \qquad (7.1.56)$$

If the system were at equilibrium in contact with a bath at temperature $T^*$, $C_{\text{st}}(\tau)$ and $R_{\text{st}}(\tau)$ would be related via the relation $R(\tau) = -\frac{1}{T^*}\frac{dC(\tau)}{d\tau}$, which corresponds to the FDT. To quantify the violation of FDT, it is convenient to introduce an effective temperature via the relation [68]

$$T_{\text{eff}}(\tau) = -\frac{1}{R_{\text{st}}(\tau)}\frac{dC(\tau)}{d\tau}\,. \tag{7.1.57}$$

Inserting the explicit expressions from (7.1.56) in Eq. (7.1.57), one finds

$$\boxed{T_{\text{eff}}(\tau) = \frac{2D\,r^2}{r^2 - \mu^2}\left[1 - e^{-(r-\mu)\tau}\right].} \tag{7.1.58}$$

When $r < \mu$, the effective temperature grows exponentially with $\tau$, which indicates a highly out-of-equilibrium situation. On the other hand, when $r > \mu$, the effective temperature $T_{\text{eff}}(\tau)$ converges to a finite value $T_{\text{eff}}(\tau) \to T_{\text{eff}} = 2D\,r^2/(r^2 - \mu^2)$. In particular, in the passive limit $r \to \infty$, one finds $T_{\text{eff}} \to 2D$ which, given the relation in Eq. (7.1.44), can be considered as the equilibrium temperature. Finally, when $r = \mu$, $T_{\text{eff}}(\tau) = D\,r\,\tau$, which also grows unboundedly with $\tau$.

In conclusion, this computation of $T_{\text{eff}}(\tau)$ indicates that this simple active dynamics is always out-of-equilibrium (see Eq. (7.1.58)). In addition, this effective temperature is a growing function of the activity in the system (quantified here by $\mu/r = 1/\beta$). In fact, the equilibrium situation is recovered only in the passive limit $r \to \infty$.

**Remark:** One can easily show that the relation obtained here for the dynamics described by Eq. (7.1.1) also holds for the RTP dynamics with a telegraphic noise with rate $\gamma$, as studied e.g. in [216]. In fact the results for the RTP are given by Eqs. (7.1.56) and (7.1.58) with the substitution $\sqrt{2\,D\,r} \to v_0$, and $r \to 2\gamma$.

## 7.2 Run-and-Tumble Particle with a General Velocity Distribution

In this section, we present unpublished results on the dynamics of a generalized run-and-tumble particle evolving in a harmonic trap with initial position $x(0) = 0$. At each tumble, the speed $v$ is drawn from an arbitrary distribution $w(v)$. The equation of motion thus reads

$$\dot{x}(t) = -\mu\,x(t) + v(t)\,, \tag{7.2.1}$$

where $v(t)$ is piecewise constant and changes value only at the tumbling times. Note that it is straightforward to show that the PDF of $x(t)$ in this model satisfies the same Fokker–Planck equation as that of the standard RTP model, in which the noise is a telegraphic process with tumbling events occurring at a fixed rate $\gamma$, provided the substitution $r \to 2\gamma$ is made.

Another way to describe the motion is to consider the deterministic motion between the different resetting epochs $\{t_1, t_2, ...t_n\}$ such that if we integrate between $t_n$ and $t_{n-1}$, and we note $\tau_n = t_n - t_{n-1}$, we have

$$x_n = x_{n-1}e^{-\mu\tau_n} + \frac{v_i}{\mu}\left[e^{-\mu\tau_n} - 1\right]\,, \tag{7.2.2}$$

with $p_{\text{int}}(\tau_n) = r\,e^{-r\tau_n}$. Again, this recursion relation is of the generalised Kesten form $x_n = U_n x_{n-1} + V_n$, with $U_n = e^{-\mu\tau_n}$ and $V_n = \frac{v_i}{\mu}\left[e^{-\mu\tau_n} - 1\right] = \frac{v_i}{\mu}[U_n - 1]$. The joint law of $U$'s and $V$'s can be written as follows

$$P(U,V) = P(U)P(V|U) = \beta\,U^{\beta-1}\,\frac{\mu}{U-1}\,w\left(\frac{\mu V}{U-1}\right)\,, \tag{7.2.3}$$

where $\beta = r/\mu$. From Eq. (5.0.6), we deduce that the stationary distribution satisfies the integral equation

$$p(x) = \int_0^1 dU \int_{-\infty}^{\infty} dV \int_{-\infty}^{\infty} dx' \, \beta \, U^{\beta-1} \, \frac{\mu}{U-1} \, w\left(\frac{\mu V}{U-1}\right) p(x') \, \delta(x - U x' - V). \tag{7.2.4}$$

Performing a Fourier transform with respect to the variable $x$, one gets

$$\hat{p}(k) = \int_0^1 dU \int_{-\infty}^{\infty} dV \int_{-\infty}^{\infty} dx' \, \beta \, U^{\beta-1} \, \frac{\mu}{U-1} \, w\left(\frac{\mu V}{U-1}\right) p(x') \, e^{ik(Ux'+V)}. \tag{7.2.5}$$

Next, we do the integration on the variable $x'$ and obtain

$$\hat{p}(k) = \int_0^1 dU \int_{-\infty}^{\infty} dV \, \beta \, U^{\beta-1} \, \frac{\mu}{U-1} \, w\left(\frac{\mu V}{U-1}\right) \hat{p}(kU) \, e^{ikV}. \tag{7.2.6}$$

We then apply the change of variables $V' = \frac{\mu V}{U-1}$ and we integrate over the new variable $V'$. It leads to

$$\hat{p}(k) = \int_0^1 dU \, \beta \, U^{\beta-1} \, \hat{w}\left(\frac{U-1}{\mu} k\right) \hat{p}(kU). \tag{7.2.7}$$

Now, we introduce $q = k U$. This change of variables gives

$$k \left( k^{\beta-1} \hat{p}(k) \right) = \beta \int_0^k dq \, \hat{w}\left(\frac{q-k}{\mu}\right) \left( q^{\beta-1} \hat{p}(q) \right). \tag{7.2.8}$$

If we write $f_\beta(k) = k^{\beta-1} \hat{p}(k)$, we can take the Laplace transform with respect to the variable $k$ and take advantage of the convolution structure of the integral. We obtain

$$\mathcal{L}_{k \to s}\left[k \, f_\beta(k)\right] = \beta \, \mu \, \tilde{w}(\mu s) \, \tilde{f}_\beta(s), \tag{7.2.9}$$

Where the tilde means that we took the Laplace transform, i.e., $\tilde{g}(s) = \int_0^{+\infty} dk \, e^{-ks} g(k)$. As $\mathcal{L}_{k \to s}\left[k \, f_\beta(k)\right] = -\tilde{f}'_\beta(s)$, we obtain a first order differential equation for $\tilde{f}_\beta(s)$

$$\tilde{f}'_\beta(s) = -\beta \, \mu \, \tilde{w}(\mu s) \, \tilde{f}_\beta(s). \tag{7.2.10}$$

The solution is

$$\tilde{f}_\beta(s) = A \exp\left[-\beta \, \mu \int_0^s du \, \tilde{w}(\mu u)\right], \tag{7.2.11}$$

and a final change of variables gives

$$f_\beta(k) = k^{\beta-1} \hat{p}(k) \quad , \quad \tilde{f}_\beta(s) = A \, e^{-\beta W(\mu s)}, \tag{7.2.12}$$

where $W'(s)$ is the Laplace transform, of the Fourier transform of the distribution of the velocity $w(v)$. The integration constant $A$ can be fixed with the normalisation condition $\hat{p}(0) = 1$. Following from Eq. (7.2.12), and using properties of the Laplace transform, one can show that

$$\boxed{\hat{p}(k) = A \, k^{1-\beta} \, \mathcal{L}^{-1}_{s \to \frac{k}{\mu}}\left[e^{-\beta W(s)}\right],} \tag{7.2.13}$$

where we changed $A \mu^{-\beta} \to A$.

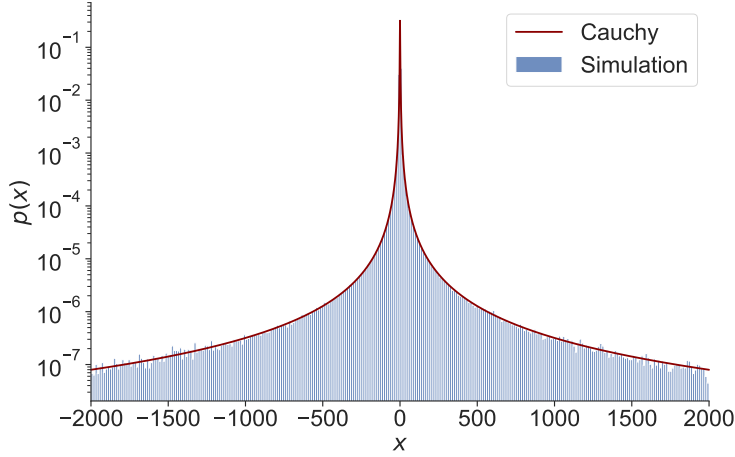

**Figure 7.6:** Plot of the stationary distribution of the position $p(x)$ for a generalised run-and-tumble particle where the velocity has a Cauchy distribution. The blue histogram is the result of numerical simulations while the solid line indicates our analytical prediction in (7.2.18), showing an excellent agreement. Here $\lambda = \mu = 1$.

### 7.2.1 Examples

**The standard RTP model:** $w(v) = 1/2\,\delta(v-v_0) + 1/2\,\delta(v+v_0)$

Here it is easy to show that

$$\tilde{f}_\beta(s) = A\,(s^2 + v_0^2/\mu^2)^{-\beta/2}\,. \tag{7.2.14}$$

Fixing the constant $A$ with $\hat{p}(0) = 1$, and going back to Fourier space gives

$$\hat{p}(k) = {}_0F_1\left(\frac{1+\beta}{2}, -\frac{k^2 v_0^2}{4\mu^2}\right)\,. \tag{7.2.15}$$

In real space it gives the expected distribution for $-v_0/\mu < x < v_0/\mu$ (see e.g. [216])

$$p(x) = \frac{\mu}{\sqrt{\pi}\,v_0} \frac{\Gamma\left(\frac{1}{2}+\frac{\gamma}{\mu}\right)}{\Gamma(\frac{\gamma}{\mu})} \left(1 - \frac{\mu^2 x^2}{v_0^2}\right)^{\frac{\gamma}{\mu}-1} \tag{7.2.16}$$

To retrieve the standard RTP, we have written $\beta = r/\mu = 2\gamma/\mu$.

**Cauchy distribution:** $\hat{w}(k) = e^{-\lambda|k|}$

In this case, it is easy to show that

$$\tilde{f}_\beta(s) = A\,(\lambda + \mu\,s)^{-\beta/\mu}\,. \tag{7.2.17}$$

Using Eq. (7.2.12), we obtain that the stationary distribution is a Cauchy distribution with parameter $\lambda/\mu$, namely

$$\boxed{p(x) = \frac{\lambda\,\mu}{\pi(\lambda^2 + \mu^2\,x^2)}}\,. \tag{7.2.18}$$

We have first found this result in [6]. In Fig. 7.6 we compare this analytical prediction (7.2.18) to numerical simulations, showing a very good agreement.

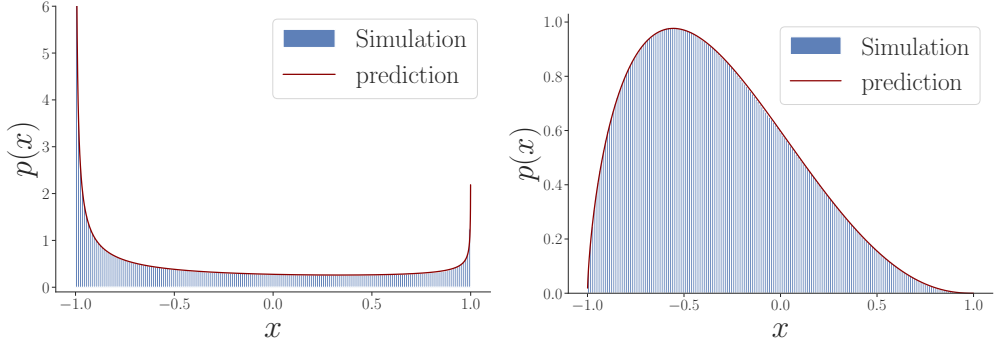

**Figure 7.7:** Simulation results for the asymmetric two-state RTP model with parameters $p = 1/3$, $\mu = 1$, and $v_0 = 1$. The left panel corresponds to $r = 2\gamma = 1$, while the right panel shows the case $r = 10$.

**The asymmetric 2 states RTP:** $w(v) = p\,\delta(v - v_0) + (1 - p)\,\delta(v + v_0)$

Here $0 < p < 1$ and if $p = 1/2$ we retrieve the standrad RTP. One can show that

$$\tilde{f}_\beta(s) = A\,(\mu s - i\,v_0)^{-\beta p}\,(\mu s + i\,v_0)^{-\beta(p-1)}\,. \tag{7.2.19}$$

Further calculations leads to

$$\hat{p}(k) = e^{\frac{ikv_0}{\mu}}\,{}_1F_1\left[\beta(1-p), \beta, -i\frac{2kv_0}{\mu}\right]\,. \tag{7.2.20}$$

To go to real space, we can use the integral representation of the hypergeometric function. We find

$$p(x) = \frac{1}{2\pi}\frac{\Gamma(\beta)}{\Gamma(\beta p)\Gamma(\beta(1-p))}\int_0^1 t^{\beta(1-p)-1}(1-t)^{\beta p-1}\int_{-\infty}^{+\infty} e^{-ik\left(x - \frac{v_0}{\mu} + \frac{2v_0}{\mu}t\right)}dk\,dt\,. \tag{7.2.21}$$

We can perform the integration with respec to $k$ and obtain

$$p(x) = \frac{\Gamma(\beta)}{\Gamma(\beta p)\Gamma(\beta(1-p))}\int_0^1 t^{\beta(1-p)-1}(1-t)^{\beta p-1}\delta\left(x - \frac{v_0}{\mu} + \frac{2v_0}{\mu}t\right)dt\,. \tag{7.2.22}$$

In the end, the distribution has a support over $[-v_0/\mu, v_0/\mu]$ and is given by

$$p(x) = \frac{\Gamma(\beta)}{\Gamma(\beta p)\Gamma(\beta(1-p))}\frac{2^{1-\beta}\mu}{v_0}\left(1 - \frac{\mu x}{v_0}\right)^{\beta(1-p)-1}\left(1 + \frac{\mu x}{v_0}\right)^{\beta p-1} \tag{7.2.23}$$

I have check that when $p = 1/2$ we retrieve the standard RTP.

We can actually solve a more general case where the RTP has a positive velocity $v_+ > 0$ and a different negative one $-v_- < 0$. The distribution of the velocities is then $w(v) = p\,\delta(v - v_+) + (1 - p)\,\delta(v + v_-)$, and for $x \in [-v_-/\mu, v_+/\mu]$ a similar computation leads to

$$\boxed{p(x) = \frac{\Gamma(\beta)}{\Gamma(\beta p)\Gamma(\beta(1-p))}\frac{\mu}{v_+ + v_-}\left(\frac{v_+ - \mu x}{v_+ + v_-}\right)^{\beta(1-p)-1}\left(\frac{v_- + \mu x}{v_+ + v_-}\right)^{\beta p-1}}\,. \tag{7.2.24}$$

In Fig. 7.7, we show that our analytical result is in excellent agreement with the simulations.

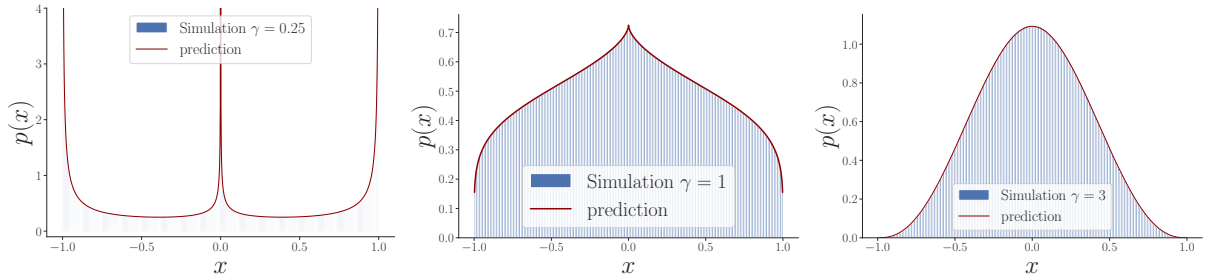

**Figure 7.8:** Simulation results for the three-state RTP model with $\mu = 1$, $v_0 = 1$, $p = 0.4$, and $r = 2\gamma$. The panels correspond to $\gamma = 0.25$ (left), $\gamma = 1$ (middle), and $\gamma = 3$ (right), illustrating different behaviors of the PDF, including eventual divergences at the edges of the finite support.

**The 3 states RTP:** $w(v) = p\,\delta(v - v_0) + p\,\delta(v + v_0) + (1 - 2p)\,\delta(v)$

Interestingly, this model closely resembles the three-state run-and-tumble model studied in [303], in which the particle cannot transition directly from velocity state $-v_0$ to $+v_0$, but instead must pass through the intermediate immobile state. It has been solved with the use of Fokker-Planck equations while here we use a completely different approach via Kesten variables. One can show that we have

$$\tilde{f}_\beta(s) = (\mu\,s)^{\beta(2p-1)} \left(\mu^2 s^2 + v_0^2\right)^{-\beta p}. \tag{7.2.25}$$

The Fourier transform of the PDF in the NESS is thus given by

$$\boxed{\hat{p}(k) = {}_1F_2\left(\beta p, \frac{1}{2}(1+\beta), \frac{\beta}{2}, -\frac{k^2 v_0^2}{4\mu^2}\right).} \tag{7.2.26}$$

Although the inverse Fourier transform can be computed using Mathematica, the resulting expression is cumbersome and therefore not presented here. Figure 7.8 shows excellent agreement between our analytical predictions and numerical simulations.

## 7.3 Conclusion

In this part of the thesis, we have introduced and studied a class of models of an active particle where the dynamics is driven by a resetting noise in the presence of an external quadratic potential. We first focused on the stationary state distribution of the position of the particle $p(x)$ and showed that it can be studied within the framework of *Kesten variables*. This allowed us to derive an integral equation satisfied by $p(x)$, which can be explicitly written for different types of resetting noises. This includes in particular the telegraphic noise, in which case the model studied here coincides with the well known run-and-tumble particle. In this case, the integral equation satisfied by $p(x)$ can be solved explicitly and allows us to recover, by a completely different method, the well known result for the corresponding stationary distribution $p(x)$. We also showed that this integral equation can be solved explicitly for a wide class of resetting noises when the resetting protocol is periodic – sometimes called "sharp resetting" or "stroboscopic resetting" [304]. For other resetting protocols, in particular in the case where the noise is a Brownian motion in the presence of Poissonian resetting – i.e., the standard resetting Brownian motion – it is very hard to solve exactly this integral equation. Nevertheless, we showed that it can be used to compute the moments of $p(x)$, as well as its asymptotic behavior for large $x$. Interestingly, we showed that $p(x)$ exhibits an exponential tail, which is markedly different form the Gaussian tail obtained in the case of a white noise – instead of the resetting noise considered here.

# Part III

# Switching Diffusion: Large Deviations and Free Cumulants

**Abstract**

We study the diffusion of a particle with a time-dependent diffusion constant $D(t)$ that switches between random values drawn from a distribution $W(D)$ at a fixed rate $r$. Using a renewal approach, we compute exactly the moments of the position of the particle $\langle x^{2n}(t) \rangle$ at any finite time $t$, and for any $W(D)$ with finite moments $\langle D^n \rangle$. For $t \gg 1$, we demonstrate that the cumulants $\langle x^{2n}(t) \rangle_c$ grow linearly with $t$ and are proportional to the free cumulants of a random variable distributed according to $W(D)$. For specific forms of $W(D)$, we compute the large deviations of the position of the particle, uncovering rich behaviors and dynamical transitions of the rate function $I(y = x/t)$. Our analytical predictions are validated numerically with high precision, achieving accuracy up to $10^{-2000}$. We then study the non-equilibrium steady state of the same process in the presence of a harmonic potential. We derive the exact stationary distribution in Fourier space for an arbitrary diffusion coefficient distribution $W(D)$. Once again, a connection to free cumulants emerges. To illustrate our exact results, we consider the case of randomly interrupted diffusion, where the particle alternates between two states: $D > 0$ and $D = 0$.

# Chapter 8

# Model and Main Results

Switching diffusion is a model in which the diffusion coefficient changes randomly over time, according to a given probability distribution. This framework has been used to describe "diffusive yet non-Brownian" motion observed in certain experiments, where the mean-squared displacement grows linearly with time – similar to classical Brownian motion – yet the displacement probability distribution exhibits distinctly non-Gaussian fluctuations [117–119]. We have already introduced and motivated this class of random diffusion models in Chapter 2, and we refer the reader to that chapter for a comprehensive overview and contextual background. In this part of the thesis, we present exact analytical results for the switching diffusion model, which was presented in the preprint [1]. Here, we also include new, unpublished findings for the same model when subject to confinement in a harmonic potential. Despite the growing interest in random diffusivity models, precise analytical results concerning their large deviation properties remain scarce. Our work provides a detailed and fully analytic characterization of these properties.

The free dynamics of the switching diffusion is given by the following Langevin equation

$$\dot{x}(t) = \sqrt{2D(t)}\,\eta(t) \quad , \quad x(0) = x_0 \quad , \tag{8.0.1}$$

where $\eta(t)$ is a Gaussian white noise with zero mean $\langle \eta(t) \rangle = 0$ and unit variance $\langle \eta(t)\eta(t') \rangle = \delta(t-t')$. For a time $\tau_1$, the process $x(t)$ diffuses with a diffusion constant $\mathsf{D}_1$, where $(\mathsf{D}_1, \tau_1)$ are drawn from a joint distribution $P_{\text{joint}}(D, \tau)$. After the duration $\tau_1$, a new pair $(\mathsf{D}_2, \tau_2)$ is drawn independently from $P_{\text{joint}}$, and this selection process is repeated at each subsequent renewal time. In our case, we assume that the renewal times and diffusion constants are independently selected, such that $P_{\text{joint}}(D, \tau) = W(D)\,p(\tau)$, where $W(D)$ is an arbitrary distribution with finite moments, and the renewal times are exponentially distributed as $p(\tau) = r\,e^{-r\tau}$. As a result, all values $\mathsf{D}_i$'s and $\tau_i$'s are independent and identically distributed (i.i.d.) random variables. For simplicity, we set the initial position to $x_0 = 0$.

As we will show, it is also interesting to study switching diffusion in the presence of an external potential $V(x)$. The dynamics then goes as follows

$$\dot{x}(t) = -V'(x) + \sqrt{2D(t)}\,\eta(t) \quad . \tag{8.0.2}$$

In the case of a harmonic potential, $V(x) = \mu x^2/2$, where $\mu$ denotes the stiffness of the trap, the confinement drives the particle toward a stationary state that exhibits deviations from the classical Boltzmann distribution. In this setting, we are able to solve the model exactly.

In the following sections, we first introduce the concept of free cumulants, followed by a presentation of the main results of our study. Chapter 9 focuses on the free switching diffusion model, while Chapter 10 presents the results obtained in the presence of a harmonic potential, including the derivation of the exact non-equilibrium steady state in Fourier space. Interestingly, both models – whether confined or free – exhibit an unexpected connection to free cumulants, which we characterize in detail.

## 8.1 Preliminaries: from cumulants to free cumulants

### 8.1.1 Bell Polynomials

As we will see, a central mathematical object emerges naturally in the analysis of switching diffusion: the partial exponential Bell polynomials $B_{n,k}$, which are fundamental in the study of set partitions [305, 306]. These polynomials are functions of several variables $x_i$ and are defined as

$$B_{n,k}(x_1, x_2, \ldots, x_{n-k+1}) = \sum_{\vec{j}} \frac{n!}{j_1! j_2! \cdots j_{n-k+1}!} \left(\frac{x_1}{1!}\right)^{j_1} \left(\frac{x_2}{2!}\right)^{j_2} \cdots \left(\frac{x_{n-k+1}}{(n-k+1)!}\right)^{j_{n-k+1}} \quad (8.1.1)$$

where the summation $\sum_{\vec{j}}$ over $\vec{j} = (j_1, j_2, \ldots, j_{n-k+1})$ denotes a sum over all non-negative integers $j_i$ subject to the following constraints

$$j_1 + j_2 + \cdots + j_{n-k+1} = k, \quad (8.1.2)$$
$$j_1 + 2j_2 + 3j_3 + \cdots + (n-k+1)j_{n-k+1} = n. \quad (8.1.3)$$

It is also useful to define the ordinary Bell polynomial $\hat{B}_{p,m}$

$$\hat{B}_{p,m}(x_1, \ldots, x_{p-m+1}) = \frac{m!}{p!} B_{p,m}(x_1, 2! \, x_2, \ldots, (p-m+1)! \, x_{p-m+1}), \quad (8.1.4)$$

and the $n^{\text{th}}$ complete exponential Bell polynomial

$$B_n(x_1, \ldots, x_n) = \sum_{m=1}^{n} B_{n,m}(x_1, x_2, \ldots, x_{n-m+1}). \quad (8.1.5)$$

### 8.1.2 From Cumulants to Free Cumulants

For a random variable $D$ with distribution $W(D)$, the classical cumulants are related to the moments via the following explicit formula

$$\langle D^n \rangle_c = \sum_{k=1}^{n} (-1)^{k-1} (k-1)! \, B_{n,k}\left(\langle D \rangle, \ldots, \langle D^{n-k+1} \rangle\right), \quad (8.1.6)$$

where $B_{n,k}$ are the partial exponential Bell polynomials defined in Eq. (8.1.1). We give below the first few classical cumulants

$$\langle D \rangle_c = \langle D \rangle, \quad (8.1.7)$$
$$\langle D^2 \rangle_c = \langle D^2 \rangle - \langle D \rangle^2, \quad (8.1.8)$$
$$\langle D^3 \rangle_c = \langle D^3 \rangle - 3\langle D^2 \rangle \langle D \rangle + 2\langle D \rangle^3, \quad (8.1.9)$$
$$\langle D^4 \rangle_c = \langle D^4 \rangle - 4\langle D^3 \rangle \langle D \rangle - 3\langle D^2 \rangle^2 + 12\langle D^2 \rangle \langle D \rangle^2 - 6\langle D \rangle^4, \quad (8.1.10)$$
$$\langle D^5 \rangle_c = \langle D^5 \rangle - 5\langle D^4 \rangle \langle D \rangle - 10\langle D^3 \rangle \langle D^2 \rangle + 20\langle D^3 \rangle \langle D \rangle^2 + 30\langle D^2 \rangle^2 \langle D \rangle$$
$$\qquad - 60\langle D^2 \rangle \langle D \rangle^3 + 24\langle D \rangle^5. \quad (8.1.11)$$

The free cumulants, on the other hand, can be computed using the following explicit formula in terms of the moments of $D$, which reads [307, 308]

$$\kappa_n(D) = \sum_{j=1}^{n} \frac{(-1)^{j-1}}{j} \binom{n+j-2}{j-1} \widetilde{\sum_{\vec{q}}} \prod_{k=1}^{j} \langle D^{q_k} \rangle, \quad (8.1.12)$$

where $\widetilde{\sum_{\vec{q}}}$, with $\vec{q} = (q_1, \cdots, q_j)$ denotes a constrained sum such that $q_1 + q_2 + \ldots + q_j = n$ with integers $q_k \geq 1$. Note that we have corrected a typo compared to [307, 308], where instead $q_k \geq 0$. The first few free cumulants are given by

$$\kappa_1(D) = \langle D \rangle, \tag{8.1.13}$$
$$\kappa_2(D) = \langle D^2 \rangle - \langle D \rangle^2, \tag{8.1.14}$$
$$\kappa_3(D) = \langle D^3 \rangle - 3\langle D^2 \rangle \langle D \rangle + 2\langle D \rangle^3, \tag{8.1.15}$$
$$\kappa_4(D) = \langle D^4 \rangle - 4\langle D^3 \rangle \langle D \rangle - 2\langle D^2 \rangle^2 + 10\langle D^2 \rangle \langle D \rangle^2 - 5\langle D \rangle^4, \tag{8.1.16}$$
$$\kappa_5(D) = \langle D^5 \rangle - 5\langle D^4 \rangle \langle D \rangle - 5\langle D^3 \rangle \langle D^2 \rangle + 15\langle D^3 \rangle \langle D \rangle^2 + 15\langle D^2 \rangle^2 \langle D \rangle$$
$$- 35\langle D^2 \rangle \langle D \rangle^3 + 14\langle D \rangle^5. \tag{8.1.17}$$

The expressions for classical and free cumulants differ starting from $n > 3$. This is because the computation of the classical cumulant $\langle D^n \rangle_c$ usually involves a sum over all possible partitions of $\{1, 2, \ldots, n\}$, including both crossing and non-crossing partitions. In contrast, free cumulants are computed using only non-crossing partitions of the indices [309].

For $n = 1, 2,$ and $3$, the possible partitions of the sets $\{1\}$, $\{1, 2\}$, and $\{1, 2, 3\}$ do not include any crossing partitions. Consequently, for $n \leq 3$, the set of all partitions coincides with the set of *non-crossing partitions* [309], meaning the first three free cumulants are identical to the corresponding classical cumulants.

### 8.1.3 The *R-transform* of a Distribution $W(D)$

The *R-transform* of a probability density function $W(D)$ is defined as the generating function of the free cumulants

$$R(z) = \sum_{n \geq 1} z^{n-1} \kappa_n(D). \tag{8.1.18}$$

It is a central object in free probability theory [151, 309–312], and is related to the Cauchy–Stieltjes transform $g(x)$ via the equation

$$g\left(R(z) + \frac{1}{z}\right) = z, \quad \text{where} \quad g(x) = \int_0^{D_{\max}} dD \, \frac{W(D)}{x - D}. \tag{8.1.19}$$

## 8.2 Main Results for the Free Case

First, for the class of switching diffusion models illustrated in Fig. 8.1, we have obtained an exact analytical expression for the moments of the positions $\langle x^{2n}(t) \rangle$ of arbitrary order $n$ at any finite time $t$ and for any distribution $W(D)$ which we assume to have all its moments well defined (note that the odd moments of $x(t)$ vanish by symmetry $x \to -x$). Here the notation $\langle \cdots \rangle$ means a simultaneous average over all the sources of randomness on the same footing (i.e., in the language of disordered systems, we consider here an "annealed" average). They read

$$\boxed{\langle x^{2n}(t) \rangle = \frac{(2n)!}{r^n} \sum_{m=1}^{n} \frac{(rt)^{m+n-1}}{(m+n-1)!} M(n-1, m+n, -rt) \, \hat{B}_{n,m}\left(\langle D \rangle, \ldots, \langle D^{n-m+1} \rangle\right)}, \tag{8.2.1}$$

where $\hat{B}_{n,m}$ is the ordinary bell polynomial – see Section 8.1.1. In Eq. (9.2.10), the function $M(a, b, x)$ denotes the Kummer's function. Of course the $2n$-th cumulant, denoted here as $\langle x^{2n}(t) \rangle_c$, can be formally obtained from (8.2.1). However, their large time behavior is more conveniently extracted from the cumulant generating function, which, as shown below, can be computed explicitly. One finds that their asymptotic behaviors at small and large time read

$$\langle x^{2n}(t)\rangle_c \simeq \begin{cases} \frac{(2n)!}{n!}\langle D^n\rangle_c\, t^n & , \quad r\,t \ll 1\,, \\ \frac{(2n)!}{r^{n-1}}\kappa_n(D)\,t & , \quad r\,t \gg 1\,. \end{cases} \tag{8.2.2}$$

In the first line, $\langle D^n\rangle_c$ denotes the (standard) cumulant of $D$, while the coefficients $\kappa_n(D) \neq \langle D^n\rangle_c$ also depend in a nontrivial way on the moments of $D$. This result (8.2.2) clearly shows that, even at large times $t$, the higher cumulants of $x(t)$ grow linearly with time, revealing the presence of non-Gaussian fluctuations in this model. As we will show, it turns out that $\kappa_n(D)$ are none other than the *free cumulants of $D$*, a class of combinatorial objects central to the field of free probability theory, which was originally developed to study non-commuting random variables [151] – see Section 8.1.2. Free probability has become crucial in random matrix theory (RMT) [151, 312–314], with applications that have sparked significant interest in both mathematics [309, 310, 315, 316] and physics [317, 318], in particular in quantum mechanics [319, 320]. While such free cumulants appeared before in more complicated classical models of *interacting* particles [321, 322], their appearance in such a simple *single* particle model here is highly surprising and intriguing. Although conventional cumulants $\langle D^n\rangle_c$ are related to the moments $\langle D^p\rangle$, with $p = 1, \ldots, n$ via Bell polynomials (8.1.6), free cumulants have a fairly explicit expression in terms of these moments (8.1.12). This enables us to compute them explicitly for various distributions of interest. For instance, for the two-state model $W(D) = p\delta(D - D_1) + (1-p)\delta(D)$ with $0 \leq p \leq 1$, one has $\langle x^2(t)\rangle_c \sim 2(pD_1 + (1-p)D_2)t$ while the higher cumulants behave for large time as (for $n \geq 2$)

$$\langle x^{2n}(t)\rangle_c \approx -(\Delta D)^n \frac{(2n)!}{r^{n-1}} \frac{\sqrt{p(1-p)}}{n(n-1)} P_{n-1}^1(1-2p)\,t, \tag{8.2.3}$$

where $\Delta D = (D_1 - D_2)$ and $P_n^m(z)$ denotes the associated Legendre polynomial of degree $n$ and parameter $m$. It is also interesting to study the case where $W(D)$ is a continuous PDF with a finite support, as discussed e.g. in [117, 142]. For example, we consider the case where $W(D)$ is given by the Wigner semi-circle on $[0, D_{\max}]$, i.e., $W(D) = 8\sqrt{D(D_{\max} - D)}/(\pi D_{\max})$ for which it is well known, from RMT, that the corresponding free cumulants are quite simple, i.e., $\kappa_n(D) = 0$ for $n \geq 3$. In this case one finds

$$\langle x^2(t)\rangle_c \approx D_{\max}t \quad, \quad \langle x^4(t)\rangle_c \approx \frac{3}{2}\frac{D_{\max}^2}{r}t\,, \tag{8.2.4}$$

while higher order cumulants vanish to leading order in $t$ [see Eq. (8.2.2)]. In fact, for $n \geq 3$, $\langle x^{2n}(t)\rangle_c = O(1)$ can also be computed – see Section 9.5.3.

What about the full distribution of $x(t)$, both at short and large times? At short time $rt \ll 1$, the particle does not have enough time to switch states and hence diffuses freely with a propagator $e^{-x^2/(4\mathsf{D}_1\tau)}/\sqrt{4\pi \mathsf{D}_1\tau}$. Averaging over $\mathsf{D}_1$ leads to

$$p_r(x,t) \approx \frac{1}{\sqrt{t}}f\left(\frac{x}{\sqrt{t}}\right)\,,\, f(u) = \int \frac{e^{-\frac{u^2}{4D}}}{\sqrt{4\pi D}}W(D)\,dD\,. \tag{8.2.5}$$

Not surprisingly, this PDF (8.2.5) has exactly the form found for diffusing diffusivity model [19, 117]. On the other hand, at large time $rt \gg 1$, one finds that the PDF of the position takes a large deviation form

$$p_r(x,t) \approx e^{-t\,I\left(y=\frac{x}{t}\right)}\,, \tag{8.2.6}$$

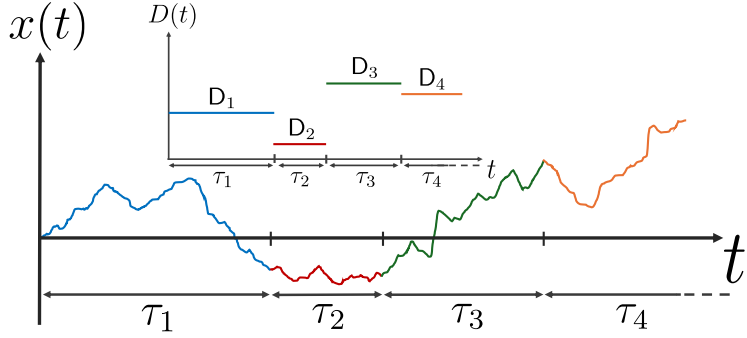

**Figure 8.1:** Trajectory of a switching diffusion process in one-dimension. During each time interval $\tau_i$, the particle performs an independent Brownian motion with a diffusion constant $\mathsf{D}_i$. In the simplest case, the $\tau_i$'s are independent exponential random variables, while the $\mathsf{D}_i$'s, which are also independent, are drawn from an arbitrary distribution $W(D)$.

where $I(y)$ is a large deviation function (LDF), whose precise shape depends on $W(D)$. However, its asymptotic behaviors for small and large arguments are universal and are given by

$$I(y) \approx \begin{cases} \frac{y^2}{4\langle D \rangle} & , \quad y \to 0 \,, \\[6pt] r + \frac{y^2}{4D_{\max}} & , \quad y \to \infty \,. \end{cases} \tag{8.2.7}$$

Here, $D_{\max}$ denotes the right edge of the support of $W(D)$. These two asymptotic behaviors can be physically understood as follows. When $y \to 0$, i.e. $x \ll t$, the Gaussian behavior near the center of the PDF picks up the average $\langle D \rangle$ (since there are many switchings, the particle samples the average of $D$). On the other hand, for $y \to \infty$, i.e., $x \gg t$, this behavior is due to very rare trajectories where the particle diffuses with the largest diffusion constant $D_{\max}$ without undergoing any switch, which occurs with a probability $e^{-rt}$. Hence, $x \sim O(t)$ acts like a separatrix between typical and atypical trajectories. As we show below, for some distributions $W(D)$, these two asymptotic behaviors (8.2.7) are separated by a transition point where the function $I(y)$ displays a singularity, signaling the presence of a dynamical transition. In some cases, like in the two-state model (i.e., when $W(D) = p\delta(D - D_1) + (1-p)\delta(D - D_2)$ with $0 \le p \le 1$), there may even be two transitions (see Fig. 9.6).

## 8.3 Main results for the confined case

In the presence of a harmonic potential, we show that the Fourier transform of the non-equilibrium steady state satisfies an integral equation, which we derive using the Kesten approach introduced in Part II. The equation reads

$$\hat{p}(k) = \int_0^{D_{\max}} dD\, W(D) \int_0^1 dU\, \beta\, U^{\beta-1}\, e^{-\frac{k^2 \alpha(U)}{2}}\, \hat{p}(kU) \quad , \quad \alpha(U) = \frac{D}{\mu}(1 - U^2). \tag{8.3.1}$$

We solve this integral equation exactly for arbitrary distributions $W(D)$ and show that the solution is given by

$$\hat{p}(k) = \sum_{n=0}^{+\infty} \frac{(ik)^{2n}}{(2\mu)^n} \frac{\Gamma\left(\frac{\beta}{2}\right)}{\Gamma\left(\frac{\beta}{2} + n\right)} \frac{B_n\left(1!\,\frac{\beta}{2}\,\frac{\langle D \rangle}{1}, \ldots, n!\,\frac{\beta}{2}\,\frac{\langle D^n \rangle}{n}\right)}{n!}. \tag{8.3.2}$$

From this solution, we can extract the exact moments of the stationary state:

$$\langle x^{2n} \rangle = \frac{(2n)!}{(2\mu)^n} \frac{\Gamma\left(\frac{\beta}{2}\right)}{\Gamma\left(\frac{\beta}{2}+n\right)} \frac{1}{n!} B_n\left(1! \frac{\beta}{2} \frac{\langle D \rangle}{1}, \ldots, n! \frac{\beta}{2} \frac{\langle D^n \rangle}{n}\right). \qquad (8.3.3)$$

We then show that the small and large $\beta = r/\mu$ behaviors of the stationary state are given by

$$p(x) \simeq \begin{cases} \int_0^{D_{\max}} dD\, W(D) \sqrt{\frac{\mu}{2\pi D}}\, e^{-\frac{\mu x^2}{2D}} & , \quad \beta \ll 1, \\ \sqrt{\frac{\mu}{2\pi \langle D \rangle}}\, e^{-\frac{\mu x^2}{2\langle D \rangle}} & , \quad \beta \gg 1. \end{cases} \qquad (8.3.4)$$

The asymptotic behaviors described are similar to a free switching diffusion process, where $\beta = r/\mu$ plays the role of $r\,t$, i.e. the mean number of switches in a period $t$. In the limit $\beta \to 0$ (low switching frequency, $r \to 0$), the diffusion remains in its initial state $\mathsf{D}_1$, yielding the Ornstein-Uhlenbeck stationary state averaged over all possible initial conditions. Conversely, in the large $\beta$ limit ($r \to +\infty$), frequent switching leads to an effective diffusion constant equal to the average value $\langle D \rangle$, resulting in the OU stationary state characterized by this averaged diffusion constant.

Similar to the free-switching diffusion process, we also show that the cumulants undergo a crossover: at small $\beta$, they are proportional to the cumulants of $D$, while at large $\beta$, they are proportional to the free cumulants of $D$. The asymptotic behavior can be expressed as follows

$$\langle x^{2n} \rangle_c \simeq \begin{cases} \frac{(2n-1)!!}{r^n} \langle D^n \rangle_c \, \beta^n & , \quad \beta \ll 1, \\ \frac{(2n-1)!}{r^n} \kappa_n(D)\, \beta & , \quad \beta \gg 1. \end{cases} \qquad (8.3.5)$$

Finally, we apply our result to obtain the exact NESS of the two-state model with $W(D) = p\, \delta(D - D_1) + (1-p)\, \delta(D - D_2)$ in the special case $D_2 = 0$. This model is called *randomly interrupted diffusion* and was studied in [138] in the special case $p = 1/2$. For all values of $p$, we find

$$p(x) = \sqrt{\frac{\mu}{2\pi D}} \frac{\Gamma\left(\frac{\beta}{2}\right)}{\Gamma\left(\frac{\beta p}{2}\right)} e^{-\frac{\mu x^2}{2D}} U\left(\frac{\beta}{2}(1-p), \frac{3}{2} - \frac{\beta p}{2}, \frac{\mu x^2}{2D}\right), \qquad (8.3.6)$$

where $U$ is Tricomi's confluent hypergeometric function. We also compare this result with numerical simulations and observe excellent agreement – see Fig. 10.1.

# Chapter 9

# Switching Diffusion: The Free Case

## 9.1 Renewal Equation and Explicit Solution

We first derive a renewal equation for the joint distribution $p_r(x, D, t|\mathsf{D}_1)$ of the position $x(t)$ and the diffusing coefficient $D(t)$ at time $t$ conditioned on the first diffusion constant value (see Fig. 9.1 for an illustration)

$$p_r(x, D, t|\mathsf{D}_1) = e^{-rt} \frac{e^{-\frac{x^2}{4\mathsf{D}_1 t}}}{\sqrt{4\pi \mathsf{D}_1 t}} \delta(D - \mathsf{D}_1) +$$

$$\int_0^t d\tau\, r\, e^{-r\tau} \int_{-\infty}^{+\infty} dz \int_0^{+\infty} dD'\, p_r(z, D', t-\tau|\mathsf{D}_1)\, W(D) \frac{e^{-\frac{(x-z)^2}{4D\tau}}}{\sqrt{4\pi D\tau}}. \quad (9.1.1)$$

The first contribution comes from the event, that occurs with probability $e^{-rt}$, where there is no reset up to time $t$ and the dynamics follow a simple Brownian motion with diffusion constant $\mathsf{D}_1$. The second term accounts for the event where the last reset occurred at time $t - \tau$, at which point the particle was at position $z$ with diffusion constant $D'$. The probability that no reset occurred between $t - \tau$ and $t$ is $e^{-r\tau}$, while the probability of a reset occurring within the small time interval $[t - \tau, t - \tau + d\tau]$ is $r d\tau$. To account for all possible reset times, we integrate over $\tau$. Next, we integrate over $z$ and $D'$, taking into account the propagator $p_r(z, D', t - \tau)$ that describes the paths from the origin $x = 0$ at $t = 0$ to position $z$ at time $t - \tau$. We also include the Gaussian propagator that governs the motion from $z$ at time $t - \tau$ to $x$ at time $t$. Finally, we need to account for the transition probability of the diffusion constant changing from $D'$ (the value just before the reset at time $t - \tau$) to $D$ (the value immediately after the reset). Since the diffusion constants are i.i.d., this is simply given by the distribution $W(D)$. If we integrate the renewal equation (9.1.1) over all values of $D$ and perform the integral over $D'$ in (9.1.1), we obtain an integral equation for the propagator of $x(t)$ conditioned on the value $\mathsf{D}_1$. This is given by

$$p_r(x, t|\mathsf{D}_1) = e^{-rt} \frac{e^{-\frac{x^2}{4\mathsf{D}_1 t}}}{\sqrt{4\pi \mathsf{D}_1 t}}$$

$$+ \int_0^t d\tau\, r e^{-r\tau} \int_{-\infty}^{+\infty} dz\, p_r(z, t-\tau|\mathsf{D}_1) \int_0^{+\infty} dD\, W(D) \frac{e^{-\frac{(x-z)^2}{4D\tau}}}{\sqrt{4\pi D\tau}}. \quad (9.1.2)$$

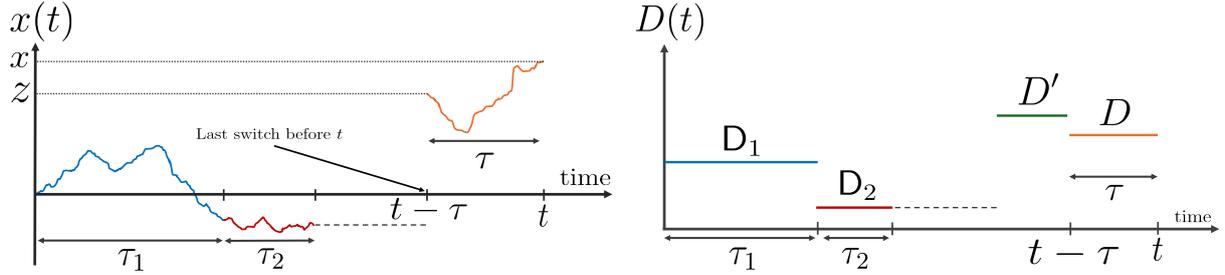

**Figure 9.1:** Schematic description of a typical time evolution of $x(t)$ and $D(t)$. In the model studied here, the $\tau_i$'s are independent exponential random variables, while the $\mathsf{D}_i$'s, which are also independent, are drawn from an arbitrary distribution $W(D)$. As considered in the renewal equation (9.1.1), two situations are considered. The first one is when there is no switch, which happens with probability $e^{-rt}$, and we do not represent it on the figure. Here, we show the second situation where the last switch happens at time $t - \tau$. At this specific time, the particle is at position $z$, and right before the switch, its diffusion constant is $D'$. At time $t$, the particle is at position $x$ with diffusion constant $D$.

A more symmetrized expression can be obtained by averaging over the values of $\mathsf{D}_1$, resulting in

$$p_r(x,t) = e^{-rt} \int_0^{+\infty} dD\, W(D) \frac{e^{-\frac{x^2}{4Dt}}}{\sqrt{4\pi Dt}}$$
$$+ \int_0^t d\tau\, r e^{-r\tau} \int_{-\infty}^{+\infty} dz\, p_r(z, t-\tau) \int_0^{+\infty} dD\, W(D) \frac{e^{-\frac{(x-z)^2}{4D\tau}}}{\sqrt{4\pi D\tau}}\,. \quad (9.1.3)$$

We now assume in the remainder of this section that $W(D)$ has a finite support on $[0, D_{\max}]^7$. We first note that the second term exhibits a convolution structure in space. To leverage this property, we introduce the generating function of $x$ which is defined as the bilateral Laplace transform (BLT) of $p_r(x,t)$. It is given by

$$\hat{p}_r(q,t) = \langle e^{qx}\rangle = \int_{-\infty}^{+\infty} dx\, e^{qx}\, p_r(x,t)\,. \quad (9.1.4)$$

Note that the normalisation of the PDF of $p_r(x,t)$ implies

$$\hat{p}_r(q=0, t) = \int_{-\infty}^{+\infty} dx\, p_r(x,t) = 1 \quad,\quad \text{for all } t\,. \quad (9.1.5)$$

We also recall that the inversion formula is given by

$$p_r(x,t) = \frac{1}{2i\pi} \int_{\gamma - i\infty}^{\gamma + i\infty} dq\, e^{-qx}\, \hat{p}_r(q,t)\,, \quad (9.1.6)$$

where the integral runs over the Bromwich contour which, in this case of a bilateral Laplace transform, lies within the region of convergence of $\hat{p}_r(q,t)$ in the complex $q$-plane. More precisely, the real $\gamma$ in (9.1.6) is such that $\gamma \in ]s_0, s_1[$ where $]s_0, s_1[$ is the maximal real interval such that $\hat{p}_r(q,t)$ is an analytic function in the vertical strip delimited by $]s_0, s_1[$ in the complex $q$-plane. Taking the BLT of Eq. (9.1.3) yields

$$\hat{p}_r(q,t) = e^{-rt} \int_0^{D_{\max}} dD\, W(D)\, e^{Dq^2 t} + \int_0^{D_{\max}} dD\, W(D) \int_0^t d\tau\, r\, e^{-r\tau}\, e^{Dq^2\tau}\, \hat{p}_r(q, t-\tau)\,. \quad (9.1.7)$$

Similarly, the convolution structure in time can be exploited to obtain a closed equation via the Laplace transformation with respect to the time variable $t$. It is thus useful to introduce the Laplace transform (with respect to time) defined as

$$\tilde{p}_r(q,s) = \int_0^{+\infty} dt\, e^{-st}\, \hat{p}_r(q,t)\,. \quad (9.1.8)$$

---

[7] The case where $W(D)$ has support on $[0, +\infty)$ is discussed in the supplementary material of [1].

Taking the Laplace transform of Eq. (9.1.7) yields

$$\tilde{p}_r(q,s) = \int_0^{D_{\max}} dD\, \frac{W(D)}{r+s-Dq^2} + r \int_0^{D_{\max}} dD\, \frac{W(D)}{r+s-Dq^2}\, \tilde{p}_r(q,s)\,. \qquad (9.1.9)$$

Ultimately, the explicit solution takes the form

$$\boxed{\tilde{p}_r(q,s) = \frac{J_r(q,s)}{1 - r\, J_r(q,s)}\,,\ \ J_r(q,s) = \int_0^{D_{\max}} dD\, \frac{W(D)}{r+s-Dq^2}}\,. \qquad (9.1.10)$$

When $q=0$, one has $J_r(0,s) = 1/(r+s)$ and we easily check from (9.1.10) that $\tilde{p}_r(0,s) = \frac{1}{s}$, which is consistent with the normalisation condition (9.1.5). Note that these functions in the $(q,s)$-plane are defined in the region $r+s > D_{\max} q^2$, such that the integrand defining $J_r(s,q)$ in (9.1.10) is free of any singularity.

**Remark.** It is also possible to derive an integral equation for the Fourier transform of $p(x,t)$ [1]. It reads

$$\hat{p}_r(k,t) = e^{-rt} \int dD\, W(D) e^{-Dk^2 t} + \int dD\, W(D) \int_0^t d\tau\, r e^{-r\tau} e^{-Dk^2 \tau} \hat{p}_r(k, t-\tau)\,, \qquad (9.1.11)$$

$$\hat{p}_r(k,t) = \int_{-\infty}^{+\infty} dx\, e^{ikx}\, p_r(x,t)\,. \qquad (9.1.12)$$

This integral equation is also well-defined for distributions $W(D)$ with infinite support. It will be particularly useful for numerically computing the distribution $p(x,t)$ via numerical inverse Fourier transform methods — see Appendix E.

## 9.2 Exact Expressions for the Moments

Deriving the formula for the moments given in Eq. (8.2.1) was a key step in advancing our understanding of the switching diffusion model and in establishing its connection to free cumulants. In fact, this expression enabled us to compute the first few cumulants using Eq. (8.1.6), revealing a clear correspondence with the structure of free cumulants in the large time limit.

The derivation of Eq. (8.2.1), as announced in the main results, is technically involved and presented in the supplementary material of [1]. In this section, we outline the main steps of the derivation to provide an overall idea of the approach. Since the distribution $p_r(x,t)$ is symmetric, i.e., $p_r(x,t) = p_r(-x,t)$, all odd moments vanish. To begin, we derive a recursive relation for the moments of the probability distribution function $p_r(x,t)$. This relation can be obtained by expanding $\hat{p}_r(q,t)$ within Eq. (9.1.7), using the series expansion

$$\hat{p}_r(q,t) = \sum_{n=0}^{+\infty} \frac{q^{2n}}{(2n)!} \langle x^{2n}(t) \rangle\,. \qquad (9.2.1)$$

First, we substitute this power series expansion in Eq. (9.1.7)

$$\sum_{n=0}^{+\infty} \frac{q^{2n}}{(2n)!} \langle x^{2n}(t) \rangle = e^{-rt} \sum_{n=0}^{+\infty} \langle D^n \rangle \frac{q^{2n} t^n}{n!}$$
$$+ \int_0^t d\tau\, r e^{-r\tau} \sum_{p=0}^{+\infty} \langle D^p \rangle \frac{q^{2p} \tau^p}{p!} \sum_{l=0}^{+\infty} \frac{q^{2l}}{(2l)!} \langle x^{2l}(t-\tau) \rangle\,, \qquad (9.2.2)$$

and we select the term of order $q^{2n}$ on both sides of to get

$$\langle x^{2n}(t)\rangle = e^{-rt}\langle D^n\rangle t^n \frac{(2n)!}{n!} + \int_0^t d\tau\, r\, e^{-r\tau} \sum_{\substack{p,l=0\\p+l=n}}^{+\infty} \langle D^p\rangle \tau^p \frac{(2n)!}{p!(2l)!}\langle x^{2l}(t-\tau)\rangle. \quad (9.2.3)$$

Further simplifications leads to

$$\frac{(r+s)^n}{(2n)!}\mathcal{L}_{t\to s}\left[\langle x^{2n}(t)\rangle\right] = \frac{1}{s}\left[\langle D^n\rangle + \sum_{l=0}^{n-1} r\,\langle D^{n-l}\rangle \left(\frac{(r+s)^l}{(2l)!}\mathcal{L}_{t\to s}\left[\langle x^{2l}(t)\rangle\right]\right)\right], \quad (9.2.4)$$

where $\mathcal{L}_{t\to s}[f(t)]$ denotes the Laplace transform of $f(t)$. We can now define a recursive sequence $u_n$ as follows

$$u_n = \frac{\langle D^n\rangle}{s} + \sum_{l=0}^{n-1}\frac{r}{s}\langle D^{n-l}\rangle u_l. \quad (9.2.5)$$

$$u_n = \frac{(r+s)^n}{(2n)!}\mathcal{L}_{t\to s}\left[\langle x^{2n}(t)\rangle\right], \quad u_0 = \frac{1}{s}. \quad (9.2.6)$$

The nontrivial part of the derivation lies in solving the recursive relation for $u_n$, which we carry out in detail in [1]. We show that

$$u_n = \frac{r+s}{r\,s}\sum_{m=1}^n \left(\frac{r}{s}\right)^m \hat{B}_{n,m}\left(\langle D\rangle,\ldots,\langle D^{n-m+1}\rangle\right), \quad (9.2.7)$$

where $\hat{B}_{p,m}$ is the ordinary Bell polynomial – see Eq. (8.1.4). From the definition of $u_n$ given in Eq. (9.2.6), we deduce

$$\mathcal{L}_{t\to s}\left[\langle x^{2n}(t)\rangle\right] = \frac{(2n)!}{r\,s\,(r+s)^{n-1}}\sum_{m=1}^n \left(\frac{r}{s}\right)^m \hat{B}_{n,m}\left(\langle D\rangle,\ldots,\langle D^{n-m+1}\rangle\right). \quad (9.2.8)$$

Since we have

$$\mathcal{L}_{s\to t}^{-1}\left[s^{-(m+1)}(r+s)^{-(n-1)}\right] = t^{m+n-1}\frac{M(n-1,m+n,-rt)}{(m+n-1)!}, \quad (9.2.9)$$

where $M(a,b,x)$ denotes the Kummer's function, taking the inverse Laplace transform of Eq. (9.2.8) using Eq. (9.2.9) then leads to the final result

$$\boxed{\langle x^{2n}(t)\rangle = \frac{(2n)!}{r^n}\sum_{m=1}^n \frac{(rt)^{m+n-1}}{(m+n-1)!} M(n-1,m+n,-rt)\, \hat{B}_{n,m}\left(\langle D\rangle,\ldots,\langle D^{n-m+1}\rangle\right)} \quad (9.2.10)$$

We show below the first three non-zero moments of the position of the switching diffusion process. These moments are computed directly from the exact expression provided in Eq. (9.2.10). They are as follows

$$\langle x^2(t)\rangle = 2\langle D\rangle t, \quad (9.2.11)$$

$$\langle x^4(t)\rangle = \frac{e^{-rt}}{r^2}\left[12(-2+e^{rt}(2+rt(-2+rt)))\langle D\rangle^2 + 24(1+e^{rt}(-1+rt))\langle D^2\rangle\right], \quad (9.2.12)$$

$$\langle x^6(t)\rangle = \frac{120\,e^{-rt}}{r^3}\left\{\left[6(4+rt)+e^{rt}\left(-24+rt(18+rt(-6+rt))\right)\right]\langle D\rangle^3\right.$$

$$\left.+6\left[-2(3+rt)+e^{rt}(6+rt(-4+rt))\right]\langle D\rangle\langle D^2\rangle + 6\left(2+rt+e^{rt}(-2+rt)\right)\langle D^3\rangle\right\}. \quad (9.2.13)$$

These expressions are valid for any distribution $W(D)$ for which the first three moments are well defined.

From Eq. (9.2.10), and using Eq. (8.1.6), it is possible to compute the cumulants for specific distributions $W(D)$ – see Fig. 9.2 for a numerical check in the case of the two-state model (left panel), and for the Wigner semi-circle distribution (right panel).

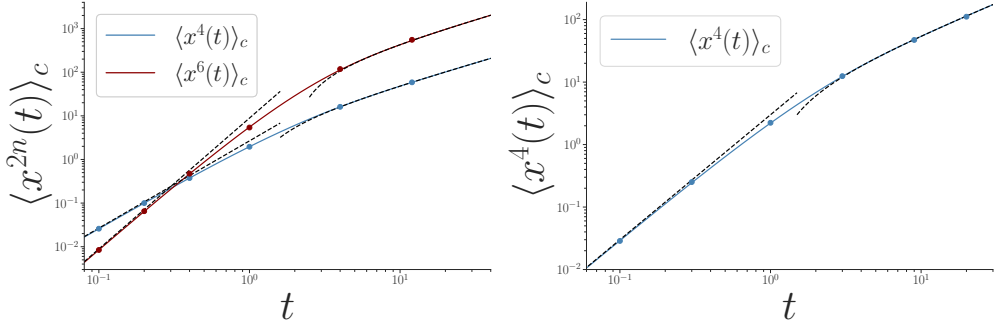

**Figure 9.2:** Using the explicit formulas for the moments given in Eq (9.2.10), we calculate and plot the cumulants (solid lines) and compare them with numerical simulations (dots). The left panel shows results for $W(D)$ for the two-state model – i.e., $W(D) = p\delta(D - D_1) + (1-p)\delta(D - D_2)$, while the right panel corresponds to the Wigner semi-circle law – i.e., $W(D) = 8/(\pi D_{\max}^2)\sqrt{D(D_{\max} - D)}$. The black dotted lines represent the asymptotic predictions given in Eq (8.2.2). Our theoretical predictions show excellent agreement with the simulations for both models. The parameters for the two-state model are: $r = 1$, $p = 1/3$, $D_2 = 1$, $D_1 = 2$. For the Wigner semi-circle law, the parameters are: $r = 1$, $D_{\max} = 1$.

## 9.3 Small Time Limit

It is possible to show from Eq. (9.2.10) that the short-time behavior of the moments is given by

$$\langle x^{2n}(t) \rangle \underset{rt \ll 1}{\approx} \frac{(2n)!}{(n!)^2} B_{n,1}\left(1!\langle D \rangle, \ldots, n!\langle D^n \rangle\right) t^n = \frac{(2n)!}{n!} \langle D^n \rangle \, t^n, \quad (9.3.1)$$

where we have used the identity $B_{n,1}(x_1, \ldots, x_n) = x_n$. The corresponding Fourier transform then reads

$$\hat{p}(k,t) \underset{rt \ll 1}{\approx} \sum_{n=0}^{+\infty} \frac{(ik)^{2n}}{n!} t^n \langle D^n \rangle = \int_0^{D_{\max}} dD \, W(D) \, e^{-Dk^2 t} \, . \quad (9.3.2)$$

Therefore, for $rt \ll 1$ and $x \sim O\left(\sqrt{t}\right)$, the distribution in real space is

$$\boxed{p(x,t) \underset{rt \ll 1}{\approx} \int_0^{D_{\max}} dD \, W(D) \, \frac{e^{-\frac{x^2}{4Dt}}}{\sqrt{4\pi Dt}}}, \quad (9.3.3)$$

which is indeed the result given in Eq. (8.2.5).

We can also extract the small-time cumulants from Eq. (9.3.3). In fact, the series expansion of the logarithm of the moment-generating function of $W(D)$ can be written in terms of the cumulants of $D$ as

$$\log\left[\langle e^{sD} \rangle\right] = \log\left[\int_0^{D_{\max}} dD \, W(D) e^{sD}\right] = \sum_{n=1}^{+\infty} \frac{s^n}{n!} \langle D^n \rangle_c \, . \quad (9.3.4)$$

Similarly, we have for small times

$$\log\left[\hat{p}(k,t)\right] \underset{rt \ll 1}{\approx} \log\left[\int_0^{D_{\max}} dD \, W(D) \, e^{-Dk^2 t}\right] = \sum_{n=1}^{+\infty} \frac{(ik)^{2n}}{(2n)!} \langle x^{2n}(t) \rangle_c \, . \quad (9.3.5)$$

Taking $s = -tk^2$ allows us to identify the cumulants of the position as

$$\boxed{\langle x^{2n}(t) \rangle_c \underset{rt \ll 1}{\approx} \frac{(2n)!}{n!} \langle D^n \rangle_c \, t^n}, \quad (9.3.6)$$

as stated in the first line of Eq. (8.2.2).

## 9.4 Large Time Limit and Large Deviations

In this section, we derive the large time asymptotic behavior of the BLT. Specifically, we show that

$$\hat{p}_r(q,t) \underset{t \to +\infty}{\approx} e^{t\Psi(q)}, \qquad (9.4.1)$$

where $\Psi(q)$ is the Scaled Cumulant Generating Function (SCGF) which encapsulate information about the cumulants and characterizes the non-Gaussian fluctuations present in the probability distribution $p(x,t)$. Moreover, we highlight how, within this model, the connection to free cumulants emerges naturally from the small-argument behavior of the SCGF. At large times, the position distribution takes the form

$$p_r(x,t) \underset{t \to +\infty}{\approx} e^{tI\left(y=\frac{x}{t}\right)}, \qquad (9.4.2)$$

where $I(y)$ is identified as the large deviation function (LDF). The large deviation function $I(y)$ and the SCGF $\Psi(q)$ are connected through a Legendre transform, provided the maximum in the following equation exists

$$I(y) = \max_{q \in \mathbb{R}} \left( q\, y - \Psi(q) \right). \qquad (9.4.3)$$

We then provide the details of the study of the scaled cumulant generating function (SCGF) $\Psi(q)$ and the rate function $I(y)$ in the case where $W(D)$ has a finite support $[0, D_{\max}]$ such that $W(D) \sim (D_{\max} - D)^\nu$ when $D \to D_{\max}$, $\nu > -1$. This includes the case where $W(D)$ follows a semi-circular distribution (corresponding to $\nu = 1/2$), a uniform distribution (for which $\nu = 0$). These cases, as well as the two-state distribution, are analyzed in the next section.

### 9.4.1 The Scaled Cumulant Generating Function $\Psi(q)$

**Behavior at Small Argument**

The large time behavior of the Fourier transform can be obtained by taking the inverse Laplace transform of Eq. (9.1.10) as follows

$$\hat{p}_r(q,t) = \frac{1}{2i\pi} \int_\Gamma ds\, e^{st}\, \tilde{p}_r(q,s) = \frac{1}{2i\pi} \int_\Gamma ds\, e^{st}\, \frac{J_r(q,s)}{1 - r\, J_r(q,s)} \underset{t \to \infty}{\approx} e^{t\Psi(q)}, \qquad (9.4.4)$$

$$J_r(q,s) = \int_0^{D_{\max}} dD\, \frac{W(D)}{r + s - Dq^2}. \qquad (9.4.5)$$

In Eq. (9.4.4), $\Gamma$ is a Bromwich contour passing to the right of all the singularities of $\tilde{p}_r(q,s)$ in the complex $s$-plane, and $\Psi(q)$ is the singularity of $\tilde{p}_r(q,s)$ with the largest real part. In general, $\tilde{p}_r(q,s)$ admits two types of singularities: (a) the one arising from $J_r(q,s)$ itself – which is a branch cut – that exists for all values of $\nu > -1$ and (b) poles – which are the roots of the denominator in $\tilde{p}_r(q,s)$ – which exists only for certain values of $\nu$ and $q$ (see below).

**(a) The branch cut**: Since the integral over $D$ defining $J_r(q,s)$ in Eq. (9.4.5) has a non-integrable singularity for $D = (r+s)/q^2$, the function $J_r(q,s)$, and hence $\tilde{p}_r(q,s)$ has a branch-cut on the real axis $[-r, -r + D_{\max}q^2]$. More precisely, $J_r(q,s)$ can be written as,

$$J_r(q,s) = \frac{1}{q^2} g\left(\tilde{s} = \frac{r+s}{q^2}\right) \quad , \quad g(z) = \int_0^{D_{\max}} \frac{W(D)}{z - D}\, ,\ z \notin [0, D_{\max}]\, , \qquad (9.4.6)$$

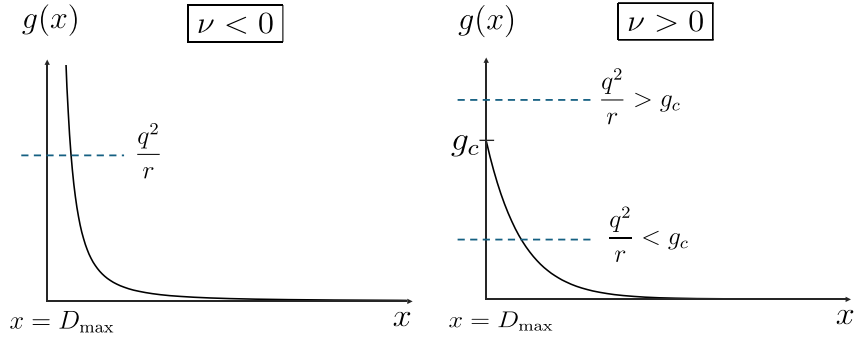

**Figure 9.3:** We illustrate here the two situations described in Eqs. (9.4.8) and (9.4.9). For a distribution $W(D) \sim (D_{\max} - D)^\nu$, the Cauchy-Stieltjes transform of $W(D)$, denoted by $g(x)$ for $x \in (D_{\max}, +\infty)$, exhibits two different behaviors for $x \to D_{\max}$. If $\nu < 0$ (left panel), $g(x \to D_{\max}) \sim (D_{\max} - x)^\nu$ diverges as $x \to D_{\max}$, while if $\nu > 0$ (right panel), $g(x \to D_{\max}) = g_c$ which is finite.

where $g(z)$ is the Cauchy-Stieltjes transform of $W(D)$.

**(b) The pole.** To find the pole of $\tilde{p}_r(q, s)$ in the complex $s$-plane, we have to solve (for $s$) the equation

$$1 - r\, J_r(q, s^*) = 0 \iff 1 = r \int_0^{D_{\max}} dD \, \frac{W(D)}{r + s^* - D\,q^2} \iff \frac{q^2}{r} = g\left(\tilde{s} = \frac{r + s^*}{q^2}\right), \quad (9.4.7)$$

where $g(x)$ is defined in (9.4.6). The pole $s^*$ must satisfy two conditions. First, the point $s^*$ does not lie on the branch cut, i.e., $s^* \notin [-r, -r + D_{\max}q^2]$. Additionally, it must satisfy $J_r(q, s^*) > 0$, since by definition $J_r(q, s^*) = 1/r > 0$. This implies that $s^*$ must lie to the right of the branch cut, i.e., $s^* > -r + D_{\max}q^2$. Therefore, whenever this pole exists, it corresponds to the singularity with the largest real part. Note that the discussion of a similar equation appeared in the study of spherical integrals (i.e., Harish Chandra/Itzykson-Zuber integrals) in random matrix theory [27].

We now argue that such a pole always exists for sufficiently small values of $q$. To proceed, we analyse the behavior of this function $g(x)$ for real $x \in (D_{\max}, +\infty)$. Note that as $g'(x) < 0$, it is a decreasing function of $x$, and we have $g(x) \sim 1/x$ when $x \to \infty$. To investigate the behavior of $g(x)$ when $x \to D_{\max}$, we need to distinguish two cases (see Fig. 9.3):

$$\text{(i)} \quad \nu > 0, \quad g(x \to D_{\max}) = g_c = \int_0^{D_{\max}} dD \, \frac{W(D)}{D_{\max} - D}, \quad (9.4.8)$$

$$\text{(ii)} \quad \nu < 0, \quad g(x \to D_{\max}) \underset{D \to D_{\max}}{\propto} (D_{\max} - x)^\nu. \quad (9.4.9)$$

While, for $\nu > 0$ the function $g(x)$ is bounded from above (since $g_c$ is finite), in the case $\nu < 0$, the function $g(x)$ diverges as $x \to D_{\max}$ (see Fig. 9.3). Thus, for sufficiently small values of $q$, $\Psi(q)$ is given by the unique solution of Eq. (9.4.7) (which exists for all $\nu > -1$).

In such a case, using the fact that for a real probability measure $W(D)$, we have the following relation between the *R-transform* and the Cauchy-Stieltjes transform (see e.g. Theorem 9.23 of [323] and Eq. (8.1.19))

$$g\left[R(z) + \frac{1}{z}\right] = z, \quad (9.4.10)$$

then, one can show that

$$\boxed{\Psi(q) = \lim_{t \to \infty} \frac{\ln \hat{p}_r(q, t)}{t} = q^2 \, R\left(\frac{q^2}{r}\right), \quad \text{when } q \to 0}. \quad (9.4.11)$$

**Large Time Behavior of the Cumulants**

We have shown that at small $q$, the following relation holds

$$\Psi(q) = q^2 R\left(\frac{q^2}{r}\right) = r \sum_{n=1}^{+\infty} \left(\frac{q^2}{r}\right)^n \kappa_n(D), \quad (9.4.12)$$

where $R(z)$ is the $R$-transform (i.e., the generating function of the free cumulants $\kappa_n(D)$). On the other hand, we also have

$$\ln \hat{p}_r(q,t) = \sum_{n=1}^{+\infty} \frac{q^{2n}}{(2n)!} \langle x^{2n}(t) \rangle_c \underset{t \to +\infty}{\approx} t \Psi(q). \quad (9.4.13)$$

By identifying terms in the power series expansions, we obtain

$$\boxed{\langle x^n(t) \rangle_c \underset{t \to \infty}{\approx} \frac{(2n)!}{r^{n-1}} \kappa_n(D) \, t}. \quad (9.4.14)$$

As we show in [1], it is also possible to calculate the pre-exponential factor of $\hat{p}_r(q,t)$ and the $O(1)$ corrections to the cumulants. This term can be expressed as a sum involving Bell polynomials, whose arguments are the free cumulants of the random variable $D$, which is distributed according to $W(D)$. The resulting expression is given by

$$\langle x^n(t) \rangle_c \underset{t \to +\infty}{\approx} \frac{(2n)!}{r^{n-1}} \kappa_n(D) t - \frac{(2n)!}{r^n} \sum_{m=1}^{n} \frac{1}{m} \hat{B}_{n,m}\left(\tilde{\kappa}_1(D), \cdots, \tilde{\kappa}_{n-m+1}(D)\right) + O(e^{-rt}) \quad (9.4.15)$$

$$\tilde{\kappa}_n(D) = (n-1)\kappa_n(D). \quad (9.4.16)$$

With the pre-exponential factor, the BLT reads

$$\hat{p}_r(q,t) \underset{t \to +\infty}{\approx} \left(1 - \frac{q^4}{r^2} R'\left(\frac{q^2}{r}\right)\right) e^{tq^2 R\left(\frac{q^2}{r}\right)}, \quad \text{when } q \to 0. \quad (9.4.17)$$

**Behavior for All Values of $q$**

The full derivation of the behavior of $\Psi(q)$ is technically involved and is presented in detail in [1]. Here, we provide a summary of the main results.

For any discrete or continuous distribution $W(D)$ with a finite support on $[0, D_{\max}]$, the asymptotic behaviors of the SCGF are

$$\boxed{\Psi(q) = \begin{cases} \langle D \rangle q^2 & , \quad q \to 0, \\ D_{\max} q^2 - r & , \quad q \to \infty. \end{cases}} \quad (9.4.18)$$

What happens between these two limits depends essentially on the behavior of $W(D)$ near $D_{\max}$, as in the extreme value statistics in the Weibull universality class [61]. Let us assume that $W(D)$ behaves as $W(D) \sim (D_{\max} - D)^\nu$ when $D \to D_{\max}$ with $\nu > -1$.

- For $-1 < \nu \leq 0$, $\Psi(q)$ is given by Eq. (9.4.11) for all $q$ and it is an analytic function of all $q \in \mathbb{R}$.

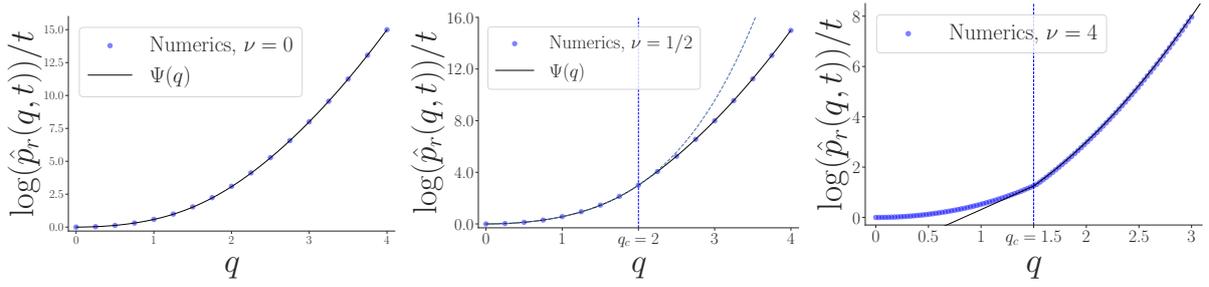

**Figure 9.4:** We show here theoretical predictions (solid lines) against numerical results (dots) for $\Psi(q)$ when $W(D) \sim (D_{\max} - D)^\nu$ on $[0,1]$ with $r = 1$ and $t = 1000$. The numerical procedure used to compute $\hat{p}_r(q,t)$ is detailed in Appendix E. **Left:** $\nu = 0$. Uniform distribution (no transition). **Middle:** $\nu = 1/2$. Wigner distribution (second order transition). The blue dotted line corresponds to the first line of Eq. (9.5.15) to emphasize the change of behavior at $q_c$. **Right:** $\nu = 4$. Beta distribution such that $W(D) \sim (1-D)^4$ (first order transition). For $q < q_c$, the linear behavior close to $q_c$ is given in [1]. For $q > q_c$, we have $\Psi(q) = D_{\max} q^2 - r$.

- Instead, for $\nu > 0$, (9.4.11) only holds for small $q$, i.e.,

$$\Psi(q) = \begin{cases} q^2 \, R\left(q^2/r\right) &, \quad q < q_c \,, \\ D_{\max} q^2 - r &, \quad q > q_c \,, \end{cases} \qquad (9.4.19)$$

where the SCGF undergoes a transition at $q = q_c$, with $q_c^2 = r g_c$, – where $g_c$ is given in Eq. (10.1.47). While $\Psi(q)$ is continuous, its higher derivatives display singularities at $q = q_c$ (see [1] for details). In particular, for $\nu > 1$ (as well as for the two-sate model in the limit $p \to 0^+$), the first derivative of $\Psi(q)$ is discontinuous – see the right panel of Fig. 9.4.

### 9.4.2 The Large Deviation Function $I(y = x/t)$

From the standard theory of large deviations [22, 24], namely the Gärtner-Ellis theorem, the exponential form of the SCGF in (9.4.1) implies the following large deviation form of $p_r(x,t)$

$$p_r(x,t) \underset{t \to +\infty}{\approx} e^{-t\, I(y=x/t)} \,, \qquad (9.4.20)$$

where the LDF $I(y)$ is given by the Legendre transform of $\Psi(q)$ (whenever $\Psi(q)$ is differentiable for all $q \in \mathbb{R}$), namely

$$I(y) = \max_{q \in \mathbb{R}}(q\,y - \Psi(q)) \,. \qquad (9.4.21)$$

Using this formula and the asymptotics of $\Psi(q)$ from Eq. (9.4.18), we find that $I(y)$ behaves as in Eq. (9.4.22). Since $I(y)$ is symmetric, we study it only for $y \geq 0$.

$$I(y) \approx \begin{cases} \frac{y^2}{4 \langle D \rangle} &, \quad y \to 0 \,, \\ r + \frac{y^2}{4 D_{\max}} &, \quad y \to \infty \,. \end{cases} \qquad (9.4.22)$$

For a distribution $W(D)$ with a finite support $[0, D_{\max}]$ as discussed above with $-1 < \nu \leq 0$, the LDF $I(y)$ is regular and crosses over smoothly between these two asymptotic behaviors in (9.4.22). This is, for instance, the case of a uniform distribution 9.5.2. However, for $0 < \nu < 1$, the LDF exhibits a dynamical transition of the form

$$I(y) = \begin{cases} \phi_\nu(y) &, \quad y \leq y_c = 2D_{\max} q_c\,, \\ r + \frac{y^2}{4D_{\max}} &, \quad y \geq y_c\,, \end{cases} \qquad (9.4.23)$$

where $\phi_\nu(y)$ is the Legendre transform of Eq. (9.4.11) – which we can compute explicitly in the case of the Wigner semi-circle law ($\nu = 1/2$) – see Section 9.5.3. This transition for the case when $W(D)$ vanishes as $D \to D_{\max}$ has an interesting physical implication. The sharp dynamical transition at $y = y_c$ implies the existence of a "light cone" $x = \pm y_c t$ in the space-time plane (see the left panel of Fig. 9.5). This light cone acts like a separatrix between rare atypical trajectories and the typical trajectories, as seen in models of diffusion with resetting [35, 38, 54, 324]. Trajectories that stay outside the light cone up to time $t$ are the ones which undergo very few switchings in time $t$, while those inside the light cone are the typical trajectories that experience a large number of switching events. However this sharp light cone and its associated sharp transition disappear when $W(D)$ does not vanish as $D \to D_{\max}$ (i.e., when $-1 < \nu \leq 0$). This is because, in that case, there is a nonzero probability for realizing many switching events but with a large fraction of them close to $D_{\max}$. As $x$ decreases, for a fixed $t$, such trajectories smoothly interpolate between atypical and typical trajectories, leading to the disappearance of the sharp transition.

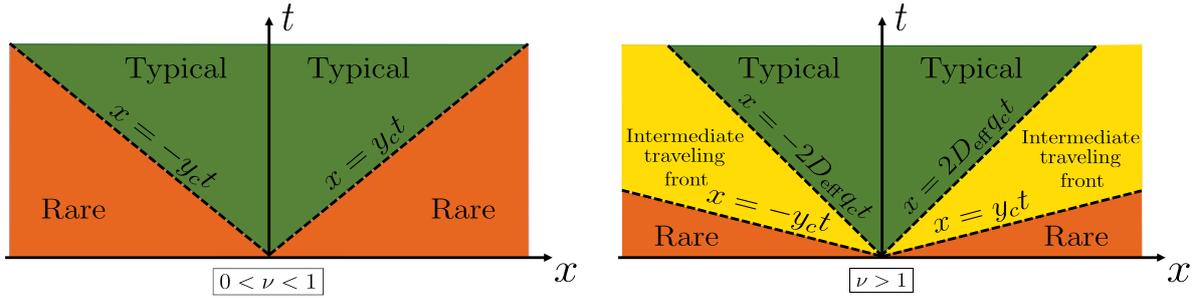

**Figure 9.5: Left:** For $0 < \nu < 1$, a light cone at $x = \pm y_c t$ separates two types of trajectories: typical ones, which switch frequently, and rare ones, which undergo almost no switches and spend most of the time in the $D_{\max}$ state. These rare trajectories dominate the large $x$ tail (see Eq. (9.4.23)). **Right:** For $\nu > 1$, a new exponential regime appears between the two existing for $0 < \nu < 1$. This exponential regime manifests as a traveling front (see Eq. (9.4.24)). In both cases, the order of the transition depends on the specific value of $\nu$ [1].

Finally, when $\nu > 1$, the LDF $I(y)$ exhibits two singular points between which its behavior is linear in $y$, namely

$$I(y) = \begin{cases} \phi_\nu(y) &, \quad 0 < y < 2D_{\text{eff}} q_c\,, \\ q_c y - \gamma &, \quad 2D_{\text{eff}} q_c < y < y_c\,, \\ r + \frac{y^2}{4D_{\max}} &, \quad y > y_c\,, \end{cases} \qquad (9.4.24)$$

where $D_{\text{eff}} < D_{\max}$ and $\gamma = D_{\max} q_c^2 - r > 0$ can be computed explicitly – see [1]. In Section 9.5.1, we show that the two-state model exhibits the same transitions in the limit $p \to 0^+$ – see also the right panel of Fig. 9.6. While $I(y)$ and $I'(y)$ are continuous across the two transitions, the second derivative $I''(y)$ is generically discontinuous at these two points – and similarly at $y = y_c$ in Eq. (9.4.23) (see [1] for more details). Thus in this case, there are two transitions as a function of the scaled distance $y$, with a new intermediate phase for $2D_{\text{eff}} q_c < y < y_c = 2D_{\max} q_c$, sandwiched between the atypical and typical regimes[8]. In this new intermediate phase, the PDF takes the form $p_r(x,t) \sim e^{-q_c(x-vt)}$ where $v = y_c - rD_{\max}/y_c > 0$. Thus, in this intermediate phase, the position distribution has the shape of a traveling front, with a nontrivial velocity $v$ [325]. Hence, in the space-time plane, we now have two light cones respectively with slopes $2D_{\text{eff}}q_c$ and $y_c$ that separate three regimes of trajectories [1] (see the right panel of Fig. 9.5). Note that a similar exponential behavior was also found for the large deviation behavior of the PDF of the position of a CTRW [112–116] – see Section 2.2. It also bears some similarities with the one found in the context of resetting Brownian motion [324].

---

[8] The existence of this new intermediate regime can be traced back to the fact that for $\nu > 1$, the derivative $W'(D)$ vanishes as $D \to D_{\max}$: this corresponds to trajectories with a typical $D$ close to $D_{\max}$ which however

## 9.5 Large Deviations for Specific Examples of $W(D)$

In this section, we show explicit results for the scaled cumulant generating function (SCGF) $\Psi(q)$, and the rate function $I(y)$, in three specific cases: (i) for the two-state model $W(D) = p\,\delta(D-D_1) + (1-p)\delta(D-D_2)$, (ii) when $W(D)$ is a uniform distribution ($\nu = 0$), and (iii) when it is a semi-circle distribution ($\nu = 1/2$).

### 9.5.1 The two-state model $W(D) = p\,\delta(D-D_1) + (1-p)\delta(D-D_2)$

The detailed analysis of the scaled cumulant generating function $\Psi(q)$ and the rate function $I(y)$ for the two-state model, corresponding to $W(D) = p\,\delta(D-D_1) + (1-p)\,\delta(D-D_2)$, is presented in [1]. Here, we summarize the main results.

For this model, the function $\tilde{p}_r(q,s)$ in Eq. (9.1.10) can be computed exactly. From this expression, one can also obtain $\hat{p}_r(q,t)$ by performing the inverse Laplace transform. The large-time exponential behavior can then be extracted, allowing for an explicit computation of the scaled cumulant generating function (SCGF) $\Psi(q)$, which leads to

$$\Psi(q) = \frac{1}{2}\left((D_1 + D_2)q^2 - r + \Delta(q)\right), \quad \Delta(q) = \sqrt{\left((D_2-D_1)q^2 + r\right)^2 + 4(D_1-D_2)prq^2}\,. \quad (9.5.1)$$

It is also possible to compute the free cumulants associated with $W(D)$ using Eq. (8.1.12), and then, by applying Eq. (9.4.14), deduce the long-time behavior of the cumulants of the position distribution

$$\langle x^{2n}(t)\rangle_c \underset{t\to+\infty}{\approx} \begin{cases} 2(pD_1 + (1-p)D_2)t & ,\quad n = 1\,, \\[6pt] -(\Delta D)^n \frac{(2n)!}{r^{n-1}} \frac{\sqrt{p(1-p)}}{n(n-1)} P_{n-1}^1(1-2p)\,t & ,\quad n > 1\,. \end{cases} \quad (9.5.2)$$

where $\Delta D = (D_1 - D_2)$ and $P_n^m(z)$ denotes the associated Legendre polynomial of degree $n$ and parameter $m$.

Interestingly, in the limit $p \to 0^+$, $\Psi(q)$ exhibits a transition as $q$ crosses some value $q_c = \sqrt{r/(D_1 - D_2)}$ (see Fig. 9.6), namely

$$\lim_{p\to 0^+} \Psi(q) = \begin{cases} D_2\,q^2 & ,\quad q < q_c\,, \\ D_1\,q^2 - r & ,\quad q > q_c\,. \end{cases} \quad (9.5.3)$$

---

undergo many switchings compared to the extreme trajectories. As a result of that, the factor $e^{-rt}$ that appears in the extreme tail is absent in this intermediate regime. Thus in the typical regime $|x| < 2D_{\text{eff}}q_c t$, the trajectories are associated with "small" values of $D$ but with a large number of switchings. In the rare regime, when $|x| > y_c t$ the trajectories up to time $t$ have a $D$ close to $D_{\max}$ and undergo almost no switchings. In the intermediate regime, when $2D_{\text{eff}}q_c t < |x| < y_c t$, the trajectories up to time $t$ have typically $D$ close to $D_{\max}$ but undergo a large number of switchings.

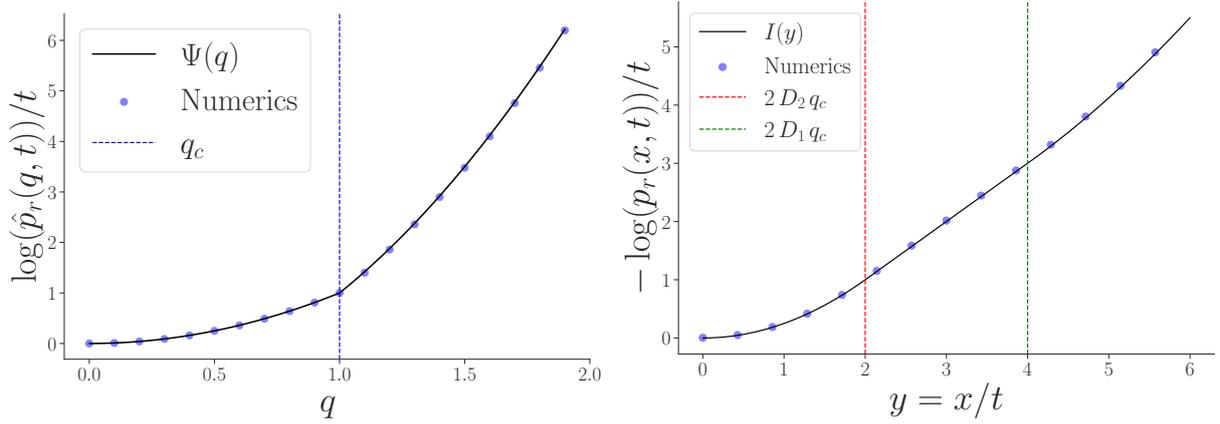

**Figure 9.6:** Plot of $\log(\hat{p}_r(q,t))/t$ vs $q$ (left panel) and of $-\log(p_r(x,t))/t$ vs $y=x/t$ (right panel) for the two-state model in the limit $p \to 0^+$. The symbols correspond to numerical results (see Appendix E for details). **Left:** the solid line shows the exact analytical result $\Psi(q)$ – see Eq. (9.5.3). **Right:** the solid line shows the rate function $I(y)$ given in Eq. (9.5.4), displaying two transition points indicated by the dotted lines. The corresponding values of the probabilities are as small as $10^{-2000}$. Here we used $r=1$, $p=10^{-10}$, $D_1=2$, $D_2=1$ and $t=1000$.

One also finds that $I(y)$ has a nontrivial limit $p \to 0^+$ where it takes the form (see [1] for details)

$$I(y) = \begin{cases} \frac{y^2}{4D_2} & , \quad |y| \leq 2D_2 q_c, \\ q_c |y| - D_2 q_c^2 & , \quad 2D_2 q_c \leq |y| \leq 2D_1 q_c, \\ r + \frac{y^2}{4D_1} & , \quad |y| \geq 2D_1 q_c. \end{cases} \quad (9.5.4)$$

These behaviors are similar to those given in Eq. (9.4.24) for $W(D) \sim (D_{\max} - D)^\nu$ when $D \to D_{\max}$ and $\nu > 1$. Interestingly, although $I(y)$ as well as its first derivative $I'(y)$ are continuous at $y = 2D_2 q_c$ and $y = 2D_1 q_c$, the second derivative $I''(y)$ is discontinuous, signaling second order dynamical transitions at these two points. A comparison with numerical results is shown in Fig. 9.6.

### 9.5.2 Uniform distribution - Case $\nu = 0$

Let us consider the case where the diffusion constants are uniformly distributed such as

$$W(D) = \frac{1}{D_{\max}}, \quad \text{for} \quad D \in [0, D_{\max}]. \quad (9.5.5)$$

To compute the SCGF, it suffices to determine the *R-transform* associated to the distribution $W(D)$ and use the relation $\Psi(q) = q^2 R\left(\frac{q^2}{r}\right)$. To proceed, we first calculate the Cauchy-Stieltjes transform of $W(D)$ which is given by

$$g(z) = \frac{1}{D_{\max}} \int_0^{D_{\max}} dD \frac{1}{z-D} = -\frac{1}{D_{\max}} \log\left(1 - \frac{D_{\max}}{z}\right), \quad (9.5.6)$$

and we have

$$g(z) = w \iff z = \frac{D_{\max}}{1 - e^{-D_{\max} w}}. \quad (9.5.7)$$

We can then use the identity $R(g(z)) + \frac{1}{g(z)} = z$ to find that the *R-transform* of a uniform distribution is given by

$$R(w) = \frac{D_{\max}}{1 - e^{-D_{\max} w}} - \frac{1}{w}. \quad (9.5.8)$$

Hence,
$$\Psi(q) = \frac{D_{\max}q^2}{1 - e^{\frac{-D_{\max}q^2}{r}}} - r. \tag{9.5.9}$$

To obtain the rate function, we need to solve the maximization problem
$$I(y) = \max_{q \in \mathbb{R}} \left( qy - \Psi(q) \right) = q^*y - \Psi(q^*), \tag{9.5.10}$$

where $q^* \equiv q^*(y)$ is a function of $y$ which is implicitly defined as the solution of
$$y = -\frac{q^* D_{\max}\left(r - e^{\frac{q^{*2} D_{\max}}{r}} r + q^{*2} D_{\max}\right)}{r\left(-1 + \cosh\left(\frac{q^{*2} D_{\max}}{r}\right)\right)}. \tag{9.5.11}$$

Unfortunately, we cannot solve this equation but we can extract the asymptotic behavior of $q^*(y)$. As the right-hand side is a monotonically increasing function of $q^*$, we can extract the small (resp. large) $y$ behavior of $q^*(y)$ by expanding the right-hand side at small (resp. large) $q^*$ values. Doing so leads to
$$I(y) = \begin{cases} \frac{y^2}{2D_{\max}} + o(y^2) & , \quad y \to 0 \\ \\ \frac{y^2}{4D_{\max}} + r + o(y^2) & , \quad y \to \infty \end{cases}. \tag{9.5.12}$$

The behavior $y \to 0$ in (9.5.12) simply corresponds to the typical fluctuations of the Gaussian where $I(y) = \frac{y^2}{4\langle D \rangle t}$, with $\langle D \rangle = D_{\max}/2$. On the other hand, the $y \to \infty$ limit corresponds to trajectories that have not experienced any switches (with probability $e^{-rt}$), and have diffused with the maximum diffusion constant $D_{\max}$. As predicted in Section 9.4.2, for $-1 < \nu \leq 0$, the function $I(y)$ interpolates smoothly between the two regimes. Note that the result in Eq. (9.5.12) obtained for the uniform distribution is in perfect agreement with the general result given in Eq. (9.4.22) which is valid for any distribution $W(D)$ with a finite support.

### 9.5.3 Wigner semi-circle distribution - $\nu = 1/2$

The Wigner distribution corresponds to the case $W(D) \sim (D_{\max} - D)^\nu$, with $\nu = 1/2$. The PDF is indeed given by
$$W(D) = \frac{8}{\pi D_{\max}^2}\sqrt{D(D_{\max} - D)} \quad , \quad 0 \leq D \leq D_{\max} \tag{9.5.13}$$

As explained in section 9.4.1, when $\nu = 1/2$, the SCGF has a transition at $q_c$ and it is given by
$$\Psi(q) = \begin{cases} q^2 R\left(\frac{q^2}{r}\right) = r \sum_{n \geq 1} \left(\frac{q^2}{r}\right)^n \kappa_n(D), & q < q_c, \\ D_{\max}q^2 - r, & q > q_c, \end{cases} \tag{9.5.14}$$

where $q_c = \sqrt{4r/D_{\max}}$. It is well known that in free probability theory, the Wigner semi-circle distribution plays the same role as the Gaussian distribution in classical probability theory in the sense that all its free cumulants $\kappa_n(D)$ vanish for $n > 2$. It is easy to compute the first free cumulants, for instance using the formulae given in the End Matter of the letter. They are given by $\kappa_1(D) = D_{\max}/2$ and $\kappa_2(D) = D_{\max}^2/16$ such that

$$R(z) = \frac{D_{\max}}{2} + z\frac{D_{\max}^2}{16} \implies \Psi(q) = \begin{cases} \frac{D_{\max}}{2}q^2 + \frac{D_{\max}^2}{16}\frac{q^4}{r}, & q < q_c, \\ D_{\max}q^2 - r, & q > q_c. \end{cases} \quad (9.5.15)$$

In the middle panel of Fig. 9.4, we have checked the equation above numerically. As shown in Section 9.4.2 for $\nu = 1/2$, the rate function $I(y)$ exhibits a second order transition at $y_c = 2D_{\max}q_c = 4\sqrt{rD_{\max}}$. We have

$$I(y) = \begin{cases} \phi_{1/2}(y) & , \quad y \leq y_c \\ r + \frac{y^2}{4D_{\max}} & , \quad y \geq y_c, \end{cases} \quad (9.5.16)$$

where

$$\phi_{1/2}(y) = \max_{q \in \mathbb{R}} \left( qy - \frac{D_{\max}}{2}q^2 - \frac{D_{\max}^2}{16}\frac{q^4}{r} \right). \quad (9.5.17)$$

Hence, one needs to solve the following equation for $q^*$

$$y = D_{\max}q + \frac{D_{\max}^2}{4}\frac{q^{*3}}{r}. \quad (9.5.18)$$

This equation has two complex roots and one real root. The large deviation function is real only for the real root, which is given by

$$q^*(y) = \frac{2 \cdot 6^{\frac{2}{3}} D_{\max}^3 r - 6^{\frac{1}{3}}\left(rD_{\max}^4\Delta\right)^{\frac{2}{3}}}{3D_{\max}^2\left(rD_{\max}^4\Delta\right)^{\frac{1}{3}}}, \quad \Delta = -9y + \sqrt{48D_{\max}r + 81y^2}. \quad (9.5.19)$$

such that we obtain

$$\phi_{1/2}(y) = \frac{r^{\frac{1}{3}}\left(-2 \cdot 6^{\frac{2}{3}} \cdot (D_{\max}r)^{\frac{1}{3}} + 6^{\frac{1}{3}}\Delta^{\frac{2}{3}}\right)}{36 \cdot D_{\max}^{\frac{2}{3}}\Delta^{\frac{4}{3}}} \times \left[81y^2 - 9y\sqrt{48D_{\max}r + 81y^2} + 9 \cdot 6^{\frac{1}{3}}y\left(D_{\max}r\Delta\right)^{\frac{1}{3}}\right.$$
$$\left. -2^{\frac{1}{3}} \cdot 3^{\frac{5}{6}}\sqrt{16D_{\max}r + 27y^2}\left(D_{\max}r\Delta\right)^{\frac{1}{3}} + 2 \cdot 6^{\frac{2}{3}}\left(D_{\max}r\Delta\right)^{\frac{2}{3}}\right]. \quad (9.5.20)$$

As expected, one can check that at small argument, we retrieve the Gaussian fluctuations $\phi_{1/2}(y) = \frac{y^2}{4\langle D\rangle} + o(y^2)$, where $\langle D\rangle = D_{\max}/2$. In Fig. 9.7, we numerically verify our prediction for the rate function in Eq. (9.5.16) with an accuracy up to $10^{-200}$. From these explicit expressions one can check that the rate function $I(y)$ as well as its derivative are continuous at $y = y_c$. However, $\phi_{1/2}''(y_c) = 1/(4D_{\max})$ while $I''(y \to y_c^-) = 1/(2D_{\max})$, hence clearly the second derivative is discontinuous at $y_c$.

**Correction of order $O(1)$ to the cumulants**

We can use the formula (9.4.17) to compute the $O(1)$ corrections to the cumulants in the large time limit. Indeed using that the $R$-transform of the Wigner distribution on $[0, D_{\max}]$ is given by

$$R(z) = \frac{D_{\max}}{2} + z\frac{D_{\max}^2}{16}, \quad (9.5.21)$$

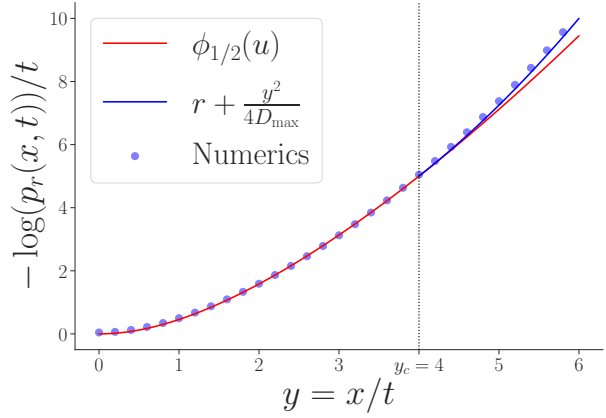

**Figure 9.7:** Plot of $-\log(p_r(x,t))/t$ vs $y = x/t$ when $W(D)$ is the Wigner semi-circle defined in Eq. (9.5.13). For $y < y_c = 4\sqrt{rD_{\max}}$, the rate function is given by $\phi_{1/2}(y)$ – see Eq. (9.5.20). For $y > y_c$, it is simply given by $r + y^2/(4D_{\max})$. The solid line corresponds to the analytic prediction, while the dots are numerical results – see Appendix E. The agreement with numerics is very good. The corresponding values of the probabilities are as small as $10^{-200}$. The parameters are $D_{\max} = 1$, $r = 1$, and $t = 50$.

one obtains from (9.4.17)

$$\tilde{p}_r(q,t) \approx \left(1 - \frac{q^4}{r^2}\frac{D_{\max}^2}{16}\right) e^{t\left(q^2 \frac{D_{\max}}{2} + q^4 \frac{D_{\max}^2}{16\,r}\right)} . \tag{9.5.22}$$

This allows us to compute the cumulants up to order $O(1)$ in the large $t$ limit, leading to

$$\langle x^2(t) \rangle_c = D_{\max}\, t \quad , \quad \langle x^{4n}(t) \rangle_c \approx \begin{cases} \frac{3}{2}\frac{D_{\max}^2}{r}(t - 1/r) + O(e^{-rt})\, , \, n=1 \\ -\frac{(4n)!}{n}\left(\frac{D_{\max}^2}{16r^2}\right)^n + O(e^{-rt})\, , \, n>1 \end{cases} \tag{9.5.23}$$

while the other cumulants are exponentially small, i.e., of order $O(e^{-rt})$ or smaller.

## 9.6 Conclusion

In summary, we have investigated the dynamics of a Brownian particle with a switching diffusion constant, obtaining the exact expression of the moments at any finite time $t$ and for any $W(D)$ with finite moments. At large times, our analysis of the cumulants and the large deviation function reveals significant deviations from Gaussian behavior in the position distribution of the particle, with intermediate exponential decay emerging in certain cases (9.4.24). Remarkably, we uncovered a surprising connection between switching diffusion and free probability theory, an unexpected link in such a classical single particle diffusion model. The origin of this connection remains a challenging and intriguing question for further investigation.

In a recent paper [318], a generic random multiplicative growth model was studied in the mean-field limit, revealing striking connections to our work, which are explicitly discussed in that preprint—notably the emergence of free cumulants. It is likely that further studies will uncover additional links between physically relevant models and free cumulants. For example, as we describe in the next chapter, switching diffusion confined within a harmonic potential also displays such a connection.

# Chapter 10

# Switching Diffusion within a Harmonic Potential

In this chapter, we consider the same switching diffusion model introduced in Eq. (8.0.1), now evolving within a harmonic potential $V(x) = \mu x^2/2$. The results presented below are original and currently unpublished – they will appear in an upcoming publication. The particle evolves under the following Langevin equation

$$\dot{x}(t) = -\mu x(t) + \sqrt{2D(t)}\,\eta(t) \quad, \tag{10.0.1}$$

where $\mu$ denotes the stiffness of the harmonic potential. As before, $D(t)$ is a switching process that takes constant values $D_i$, drawn from a distribution $W(D_i)$, and switches between them at exponentially distributed times with a constant rate $r$. In the limit $r \to 0$, the particle dynamics reduce to the standard Ornstein-Uhlenbeck process, introduced in Section 1.4. To analyze this process, it is useful to introduce the dimensionless parameter $\beta = r/\mu$, which represents the ratio of the two characteristic timescales of the system: $\tau_\mu = 1/\mu$, the average relaxation time of the particle in the harmonic potential, and $\tau_r = 1/r$, the average time between two switches of the diffusion coefficient.

In the first section, we derive the exact stationary distribution of the position of the particle in Fourier space, for an arbitrary distribution $W(D)$, using a Kesten approach — see Part II. We then show that the non-equilibrium steady state satisfies a large deviation principle in the limit of large $\beta$, once again revealing a connection to free cumulants. Finally, we apply our results to a specific case known as *randomly interrupted diffusion*, where the diffusion coefficient alternates between a fixed value $D > 0$ and zero. We demonstrate through simulations that our theoretical predictions are in excellent agreement with numerical results.

## 10.1 Non-Equilibrium Steady State

### 10.1.1 An Integral Equation from a Kesten Approach

Here, we use Kesten variables and the tools developed in [6] to derive an integral equation for the stationary distribution of the position $p(x)$ – see also Part II. In this model, there are three sources of randomness. Considering the motion between two resetting times $t_{n-1}$ and $t_n$, then the interval duration $\tau_n = t_n - t_{n-1}$ is distributed as $p(\tau_n) = re^{-r\tau_n}$. The diffusion constant $\mathsf{D}_n$ between two switches is also random and distributed according to $W(D)$, and finally, the white noise is another source of randomness. Between $t_{n-1}$ and $t_n$, it is possible to write the explicit solution of the differential equation (8.0.2). It is given by

$$x_n = x_{n-1}\,e^{-\mu\tau_n} + \sqrt{2\,\mathsf{D}_n}\,e^{-\mu\tau_n}\int_0^{\tau_n} d\tau\,\eta(\tau)\,e^{\mu\tau}\,. \tag{10.1.1}$$

As noticed in [6], Eq. (10.1.1) is of the generalised Kesten form

$$x_n = U_n \, x_{n-1} + V_n \,, \qquad (10.1.2)$$

with $U_n = e^{-\mu \tau_n}$ and $V_n = \sqrt{2 \mathsf{D}_n} \, e^{-\mu \tau_n} \int_0^{\tau_n} d\tau \, \eta(\tau) \, e^{\mu \tau}$. The distribution of the position at the $n^{\text{th}}$ switching time is given by the following integral equation

$$p(x, n) = \int_0^{D_{\max}} d\mathsf{D}_n \, W(\mathsf{D}_n) \int_0^1 dU_n \int_{-\infty}^{\infty} dV_n \int_{-\infty}^{\infty} dx' \, P(U_n, V_n) p(x', n-1) \delta(x - U_n x' - V_n) \,, \qquad (10.1.3)$$

Because the particle is trapped in a harmonic potential, we expect the particle to reach eventually a stationary state such as in the large $n$ limit, both distributions $p(x, n)$ and $p(x, n-1)$ converge to a fixed-point distribution, which we denote by $p(x)$. Therefore, when $n \to \infty$, we have

$$p(x) = \int_0^{D_{\max}} dD \int_0^1 dU \int_{-\infty}^{+\infty} dV \int_{-\infty}^{+\infty} dx' \, W(D) \, P(U, V) \, p(x') \delta(x - U \, x' - V) \,, \qquad (10.1.4)$$

and $P(U, V)$ is the joint distribution of $U_n$ and $V_n$. By definition of the conditional probability,

$$P(U, V) = P(U) P(V|U) \,. \qquad (10.1.5)$$

Based on the fact that $p(\tau) = r \, e^{-r \tau}$, we deduce that $U_n$'s are distributed over the interval $U \in [0, 1]$ with the following PDF

$$P(U) = \beta \, U^{\beta - 1} \quad \text{with} \quad \beta = \frac{r}{\mu} \,. \qquad (10.1.6)$$

At fixed $\tau_n$ and $\mathsf{D}_n$, $V_n$ is a linear transformation of Gaussian white noises and is therefore Gaussian. Hence, $P(V|U)$ depends only on the variance of $V_n$'s.

$$\begin{aligned}
\langle V_n^2 \rangle &= 2 \mathsf{D}_n \, e^{-2\mu \tau_n} \int_0^{\tau_n} d\tau \int_0^{\tau_n} d\tau' \, \langle \eta(\tau) \eta(\tau') \rangle \, e^{\mu (\tau + \tau')} \\
&= 2 \mathsf{D}_n \, e^{-2\mu \tau_n} \int_0^{\tau_n} d\tau \, e^{2\mu \tau} \\
&= \frac{\mathsf{D}_n}{\mu} \left(1 - U_n^2\right) \,. \end{aligned} \qquad (10.1.7)$$

The joint law of $U$ and $V$ is thus given by

$$P(U, V) = \beta \, U^{\beta - 1} \frac{1}{\sqrt{2\pi \alpha(U)}} e^{\frac{-V^2}{2\alpha(U)}} \quad , \quad \alpha(U) = \frac{D}{\mu} (1 - U^2) \,. \qquad (10.1.8)$$

The explicit integral equation for the stationary distribution is therefore

$$p(x) = \int_0^{D_{\max}} dD \int_0^1 dU \int_{-\infty}^{+\infty} dV \int_{-\infty}^{+\infty} dx' \, W(D) \, \beta \, U^{\beta - 1} \frac{1}{\sqrt{2\pi \alpha(U)}} e^{\frac{-V^2}{2\alpha(U)}} p(x') \delta(x - U \, x' - V) \,. \qquad (10.1.9)$$

Going to Fourier space, we obtain

$$\hat{p}(k) = \int_0^{D_{\max}} dD \int_0^1 dU \int_{-\infty}^{+\infty} dV \int_{-\infty}^{+\infty} dx' \, W(D) \, \beta \, U^{\beta - 1} \frac{1}{\sqrt{2\pi \alpha(U)}} e^{\frac{-V^2}{2\alpha(U)}} p(x') e^{ik(Ux' + V)} \,, \qquad (10.1.10)$$

where $\hat{p}(k)$ is the Fourier transform of $p(x)$. Performing the integral over $x'$ leads to

$$\hat{p}(k) = \int_0^{D_{\max}} dD \int_0^1 dU \int_{-\infty}^{+\infty} dV \, W(D) \, \beta \, U^{\beta - 1} \, e^{ikV} \frac{1}{\sqrt{2\pi \alpha(U)}} e^{\frac{-V^2}{2\alpha(U)}} \hat{p}(kU) \,. \qquad (10.1.11)$$

A last integration over $V$ yields

$$\boxed{\hat{p}(k) = \int_0^{D_{\max}} dD\, W(D) \int_0^1 dU\, \beta\, U^{\beta-1}\, e^{-\frac{k^2 \alpha(U)}{2}}\, \hat{p}(kU)}, \quad \alpha(U) = \frac{D}{\mu}(1-U^2)\,. \qquad (10.1.12)$$

which is exactly the result given in Eq. (7.1.20), with another function $\alpha(U)$, and an additional average over $W(D)$.

Note that it is also possible to derive a similar integral equation for the bilateral Laplace transform of $p(x)$ which we denote $\tilde{p}(q)$. It reads

$$\tilde{p}(q) = \int_0^{D_{\max}} dD\, W(D) \int_0^1 dU\, \beta\, U^{\beta-1}\, e^{\frac{q^2 \alpha(U)}{2}}\, \hat{p}(qU) \quad , \quad \tilde{p}(q) = \int_{-\infty}^{+\infty} dx\, e^{qx}\, p(x)\,. \qquad (10.1.13)$$

It turns out that the BLT is more convenient to use when studying the large deviations of $p(x)$.

### 10.1.2 Exact Expressions for the Moments and Series Expansion of $\hat{p}(k)$

We first write the series expansion of the Fourier transform of the distribution as

$$\hat{p}(k) = \sum_{n=0}^{+\infty} \frac{(ik)^{2n}}{(2n)!} \left\langle x^{2n} \right\rangle, \qquad (10.1.14)$$

where odd moments vanish by symmetry. Injecting the series expansion of $\hat{p}(k)$ in Eq. (10.1.12) leads to a recursive formula for the moments that was already derived in Section 7.1.2 for any $\alpha(U)$. Here, we have an additional average over the random variable $D$, and the moments are given by the following recursive relation

$$\left\langle x^{2n} \right\rangle = \beta\,(\beta+2n)\,(2n-1)! \sum_{p=0}^{n-1} \frac{\left\langle x^{2p} \right\rangle}{2^{n-p}\,(n-p)!\,(2p)!} \int_0^{D_{\max}} dD\, W(D) \int_0^1 dU\, U^{\beta-1+2p}\, [\alpha(U)]^{n-p}\,, \qquad (10.1.15)$$

where we recall that here $\alpha(U) = \frac{D}{\mu}(1-U^2)$. It is possible to perform the integrations over $U$ and $D$ to obtain

$$\left\langle x^{2n} \right\rangle = \beta\,(2n-1)! \sum_{p=0}^{n-1} \frac{\left\langle x^{2p} \right\rangle}{(2p)!} \langle D^{n-p}\rangle (2\mu)^{p-n} \frac{\Gamma\left(\frac{\beta}{2}+p\right)}{\Gamma\left(\frac{\beta}{2}+n\right)}\,. \qquad (10.1.16)$$

When regrouping terms together, the above recursive relation reads

$$\left[\frac{(2\mu)^n}{(2n)!}\Gamma\left(\frac{\beta}{2}+n\right)\left\langle x^{2n}\right\rangle\right] = \frac{\beta}{2n}\sum_{p=0}^{n-1}\langle D^{n-p}\rangle \left[\frac{(2\mu)^p}{(2p)!}\Gamma\left(\frac{\beta}{2}+p\right)\left\langle x^{2p}\right\rangle\right]\,. \qquad (10.1.17)$$

Hence, it can be written in a simpler way as follows

$$a_n = \frac{\beta}{2n}\sum_{p=0}^{n-1}\langle D^{n-p}\rangle\, a_p\,, \qquad (10.1.18)$$

$$a_n = \frac{(2\mu)^n}{(2n)!}\Gamma\left(\frac{\beta}{2}+n\right)\left\langle x^{2n}\right\rangle \quad , \quad a_0 = \Gamma\left(\frac{\beta}{2}\right)\,. \qquad (10.1.19)$$

Introducing $\tilde{a}_n = a_n/\Gamma\left(1 + \frac{\beta}{2}\right)$, it is possible to compute the first terms of the sequence. They are given by

$$\tilde{a}_1 = \langle D \rangle, \tag{10.1.20}$$

$$\tilde{a}_2 = \frac{1}{4}\left(\beta\langle D\rangle^2 + 2\langle D^2\rangle\right), \tag{10.1.21}$$

$$\tilde{a}_3 = \frac{1}{24}\left(\beta^2\langle D\rangle^3 + 6\beta\langle D\rangle\langle D^2\rangle + 8\langle D^3\rangle\right), \tag{10.1.22}$$

$$\tilde{a}_4 = \frac{1}{192}\left(\beta^3\langle D\rangle^4 + 12\beta^2\langle D\rangle^2\langle D^2\rangle + 12\beta\langle D^2\rangle^2 + 32\beta\langle D\rangle\langle D^3\rangle + 48\langle D^4\rangle\right), \tag{10.1.23}$$

$$\tilde{a}_5 = \frac{1}{1920}\Big(\beta^4\langle D\rangle^5 + 20\beta^3\langle D\rangle^3\langle D^2\rangle + 60\beta^2\langle D\rangle\langle D^2\rangle^2 + 80\beta^2\langle D\rangle^2\langle D^3\rangle$$
$$+ 160\beta\langle D^2\rangle\langle D^3\rangle + 240\beta\langle D\rangle\langle D^4\rangle + 384\langle D^5\rangle\Big). \tag{10.1.24}$$

The sequence on OEIS [326] reveals the polynomial structure of the $\tilde{a}_n$'s. They are given by

$$\tilde{a}_n = [t^n]\left[\frac{2}{\beta}\left(\exp\left[\frac{\beta}{2}\sum_{m=1}^{+\infty}\langle D^m\rangle\frac{t^m}{m}\right] - 1\right)\right], \tag{10.1.25}$$

where $[t^n]f(t)$ is the coefficient at order $t^n$ of the series expansion of $f$ with respect to $t$. At this stage, Eq. (10.1.25) is an analytical guess. One can identify the coefficients with respect to partial exponential Bell polynomials $B_{n,k}$ (see [305, 306])

$$\tilde{a}_n = \frac{2}{\beta\, n!}\sum_{m=1}^{n} B_{n,m}\left(1!\frac{\beta}{2}\frac{\langle D\rangle}{1}, \ldots, (n-m+1)!\frac{\beta}{2}\frac{\langle D^{n-m+1}\rangle}{n-m+1}\right). \tag{10.1.26}$$

Using the definition of $B_n(x_1,\ldots,x_n)$, the $n^{\text{th}}$ complete exponential Bell polynomial, given in Eq. (8.1.5), we obtain

$$\tilde{a}_n = \frac{2}{\beta\, n!}B_n\left(1!\frac{\beta}{2}\frac{\langle D\rangle}{1}, \ldots, n!\frac{\beta}{2}\frac{\langle D^n\rangle}{n}\right). \tag{10.1.27}$$

Hence, using Eq. (10.1.19), the final expression for the moments reads

$$\langle x^{2n}\rangle = \frac{(2n)!}{(2\mu)^n}\frac{\Gamma\left(\frac{\beta}{2}\right)}{\Gamma\left(\frac{\beta}{2}+n\right)}\frac{1}{n!}B_n\left(1!\frac{\beta}{2}\frac{\langle D\rangle}{1}, \ldots, n!\frac{\beta}{2}\frac{\langle D^n\rangle}{n}\right). \tag{10.1.28}$$

Using $B_0 = 1$, it leads to

$$\boxed{\hat{p}(k) = \sum_{n=0}^{+\infty}\frac{(ik)^{2n}}{(2\mu)^n}\frac{\Gamma\left(\frac{\beta}{2}\right)}{\Gamma\left(\frac{\beta}{2}+n\right)}\frac{B_n\left(1!\frac{\beta}{2}\frac{\langle D\rangle}{1}, \ldots, n!\frac{\beta}{2}\frac{\langle D^n\rangle}{n}\right)}{n!}}, \tag{10.1.29}$$

This formula is based on the guess (10.1.25), but in Appendix F, we show that Eq. (10.1.29) is indeed a solution of the integral equation (10.1.12), thus completing the proof.

### 10.1.3 Large and Small $\beta$ Asymptotics of the Distribution

We find that the small and large $\beta$ behaviors of the stationary state are given by

$$\boxed{p(x) \simeq \begin{cases} \int_0^{D_{\max}} dD\, W(D)\sqrt{\frac{\mu}{2\pi D}}e^{-\frac{\mu x^2}{2D}} & , \quad \beta \ll 1, \\ \\ \sqrt{\frac{\mu}{2\pi\langle D\rangle}}e^{-\frac{\mu x^2}{2\langle D\rangle}} & , \quad \beta \gg 1. \end{cases}} \tag{10.1.30}$$

These asymptotics are similar to the ones derived in [1] for the free switching diffusion process where $r \times t$, i.e. the mean number of switches in a period $t$, plays the role of $\beta = r/\mu$. In the limit $\beta \to 0$, which can be seen as a limit $r \to 0$ with $\mu$ fixed, the diffusion constant of the particles does not switch and remains in the initial state $\mathsf{D}_1$. Hence, the stationary state is that of an Ornstein-Uhlenbeck process with diffusion constant $\mathsf{D}_1$. Averaging over all possible $\mathsf{D}_1$ leads to the first line of Eq. (10.1.30). On the other hand, the large beta limit, i.e. $r \to +\infty$ while $\mu$ remains fixed, corresponds to the OU stationary state with diffusion constant $\langle D \rangle$. In this regime, due to the high frequency of switches, the particle effectively samples the average value of $D$, hence the second line of Eq. (10.1.30).

Consequently, in the limit of small and large $\beta$, the moments are given by

$$\langle x^{2n} \rangle \simeq \begin{cases} \frac{(2n-1)!!}{\mu^n} \langle D^n \rangle &, \quad \beta \ll 1, \\ \\ \frac{(2n-1)!!}{\mu^n} \langle D \rangle^n &, \quad \beta \gg 1. \end{cases} \quad (10.1.31)$$

**Remark.** It is possible to analyze the small $\beta$ limit of the integral equation for $\hat{p}(k)$. Indeed, the second integral in the right-hand side of Eq. (10.1.12) is

$$\int_0^1 dU \, \beta \, U^{\beta-1} \, e^{-\frac{k^2 \alpha(U)}{2}} \hat{p}(kU) = \int_0^{\beta^{-1}} dz \, \beta^{\beta+1} \, z^{\beta-1} \, e^{\frac{-k^2 D}{2\mu}(1-z^2\beta^2)} \hat{p}(kz\beta), \quad (10.1.32)$$

where we have made the change of variable $U = z\beta$ in the second equality. In the limit $\beta \to 0$, using that $\hat{p}(0) = 1$, which is the normalization of $p(x)$, and performing the integral over $z$, we obtain

$$\hat{p}(k) \underset{\beta \to 0}{\approx} \int_0^{D_{\max}} dD \, W(D) e^{\frac{-k^2 D}{2\mu}}. \quad (10.1.33)$$

Going back in real space leads directly to the first line of Eq. (10.1.30).

In the large $\beta$ limit, it is possible to show that

$$B_n\left(1! \frac{\beta}{2} \frac{\langle D \rangle}{1}, \ldots, n! \frac{\beta}{2} \frac{\langle D^n \rangle}{n}\right) \underset{\beta \to +\infty}{\approx} \frac{\beta^n}{2^n} \langle D \rangle^n. \quad (10.1.34)$$

Then, using Eq. (10.1.27), we have

$$\tilde{a}_n \underset{\beta \to +\infty}{\approx} \frac{\beta^{n-1}}{2^{n-1} n!} \langle D \rangle^n, \quad \tilde{a}_n = a_n / \Gamma\left(1 + \frac{\beta}{2}\right), \quad (10.1.35)$$

which, when substituted into the definition (10.1.19), leads to the second line of Eq. (10.1.31).

### 10.1.4 Large and Small $\beta$ Asymptotics of the Cumulants: From Classical to Free Cumulants

It is interesting to compute the cumulants of this process, as higher-order cumulants play a crucial role in characterizing the non-Gaussian nature of the non-equilibrium steady state. It is also worth investigating whether the connection with free cumulants persists when the switching diffusion occurs in the presence of an external potential.

Interestingly, the large and small $\beta$ limits reveal a distinct polynomial structure of the cumulants. Similar to the free-switching diffusion process, the cumulants undergo a crossover: at small $\beta$, they are proportional to the cumulants of $D$ (denoted as $\langle D^n \rangle_c$), while at large $\beta$, they are proportional to the free cumulants of $D$ (denoted as $\kappa_n(D)$). In this context, $\beta$ plays a role analogous to $rt$ in the free case – see Eq. (8.2.2). The asymptotic behavior can be expressed as follows

$$\boxed{\langle x^{2n}\rangle_c \simeq \begin{cases} \frac{(2n-1)!!}{r^n} \langle D^n\rangle_c \beta^n &, \quad \beta \ll 1, \\ \\ \frac{(2n-1)!}{r^n} \kappa_n(D)\beta &, \quad \beta \gg 1 \, . \end{cases}} \tag{10.1.36}$$

The small $\beta$ limit of the cumulants is proved in Appendix G.

**Large $\beta$ limit of the cumulants**

To show the second line of Eq. (10.1.36), let us begin by assuming that it holds. It is natural here to try to relate the cumulant generating function (CGF)[9] in the large $\beta$ limit to the *R-transform*, i.e, the generating function of the free cumulants $R(z) = \sum_{n\geq 1} z^{n-1}\kappa_n(D)$. For sufficiently small values of $q$, The CGF reads

$$\log \tilde{p}(q) = \sum_{n=1}^{+\infty} \frac{q^{2n}}{(2n)!}\langle x^{2n}\rangle_c \underset{\beta\to+\infty}{\approx} \beta \sum_{n=1}^{+\infty} \frac{\left(\frac{q}{\sqrt{\mu\beta}}\right)^{2n}}{2n}\kappa_n(D) \, , \tag{10.1.37}$$

such that taking a derivative with respect to $q$ leads to

$$\frac{d\log\tilde{p}(q)}{dq} \underset{\beta\to+\infty}{\approx} \sqrt{\frac{\beta}{\mu}} \sum_{n=1}^{+\infty} \left(\frac{q}{\sqrt{\mu\beta}}\right)^{2n-1}\kappa_n(D) = \frac{q}{\mu}\sum_{n=1}^{+\infty}\left(\frac{q^2}{\mu\beta}\right)^{n-1}\kappa_n(D) = \beta\frac{q}{r}R\left(\frac{q^2}{r}\right). \tag{10.1.38}$$

Therefore, we have

$$\lim_{\beta\to+\infty}\frac{1}{\beta}\log\tilde{p}(q) = \int_0^q dz\, \frac{z}{r}R\left(\frac{z^2}{r}\right). \tag{10.1.39}$$

A change of variable $u = z^2/r$ leads to

$$\boxed{\lim_{\beta\to+\infty}\frac{1}{\beta}\log\tilde{p}(q) = \frac{1}{2}\int_0^{q^2/r} du\, R(u)}. \tag{10.1.40}$$

This result, which is valid for sufficiently small values of $q$, is interestingly quite similar to the one obtained for asymptotics of spherical integrals of random matrices [27].

The large deviation form in Eq. (10.1.40) suggests the following ansatz for the bilateral Laplace transform (BLT) of the position distribution

$$\tilde{p}(q) \underset{\beta\to+\infty}{\approx} e^{\beta\Psi\left(\frac{q^2}{r}\right)}. \tag{10.1.41}$$

This expression can be inserted into Eq. (10.1.13) to verify self-consistently that Eq. (10.1.40) indeed provides the correct result in the large $\beta$ limit. It yields

$$e^{\beta\Psi\left(\frac{q^2}{r}\right)} = \int_0^{D_{\max}} dD\, W(D)\int_0^1 dU\, \beta U^{\beta-1}\, e^{\frac{q^2 D}{2\mu}(1-U^2)}\, e^{\beta\Psi\left(\frac{q^2}{r}U^2\right)}. \tag{10.1.42}$$

Performing a change of variables $U = 1 - v/\beta$, with $v > 0$ leads to

$$e^{\beta\Psi\left(\frac{q^2}{r}\right)} = \int_0^{D_{\max}} dD\, W(D)\int_0^\beta dv\, \frac{1}{1-\frac{v}{\beta}}\, e^{\beta\log\left(1-\frac{v}{\beta}\right)+\beta\frac{q^2}{r}\frac{D}{2}\frac{v}{\beta}\left(2-\frac{v}{\beta}\right)}\, e^{\beta\Psi\left(\frac{q^2}{r}\left(1-\frac{v}{\beta}\right)^2\right)}, \tag{10.1.43}$$

---

[9] Here, we will use the BLT $\tilde{p}(q)$ instead of the Fourier transform.

where we recall that $\beta = r/\mu$. For large $\beta$, it gives

$$e^{\beta \Psi\left(\frac{q^2}{r}\right)} = \int_0^{D_{\max}} dD\, W(D) \int_0^{+\infty} dv\, e^{-v + v\frac{q^2}{r}D + \beta \Psi\left(\frac{q^2}{r}\right) - \frac{2q^2 v}{r}\Psi'\left(\frac{q^2}{r}\right)}. \quad (10.1.44)$$

Hence

$$1 = \int_0^{D_{\max}} dD\, W(D) \int_0^{+\infty} dv\, e^{-v + v\frac{q^2}{r}D - \frac{2q^2 v}{r}\Psi'\left(\frac{q^2}{r}\right)}. \quad (10.1.45)$$

One can perform the integral over $v$, such that the ansatz solves the integral equation for the following constraint on $\Psi$

$$\frac{q^2}{r} = \int_0^{D_{\max}} dD\, \frac{W(D)}{\frac{r}{q^2} + 2\Psi'\left(\frac{q^2}{r}\right) - D}. \quad (10.1.46)$$

This equation can also be written in terms of the Cauchy-Stieltjes transform $g(x)$ as

$$\frac{q^2}{r} = g\left[\frac{r}{q^2} + 2\Psi'\left(\frac{q^2}{r}\right)\right] \quad \text{where} \quad g(x) = \int_0^{D_{\max}} dD\, \frac{W(D)}{x - D}. \quad (10.1.47)$$

Now using the relation (8.1.19) for a real probability measure $W(D)$, we conclude that, for sufficiently small values of $q$, we have

$$\Psi'\left(\frac{q^2}{r}\right) = \frac{1}{2}R\left(\frac{q^2}{r}\right) \quad (10.1.48)$$

which is consistent with Eq. (10.1.40) and the second line of Eq. (10.1.36).

Following Eq. (10.1.46), one expects interesting transitions in the scaled cumulant generating function for certain classes of $W(D)$, as well as for the NESS in the large $\beta$ limit (see Section 9.4). These aspects will be explored in greater detail in an upcoming publication.

## 10.2 Randomly Interrupted Diffusion

To conclude this chapter, we consider as an illustrative example the *randomly interrupted diffusion*[10] model, defined by the distribution $W(d) = p\delta(d - D) + (1-p)\delta(d)$. This model was originally introduced in [138] for the case $p = 1/2$, and we extend the analysis here to arbitrary values of $p$[11]. This model describes the motion of a particle that diffuses with a diffusion coefficient $D$ for a random duration, after which it switches to a deterministic phase where it relaxes in the harmonic potential. Both phases last for random durations that are exponentially distributed, with switching rates controlled by the parameter $p$: the transition rate from the diffusive state $(D \to 0)$ is $r(1-p)$, while the transition rate from the deterministic state $(0 \to D)$ is $rp$[12].

For this specific distribution of diffusion coefficients, we have $\langle D^n \rangle = pD^n$. Starting from Eq. (10.1.29) and using the identities listed below, we can simplify the expression

$$B_n\left(\alpha\beta x_1, \alpha\beta^2 x_2, \ldots, \alpha\beta^n x_n\right) = \beta^n \sum_{k=1}^n \alpha^k B_{n,k}\left(x_1, x_2, \ldots, x_{n-k+1}\right), \quad (10.2.1)$$

$$B_{n,k}(0!, 1!, \ldots, (n-k)!) = \begin{bmatrix} n \\ k \end{bmatrix}, \quad (10.2.2)$$

$$x^{(n)} = \sum_{k=0}^n \begin{bmatrix} n \\ k \end{bmatrix} x^k, \quad (10.2.3)$$

---

[10] This model is also sometimes referred to as *randomly flashing diffusion* [327].
[11] We have verified that the results derived in this section are fully consistent with those obtained in [138] when specialized to $p = 1/2$. To make the correspondence with their notation, one should perform the substitutions $\nu \to r/2$, $a \to \mu$, and $p \to 1/2$.
[12] These rates can be derived by writing the Fokker–Planck equation associated with the dynamics.

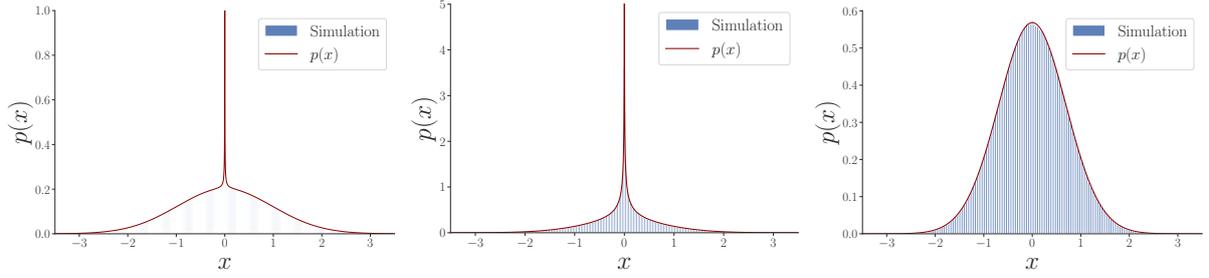

**Figure 10.1:** Comparison between simulation results and the analytical prediction given by Eq. (10.2.7). We show the steady-state solution for the randomly interrupted diffusion model $W(d) = p\delta(d - D) + (1 - p)\delta(d)$ with parameters $p = 0.5$, $D = 1$, and three different values of $\beta$: $\beta = 0.01$ (left), $\beta = 1$ (middle), and $\beta = 100$ (right). The asymptotic behaviors corresponding to these values of $\beta$ are provided in Eq. (10.2.8).

where $\begin{bmatrix} n \\ k \end{bmatrix}$ are the Stirling numbers of the first kind. Using these properties, one can show that

$$\hat{p}(k) = 1 + \sum_{n=1}^{+\infty} \frac{\left(-\frac{k^2 D}{2\mu}\right)^n}{\left(\frac{\beta}{2}\right)^{(n)} n!} \left(p\frac{\beta}{2}\right)^{(n)}. \tag{10.2.4}$$

This series simplifies to a closed-form expression

$$\hat{p}(k) = {}_1F_1\left(\frac{\beta p}{2}, \frac{\beta}{2}, -\frac{Dk^2}{2\mu}\right), \tag{10.2.5}$$

where ${}_1F_1(a, b, z)$ denotes Kummer's function. When $p = 1$, we recover

$$\hat{p}(k) = 1 + \sum_{n=1}^{+\infty} \frac{\left(-\frac{k^2 D}{2\mu}\right)^n}{n!} = e^{-\frac{k^2 D}{2\mu}}, \tag{10.2.6}$$

which is exactly what we expect: the Fourier transform of the steady state of an Ornstein–Uhlenbeck process with diffusion coefficient $D$ and stiffness $\mu$.

It is possible to invert Eq. (10.2.5) into real space. Through nontrivial manipulations involving hypergeometric functions, one can show that the NESS takes the following simple form

$$\boxed{p(x) = \sqrt{\frac{\mu}{2\pi D}} \frac{\Gamma\left(\frac{\beta}{2}\right)}{\Gamma\left(\frac{\beta p}{2}\right)} e^{-\frac{\mu x^2}{2D}} U\left(\frac{\beta}{2}(1-p), \frac{3}{2} - \frac{\beta p}{2}, \frac{\mu x^2}{2D}\right),} \tag{10.2.7}$$

where $U(a, b, c)$ is Tricomi's (confluent hypergeometric) function.

It is also possible to derive the asymptotic behavior of $p(x)$ in the limits of small and large $\beta$, as well as to compute it explicitly for $\beta = 1$. We obtain

$$p(x) \simeq \begin{cases} (1-p)\,\delta(x) + p\sqrt{\frac{\mu}{2\pi D}} e^{-\frac{\mu x^2}{2D}} & ,\quad \beta = 0, \\[2mm] \frac{1}{\Gamma\left(\frac{p}{2}\right)} \sqrt{\frac{\mu}{2D}} \left(\frac{\mu x^2}{2D}\right)^{\frac{1}{2}(p-1)} e^{-\frac{\mu x^2}{2D}} & ,\quad \beta = 1, \\[2mm] \sqrt{\frac{\mu}{2\pi Dp}} e^{-\frac{\mu x^2}{2Dp}} & ,\quad \beta \to +\infty. \end{cases} \tag{10.2.8}$$

When $\beta \to 0$, the mean time between two switches, $\tau_r$, greatly exceeds the relaxation time within the harmonic trap, $\tau_\mu$. Consequently, particles in the non-diffusive state quickly relax toward the origin, resulting in a sharply peaked distribution that becomes a delta function at $\beta = 0$. Conversely, as $\beta \to \infty$, the situation reverses: the system reaches a steady state characteristic of an Ornstein–Uhlenbeck particle, where the effective diffusion coefficient corresponds to the mean value of $W(D)$, which, in this context, is given by $\langle D \rangle = Dp$. Note also that the small and large $\beta$ limits are consistent with Eq. (10.1.30). We compare the exact result in Eq. (10.2.7) with numerical simulations in Figure 10.1, which shows excellent agreement.

## 10.3 Summary

In this chapter, we developed a Kesten approach to obtain an explicit integral equation for the non-equilibrium steady state of a switching diffusion in a harmonic trap. We managed to solve this integral equation in Fourier space for any distribution of the diffusion coefficient $W(D)$ (with finite moments). We also derived the small $\beta = r/\mu$ (slow-switching) and large $\beta$ (fast-switching) asymptotics, revealing a crossover from a mixture of Ornstein–Uhlenbeck Boltzmann distributions to an effective Gaussian with a variance being the mean of $W(D)$. In the large $\beta$ limit, we also show the cumulants of the position of the particle are proportional to the free cumulants of $W(D)$ – as is the case in the free (unconfined) setting in the large-time limit. We then specialized to the "randomly interrupted" case where the diffusion coefficient alternates between a fixed value and zero and confirmed our exact theoretical predictions through numerical simulations. These results are the subject of a forthcoming publication currently in preparation.

# Part IV

# Run-and-Tumble Particle in One-Dimensional Potentials: Mean First-Passage Time and Applications

**Abstract**

We study a one-dimensional run-and-tumble particle (RTP), which is a prototypical model for active system, moving within an arbitrary external potential. Using backward Fokker-Planck equations, we derive the differential equation satisfied by its mean first-passage time (MFPT) to an absorbing target, which, without any loss of generality, is placed at the origin. Depending on the shape of the potential, we identify four distinct "phases", with a corresponding expression for the MFPT in every case, which we derive explicitly. We study in particular confining potential of the type $V(x) = \alpha |x|^p$ with $p \geq 1$. We obtain a closed form expression for the MFPT $\tau_\gamma(x_0)$ for all $p \geq 1$, which becomes fully explicit in the case $p = 1$, $p = 2$ and in the limit $p \to \infty$. For generic $p > 1$ we find that there exists an optimal rate $\gamma_\text{opt}$ that minimizes the MFPT and we characterize in detail its dependence on $x_0$. We find that $\gamma_\text{opt} \propto 1/x_0$ as $x_0 \to 0$, while $\gamma_\text{opt}$ converges to a nontrivial constant as $x_0 \to \infty$. In contrast, for $p = 1$, there is no finite optimum and $\gamma_\text{opt} \to \infty$ in this case. These analytical results are confirmed by our numerical simulations. We then present different applications of these general formulae to (i) the generalization of the Kramer's escape law for an RTP in the presence of a potential barrier, (ii) the "trapping" time of an RTP moving in a harmonic well and (iii) characterizing the efficiency of the optimal search strategy of an RTP subjected to stochastic resetting. Our results reveal that the MFPT of an RTP in an external potential exhibits a far more complex and, at times, counter-intuitive behavior compared to that of a passive particle (e.g., Brownian) in the same potential.

# Chapter 11

# Exact Mean First-Passage Time

## 11.1 Introduction

First-passage time (FPT) properties constitute a classical and notoriously difficult problem in the study of stochastic processes [40, 43, 63, 69]. During the last decades they have found myriad of applications ranging from chemical reaction kinetics [328, 329], intracellular transport [330] and slow dynamics in complex and disordered systems [63, 82] all the way to animals searching for foods [331, 332], or more generally random search strategies [40, 333]. When the stochastic dynamics under study takes place in a confined geometry, such as in a bounded domain or in the presence of an external potential, the main features of the full distribution of the FPT are well captured by its average value, namely the mean first-passage time (MFPT) – see e.g. [334].

While the MFPT has been widely studied for passive particles, such as a Brownian particle in an external potential, much less is known for active particles. At variance with the passive particles, active particles are able to consume energy from their environment, leading to self-propelled motion [47–50]. In the absence of any confining potential, the first-passage properties of a single run-and-tumble active particle in one dimension has been computed exactly with many interesting results [222, 225, 239, 240, 270, 335–338]. However, a central topic of current active matter research dwells on the effect of confinement, either in a finite system or in the presence of a confining trap (for example optical traps [339, 340]). It is well known that, with increasing activity, the position distribution of an active particle undergoes a cross-over from a bell-shaped Gaussian distribution concentrated around the trap center to a structure with peaks near the edge of the trap signalling accumulation of the particles there [47–50, 215, 216, 241, 339–341]. In contrast, the first-passage properties of active particles in a confined geometry are much less explored. A quantity of prime interest is the MFPT, which has been studied for active particles in a finite domain, and in the absence of any potential, in dimensions up to $d = 3$ [217, 235, 236, 336, 337]. However, most experiments are performed in the presence of an optical trap, both harmonic and non-harmonic [339, 340]. In Ref. [216] the late time decay of the full first-passage probability of an RTP in a harmonic potential in $d = 1$ was studied, though the MFPT remained unknown even in this case prior to our work.

In [5], we derived the MFPT of a one-dimensional RTP in the presence of a confining potential of the form $V(x) = \alpha |x|^p$ with $p \geq 1$ and $\alpha > 0$. The case $p = 2$ corresponds to the harmonic potential. On the other hand, the limit $p \to \infty$ corresponds to an RTP in a finite domain $|x| < 1$ with reflecting boundary conditions at $x = 1$. One of our principal results here is that, for $p > 1$, there exists an optimal tumbling rate $\gamma$ that minimizes the MFPT. This has important implication for finding optimal strategies for active particles to navigate in a noisy environment [342–345]. In fact, this question of optimizing the MFPT with respect to some parameters of the underlying dynamics is of paramount importance in many search strategies involving Brownian motion, Lévy flights [346, 347], stochastic resetting [35, 54, 155, 169, 348, 349], etc. For active particles, the optimization of MFPT has been studied mostly for systems in a finite domain [35, 158, 235, 236, 239, 240, 255, 350–353], but very little is known about such optimization in the presence of an external confining potential. Our exact results, in particular the existence of an optimal tumbling rate, valid for a wide class of one-dimensional potentials, will serve as important benchmarks for future studies of the MFPT in more complex environments. We show these results in Chapter 12. Based on the exact expressions derived for the MFPT, we also present several applications, including the derivation of Kramers' law for RTPs, in Chapter 13.

In [2] we derived exact analytical results for the MFPT of an RTP in $d = 1$ in the presence of an arbitrary potential. We showed that the MFPT satisfies a second-order differential equation with non-trivial boundary conditions that depend on the specific shape of the potential. These results are presented in Section 11.2 below.

## 11.2 Model and Differential Equations for the MFPT

We consider a run-and-tumble particle whose position is denoted by $x(t) \in [0, +\infty[$. The RTP starts from $x(t = 0) = x_0 \geq 0$ and its equation of motion reads

$$\frac{dx(t)}{dt} = \dot{x}(t) = f(x) + v_0\, \sigma(t)\,, \qquad (11.2.1)$$

where $v_0$ represents a constant speed, and $f(x)$ denotes the force acting on the RTP.

This force derives from a potential $V(x)$, i.e., $f(x) = -V'(x)$. The stochastic part of the dynamics is driven by a telegraphic noise $\sigma(t)$, which alternates between the values $\pm 1$. This noise mimics the tumblings of the particle, i.e., the random changes in direction. Here, we consider Poissonian tumblings, where the time intervals $\tau$ between successive tumbling events are exponentially distributed with rate $\gamma > 0$, such that $p(\tau) = \gamma\, e^{-\gamma \tau}$. Alternatively, the dynamics of the telegraphic noise $\sigma(t)$ can be described as follows. Within the infinitesimal time interval $[t, t + dt]$, the noise $\sigma(t)$ evolves via the rule

$$\sigma(t + dt) = \begin{cases} \sigma(t) & \text{, with probability } (1 - \gamma\, dt) \\ -\sigma(t) & \text{, with probability } \gamma\, dt \end{cases}. \qquad (11.2.2)$$

Initially, we have $\sigma(0) = +1$ or $\sigma(0) = -1$, with equal probability $1/2$. Here we are interested in the calculation of $\tau_\gamma(x_0)$ which is the mean first-passage time at the origin $x = 0$, i.e., the average time it takes for an RTP to reach the origin for the first time, starting from $x_0 \geq 0$. Of course $\tau_\gamma(x_0)$ is also a function of the speed $v_0$ and of the force $f(x)$. Note that the MFPT to a target located at $X > 0$ can be easily obtained from the expression of the MFPT to the origin by shifting the potential $V(x) \to V(x + X)$.

To compute the MFPT, it is useful to first consider the survival probabilities $Q^\pm(x_0, t)$. They are the probabilities that the particle remains in the positive region of the real line up to time $t$, with initially $\sigma(0) = \pm 1$. We have shown in Section 4.3.1 that these probabilities obey the following coupled differential equations

$$\partial_t Q^+(x_0, t) = [f(x_0) + v_0]\, \partial_{x_0} Q^+(x_0, t) - \gamma\, Q^+(x_0, t) + \gamma\, Q^-(x_0, t)\,, \qquad (11.2.3)$$

$$\partial_t Q^-(x_0, t) = [f(x_0) - v_0]\, \partial_{x_0} Q^-(x_0, t) - \gamma\, Q^-(x_0, t) + \gamma\, Q^+(x_0, t)\,. \qquad (11.2.4)$$

If the initial state of the particle is chosen with equal probability, the "average" survival probability of an RTP is $Q(x_0, t) = (Q^+(x_0, t) + Q^-(x_0, t))/2$. From Eqs. (11.2.3) and (11.2.4), we will now derive coupled differential equations for the MFPTs $\tau_\gamma^\pm(x_0)$, i.e., the MFPT starting from $x_0 \geq 0$ with $\sigma(0) = \pm 1$. To proceed, it is useful to recall that we have the important relation $F_{\text{fp}}(x_0, t) = -\partial_t Q(x_0, t)$, where $F_{\text{fp}}(x_0, t)$ is the probability density function of the first-passage time $T$ – see Section 1.5.1. This means that $F_{\rm fp}(x_0, t)\, dt$ is the probability that an RTP initially located at $x_0$ reaches the origin for the first time in the interval $[t, t + dt]$. The MFPT $\tau_\gamma(x_0)$ is thus given by

$$\tau_\gamma(x_0) = \int_0^{+\infty} dt\, t\, F_{\rm fp}(x_0, t) = -\int_0^{+\infty} dt\, t\, \partial_t Q(x_0, t)\,, \qquad (11.2.5)$$

and similarly

$$\tau_\gamma^\pm(x_0) = -\int_0^{+\infty} dt\, t\, \partial_t Q^\pm(x_0, t)\,. \qquad (11.2.6)$$

To find the coupled differential equations for $\tau_\gamma^\pm(x_0)$ (assuming that both MFPTs are finite), we differentiate Eqs. (11.2.3) and (11.2.4) with respect to $t$, multiply by $t$, and integrate them over $t$ from 0 to $+\infty$. Using the initial conditions $Q^\pm(x_0 > 0, t = 0) = 1$ and assuming that $Q^\pm(x_0, t = +\infty) = 0$ (otherwise the MFPT is infinite)[13] one obtains (see Appendix H)

$$[f(x_0) + v_0]\, \partial_{x_0} \tau_\gamma^+(x_0) - \gamma \tau_\gamma^+(x_0) + \gamma \tau_\gamma^-(x_0) = -1\,, \qquad (11.2.7)$$
$$[f(x_0) - v_0]\, \partial_{x_0} \tau_\gamma^-(x_0) + \gamma \tau_\gamma^+(x_0) - \gamma \tau_\gamma^-(x_0) = -1\,. \qquad (11.2.8)$$

From the coupled equations (11.2.7) and (11.2.8) one can show that the MFPT

$$\tau_\gamma(x_0) = \left(\tau_\gamma^+(x_0) + \tau_\gamma^-(x_0)\right)/2\,, \qquad (11.2.9)$$

obeys a closed ordinary differential equation

$$\boxed{\left[v_0^2 - f(x_0)^2\right] \partial_{x_0}^2 \tau_\gamma(x_0) + 2 f(x_0) \left[\gamma - f'(x_0)\right] \partial_{x_0} \tau_\gamma(x_0) = f'(x_0) - 2\gamma}\,. \qquad (11.2.10)$$

The computation of $\tau_\gamma(x_0)$ gives us also the full solution for $\tau_\gamma^\pm(x_0)$. Indeed, by manipulating Eqs. (11.2.7) and (11.2.8), one can express $\tau_\gamma^\pm(x_0)$ in terms of $\tau_\gamma(x_0)$ and its first derivative as (see again Appendix H)

$$\tau_\gamma^-(x_0) = \frac{f(x_0)}{2\gamma\, v_0} - \frac{1}{2\gamma v_0} \left[v_0^2 - f(x_0)^2\right] \partial_{x_0} \tau_\gamma(x_0) + \tau_\gamma(x_0)\,, \qquad (11.2.11)$$

$$\tau_\gamma^+(x_0) = -\frac{f(x_0)}{2\gamma\, v_0} + \frac{1}{2\gamma v_0} \left[v_0^2 - f(x_0)^2\right] \partial_{x_0} \tau_\gamma(x_0) + \tau_\gamma(x_0)\,. \qquad (11.2.12)$$

Interestingly, the differential equation satisfied by $\tau_\gamma(x_0)$ is a second order differential equation but it is a first order differential equation for $\partial_{x_0} \tau_\gamma(x_0)$. It can thus be solved explicitly for any $f(x)$ up to two integration constants, which have to be fixed by appropriate boundary conditions. We will see below that these conditions depend in a rather subtle way on the force $f(x)$, which eventually lead to the different "phases" discussed below (see Figs. 11.2 and 11.3).

We conclude this subsection by a comment on the higher moments of the PDF $F_{\rm fp}(x_0, t)$. Here we focus on the MFPT, which is the first moment of this PDF. In fact, starting from Eqs. (11.2.3) and (11.2.4) and performing manipulations very similar to the ones presented here and leading to Eq. (11.2.10) – see Appendix I – one can in principle compute recursively the higher moments $\langle T^n \rangle$ of this PDF with $n \geq 2$. These higher moments can be obtained from the following differential equation

$$\left(v_0^2 - f(x_0)^2\right) \partial_{x_0}^2 \langle T^n \rangle + 2 f(x_0) \left[\gamma - f'(x_0)\right] \partial_{x_0} \langle T^n \rangle$$
$$= n\, (f'(x_0) - 2\gamma) \langle T^{n-1} \rangle + 2 n f(x_0) \partial_{x_0} \langle T^{n-1} \rangle + n^2 \langle T^{n-2} \rangle \quad,\quad n \geq 2\,, \qquad (11.2.13)$$

---
[13] It is also possible to compute the MFPT conditioned on trajectories that reach the origin in finite time, as was done in [354] for the case of periodic forces.

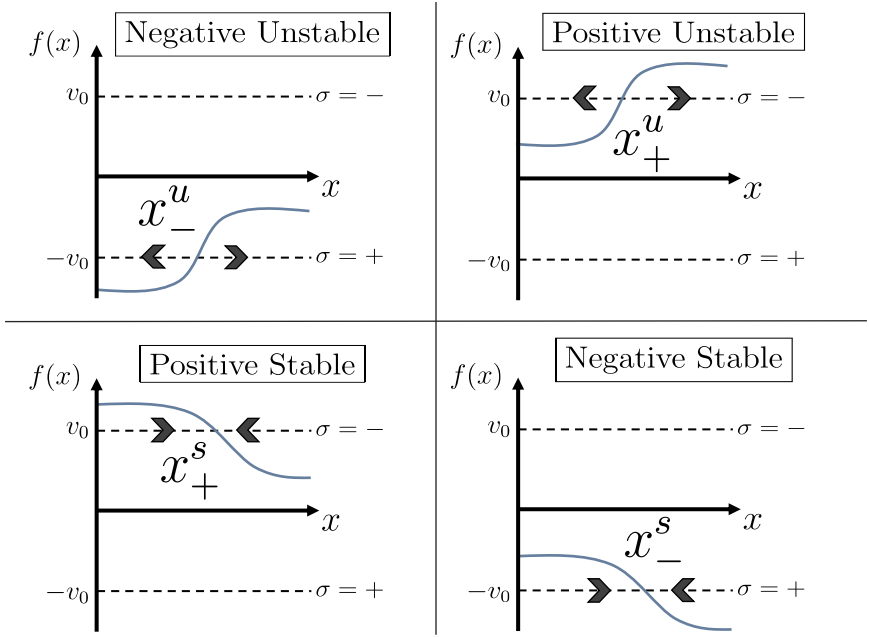

**Figure 11.1:** Schematic of the four possible turning points for a one-dimensional run-and-tumble particle subjected to a force $f(x)$. We refer to Table 11.1 for a precise definition of the turning points. In each panel, the solid blue curve shows $f(x)$ versus position $x$, the dashed horizontal lines indicate the two self-propulsion speeds $\pm v_0$ (with corresponding internal state $\sigma = \pm$), and the black arrows denote the drift direction in each state on the left or right of the turning point. The panels represent (top left) the *negative unstable* turning point $x^u_-$, (top right) the *positive unstable* turning point $x^u_+$, (bottom left) the *positive stable* turning point $x^s_+$, and (bottom right) the *negative stable* turning point $x^s_-$.

which generalizes the recursion relations found in the passive case [39, 77] – see also Appendix I. Interestingly, the differential operator acting on $\langle T^n \rangle$ on the left hand side of (11.2.13) does not depend on $n$ – and is thus the same as in the case $n = 1$ in Eq. (11.2.10). However the right hand side depends explicitly on $n$ and involves the lower moments of $T$. Thus, in principle, these differential equations can be solved recursively. Here, we provide a detailed analysis of the first moment ($n = 1$), and it would be very interesting to extend it to higher moments $n > 1$.

## 11.3 Explicit Solutions in Different Phases

Having established the differential equations (11.2.10)-(11.2.12) satisfied by $\tau^\pm_\gamma(x_0)$ and $\tau_\gamma(x_0) = (\tau^+_\gamma(x_0) + \tau^-_\gamma(x_0))/2$, we now need to fix the appropriate boundary conditions for these functions. Here we will restrict our analysis to the case where $f(x)$ is continuous. A first boundary condition can be easily found from the following argument. First, we notice that if $f(0) \geq v_0$, the origin can never be crossed, starting from $x_0 \geq 0$ since, from Eq. (11.2.1), the force felt by the particle at 0 is always positive, namely $f(0) \pm v_0 \geq 0$ if $f(0) \geq v_0$. Therefore the computation of the MFPT for $f(0) \geq v_0$ is trivial since $\tau^\pm_\gamma(x_0) = \tau_\gamma(x_0) = +\infty$. Hence we only need to consider situations where $f(0) < v_0$. Let us then analyse the MFPT starting at the origin $x_0 = 0$. In this case, from Eq. (11.2.1), we see that $\dot{x}(0) = f(0) + \sigma(0)v_0$. Therefore, if $\sigma(0) = -1$, the initial velocity is negative: namely $\dot{x}(0) = f(0) - v_0 < 0$ since $-v_0 < f(0) < v_0$ (while the initial velocity $\dot{x}(0) > 0$ if $\sigma(0) = +1$). Thus if $\sigma(0) = -1$ and $x_0 = 0$ the position of the RTP $x(t) < 0$ for $t = 0^+$ and we thus obtain the first boundary condition

$$\boxed{\tau^-_\gamma(0) = 0}, \qquad (11.3.1)$$

while a priori $\tau^+_\gamma(0) > 0$.

The second boundary condition, which still needs to be fixed, depends crucially on the existence or not of fixed points (or turning points) of the dynamics, i.e., values of $x$ such that $f(x) \pm v_0 = 0$. We show the different turning points in Fig. 11.1. In general, we need to distinguish different types of turning points [355]: negative (respectively positive) turning points when $f(x_-) = -v_0$ (respectively $f(x_+) = v_0$), that may be stable and denoted by $x_\pm^s$ (respectively unstable and denoted by $x_\pm^u$) if $f'(x_\pm^s) < 0$ (respectively $f'(x_\pm^u) > 0$) – see Table 11.1. The stability of the turning points $x_\pm$ has a strong influence on the behavior of $\tau_\gamma(x_0)$ around $x_0 = x_\pm$ (see below). Specifically, depending on the starting position $x_0$, the particle may become permanently trapped in certain regions of space, making the origin inaccessible from $x_0$. For instance, in the presence of a positive turning point $x_+$, if $x_0 > x_+$ the RTP is unable to reach the origin and therefore in this case $\tau_\gamma(x_0) = +\infty$ (see e.g., the right panel of Fig. 11.3 when $x_0 > x_+^u$).

To cover most of the different force landscapes $f(x)$, it turns out that it is sufficient to compute the MFPT in four different situations where there are either no turning point – "Phase I" and "Phase II" – or only one turning point – "Phase III" and "Phase IV". Note that when there is one turning point, since $f(0) < v_0$, this turning point can be either negative-stable ("Phase III") or positive-unstable ("Phase IV") – corresponding to $f(0) > -v_0$ – as well as negative-unstable – corresponding to $f(0) < -v_0$. Note however that this turning point can not be positive-stable (because $f(0) < v_0$). In fact, it turns out that the MFPT in the case of a negative-unstable turning point can actually be obtained by combining the results of Phase II and Phase I, as discussed below (see also the right panel of Fig. 11.4). Hence there are only four distinct phases to consider. The formula for the MFPT in more complicated situations, i.e., with a higher number of fixed points can then be obtained by "gluing" together the results obtained from these four "building blocks", as discussed and illustrated below in some specific examples.

We now present our results for these four different phases, which lead to different expressions of the MFPT. In particular, we provide and explain the additional condition required to supplement Eq. (11.3.1). Together, these conditions allow one to solve Eq. (11.2.10). While we do not detail the solution of the differential equation here, references [2, 5] provide a thorough treatment. Note that, using the explicit expressions for $\tau_\gamma(x_0)$ obtained in the different phases, one can deduce the corresponding expressions for $\tau_\gamma^\pm(x_0)$ from Eqs. (11.2.11) and (11.2.12).

| Type of Turning Point | Notation | Condition on $f(x)$ | Stability Criterion |
|---|---|---|---|
| Negative stable | $x_-^s$ | $f(x_-^s) = -v_0$ | $f'(x_-^s) < 0$ |
| Negative unstable | $x_-^u$ | $f(x_-^u) = -v_0$ | $f'(x_-^u) > 0$ |
| Positive stable | $x_+^s$ | $f(x_+^s) = v_0$ | $f'(x_+^s) < 0$ |
| Positive unstable | $x_+^u$ | $f(x_+^u) = v_0$ | $f'(x_+^u) > 0$ |

Table 11.1: Classification of the four turning point types for a one-dimensional run-and-tumble particle (RTP) under an external force $f(x)$. Each row provides the notation for the fixed point, the condition $f(x) = \pm v_0$ that defines its location, and the sign of the derivative $f'(x)$ which determines its stability. See Fig. 11.1 for schematic illustrations. These turning points play a crucial role in the dynamics of the RTP and in formulating the boundary conditions used to solve the differential equation satisfied by the mean first-passage time – see Eq. (11.2.10).

### 11.3.1 Phase I: $|f(x)| < v_0$

In this case there is no turning point – see the left panel of Fig. 11.2 for a schematic description of this situation. To fix the second boundary condition (in addition to the first one in Eq. (11.3.1)), we impose a reflecting barrier at position $L > x_0$, and then take the limit $L \to \infty$. One can show that this leads to the following boundary condition (see Appendix J.1)

$$\boxed{\left.\partial_{x_0}\tau_\gamma^+(x_0)\right|_{x_0=L} = 0}. \qquad (11.3.2)$$

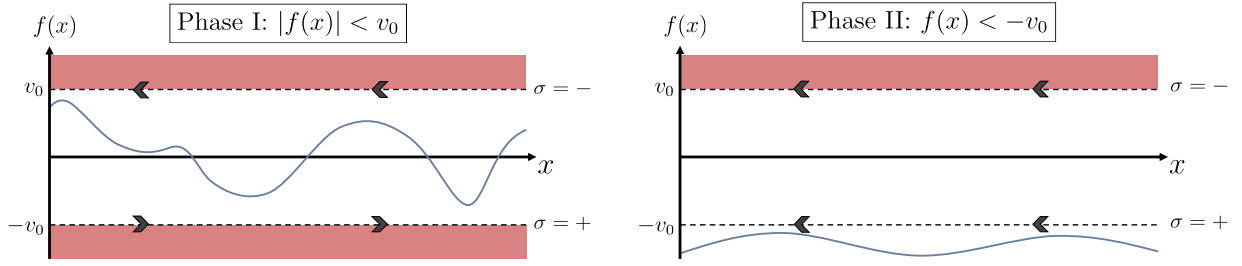

**Figure 11.2: Left panel**: We show a force such that $|f(x)| < v_0$ which correspond to phase I. **Right panel**: in phase II, the force is bounded such that $f(x) < -v_0$. The arrows on the dotted lines show the direction of the velocity of the RTP in state $\sigma = \pm$. If the arrow is directed to the right (left), the velocity is positive (negative) in this region.

One can then solve the Eqs. (11.2.10)-(11.2.12) with these two boundary conditions (11.3.1) and (11.3.2) and eventually take the limit $L \to \infty$. This leads to the following expression for the MFPT in this case (see Appendix IV in [5])

$$\tau_\gamma(x_0) = \frac{1}{2\gamma} + \int_0^{+\infty} \frac{dy}{v_0 - f(y)} \exp\left[\int_0^y du \frac{2\gamma f(u)}{v_0^2 - f(u)^2}\right]  \\ - \int_0^{x_0} \frac{dz}{v_0^2 - f(z)^2} \int_z^{+\infty} dy \left(f'(y) - 2\gamma\right) \exp\left[\int_y^z du \frac{-2\gamma f(u)}{v_0^2 - f(u)^2}\right], \tag{11.3.3}$$

while $\tau_\gamma^\pm(x_0)$ can be obtained from Eqs. (11.2.11) and (11.2.12).

A simple example belonging to this phase I is the linear potential $V(x) = \alpha|x|$, with $-v_0 < -\alpha < v_0$. For positive values of $x \geq 0$, the force is $f(x) = -\alpha$. Hence substituting this expression for $f(x)$ in Eq. (11.3.3), one obtains (see also [5, 226, 356])

$$\tau_\gamma(x_0) = \frac{x_0}{\alpha} + \frac{v_0}{2\alpha\gamma}. \tag{11.3.4}$$

### 11.3.2 Phase II: $f(x) < -v_0$

Clearly here there is no turning point – this phase is illustrated on the right panel of Fig. 11.2. Since $f(0) < -v_0$, if the particle starts at the origin $x_0 = 0$, the initial velocity $\dot{x}(0) = f(0) + \sigma(0)v_0 < 0$ in any of the two states $\sigma(0) = \pm 1$. In this case, the boundary conditions are thus

$$\boxed{\tau_\gamma^+(x_0 = 0) = 0\,,} \tag{11.3.5}$$

$$\boxed{\tau_\gamma^-(x_0 = 0) = 0\,.} \tag{11.3.6}$$

By solving Eq. (11.2.10) with these two boundary conditions (11.3.5) and (11.3.6), we show in Appendix D of [2], that the MFPT is given by

$$\tau_\gamma(x_0) = \int_0^{x_0} dz \frac{1}{v_0^2 - f(z)^2} \left[\int_0^z dy \left(f'(y) - 2\gamma\right) \exp\left(\int_y^z du \frac{-2\gamma f(u)}{v_0^2 - f(u)^2}\right) \right. \\ \left. + f(0) \exp\left(\int_0^z du \frac{-2\gamma f(u)}{v_0^2 - f(u)^2}\right)\right]. \tag{11.3.7}$$

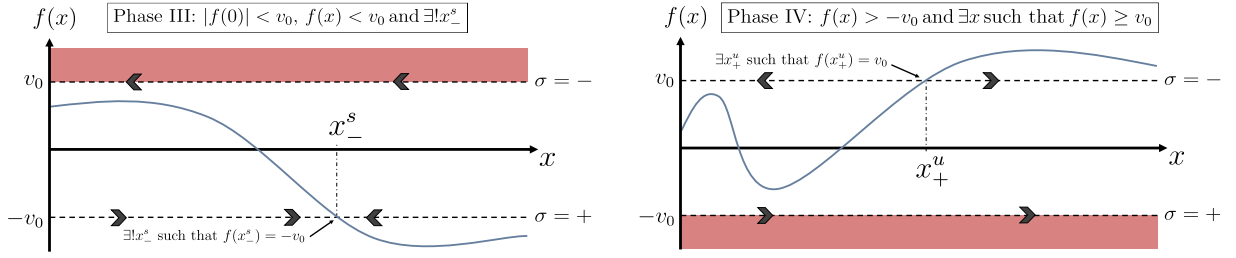

**Figure 11.3: Left panel:** we show an example of force from Phase III in which there is a unique stable negative turning point $x_-^s$ and we have $|f(0)| < v_0$ as well as $f(x) < v_0$. **Right Panel:** in Phase IV, the turning point is positive and unstable. The direction of the velocity of the RTP in state $\sigma = \pm$ is indicated by the arrows on the dotted lines. When the arrow points to the right (left), the velocity is positive (negative) in that region.

Here also, the linear potential $V(x) = \alpha|x|$ with $\alpha > v_0$ is an illustration for this phase. By substituting $f(x) = -\alpha < -v_0$ in the general expression (11.3.7) one finds

$$\tau_\gamma(x_0) = \frac{x_0}{\alpha} + \frac{v_0^2}{2\alpha^2 \gamma}\left(1 - e^{-\frac{2\alpha\gamma x_0}{\alpha^2 - v_0^2}}\right), \tag{11.3.8}$$

thus recovering the result obtained in [5, 226, 356].

### 11.3.3 Phase III

Here, we consider a force $f(x) < v_0$ such that $|f(0)| < v_0$ but, at variance with Phase I, we assume here that there is a unique stable negative turning point $x_-^s$, i.e., $f(x_-^s) = -v_0$ - see Figure 11.3. The second condition arises when writing the Fokker-Plank equations (11.2.3) and (11.2.4) at $x_-^s$ – see Appendix J.2. It reads

$$\boxed{\lim_{x_0 \to x_-^s}\left(f(x_0) + v_0\right)\partial_{x_0}\tau_\gamma^+(x_0) = 0}. \tag{11.3.9}$$

Solving Eq. (11.2.10) with boundary conditions (11.3.1) and (11.3.9) leads to (see Appendix II of [5])

$$\boxed{\begin{aligned}\tau_\gamma(x_0) &= \frac{1}{2\gamma} + \int_0^{x_-^s}\frac{dy}{v_0 - f(y)}\exp\left[\int_0^y du\,\frac{2\gamma f(u)}{v_0^2 - f(u)^2}\right] \\ &\quad + \int_0^{x_0}dz\,\frac{1}{v_0^2 - f(z)^2}\int_{x_-^s}^z dy\,\left(f'(y) - 2\gamma\right)\exp\left[\int_y^z du\,\frac{-2\gamma f(u)}{v_0^2 - f(u)^2}\right]\end{aligned}}. \tag{11.3.10}$$

For a particle inside the interval $[0, x_-^s[$, the turning point acts as a reflective hard wall explaining the similarities in the formulae (11.3.10) and (11.3.3) (see the Supp. mat. of [5]). Phase III includes for instance the harmonic potential $V(x) = \mu x^2/2$ with $\mu > 0$ leading to a force applied on the RTP given by $f(x) = -\mu x$. Since $x > 0$ and $\mu > 0$, we indeed have $f(x) < v_0$, and there is a stable negative turning point $x_-^s = v_0/\mu$. The MFPT is given by

$$\boxed{\begin{aligned}\tau_\gamma(x_0) &= \frac{\sqrt{\pi}}{2\gamma}\frac{\Gamma\left(1 + \frac{\gamma}{\mu}\right)}{\Gamma\left(\frac{1}{2} + \frac{\gamma}{\mu}\right)}\left[1 + 2\gamma\frac{x_0}{v_0}\,{}_2F_1\left(\frac{1}{2}, 1 + \frac{\gamma}{\mu}; \frac{3}{2}; \frac{\mu^2 x_0^2}{v_0^2}\right)\right] \\ &\quad - (2\gamma + \mu)\frac{x_0^2}{2v_0^2}\,{}_3F_2\left(\left\{1, 1, \frac{3}{2} + \frac{\gamma}{\mu}\right\}; \left\{\frac{3}{2}, 2\right\}; \frac{\mu^2 x_0^2}{v_0^2}\right)\end{aligned}}, \tag{11.3.11}$$

where $_2F_1(.;z)$ and $_3F_2(.;z)$ are hypergeometric functions [357]. Another interesting example is the double-well potential $V(x) = \alpha/2\left(|x|-1\right)^2$. The force is then $f(x) = -\alpha(x-1)$ and if $0 < \alpha < v_0$, this is indeed a phase III case. The double-well potential is studied in detail in [2].

### 11.3.4 Phase IV

Consider a case where $-v_0 < f(0) < v_0$ and there is a positive unstable turning point $x_+^u$, i.e., $f(x_+^u) = v_0$ and $f'(x_+^u) > 0$. Therefore, if the position of the RTP $x(t)$ is such that $x(t) > x_+^u$ then its velocity is necessarily positive in both states $\sigma(t) = \pm 1$ (see the right panel of Fig. 11.3). Hence, if a particle reaches or surpasses this point $x_+^u$, it will never be able to return to the origin. Clearly, such a trajectory will give an infinite contribution to the MFPT – see the left panel of Figure 11.3 – and in this case we simply have

$$\tau_\gamma(x_0) = +\infty \, . \tag{11.3.12}$$

An illustration of this phase is the (inverted) double-well potential $V(x) = \alpha/2\left(|x|-1\right)^2$ when $-v_0 < \alpha < 0$ - see [2].

### 11.3.5 Combination of Phases

The four phases introduced above constitute the building blocks to derive the MFPT for a general force $f(x)$. The general strategy is to study the force $f(x)$ on distinct intervals, each corresponding to one of the four phases discussed above. The full expression of the MFPT is then obtained by ensuring the continuity at the boundaries of these intervals. Below, to illustrate this construction, we provide the full solutions in two different situations:

- *Combination of phases: Example 1.* Consider a force such that $f(0) < -v_0$ which displays an unstable negative turning point $x_-^u$, and, in addition, an unstable positive turning point $x_+^u$ such that $x_+^u > x_-^u$. This phase is illustrated on the left panel of Fig. 11.4. For $x_0 \in [0, x_-^u[$, the MFPT can be computed as a phase II case, while for $x_0 \in ]x_-^u, +\infty[$, this is a phase IV case and the MFPT diverges. We solve an example of such a force in [2]. It corresponds to the inverted double-well $V(x) = \alpha/2(|x|-1)^2$ when $\alpha < -v_0$.

    Similarly, we could also consider the case $|f(0)| < v_0$ with a stable turning point $x_-^s$, and an unstable turning point $x_-^u > x_-^s$ in addition to the positive turning point $x_+^u$. In this case, the MFPT on the interval $[0, x_-^u[$ is described by phase III, i.e., Eq. (11.3.10).

- *Combination of phases: Example 2.* Consider now another, more complicated, example where $f(0) < -v_0$ such that there exists an unstable negative turning point $x_-^u$, and $|f(x)| < v_0$, see the right panel of Fig. 11.4 for an illustration. Concrete examples include for instance $f(x) = -1/(1+x)$ deriving from a log-potential $V(x) = \log(1+|x|)$ that we study in [2]. This situation requires a careful analysis. Indeed we have to solve the MFPT separately for $x_0 \in [0, x_-^u[$, denoted by $\tau_{\text{I}\gamma}(x_0)$ and for $x_0 \in ]x_-^u, +\infty[$, denoted by $\tau_{\text{II}\gamma}(x_0)$. We have here four integration constants to fix (i.e., two for each interval). First, on $[0, x_-^u[$, as $f(0) < -v_0$, it is a phase II case, and $\tau_{\text{I}\gamma}(x_0)$ is thus given by Eq. (11.3.7), that fixes two integration constants. A third integration constant is fixed by imposing the continuity of the MFPT at $x_-^u$. To fix the remaining constant, we introduce a reflecting wall at infinity, as done in phase I. We provide the details of the derivation in Appendix E of [2] and the solution on $]x_-^u, +\infty[$ is given by

$$\tau_{\text{II}\gamma}(x_0) = \tau_{\text{I}\gamma}(x_-^u) - \int_{x_-^u}^{x_0} dz \, \frac{1}{v_0^2 - f(z)^2} \int_z^{+\infty} dy \, \left(f'(y) - 2\gamma\right) \exp\left[\int_y^z du \, \frac{-2\gamma \, f(u)}{v_0^2 - f(u)^2}\right] \, . \tag{11.3.13}$$

    This formula is valid provided the integrals over $z$ and $y$ are well defined. If this not the case, this means that the MFPT is infinite, i.e., $\tau_{\text{II}\gamma}(x_0) = +\infty$ for $x_0 > x_-^u$. For instance, if

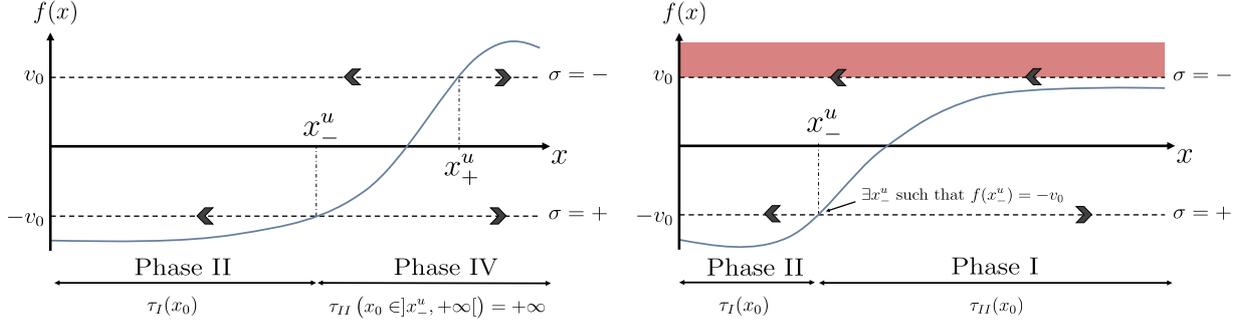

**Figure 11.4: Left panel:** the force has an unstable negative turning point $x_-^u$, and an unstable positive turning point $x_+^u$, with $x_+^u > x_-^u$. The force at the origin is such that $f(0) < -v_0$. On $[0, x_-^u[$, the MFPT is derived as in Phase II, while for $x_0 > x_-^u$, it diverges as in Phase IV. **Right panel:** the force has an unstable negative turning point $x_-^u$. We also have $f(0) < -v_0$ and $f(x) < v_0$. On $[0, x_-^u[$ the MFPT is computed as in Phase II. However, on the right of $x_-^u$, the force is such that $|f(x)| < v_0$ and we use continuity at $x_-^u$ to calculate the MFPT. The direction of the RTP's velocity in the state $\sigma = \pm$ is indicated by the arrows on the dotted lines. If you see an arrow pointing to the right (left), it means the velocity is positive (negative) in that particular region.

$f(x) \to C$ as $x \to \infty$ with $|C| < v_0$ a constant, then $\tau_{\text{II}\gamma}(x_0)$ is finite for $C < 0$ and infinite for $C > 0$. The marginal case $C = 0$ is studied in detail in [2] for the case of the logarithmic potential. This formula (11.3.13) would also hold for potentials of the form $V(x) = \alpha|x|^p$ with $p < 1$. Note that we can also consider the case $|f(0)| < v_0$ with a stable turning point $x_-^s$, and another unstable turning point $x_-^u > x_-^s$ – illustrated for instance by a potential of the form $V(x) = \log(1 + x^2)$, as studied in [355]. The reasoning would be similar, and $\tau_{\text{I}\gamma}(x_0)$ would be instead described by phase III, i.e., Eq. (11.3.10).

### 11.3.6 Special Cases

Let us briefly comment on a particular class of forces, namely $f(x)$ that vanishes identically beyond a certain value $X$, i.e., $f(x) = 0$ for $x > X$, such as a barrier of potential. In this case, if there is a non zero probability for an RTP to reach the point $X$ then, since the RTP behaves diffusively at long times, the MFPT diverges as for the free Brownian motion. As it is a rather peculiar case, we will not consider this class of force.

Finally, we will not consider peculiar forces with a singular turning point $x_\pm$ that is neither stable nor unstable, i.e., with $f'(x_\pm) = 0$. However, we have verified in some specific instances that the correct combination of the different phases discussed above yields the complete solution (for example, see the remark in Appendix D of [2]).

# Chapter 12

# Optimal MFPT of an RTP in a Class of Confining Potentials

## 12.1 Introduction

A particularly interesting class of potentials corresponds to confining potentials of the form $V(x) = \alpha |x|^p$, with $p \geq 1$. For this model, the position distribution $p_s(x)$ in the steady state is well known [216, 341] – see also Section 4.3. Indeed, $p_s(x)$ has actually a *finite* support $[-x_e, x_e]$ with two edges at $\pm x_e = \pm(v_0/(\alpha p))^{1/(p-1)}$ which are the two fixed points of the dynamics (11.2.1), i.e., $f(\pm x_e) = \mp v_0$. In addition, close to the edges at $\pm x_e$, the stationary distribution behaves as $p_s(x) \propto |x \mp x_e|^\phi$ with a nontrivial exponent $\phi$ given by

$$\phi = \frac{1}{p(p-1)} \left(\frac{v_0}{p}\right)^{\frac{2-p}{p-1}} \frac{\gamma}{\alpha^{1/(p-1)}} - 1 \,. \tag{12.1.1}$$

Thus, for fixed $\alpha, v_0$ and $p$ there exists a critical value

$$\gamma_c = \gamma_c(p) = p(p-1)\alpha^{1/(p-1)} \left(\frac{p}{v_0}\right)^{\frac{2-p}{p-1}} \,, \tag{12.1.2}$$

such that $\phi < 0$ for $\gamma < \gamma_c$ while $\phi > 0$ for $\gamma > \gamma_c$[14]. This indicates a shape transition for the distribution $p_s(x)$, from a bell shape with $\phi > 0$ (the passive-like phase) to a U-shape with $\phi < 0$ (the active-like phase where the particles accumulate at the edges) [216]. For $0 < p < 1$, the stationary distribution is trivially a delta-function centered at the origin and we will not discuss this case henceforth[15].

Given the solvable structure for the position distribution in this class of potentials, it is natural to ask whether the first-passage time distribution, or at least the MFPT, can be computed exactly for all $p \geq 1$. For $p > 1$, the force has one unique stable negative turning point $x_e$ (see Fig. 12.1), therefore, the expression of the MFPT $\tau_\gamma(x_0)$ is given by the expression in Phase III, i.e., Eq. (11.3.10) with $x_e \equiv x_-^s$.

This expression in Eq. (11.3.10) becomes fully explicit in the case of a harmonic potential (i.e., $p = 2$) – see Eq. (11.3.11) – as well as in the limiting case $p \to \infty$ – see Eq. (12.4.1). Based on this exact formula (11.3.10), together with numerical simulations, we show that there indeed exists a finite optimal tumbling rate $\gamma_{\rm opt}$ that minimizes the MFPT for all $p > 1$ (see Fig. 12.2), and we characterize its dependence on $x_0$ (see Fig. 12.3 for $p = 2$). Interestingly, we also find that the MFPT displays rather different behaviors for $p < 2$ (left panel of Fig. 12.4) and $p \geq 2$ (right panel of Fig. 12.4). Indeed, while $\tau_\gamma(x_0)$ diverges as $\gamma \to 0$ for all $p > 1$, it also diverges, as $\gamma \to \infty$, for $p \geq 2$ but remains finite for $1 < p < 2$. Below, we provide a physical explanation for these analytical results and numerical observations. We further show that only $\tau_\gamma^+(x_0)$ exhibits a minimum as a function of $\gamma$, while $\tau_\gamma^-(x_0)$ is a monotonically increasing function of $\gamma$. Finally, we study the case $p = 1$ in Section 12.5, i.e., $V(x) = \alpha|x|$, with $\alpha > 0$, for which the MFPT can also be computed exactly. In this case, we show that there is no finite optimal rate $\gamma_{\rm opt}$ that minimizes $\tau_\gamma(x_0)$. Besides, we show that $\tau_\gamma(x_0)$ exhibits very different functional form for $\alpha < v_0$ (12.5.1) and $\alpha > v_0$ (12.5.3) – see also Fig. 12.5. Establishing analytically the existence of an optimal tumbling rate $\gamma_{\rm opt}$ for all $p > 1$ is one of the central results of this chapter.

---

[14] In Ref. [216], the transition was described in the $(\alpha, p)$ plane, instead of $(\gamma, p)$, but it is of course totally equivalent.

[15] Note that we discussed in [2] the behavior of the MFPT in the case $0 < p < 1$ for which the force has an unstable negative turning point. The numerical evaluations of the exact formula (11.3.13) for $V(x) = \alpha |x|^p$ with $0 < p < 1$ seem to indicate that none of the MFPT $\tau_\gamma(x_0)$ and $\tau_\gamma^\pm(x_0)$ exhibit a minimum in contrast to the case $p > 1$ (see below). This shows that the MFPT of an RTP in a potential of the form $V(x) = \alpha |x|^p$ behaves quite differently for $p < 1$ and $p > 1$ ($p = 1$ being thus a borderline case).

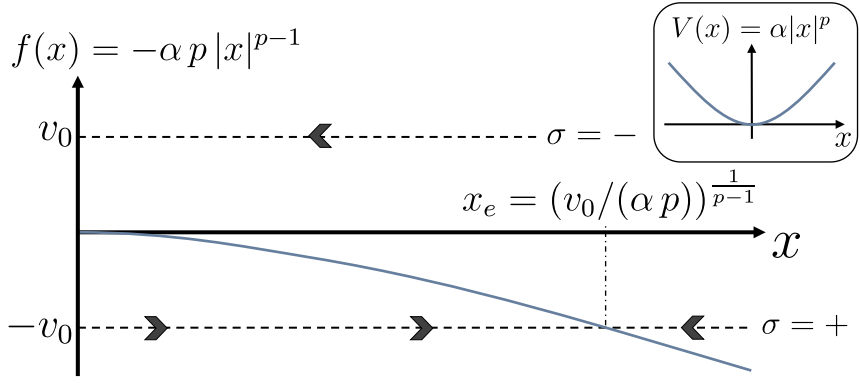

**Figure 12.1:** Schematic motion of an RTP inside a confining potential $V(x) = \alpha|x|^p$, and the associated force is $f(x) = -\alpha p x^{p-1}$, when $p > 1$. The force has a stable negative fixed point $x_e = (v_0/(\alpha p))^{\frac{1}{p-1}}$ such that $f(x_e) = -v_0$. Note that when $1 < p \leq 2$, the concavity of the force is not the same as the one represented in the figure. The arrows on the dotted lines show the direction of the velocity of the RTP in state $\sigma = \pm$. If the arrow is directed to the right (left), the velocity is positive (negative) in this region.

In the specific case $f(x) = -\alpha p x^{p-1}$, it turns out that $\tau_\gamma(x_0)$ can be written under the scaling form

$$\tau_\gamma(x_0) = \frac{1}{\gamma_c} \mathcal{F}_p\left(\beta = \frac{\gamma}{\gamma_c}, u = \frac{x_0}{x_e}\right), \qquad (12.1.3)$$

where we recall that $x_e = (v_0/(\alpha p))^{1/(p-1)}$ while $\gamma_c$ is given in Eq. (12.1.2). For general $p > 1$, the MFPT $\tau_\gamma(x_0)$ (or equivalently $\mathcal{F}_p(\beta, u)$) is given in terms of a double integral, which can however be studied in detail (see below). However for $p = 2$, on which we first focus, $\tau_\gamma(x_0)$ can be explicitly computed.

The details of the derivation of the results presented below can be found in the supplementary materials of [5].

## 12.2 The Case $p = 2$ (Harmonic Potential)

In this case where $V(x) = \mu x^2/2$ (i.e., we use $\alpha = \mu/2$), the double integral in Eq. (11.3.10) can be computed explicitly and $\tau_\gamma(x_0)$ takes the scaling form as in Eq. (12.1.3) with $\gamma_c = \mu$,

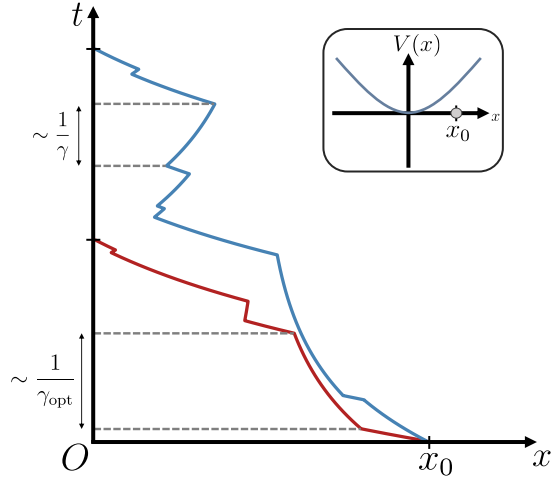

**Figure 12.2:** Schematic trajectories of an RTP in a confining potential $V(x) = \alpha|x|^p$ with $p > 1$, for two different values of the tumbling rate $\gamma$. The typical time between two consecutive tumblings scales as $1/\gamma$. The trajectory stops when the particle hits the origin for the first time, starting at $x_0$. Here $\gamma_{\text{opt}}$ denotes the optimal tumbling rate that minimizes the MFPT to the origin.

$x_e = v_0/\mu$ and

$$\begin{aligned}\mathcal{F}_2(\beta, u) &= \frac{\sqrt{\pi}}{2\beta} \frac{\Gamma(1+\beta)}{\Gamma\left(\frac{1}{2}+\beta\right)} \left[1 + 2\beta u \,_2F_1\left(\frac{1}{2}, 1+\beta, \frac{3}{2}; u^2\right)\right] \\ &\quad - (2\beta+1)\frac{u^2}{2} \,_3F_2\left(\{1,1,\frac{3}{2}+\beta\}, \{\frac{3}{2}, 2\}; u^2\right),\end{aligned} \quad (12.2.1)$$

where $_2F_1(\cdot;z)$ and $_3F_2(\cdot;z)$ are hypergeometric functions [357]. For some special values of $\beta$, it takes a simpler form. For instance $\mathcal{F}_2(1, u) = 2 - 1/(1+u) + \log(1+u)$. Note that although these two hypergeometric series have a branch cut along $[1, +\infty)$, the combination of these two that enters the expression of $\mathcal{F}_2(\beta, u)$ in Eq. (12.2.1) is, for any $\beta > 0$, a perfectly smooth function of $u$ on the whole real line, and in particular at $u = 1$ (i.e., for $x$ close to the edge $x_e$). We recall that one can straightforwardly deduce $\tau_\gamma^\pm$ from (12.1.3) and (12.2.1) using (11.2.11) and (11.2.12). In what follows we focus on the optimal value $\gamma_{\text{opt}}$ associated to $\tau_\gamma(x_0)$ (but a similar analysis can be carried out for $\gamma_{\text{opt}}^+$ associated to $\tau_\gamma^+(x_0)$).

From the explicit expression (12.2.1) one can straightforwardly obtain the behaviors of $\tau_\gamma(x_0)$ in the two opposite limits $\gamma \to 0$ and $\gamma \to \infty$. They read (returning back to dimensionful variables)

$$\tau_\gamma(x_0) \sim \begin{cases} \dfrac{1}{2\gamma} & , \quad \gamma \to 0, \\ \dfrac{1}{2\gamma_c}\log(\gamma/\gamma_c) & , \quad \gamma \to +\infty. \end{cases} \quad (12.2.2)$$

Interestingly, the behavior of $\tau_\gamma(x_0)$ in the two limits $\gamma \to 0$ and $\gamma \to \infty$ are independent of $x_0$. The divergence in the two limits are easy to understand physically. When $\gamma \to 0$, the RTP is highly persistent and if it starts with a positive velocity at $x_0 > 0$ it will converge to the fixed point at $+x_e$ (see Fig. 12.1) and it will get stuck there for a very long time before the velocity reverses direction (since $\gamma \to 0$). This makes the MFPT divergent as $\gamma \to 0$. In the opposite limit $\gamma \to \infty$, the telegraphic noise behaves as white noise with an effective diffusion constant $D = v_0^2/(2\gamma)$. Thus the diffusion constant of the effective Ornstein-Uhlenbeck (OU) particle also vanishes in this limit for fixed $v_0$. Consequently the MFPT for the effective OU process also diverges as $\gamma \to \infty$, which can be checked from Eq. (1.5.45) when $D \to 0$. Since $\tau_\gamma(x_0)$ diverges in these two limits, this shows that there exists an optimal value $\gamma_{\text{opt}}$, which we can reasonably assume to be unique. It is however difficult to compute $\gamma_{\text{opt}}$ for any finite $x_0$. To make progress we analyse the two limits $x_0 \to 0$ and $x_0 \to \infty$. For small $x_0$, the MFPT $\tau_\gamma(x_0)$ behaves as

$$\tau_\gamma(x_0) \approx \frac{1}{\gamma_c}\left(A_\gamma + B_\gamma \frac{x_0}{x_e}\right) , \qquad (12.2.3)$$

where $A_\gamma = \sqrt{\pi}\,\Gamma(\frac{\gamma}{\gamma_c})/\Gamma(\frac{1}{2} + \frac{\gamma}{\gamma_c})$ and $B_\gamma = 2(\gamma/\gamma_c)A_\gamma$. By analysing the $\gamma$-dependence of the coefficient $A_\gamma$ and $B_\gamma$, one finds that there indeed exists an optimal value $\gamma_{\text{opt}}$ which is diverging as $x_0 \to 0$. Expanding $A_\gamma$ and $B_\gamma$ for large $\gamma$ one finds that $\gamma_{\text{opt}}$ behaves, when $x_0 \to 0$, as

$$\gamma_{\text{opt}} \approx \frac{\gamma_c}{2}\frac{x_e}{x_0} . \qquad (12.2.4)$$

For large $x_0$, it is also straightforward to obtain the behavior of $\tau_\gamma(x_0)$ from Eqs. (12.1.3) and (12.2.1). It reads

$$\tau_\gamma(x_0) = \frac{1}{\gamma_c}\left(\log(x_0/x_e) + c_\gamma + o(1)\right) , \qquad (12.2.5)$$

where $c_\gamma$ is a constant (i.e., independent of $x_0$) that can be computed explicitly. Here we see that only the subleading term, namely the constant $c_\gamma$ depends on $\gamma$. We find that $c_\gamma$ indeed admits a minimum for $\gamma_{\text{opt}} = \beta^* \gamma_c$ where $\beta^* = 1.38657\ldots$ is the solution of a transcendental equation [5].

To summarize, the optimal dimensionless rate $\beta_{\text{opt}} = \gamma_{\text{opt}}/\gamma_c$, as a function of the scaled initial distance $u = x_0/x_e$, behaves asymptotically as

$$\boxed{\beta_{\text{opt}} = \frac{\gamma_{\text{opt}}}{\gamma_c} \sim \begin{cases} \dfrac{x_e}{2x_0} & , \quad u \to 0 , \\ \beta^* = 1.38657\ldots & , \quad u \to +\infty . \end{cases}} \qquad (12.2.6)$$

In Fig. 12.3 we show a plot of $\beta_{\text{opt}}$ as a function of $u = x/x_e$ which has been obtained from the numerical minimization with respect to $\beta$ (for different values of $u = x/x_e$) of the exact expression in Eq. (12.2.1). We also compare this curve with the exact asymptotic behaviors in Eq. (12.2.6), showing a very good agreement in both limits.

## 12.3 The Generic Case $p > 1$

Let us now turn to the generic case. From the formula (11.3.10), one can extract the asymptotic behaviors of $\tau_\gamma(x_0)$ in the two limits $\gamma \to 0$ and $\gamma \to \infty$. In the limit $\gamma \to 0$, one finds [5]

$$\tau_\gamma(x_0) \approx \frac{1}{2\gamma} , \qquad (12.3.1)$$

independently of $p > 1$. The mechanism responsible for this divergence as $\gamma \to 0$ is again due to the long persistence time $1/\gamma$ of the RTP, as discussed in the $p = 2$ case below Eq. (12.2.2). The large $\gamma$ behavior of $\tau_\gamma(x_0)$ is a bit more subtle. It can again be obtained from Eq. (11.3.10) and we find that, depending on $p$ it exhibits different behaviors, namely [5]

$$\boxed{\tau_\gamma(x_0) \underset{\gamma \to \infty}{\sim} \begin{cases} \dfrac{x_0^{2-p}}{\alpha p(2-p)} & , \quad 1 < p < 2 , \\ A_p\, \gamma^{1-\frac{2}{p}} & , \quad p > 2 . \end{cases}} \qquad (12.3.2)$$

where $A_p$ is a computable constant [5]. We recall that exactly at $p = 2$, we found a logarithmic behavior [see Eq. (12.2.2)]. These behaviors in (12.3.1) and (12.3.2) are in agreement with the numerical results shown in Fig. 12.4. The limit $\gamma \to \infty$ corresponds to passive motion with a diffusion coefficient $D = v_0^2/(2\gamma) \to 0$, since $v_0$ is finite here. In this limit, Eq. (11.2.1) thus reduces to the standard Langevin equation but in the absence of noise, i.e., $dx/dt = f(x)$, since $D \to 0$. Hence the MFPT is just the time to reach the origin, starting from $x_0$, which reads in this limit $\tau_\gamma(x_0) = \lim_{\epsilon \to 0} -\int_\epsilon^{x_0} dx/f(x)$. For $p < 2$, this limit is well defined and, for $f(x) = -\alpha p x^{p-1}$ reproduces exactly the first line of Eq. (12.3.2). On the other hand, for $p > 2$, the integrand $1/f(x)$ has a nonintegrable singularity at the origin and therefore $\tau_{\gamma=0}(x_0)$ diverges, which is consistent with the second line of (12.3.2). Note that these behaviors (12.3.2) can also be recovered by analysing the passive formula (1.5.45) in the limit $D \to 0$ limit.

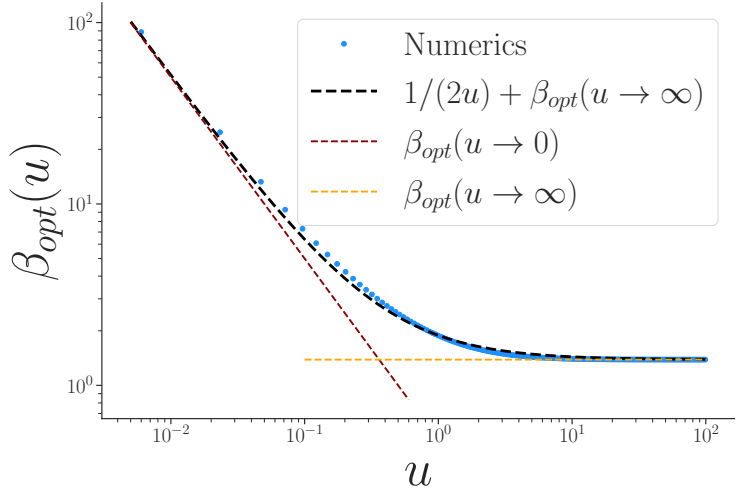

**Figure 12.3:** Plot of $\beta_{\text{opt}} = \gamma_{\text{opt}}/\gamma_c$ for the harmonic potential $V(x) = \mu x^2/2$ as a function of $u = x_0/x_e = \mu x_0/v_0$ on a log-log scale. The blue data points represent numerical computations of the minimum using Eq. (12.2.1) while the red and yellow dotted lines correspond to the asymptotic behaviors of $\beta_{\text{opt}}$ for small and large $u$ as given in Eq. (12.2.6). The black dotted line is the sum of the two asymptotic behaviors, which describes quite accurately the whole curve.

Therefore, for $p > 2$, the MFPT $\tau_\gamma(x_0)$ diverges in both limits $\gamma \to 0$ and $\gamma \to \infty$ (see the right panel of Fig. 12.4), which shows that there is an optimal rate $\gamma_{\text{opt}}$. Nevertheless, this analysis is not conclusive for $p < 2$ since $\tau_\gamma(x_0)$ remains finite for $\gamma \to \infty$ (see the left panel of Fig. 12.4). To make further progress, we considered the limit $x_0 \to 0$ where we could show that there is an optimal $\gamma_{\text{opt}}$ for all $p > 1$ given by the compact formula (see [5])

$$\boxed{\gamma_{\text{opt}} \approx \frac{\gamma_c}{2(p-1)^2} \frac{x_e}{x_0} \,, \quad \text{as } x_0 \to 0} \,, \tag{12.3.3}$$

which matches with the result in Eq. (12.2.4) for $p = 2$. Similarly, one can also analyse the limit $x_0 \to \infty$ and show that there is an optimal tumbling rate $\gamma_{\text{opt}}$ which converges to a $p$-dependent constant in that limit [5].

## 12.4 The Case $p \to \infty$

In this limiting case, one can show that the function $\exp\left[\int_0^x du \, \frac{-2\gamma f(u)}{v_0^2 - f(u)^2}\right]$ in Eq. (11.3.10) reduces simply to 1 for all $x \geq 0$. Hence the integrals in Eq. (11.3.10) can be performed explicitly. Besides $x_e \to 1$ in this limit and, interestingly, one finds that $\tau_\gamma(x_0)$ has a well defined limit as $p \to \infty$

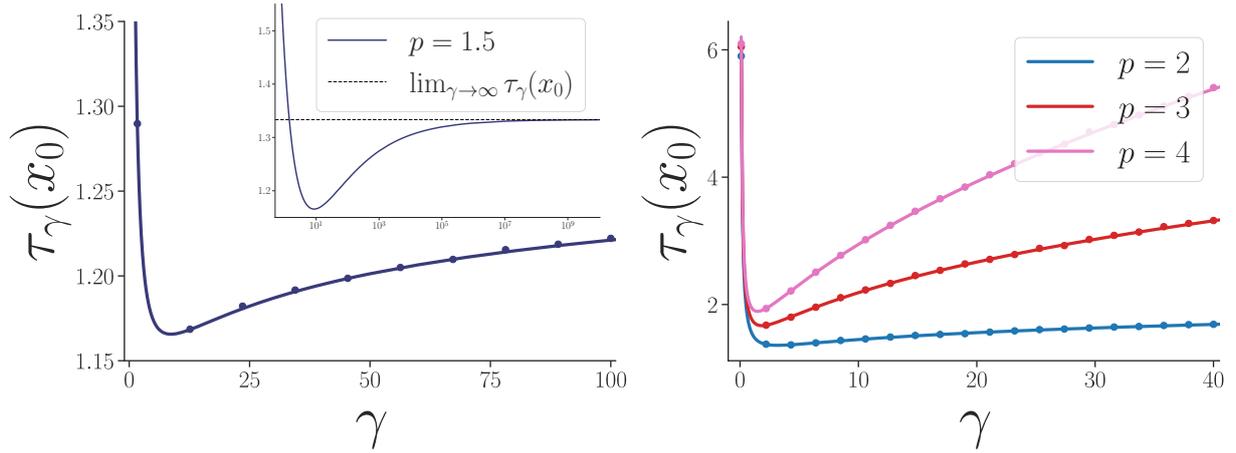

**Figure 12.4:** Plot of the MFPT $\tau_\gamma(x_0)$ as a function of the tumbling rate $\gamma$ for different values of $p$ (with $1 < p < 2$ on the left panel and $p \geq 2$ on the right panel). In both panels, the solid lines correspond to a numerical evaluation of our exact formula in Eq. (11.3.10) while the dots are the results of numerical simulations of (11.2.1), showing an excellent agreement. In all these cases, there exists an optimal rate $\gamma_{\text{opt}}$ that minimizes $\tau_\gamma(x_0)$. Here we used $v_0 = 1$, $x_0 = 1$ and $\alpha = 1$.

which reads

$$\tau_\gamma(x_0) = \begin{cases} \dfrac{1}{2\gamma} + \dfrac{x_0 + 1}{v_0} + \dfrac{\gamma x_0(2 - x_0)}{v_0^2} & , \quad x_0 \leq 1 , \\ \dfrac{1}{2\gamma} + \dfrac{2}{v_0} + \dfrac{\gamma}{v_0^2} & , \quad x_0 > 1 . \end{cases} \quad (12.4.1)$$

It is easy to see from Eq. (12.4.1) that there exists an optimal rate $\gamma_{\text{opt}}$ that minimizes $\tau_\gamma(x_0)$. This optimal rate reads

$$\gamma_{\text{opt}} = \begin{cases} \dfrac{v_0}{\sqrt{2x_0(2 - x_0)}} & , \quad x_0 \leq 1 , \\ \dfrac{v_0}{\sqrt{2}} & , \quad x_0 > 1 . \end{cases} \quad (12.4.2)$$

Hence in this case, at variance with the case $p = 2$ (see Fig. 12.3), $\gamma_{\text{opt}}$ exhibits a transition at $x_0 = 1$ where its second derivative with respect to $x_0$ is discontinuous, while $\gamma_{\text{opt}}$ itself and its first derivative are continuous. We note that the result for $x_0 < 1$ in (12.4.2) can also be obtained by noticing that in the limit $p \to \infty$ the potential $V(x) = \alpha|x|^p$ is equivalent (for particle coming from $x_0 < 1$) to a reflecting hard wall at $x = 1$. It does indeed exhibit an optimal $\gamma_{\text{opt}}$ and this result in $d = 1$ is thus the counterpart of the ones found for $d = 2, 3$ in Refs. [235, 236].

## 12.5 The Special Case $p = 1$

It turns out that for a linear potential $V(x) = \alpha |x|$ where $\alpha > 0$, the stationary state is different from the case $p > 1$ [216]. In this case, indeed, there is a nontrivial stationary state $p_s(x)$ only for $\alpha < v_0$, while $p_s(x) = \delta(x)$ for $\alpha > v_0$. In addition, for $\alpha < v_0$, in contrast to the case $p > 1$, the support of $p_s(x)$ is the whole real line, consistent with the fact that $x_e \to +\infty$ as $p \to 1$. It is thus natural to ask if there is a signature of this critical value $\alpha_c = v_0$ in the MFPT and if there still exists an optimal rate $\gamma_{\text{opt}}$ in this case.

By analysing Eq. (11.2.10), it is easy to see that the two aforementioned cases $\alpha < v_0$ and $\alpha > v_0$ need to be treated separately. For $0 < \alpha < v_0$, the force is in Phase I, and we can apply the formula given in Eq. (11.3.3). We obtain

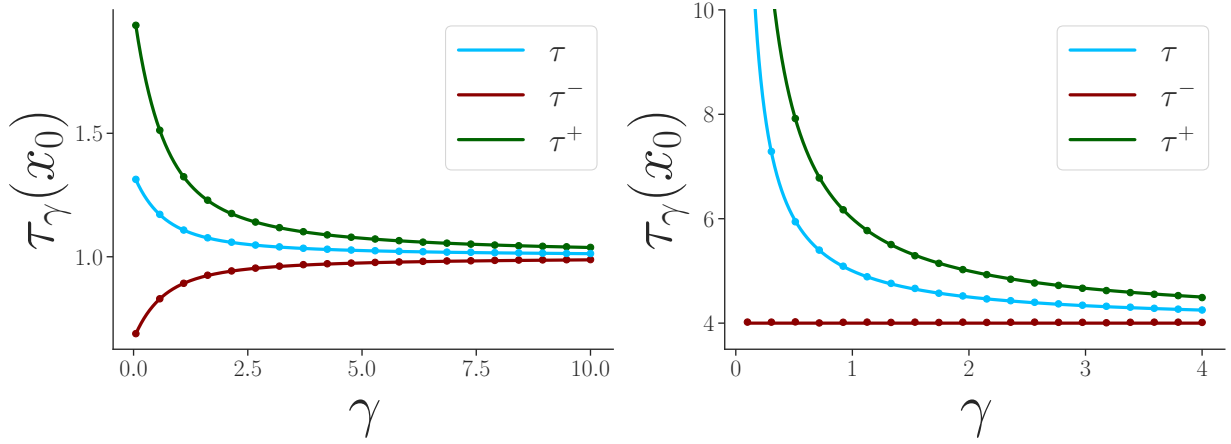

**Figure 12.5: Left panel:** Plot of the MPFT's for $V(x) = \alpha|x|$ ($p = 1$) for $\alpha < v_0$ as a function of $\gamma$. Here $v_0 = 0.5$, $\alpha = 1$ and $x_0 = 1$. **Right panel:** Same plot but for $\alpha > v_0$ – here $v_0 = 1$, $\alpha = 2$ and $x_0 = 1$. In both panels, the solid lines are the exact analytical results in Eqs. (12.5.1) and (12.5.3) while the points are the results of numerical simulations. At variance with the case $p > 1$ in Fig. 12.4, there is no finite $\gamma_{\text{opt}}$ for $p = 1$.

$$\boxed{\tau_\gamma(x_0) = \frac{x_0}{\alpha} + \frac{v_0}{2\alpha\gamma} \;,\; \alpha < v_0} \;. \tag{12.5.1}$$

From the knowledge of $\tau_\gamma(x_0)$, one can also compute $\tau_\gamma^\pm(x_0)$ using Eqs. (11.2.11) and (11.2.12). In the left panel of Fig. 12.5, we show results of simulations and compare them to the analytical results, showing a very good agreement.

In the opposite case $\alpha > v_0$, since $f(0) = -\alpha < -v_0$, the force is in Phase II and the initial velocity in the two states of the RTP $\dot x(0) = -\alpha \pm v_0$ are both negative. This implies

$$\tau_\gamma^+(x_0 = 0) = 0 \;,\; \tau_\gamma^-(x_0 = 0) = 0 \;. \tag{12.5.2}$$

We can then use Eq. (11.3.7) to obtain

$$\boxed{\tau_\gamma(x_0) = \frac{x_0}{\alpha} + \frac{v_0^2}{2\alpha^2\gamma}\left(1 - e^{-\frac{2\alpha\gamma x_0}{\alpha^2 - v_0^2}}\right) \;,\; \alpha > v_0} \;. \tag{12.5.3}$$

Again, using Eq. (11.2.11) and Eq. (11.2.12), we can compute the MFPT's $\tau_\gamma^\pm(x_0)$. For a comparison with simulations, see the right panel of Fig. 12.5.

The first observation is that for $p = 1$, there is no finite $\gamma_{\text{opt}}$ that minimizes the MFPT (see Fig. 12.5): this is in sharp contrast with the case $p > 1$ studied above. In addition, by considering $\tau_\gamma(x_0)$ in Eqs. (12.5.1) and (12.5.3) as a function of $\alpha$, we see that, although it is continuous as $\alpha$ crosses $v_0$, the first derivative of $\tau_\gamma(x_0)$ with respect to $\alpha$ is discontinuous at $\alpha = v_0$. It shows that when $p = 1$ the MFPT carries a signature of the transition found in the stationary state. This is in contrast to the case $p > 1$ where the transition from a bell-shape to a U-shape does not show up so drastically in the MFPT, except in the limit $p \to \infty$.

# Chapter 13

# Applications

## 13.1 Kramers' Law for RTP

For a passive or diffusive particle with diffusion coefficient $D$, Kramers' law gives the main contribution to the averaged time needed for a particle to cross a barrier in the weak noise limit $D \ll \Delta V$ where $\Delta V$ is the barrier height. Let us suppose that the potential has a local minimum at $x_{\min} > 0$, and it can only escape through the left where a local maximum is located at the origin (see Fig. 13.1). The barrier height is then $\Delta V = V(0) - V(x_{\min})$, and the MFPT is simply proportional to $\exp(\Delta V/D)$ (Arrhenius law) – see Section 1.5.5. It is natural to ask: how does this Arrhenius law get modified for active particles, such as an RTP?

Based on the above results, in [2] we have derived the explicit formula for the MFPT of an RTP inside a double-well $V(x) = \frac{\alpha}{2}(|x| - 1)^2$ when $0 < \alpha < v_0$. This is the relevant case since for $\alpha > v_0$ the particle cannot reach the origin. In this case, the force $f(x) = -\alpha(x-1)$ has one unique stable negative turning point $x_-^s = 1 + v_0/\alpha$ such that $f(x_-^s) = -v_0$, and of course $|f(0)| = \alpha < v_0$. This force is an example of Phase III (see Fig. 11.3). In this case, the MFPT is given by Eq. (11.3.10) which reads here

$$\tau_\gamma(x_0) = \frac{1}{2\gamma} + \int_0^{1+\frac{v_0}{\alpha}} \frac{dy}{v_0 + \alpha(y-1)} \exp\left[\int_0^y du \, \frac{-2\gamma\,\alpha(u-1)}{v_0^2 - \alpha^2(u-1)^2}\right] \\ - \int_0^{x_0} dz \, \frac{1}{v_0^2 - \alpha^2(z-1)^2} \int_{1+\frac{v_0}{\alpha}}^z dy \, (\alpha + 2\gamma) \exp\left[\int_y^z du \, \frac{2\gamma\,\alpha(u-1)}{v_0^2 - \alpha^2(u-1)^2}\right]. \tag{13.1.1}$$

We want to analyse this expression in the limit where $D_{\text{eff}} = v_0^2/(2\gamma) \to 0$, e.g., in the limit $\gamma \to \infty$ keeping $v_0$ fixed. In this limit, it is natural to expect that $\tau_\gamma(x_0)$ grows exponentially with $\gamma$ and we thus estimate the integrals in Eq. (13.1.1) by the saddle point method (note that the first term $1/(2\gamma)$ in Eq. (13.1.1) is subdominant compared to the other terms). Let us first analyse the first integral in Eq. (13.1.1), which leads to

$$\int_0^{1+v_0/\alpha} dy \, \frac{1}{v_0 + \alpha(y-1)} \exp\left[-\gamma \int_0^y du \, \frac{2\,\alpha(u-1)}{v_0^2 - \alpha^2(u-1)^2}\right] \\ \underset{\gamma \to +\infty}{\sim} \exp\left[-\gamma \min_{y \in [0, 1+v_0/\alpha]} \left(\int_0^y du \, \frac{2\,\alpha(u-1)}{v_0^2 - \alpha^2(u-1)^2}\right)\right]. \tag{13.1.2}$$

It is easy to check that, for $0 \le y \le 1 + \frac{v_0}{\alpha}$, the minimum in the argument of the exponential is reached for $y = 1$ such that we have

$$\int_0^{1+v_0/\alpha} dy \, \frac{1}{v_0 + \alpha(y-1)} \exp\left[-\gamma \int_0^y du \, \frac{2\,\alpha(u-1)}{v_0^2 - \alpha^2(u-1)^2}\right] \underset{\gamma \to +\infty}{\sim} \exp\left[\int_0^1 du \, \frac{2\gamma\,\alpha(1-u)}{v_0^2 - \alpha^2(u-1)^2}\right]. \tag{13.1.3}$$

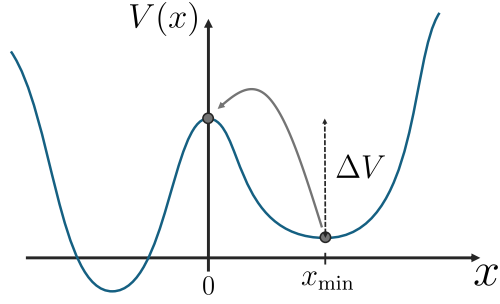

**Figure 13.1:** Consider a potential with a local minimum located at $x_\text{min} > 0$ and a local maximum at the origin $x = 0$. We want to estimate the average time needed for a particle to escape from this local minimum. This is given by the mean first-passage time to the origin. For a diffusive particle, it is simply proportional to $\exp(\Delta V/D)$ where $\Delta V$ is the barrier height and $D$ the diffusion coefficient. However, for an RTP, we show that the MFPT is approximated by $\exp(\Delta W/D)$ where $\Delta W$ is the height of an "active external potentia" given in Eq. (13.1.6).

The last term in Eq. (13.1.1) is a double integral which we rewrite as

$$\int_0^{x_0} dz \, \frac{1}{v_0^2 - \alpha^2(z-1)^2} \int_{1+\frac{v_0}{\alpha}}^{z} dy \, (\alpha + 2\gamma) \, e^{-\gamma \phi(z,y)} \quad , \quad \phi(z,y) = \int_z^y du \, \frac{2\gamma \, \alpha(u-1)}{v_0^2 - \alpha^2(u-1)^2}. \tag{13.1.4}$$

We are looking for the minimum of $\phi(z,y)$. If $x_0 > 1$, then $\partial_z \phi(z,y) = 0$ for $z^* = 1$. In that case, $z^* = 1 < y < 1 + \frac{v_0}{\alpha}$, and $\partial_y \phi(z^*, y) = 0$ for $y^* = 1$. In this case, this term is subdominant compared to the exponentially diverging log-equivalent (13.1.3). Instead, if $x_0 < 1$, separating the integral over $y$ in two regions $y > 1$ and $y < 1$ allows to show that the integral is of the same order as (13.1.3). Hence,

$$\log\left(\tau_\gamma(x_0)\right) \underset{\gamma \to +\infty}{\sim} \int_0^1 du \, \frac{2\gamma \, \alpha(1-u)}{v_0^2 - \alpha^2(1-u)^2} = \frac{W(0) - W(1)}{D_\text{eff}}, \tag{13.1.5}$$

where $W(x) = -\frac{v_0^2}{2\alpha} \log\left|1 - \frac{\alpha^2}{v_0^2}(|x|-1)^2\right|$ is an effective potential. Therefore, for an RTP, in the weak noise limit, Kramers' law is modified in the sense that the averaged time to go from one side of the double-well to the other is exponentially diverging with the height of an effective barrier $\Delta V_\text{eff} = W(0) - W(1) = -\frac{v_0^2}{2\alpha} \log\left|1 - \frac{\alpha^2}{v_0^2}\right| > \Delta E$ greater than in the diffusive case. In the full diffusive limit which is retrieved when $v_0 \to \infty$, we find back the diffusive result $\log\left(\tau_\gamma(x_0)\right) \approx \frac{\Delta E}{D_\text{eff}}$ with $D_\text{eff} = v_0^2/(2\gamma)$.

The previous analysis can actually be carried out for a general potential $V(x)$ with a single minimum located at $x_\text{min}$, and the top of the barrier at the origin, while $V(x)$ is an increasing function of $x$ for $x > x_\text{min}$ (as in Fig. 13.1). If the origin is accessible to an RTP located inside the minimum of the well, then from equation (11.3.10), it is possible to show that the modified Kramers' law reads

$$\boxed{\log\left(\tau_\gamma(x_0)\right) \underset{\gamma \to +\infty}{\sim} \exp\left(\Delta W/D_\text{eff}\right) \quad , \quad \Delta W = W(0) - W(x_\text{min}) = \int_0^{x_\text{min}} du \, \frac{f(u)}{1 - \frac{f^2(u)}{v_0^2}}}, \tag{13.1.6}$$

where $f(x)$ is the force associated to the potential, i.e., $f(x) = -V'(x)$ and $W(x)$ is called the "active external potential" [241, 355]. In the diffusive limit, we indeed obtain $\log\left(\tau_\gamma(x_0)\right) \approx (V(0) - V(x_\text{min}))/D_\text{eff}$ where the numerator is the barrier height. Note that in principle our exact formula (11.3.10) allows to compute the pre-exponential corrections to the modified Kramer's law (13.1.6), as it can be done in the passive case, see e.g. [39]. However we have not pursued this analysis here.

**Remark.** Finally, it is interesting to compare this formula (13.1.6) with the result of Ref. [242] who studied the MFPT in the weak noise limit. In that limit, the authors established a relation between the MFPT in a confining potential and the stationary distribution of the RTP $P_{\text{st}}(x)$ in the same potential, namely [242]

$$\lim_{\gamma \to \infty} \log\left(\tau_\gamma(x_0)\right) \sim - \lim_{\gamma \to \infty} \log\left(P_{st}(X)\right), \qquad (13.1.7)$$

where $X$ is the position of the absorbing state (here $X = 0$). Using the explicit expression of $P_{st}(x)$ for a generic potential (e.g., from [216, 341, 358–360]) one has for $V(x) = \frac{\alpha}{2}(|x|-1)^2$

$$\frac{1}{P_{\text{st}}(0)} = (v_0^2 - \alpha^2) \int_{-1-\frac{v_0}{\alpha}}^{1+\frac{v_0}{\alpha}} dy \, \frac{1}{v_0^2 - \alpha^2(y-1)^2} \, e^{-2\gamma \int_0^y dz \, \frac{\alpha(z-1)}{v_0^2 - \alpha^2(z-1)^2}}. \qquad (13.1.8)$$

At large $\gamma$, this integral can be evaluated by a saddle-point method. It turns out the saddle point is located at $y^* = 1$, leading to

$$\log\left(\frac{1}{P_{\text{st}}(0)}\right) \underset{\gamma \to +\infty}{\sim} \int_0^1 du \, \frac{2\gamma \, \alpha(1-u)}{v_0^2 - \alpha^2(1-u)^2}. \qquad (13.1.9)$$

By using the relation in (13.1.7), this result indeed coincides with our prediction in (13.1.5).

## 13.2 Mean Trapping Time of an RTP Inside a Harmonic Trap

When confined in a harmonic trap described by the potential $V(x) = \mu \, x^2/2$, an RTP reaches a stationary state at large time. What is remarkable is that the support of the distribution is finite and the particle is trapped in the interval $[-v_0/\mu, v_0/\mu]$ [216]. This is because $\pm v_0/\mu$ are turning points of the force $f(x) = -V'(x) = -\mu \, x$. At $v_0/\mu$, the positive state of the RTP has a zero velocity while the negative state has a negative velocity. This means that once an RTP that initiates its motion at $x_0 > v_0/\mu$ reaches for the first time $[0, v_0/\mu]$, it stays inside this interval forever. Therefore, in order for the particle to relax in the stationary state, it must first reach $v_0/\mu$. Consequently, the mean first-passage time to $v_0/\mu$, which we call the mean "trapping time" serves as a lower bound for the relaxation time of an RTP inside a harmonic trap to the stationary state.

Although, up to now, we have only computed the MFPT to the origin $x = 0$, it is possible to instead compute the MFPT to an arbitrary point by shifting the force $f(x)$. Here, we shift the potential such that the right edge of the support of the steady state is at the origin. For this purpose, the shifted potential is $V(x) = \mu \, (x + v_0/\mu)^2/2$ and $f(x) = -\mu \, x - v_0$. The limit $x^s_- \to 0$ inside Eq. (11.3.10) is well defined and the MFPT is given by

$$\tau_{\text{trap}}(x_0, \gamma) = \frac{1}{2\gamma} + \int_0^{x_0 - \frac{v_0}{\mu}} dz \, \frac{1}{v_0^2 - f(z)^2} \int_0^z dy \, (f'(y) - 2\gamma) \exp\left[\int_y^z du \, \frac{-2\gamma f(u)}{v_0^2 - f(u)^2}\right]. \qquad (13.2.1)$$

Performing the integrals explicitly leads to

$$\boxed{\tau_{\text{trap}}(x_0, \gamma) = \frac{1}{2\gamma} + \frac{(2\gamma + \mu)}{\gamma + \mu} \frac{(x_0 - v_0/\mu)}{2v_0} \, _3F_2\left(\{1, 1, 2(1+\frac{\gamma}{\mu})\}; \{2, 2+\frac{\gamma}{\mu}\}; -\frac{\mu \, x_0}{2v_0} + \frac{1}{2}\right).} \qquad (13.2.2)$$

The relaxation time of the RTP is equal to the sum of the MFPT to $v_0/\mu$ and the relaxation time of an RTP starting from $v_0/\mu$ which is of order $1/\mu$ [216]. Hence, for large $x_0$, $\tau_{\text{trap}}(x_0, \gamma) \approx \frac{1}{\mu} \log(\frac{\mu \, x_0}{v_0})$ dominates the relaxation time. In Fig. 13.2, we show a perfect agreement between simulations and our theoretical prediction (13.2.2).

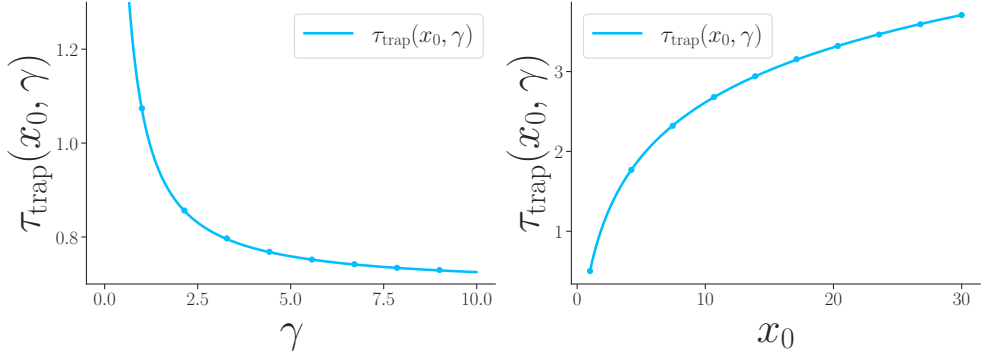

**Figure 13.2:** Plots of the mean "trapping time" of an RTP confined in a harmonic trap, i.e., the mean time needed for the particle to reach the right edge of the support of the stationary distribution located at $v_0/\mu$, and starting at $x_0 > v_0/\mu$. Here, $v_0 = 1$ and $\mu = 1$ such that the right edge is located at $v_0/\mu = 1$. On both panels, the dots show numerical results, while the solid lines are analytical results Eq. (13.2.2). On the left plot, we plot the MFPT with respect to $\gamma$ fixing $x_0 = 2$, while on the right, we plot the MFPT with respect to $x_0$ fixing $\gamma = 1$.

## 13.3 Optimal Search Strategy: Resetting RTP vs Potential Driven RTP

The last application that we want to discuss here is the characterization of the efficiency of the search strategy of a *free* RTP in the presence of stochastic resetting. Let us thus consider a single particle on an infinite line starting at the initial position $x_0$ at time $t = 0$. The position of the particle at time $t$ evolves, during an infinitesimal amount of time $dt$ according to

$$x(t+dt) = \begin{cases} x_0, & \text{with proba. } r\,dt, \\ x(t) + \eta(t)\,dt & \text{with proba. } 1 - r\,dt, \end{cases} \quad (13.3.1)$$

where $r$ is the resetting rate and $\eta(t)$ is a Gaussian white noise with zero mean $\langle \eta(t) \rangle = 0$ and delta correlations $\langle \eta(t)\eta(t') \rangle = 2D\delta(t-t')$, with $D$ the diffusion constant. In the limit $dt \to 0$, this dynamics (13.3.1) defines the *resetting Brownian motion* (rBM) (see Section 3.1). Interestingly, it was shown that, in the large time limit, the distribution of the position of the particle $p(x,t)$ at time $t$ converges to a stationary distribution given by [54, 176]

$$\lim_{t \to \infty} p(x,t) = p_{\text{st}}(x) = \frac{\alpha_0}{2} \exp\left[-\alpha_0 |x - x_0|\right], \quad \text{where} \quad \alpha_0 = \sqrt{r/D}. \quad (13.3.2)$$

Therefore, although the rBM is a non-stationary process (one can indeed show that the rules of the dynamics (13.3.1) violate detailed balance), the stationary distribution (13.3.2) can nevertheless be expressed as an effective Boltzmann weight $p_{\text{st}}(x) \propto \exp\left[-U_{\text{eff}}(x)\right]$ with the effective potential

$$U_{\text{eff}}(x) = \alpha_0 |x - x_0|. \quad (13.3.3)$$

Therefore, if we consider the following equilibrium Langevin dynamics

$$\frac{dx}{dt} = -D\partial_x U_{\text{eff}}(x) + \eta(t), \quad (13.3.4)$$

where $\eta(t)$ is the same Gaussian white noise as in (13.3.1), then the stationary state of Eq. (13.3.4) is characterized by the same stationary distribution $p_{\text{st}}(x) \propto \exp[-U_{\text{eff}}(x)]$ – although of course the two dynamics (13.3.1) and (13.3.4) are quite different.

An other remarkable property of the rBM is that the resetting parameter $r$ can be tuned to minimize the MFPT to the origin. Indeed, for any finite $r > 0$ the MFPT is finite (while it is infinite for the standard Brownian motion corresponding to $r = 0$). In addition, there exists an optimal value of the resetting rate $r_1$ that minimizes this MFPT [54, 176]. In Ref. [179], the authors asked the following question: since the non-equilibrium resetting dynamics (13.3.1) and the equilibrium Langevin process (13.3.4) lead to the same stationary state, can one compare the MFPT to the origin of these two processes? Interestingly, they showed that the optimal "nonequilibrium" MFPT (i.e., with resetting with $r = r_1$) is always smaller than the "equilibrium" MFPT (evaluated at its optimal value of $r = r_2 \neq r_1$). Loosely speaking, "nonequilibrium offers a better search strategy than equilibrium" [179] – see also Section 3.1.4. In this section, using our results derived for the MFPT of an RTP in a confining potential, we address this question of the efficiency of the search strategy offered by a free RTP subjected to resetting, in the same spirit as in Ref. [179].

We thus consider a free RTP evolving under Eq. (11.2.1) with $f(x) = 0$ where we add resetting to the dynamics – as in Eq. (13.3.1). This comprises simultaneously resetting both the position and the velocity. With rate $r$ the particle thus resets to its initial position $x_0$, while the velocity $\sigma$ is set $\pm 1$ with probability $1/2$, i.e. the velocity is randomized. This resetting protocol is referred to as position resetting and velocity randomization [157]. In this case, the system also reaches a steady state characterized by a steady state distribution of the position of the particle $P_{\text{st},r}(x)$ given by [157]

$$P_{\text{st},r}(x) = \frac{\lambda_r}{2} \exp\left[-\lambda_r |x - x_0|\right] \quad , \quad \lambda_r = \left(\frac{r(r + 2\gamma)}{v_0^2}\right)^{\frac{1}{2}}. \tag{13.3.5}$$

On the other hand, an RTP in a linear potential $V(x) = \alpha |x - x_0|$, if $\alpha < v_0$, reaches also a steady state described by (see e.g., [216])

$$P_{\text{st}}(x) = \frac{\gamma \alpha}{v_0^2 - \alpha^2} \exp\left[-\frac{2\gamma \alpha}{v_0^2 - \alpha^2} |x - x_0|\right]. \tag{13.3.6}$$

Therefore, the two steady states (13.3.5) and (13.3.6) coincide if we choose $\lambda_r = 2\gamma\alpha/(v_0^2 - \alpha^2)$, i.e.

$$\alpha = v_0 \sqrt{\frac{r}{2\gamma + r}}. \tag{13.3.7}$$

Hence, in the spirit of the work [179] described above, we can compare the optimal MFPT of the potential driven particle to the resetting driven one.

The MFPT of the resetting RTP was derived in [157] and reads

$$T_r(x_0) = \frac{1}{r}\left[-1 + e^{\frac{\sqrt{1+u}}{u}z} \frac{u}{1 + u - \sqrt{1+u}}\right] \quad , \quad u = \frac{2\gamma}{r} \quad , \quad z = \frac{2\gamma x_0}{v_0}. \tag{13.3.8}$$

To calculate the MFPT of an RTP within a linear potential centered at the particle's initial position, it is convenient to consider the potential $V(x) = \alpha |x - x_1|$ and then take the limit as $x_1$ approaches $x_0$. When $\alpha < v_0$, the force is in phase I, and the MFPT is given by Eq. (11.3.3). The force is given by $f(x) = -V'(x) = -\alpha \,\text{sign}(x - x_1)$, so it is crucial to carefully split the integrals in Eq. (11.3.3) into two separate cases to account for the regions where $x > x_1$ and $x < x_1$. Upon calculating these integrals, one finds (see also [243, 356])

$$\tau_\gamma(x_0) = -\frac{x_0}{\alpha} + \frac{v_0}{2\alpha\gamma} + \frac{v_0(v_0 + \alpha)}{\alpha^2 \gamma}\left(e^{\frac{2\alpha\gamma}{v_0^2 - \alpha^2}x_0} - 1\right) \tag{13.3.9}$$

$$= \frac{1}{r}\frac{(1+u)}{u}\left[2 e^{\frac{\sqrt{1+u}}{u}z}\left(1 + \sqrt{\frac{1}{1+u}}\right) - \sqrt{\frac{1}{1+u}}(z+1) - 2\right] \quad , \quad u = \frac{2\gamma}{r} \quad , \quad z = \frac{2\gamma x_0}{v_0}, \tag{13.3.10}$$

where we have used the relation (13.3.7) in the second equality.

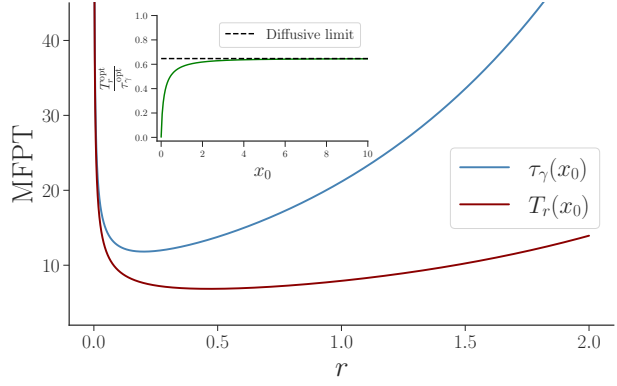

**Figure 13.3:** For $\gamma = 1$, $x_0 = 1$ and $v_0 = 1$, we plot $T_r(x_0)$ and $\tau_\gamma(x_0)$ versus $r$. Both curves exhibit a minimum at an optimal tumbling rate, with $T_r(x_0)$ always smaller than $\tau_\gamma(x_0)$. The inset shows that the ratio $T_r^{\mathrm{opt}}/\tau_\gamma^{\mathrm{opt}}$ converges to $0.646728\ldots$ for large value of $x_0$ matching the ratio of the optimal MFPTs for a diffusive particle.

We now introduce the difference between the mean first-passage times and study its sign to determine which strategy is more efficient in finding the target at the origin

$$\delta(u) \quad = r\left(T_r(x_0) - \tau_\gamma(x_0)\right) \tag{13.3.11}$$

$$= \frac{2 + u + \sqrt{1+u}(1+z)}{u} - \frac{e^{\frac{\sqrt{1+u}}{u}z}(2+u)}{(1+u) - \sqrt{1+u}}. \tag{13.3.12}$$

Note that we have multiplied the difference of the MFPTs by a factor $r$ due to the scaling form in Eqs. (13.3.8) and (13.3.10) – since $r > 0$ this does not affect the analysis of the sign of $\delta(u)$. Using the fact that $e^{\frac{\sqrt{1+u}}{u}z} \geq 1 + \frac{\sqrt{1+u}}{u}z$, it is possible to verify that $\delta(u)$ is always negative for fixed $z > 0$. This implies that the resetting RTP is always more efficient in reaching the origin than the potential-driven RTP, for any value of $r$. Hence, as in the passive case, resetting offers a better search strategy!

For $x_0 > 0$, both $T_r(x_0)$ and $\tau_\gamma(x_0)$ have a minimum with respect to $r$. In the large $x_0$ limit, with fixed $\gamma$ and $v_0$, the particles that reach the origin have experienced a large number of tumbles, and have taken a long time, so their behavior is essentially diffusive. Therefore, we expect the ratio $T_r^{\mathrm{opt}}/\tau_\gamma^{\mathrm{opt}}$ to take the diffusive value $0.646728\ldots$ found in [179] when $x_0 \to \infty$. In fact, for diffusive particles, the minimum of the MFPTs is reached at a specific rate $r^* \propto x_0^{-2}$. Using this scaling, one can solve the minimization equations $\partial_r T_r(x_0) = 0$ and $\partial_r \tau_\gamma(x_0) = 0$ and identify the optimal rate in both cases. Plugging this back into the original expression, we obtain in the large $x_0$ limit

$$T_r^{\mathrm{opt}}(x_0) \approx \left(\frac{e^{C_1}-1}{C_1^2}\right)\frac{x_0^2}{D_{\mathrm{eff}}}, \quad \text{where} \quad 2(1 - e^{C_1}) - C_1 = 0, \quad C_1 = 1.54414\ldots, \tag{13.3.13}$$

$$\tau_\gamma^{\mathrm{opt}}(x_0) \approx \left(\frac{2(e^{C_2}-1) - C_2)}{C_2^2}\right)\frac{x_0^2}{D_{\mathrm{eff}}}, \quad \text{where} \quad 4 + 2e^{C_2}(C_2 - 2) + C_2 = 0, \quad C_2 = 2.38762\ldots, \tag{13.3.14}$$

with $D_{\mathrm{eff}} = \frac{v_0^2}{2\gamma}$. These coincide with the results derived in [179] and as expected, it gives $T_r^{\mathrm{opt}}/\tau_\gamma^{\mathrm{opt}} \approx 0.646728\cdots$. In Fig. 13.3, we show a plot of $T_r(x_0)$ and $\tau_\gamma(x_0)$ when $\gamma = 1$; $x_0 = 1$ and $v_0 = 1$. We also show the ratio $T_r^{\mathrm{opt}}/\tau_\gamma^{\mathrm{opt}}$ with respect to $x_0$ when $\gamma = 1$ and $v_0 = 1$.

## 13.4 Conclusion

The mean first-passage time is a key quantity for probing the first-passage properties of stochastic processes. In general, its computation is challenging – especially for active particles, where only a few analytical results are available. In this work, we focused on a one-dimensional run-and-tumble particle evolving in an arbitrary external potential. Specifically, we considered the case where the particle is driven solely by telegraphic noise, with thermal noise absent. Depending on the shape of the potential, we derived explicit expressions for the MFPT in different phases. The approach developed here provides a systematic framework to compute the MFPT for a wide class of potentials. In particular, it includes confining potentials of the form $V(x) = \alpha |x|^p$ with $p > 1$, for which we showed that there exists an optimal tumbling rate $\gamma_{\text{opt}}$ that minimizes the MFPT. We also analyzed in detail how $\gamma_{\text{opt}}$ depends on the initial position $x_0$: for generic $p > 1$, it diverges as $\gamma_{\text{opt}} \propto 1/x_0$ when $x_0 \to 0$ and saturates to a constant as $x_0 \to \infty$. However, this behavior breaks down in the case $p = 1$, where no finite optimal rate exists, i.e., $\gamma_{\text{opt}} \to \infty$.

In addition, we proposed three applications to illustrate the physical relevance of our results. Among them, we derived Kramers' law for a one-dimensional RTP, and we studied optimal search strategies, showing that a resetting RTP can outperform a potential-driven RTP in locating a target.

The study of first-passage properties of active particles is a timely and evolving field, attracting increasing attention in recent literature. For example, in [354], the authors combined results presented in this part of the thesis (from [2, 5]) with those from [4], where we computed the exact exit probability of a confined RTP, to calculate the MFPT of an RTP subjected to a periodic force, conditioned on trajectories that reach the origin. Another recent work investigates the optimization of first-passage properties for an active Brownian particle subjected to both position and orientation resetting [260].

Another important first-passage property for stochastic processes is the splitting probability, i.e., the probability that a particle exits an interval with two absorbing boundaries on a given side. In [4], we computed this quantity exactly for a run-and-tumble particle subjected to an arbitrary force $F(x)$. Remarkably, the resulting expression coincides with the cumulative distribution of the same RTP confined between two hard walls under the reversed force $-F(x)$. This surprising identity is known as *Siegmund duality*, and it extends well beyond the case of the RTP. In the next part of this thesis, we introduce Siegmund duality in detail and explain its usefulness and relevance in physical applications.

# Part V

# Siegmund Duality: a Bridge Between Spatial and First-Passage Properties of Stochastic Processes

**Abstract**

In this part of the thesis, we introduce Siegmund duality, a concept not widely known in the physics literature. This duality is a powerful tool, as it connects the first-passage properties of a stochastic process to the spatial observables of its dual. It was initially derived by mathematicians for Brownian motion [59]. A particular challenge is that, given a stochastic process, the construction of the dual process is not always known. With Léo Touzo, we first extended Siegmund duality to the run-and-tumble particle in [4]. In a second paper [3], we developed a framework to generalize this duality to a wide range of physically relevant stochastic processes, including active particle models, random diffusion models, stochastic resetting, continuous- and discrete-time random walks, and even fractional Brownian motion. In the following chapters, we will derive most of the results obtained in [3, 4] and illustrate their usefulness. The derivation of the results, the numerical simulations and the writing of the articles were all performed by both of us in equal proportions.

# Chapter 14

# Siegmund Duality: Introduction and Main Results

## 14.1 Introduction

The study of first-passage properties of stochastic processes has a long history, both in mathematics and in physics, with applications ranging from biology to mathematical finance. Consider for instance a random walk on the interval $[a, b]$, with absorbing boundaries at both extremities. Finding its first-passage properties consists in addressing the following questions: what is the probability that it reaches $b$ before reaching $a$ (called the *exit*, *splitting* or *hitting* probability)? What is the probability that it remains in the interval up to time $t$ (the *survival* probability)? What is the distribution of first-passage time at $b$? Beyond their direct applications, these very general questions also share strong connections with the study of extreme value statistics. For general surveys on the topic, see [40, 43, 61, 63, 69, 361].

On a seemingly unrelated note, the behavior of stochastic processes confined between hard walls is an interesting subject of study. For instance, active particles exhibit peculiar dynamics under confinement. In particular, their persistent motion causes them to accumulate near boundaries [362–365]. Therefore, understanding and computing spatial observables, such as the distribution of the position of the particle between hard walls, is crucial.

Surprisingly, there is a strong connection between absorbing and hard wall (or reflective) boundary conditions. This was first pointed out by Lévy in the case of Brownian motion [75], and by Lindley for discrete random walks [76]. Siegmund later generalized this relation, which became known as *Siegmund duality*: two processes $x(t)$ and $y(t)$ (where $t$ is either discrete or continuous) such that $x(0) = x$ and $y(0) = y$ are said to be Siegmund duals if, at any time $t$,

$$\mathbb{P}(x(t) \geq y | x(0) = x) = \tilde{\mathbb{P}}(y(t) \leq x | y(0) = y) , \qquad (14.1.1)$$

where for the sake of clarity we will denote with a tilde all quantities associated to the dual. Siegmund showed the existence of a Siegmund dual for any one-dimensional stochastically monotone Markov process (meaning that $\mathbb{P}(x(t) \geq y | x(0) = x)$ is a non-decreasing function of $x$) [59]. Since then, extensions to more general Markov processes have been considered [366–369]. Depending on the setting it is not always obvious how to explicitly construct this dual.

In this part of the thesis, $x(t)$ takes values in an interval $[a, b]$ and has absorbing walls at $a$ and $b$, while the dual $y(t)$, as a consequence, has hard walls at $a$ and $b$. By *absorbing boundary conditions* we mean that, if the particle reaches $a$ or $b$, it "sticks" to the wall and remains there afterwards (i.e., it does not disappear, as it is sometimes the case in the physics literature [43, 63, 69]). Conversely, a *hard wall* can be considered as an infinite potential step, i.e. when a particle encounters it, it remains there until its total velocity changes sign (see [4, 370]). This is sometimes called a "reflecting" boundary condition, but we prefer the terminology "hard wall" in the context of active particles, to make it clear that when the particle does not instantly tumble when it reaches the wall, it can remain stuck for a certain time due to the persistence in its motion. In Figure 14.1, we show a schematic description of these boundary conditions. Our main quantity of interest is the finite time exit probability at $b$, which we define as the probability to be absorbed at $b$ before or at time $t$, when $x(0) = x$,

$$E_b(x,t) = \mathbb{P}(x(t) = b|x(0) = x) \,. \tag{14.1.2}$$

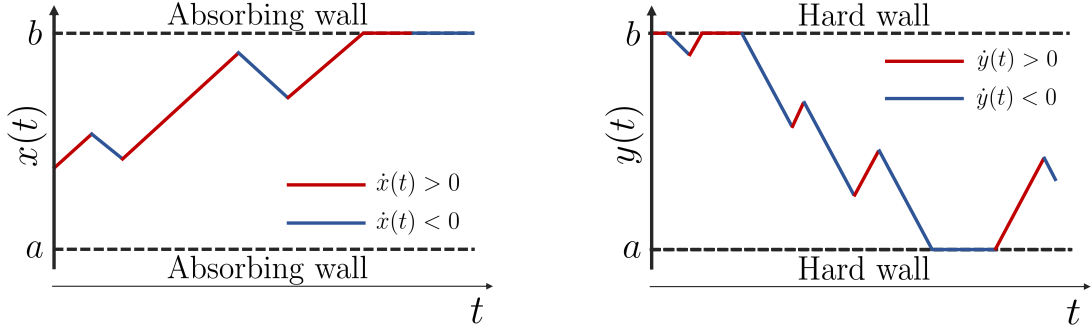

**Figure 14.1:** In this figure, we show the two types of boundaries that we consider throughout this part of the thesis, and how particles behave around them. The trajectories shown are schematics and could be for instance sampled from a run-and-tumble dynamics. **Left**: The process $x(t)$ is surrounded by two absorbing walls at $a$ and $b$. If the particle reaches one of the two walls it will be absorbed and stay stuck to the wall forever. **Right**: The process $y(t)$ is surrounded by two hard walls at $a$ and $b$. In this case, if the particle is at a wall, and if its velocity $\dot{y}(t)$ is directed toward the wall, it will stay at the wall until $\dot{y}(t)$ changes sign.

Here we mostly focus on the case where $y = b$ in (14.1.1). The identity (14.1.1) then gives us a relation between $E_b(x,t)$ and the cumulative distribution of the dual $y(t)$ denoted $\tilde{\Phi}(x,t|b)$ (since with our definition of absorbing walls, $x(t) \geq b$ is simply equivalent to $x(t) = b$)

$$E_b(x,t) = \tilde{\Phi}(x,t|b) \quad, \quad \text{where} \quad \tilde{\Phi}(x,t|b) = \tilde{\mathbb{P}}(y(t) \leq x|y(0) = b) \,. \tag{14.1.3}$$

One simple example is if $x(t)$ is a Brownian motion subjected to a force $F(x)$, with absorbing boundary conditions. In that case, the dual $y(t)$ has the exact same dynamics as the one of $x(t)$ but with hard walls and subjected to the reversed force $-F(x)$, and (14.1.3) holds. This is a general result: if $x(t)$ is subjected to an external force, the sign of the force is reversed in the case of its dual $y(t)$. This can be shown using the Fokker-Planck equation, as we have done it in Section 1.5.4 for Brownian motion.

Another class of processes for which one can apply (14.1.1) is Lévy flights. In the continuum limit, when $x(t)$ is a Lévy flight whose increments obey a Lévy stable symmetric law of index $0 < \mu \leq 2$, the probability that it exits the interval $[a,b]$ at $b$ after an infinite time reads [74, 371, 372]

$$E_b(x, t \to +\infty) = \frac{\Gamma(2\phi)}{\Gamma(\phi)^2}(b-a)^{1-2\phi} \int_a^x du \, [(u-a)(b-u)]^{\phi-1} \,, \tag{14.1.4}$$

where $\phi = \mu/2$. If $x(t)$ is a Lévy flight, then its dual $y(t)$ is also a Lévy flight with the same increments. This is confirmed by a more recent and completely independent derivation of the stationary distribution of a Lévy flight between two hard walls, $\tilde{P}_{st}(x) = \partial_x \tilde{\Phi}(x, t \to +\infty|b)$, in [373],

$$\tilde{P}_{st}(x) = \frac{\Gamma(2\phi)}{\Gamma(\phi)^2}(b-a)^{1-2\phi} [(x-a)(b-x)]^{\phi-1} = \frac{dE_b(x,t \to \infty)}{dx} \,. \tag{14.1.5}$$

As we can see, $\tilde{P}_{st}(x)$ is simply the derivative of $E_b(x)$, which is a direct consequence of the duality (14.1.3) at infinite time[16]. The possibility of deriving one quantity directly from the other is a strong analytical motivation for the usefulness of the duality presented here.

The original results by Siegmund were derived for one-dimensional Markov processes. Active particle models are by definition non-Markovian when one only considers the position of the particle, but they are Markovian when considering both the position and the driving velocity. In fact, the existence of a dual process was also shown for processes driven by a stationary process, in [374] for discrete time and in [375] for continuous time (it has even been extended to higher dimensions in [376]). In this setting, inspired by applications to finance, the exit probability (called ruin probability in this context) is related to the cumulative distribution of some dual process, defined in very general terms.

Siegmund duality can be seen as a particular case of Markov duality (see [377–380] for reviews and other examples). We say that two Markov processes $x(t)$ and $y(t)$ are *Markov duals* with respect to a function $H(x,y)$ if and only if, for all $(x,y) \in \mathbb{R}^2$ and $t \geq 0$, we have

$$\mathbb{E}\left[H(x(t),y)\right] = \tilde{\mathbb{E}}\left[H(y(t),x)\right] . \tag{14.1.6}$$

If $H(x,y) = \mathbb{1}_{x \geq y}$, this yields Siegmund duality (see equation (14.1.1)). Such duality relations are frequently used by mathematicians and have been applied in various contexts, including queuing theory, finance and population genetics, as well as interacting particle systems and systems with a reservoir of particles [381–386]. Yet, they seem to be less well-known among the physics community, although similar relations have been sometimes pointed out [74, 387–391].

There are several situations in which one may want to use Siegmund duality. First of all, analytically computing the distribution of positions and the first-passage properties of a stochastic process can be quite cumbersome, and knowing that it is possible to derive one from the other could save a lot of effort in some situations (see e.g. the example of Lévy flights mentioned above). Additionally, in numerical simulations as well as in experiments, it is often much simpler to compute spatial properties than first-passage properties. In particular, if one is interested in the stationary state of an ergodic system, the former can be computed from a single, long enough time series of data, contrary to the latter.

In Section 14.2, we consider two types of models and present the main results. The first is a one-dimensional continuous stochastic process subjected to a force and to a Gaussian white noise, with absorbing boundary conditions. Both the force and the temperature may depend on a parameter which is itself a Markov process at equilibrium. This definition was inspired by active particle models, but it also includes other models of interest such as diffusing diffusivity models or switching diffusion [111, 117, 129, 130] – see Chapter 2. The second model is a 1d random walk with stationary increments (and again absorbing boundary conditions). In both settings, we construct explicitly a dual process, with hard wall boundary conditions, such that (14.1.1) is satisfied.

In Chapter 15, we highlight the connection between absorbing boundaries and hard walls for a specific model of active particle: a run-and-tumble particle moving within an arbitrary external potential. For this system, we give an explicit formulation of the Siegmund dual and derive the relation between the exit probability and the distribution of positions of the particle at any time. In the stationary state, both quantities are computed explicitly. We also derive the duality for the persistent random walks, a discrete time version of the RTP. Then, in Chapter 16, we extend the duality to processes subjected to stochastic resetting, which is also an entirely new result. Finally, in Chapter 17, we provide numerical results which illustrate this duality for many processes, including fractional Brownian motion.

Although some general mathematical results already exist (see, e.g. [374, 375]), to our knowledge, this is the first time that a fully explicit construction of the dual process is proposed for such a wide range of physically relevant models (including active particles). We provide in [3] original and intuitive derivations of the duality relation (14.1.1) for both the continuous time and the discrete time settings, as well as analytical and numerical illustrations for some well-known models. Our goal is not to provide fully rigorous proofs, but rather to extend the use of Siegmund duality to new fields and to convince physicists of its usefulness.

---

[16] The original results by Lindley, apply in particular to Lévy flights in discrete time. Taking the continuous time limit would require a rigorous analysis on its own, but it is reasonable to assume that the duality still holds in that case.

## 14.2 Main Results

### 14.2.1 The model, the dual, and the duality

We consider a one-dimensional stochastic process $x(t)$ which evolves according to the following stochastic differential equation (SDE),

$$\dot{x}(t) = f\left(x(t), \boldsymbol{\theta}(t)\right) + \sqrt{2\mathcal{T}\left(x(t), \boldsymbol{\theta}(t)\right)}\, \xi(t)\,, \tag{14.2.1}$$

where $f(x, \boldsymbol{\theta})$ and $\mathcal{T}(x, \boldsymbol{\theta})$ are arbitrary functions which act respectively as a force and a temperature. In Eq. (14.2.1), $\xi(t)$ represents a Gaussian white noise with zero mean and unit variance, and $\boldsymbol{\theta}(t)$ is a vector of arbitrary dimension with stochastic components governed by the Markovian dynamics (independent of $x(t)$)

$$\dot{\boldsymbol{\theta}}(t) = \boldsymbol{g}\left(\boldsymbol{\theta}(t)\right) + \left[2\underline{\mathcal{D}}(\boldsymbol{\theta}(t))\right]^{1/2} \cdot \boldsymbol{\eta}(t)\,, \tag{14.2.2}$$

where $\underline{\mathcal{D}}$ is a positive matrix[17]. The $\eta_i(t)$'s are again independent Gaussian white noises with zero mean and unit variance. In the following, we use the Itō prescription for multiplicative noise [42, 361]. Additionally, we allow $\boldsymbol{\theta}(t)$ to jump from a value $\boldsymbol{\theta}$ to $\boldsymbol{\theta}'$ with a transition kernel $\mathcal{W}(\boldsymbol{\theta}'|\boldsymbol{\theta})$. More precisely, during a time interval $dt$, $\boldsymbol{\theta}(t)$ jumps to some value $\boldsymbol{\theta}'$ with probability $\mathcal{W}(\boldsymbol{\theta}'|\boldsymbol{\theta})d\boldsymbol{\theta}'\,dt$ or evolves according to (14.2.2) with probability $1 - dt \int d\boldsymbol{\theta}'\mathcal{W}(\boldsymbol{\theta}'|\boldsymbol{\theta})$. We assume that $\boldsymbol{\theta}(t)$ admits an equilibrium distribution $p_{eq}(\boldsymbol{\theta})$ which satisfies the local detailed balance conditions[18]

$$-g_i(\boldsymbol{\theta})p_{eq}(\boldsymbol{\theta}) + \sum_j \partial_{\theta_j}[\mathcal{D}_{ij}(\boldsymbol{\theta})p_{eq}(\boldsymbol{\theta})] = 0\,,\ \forall\ i, \tag{14.2.3}$$

$$\mathcal{W}(\boldsymbol{\theta}|\boldsymbol{\theta}')p_{eq}(\boldsymbol{\theta}') = \mathcal{W}(\boldsymbol{\theta}'|\boldsymbol{\theta})p_{eq}(\boldsymbol{\theta})\,. \tag{14.2.4}$$

The first equation (14.2.3) corresponds to the vanishing of the probability current in (14.2.2). Since discrete jumps are allowed, we also need a separate detailed balance condition given by (14.2.4).

This definition encompasses a wide variety of stochastic processes which are relevant to physics. At the end of this section we will show how it can be specialized to active particles, leading to the results of our previous paper in the case of RTPs [4]. To put it simply, for these models $\boldsymbol{\theta}(t) = v(t)$ corresponds to the intrinsic velocity of the active particle. In models such as the AOUP, the evolution of $v(t)$ follows a Langevin equation similar to (14.2.2) (an Ornstein-Uhlenbeck process in that case), while for other models such as the RTP, $v(t)$ takes discrete values, hence the transition kernel $\mathcal{W}(\boldsymbol{\theta}'|\boldsymbol{\theta})$. A combination of these two types of evolution is possible, as it is the case for the direction reversing active Brownian particle [392, 393]. Another important class of models included in this definition are diffusing diffusivity models [111, 117, 129, 130]: in that case, $\boldsymbol{\theta}$ is a $d$-dimensional Ornstein-Uhlenbeck process ($\underline{\mathcal{D}}$ is a constant, $\boldsymbol{g}$ is a harmonic force and $\mathcal{W} = 0$), $f(x, \boldsymbol{\theta}) = F(x)$ and $\mathcal{T}(x, \boldsymbol{\theta}) = \boldsymbol{\theta}^2$. In some situations, one may also want to consider a force $f(x, \boldsymbol{\theta}, t)$ with an explicit time dependence. We will mention this case in Sec. 17.5.

---

[17] Here we use bold letters to denote vectors and we underline matrices. In the right-hand side of Eq. (14.2.2), the dot product corresponds to the canonical product between a matrix and a vector.

[18] When initialized in its equilibrium distribution $p_{eq}(\boldsymbol{\theta})$, $\boldsymbol{\theta}(t)$ satisfies $P(\boldsymbol{\theta}(t_1),...,\boldsymbol{\theta}(t_n)) = P(\boldsymbol{\theta}(t_1+\tau),...,\boldsymbol{\theta}(t_n+\tau))$ for any times $t_1,...,t_n$ and any time-shift $\tau$, and is called a stationary process.

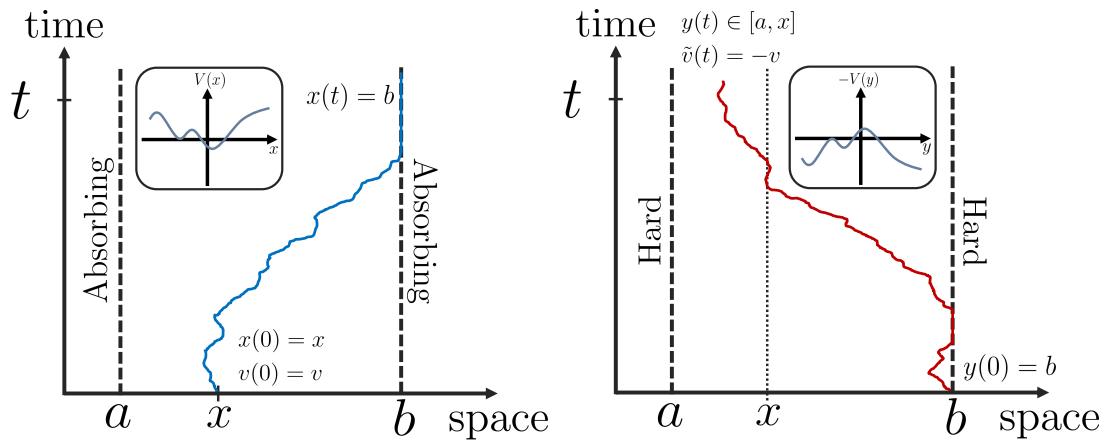

**Figure 14.2:** Schematic representation of typical trajectories which contribute to the probabilities $E_b(x, v, t)$ and $\tilde{\Phi}(x, t| - v; b)$. **Left**: The process $x(t)$ following the Langevin dynamics (14.2.15) initiates its motion at position $x(t = 0) = x$ with velocity $v(t = 0) = v$. Its motion is eventually influenced by the presence of an external potential $V(x)$. Two absorbing walls are located at $a$ and $b$. The trajectory shown is absorbed at $b$ before time $t$ and hence contributes to the exit probability $E_b(x, v, t)$. **Right**: The dual process of $x(t)$, namely $y(t)$ (with Langevin dynamics (14.2.16)), featuring hard walls at $x = a$ and $x = b$. The dual particle initiates its motion at $y(t = 0) = b$ with an initial velocity drawn from the equilibrium distribution of the process $v(t)$, and in the presence of the reversed potential $-V(y)$. The trajectory shown contributes to $\Phi(x, t|\tilde{v}(t) = -v; b)$, i.e. the probability of locating the dual particle within the interval $[a, x]$ with a velocity $-v$ at time $t$. In this paper we show that $E_b(x, v, t) = \tilde{\Phi}(x, t| - v; b)$.

We are interested in a situation where two absorbing walls are present at the positions $x = a$ and $x = b$, with $b > a$. By absorbing wall we mean that if the particle reaches one of these walls at any point in time, it remains there indefinitely (see left panel of Fig. 14.1). It is important to stress that, contrary to another common definition of absorbing boundary conditions, the particle does not disappear, such that the integral of the particle density over the interval $[a, b]$, including walls, remains equal to 1 at all times. The particle is initially placed inside the box such that $x(0) \in [a, b]$. The quantity that we want to compute is the probability that the particle reaches the wall located at $b$ before or at time $t$ (without touching the wall at $a$). This is called the *exit probability* at $b$. In the current setting it can be formally defined as

$$E_b(x, \boldsymbol{\theta}, t) = \mathbb{P}(x(t) = b | x(0) = x, \boldsymbol{\theta}(0) = \boldsymbol{\theta}). \qquad (14.2.5)$$

Here and in the rest of the paper, we use the notation $\mathbb{P}(x(t) \in I)$ to refer to the probability mass function that $x(t)$ belongs to some interval $I$, while the notation $P$ or $p$ refers to a probability density. Being able to compute this quantity for a certain process gives us a lot of information about the first-passage properties of this process. In particular, it contains the information on the survival probability, i.e. the probability that the particle is still in the interval $[a, b]$[19] after a duration $t$,

$$Q_{[a,b]}(x, \boldsymbol{\theta}, t) = 1 - E_a(x, \boldsymbol{\theta}, t) - E_b(x, \boldsymbol{\theta}, t). \qquad (14.2.6)$$

One may also take the limit $a \to -\infty$, in which case it describes the probability that the particle is still below the value $x = b$ after a duration $t$. The survival probability can in turn be used to obtain the first passage time distribution of the process. It also plays an important role in the computation of extreme value statistics [63, 71, 394].

---

[19] Eq. (14.2.6) is mostly interesting for $x \in [a, b]$, but with the definition (14.2.5) it remains valid for $x = a$ and $x = b$ since $E_a(a, \boldsymbol{\theta}, t) = E_b(b, \boldsymbol{\theta}, t) = 1$.

One important result relies on the definition of a dual process of $x(t)$, which we denote $y(t)$. Using Fokker-Planck equations, we have shown in [3] that the evolution of $y(t)$ is governed by the equation

$$\boxed{\dot{y}(t) = \tilde{f}\left(y(t), \boldsymbol{\theta}(t)\right) + \sqrt{2\mathcal{T}\left(y(t), \boldsymbol{\theta}(t)\right)}\,\xi(t) \quad , \quad \tilde{f}(x, \boldsymbol{\theta}) = -f(x, \boldsymbol{\theta}) + \partial_x \mathcal{T}(x, \boldsymbol{\theta})}. \tag{14.2.7}$$

If the temperature is independent of the position, $y(t)$ has the same dynamics as $x(t)$ but with the sign of $f$ reversed, while if the temperature is space dependent the force also has an additional term $\partial_x \mathcal{T}$. Note that the transformation $f \to -f + \partial_x \mathcal{T}$ is its own inverse. Thus if one wants to find the process $x(t)$ which is dual to $y(t)$, the force applied to $x(t)$ is also given by $f = -\tilde{f} + \partial_x \mathcal{T}$. Here, $\boldsymbol{\theta}(t)$ and $\xi(t)$ denote different realisations of the same processes as the ones appearing in (14.2.1). In addition, we assume the presence of hard walls at $a$ and $b$ instead of absorbing walls. This means that if the particle is at $y = b$ with a positive velocity, it remains at $y = b$ until its velocity changes sign (and symmetrically for $y = a$ – see right panel of Fig. 14.1).

Let us now assume that the dual process is initialized at position $y(0) = b$ and that $\boldsymbol{\theta}(0)$ is drawn from the equilibrium distribution $p_{eq}(\boldsymbol{\theta})$. Given these initial conditions, we are interested in the probability of finding the dual particle in an interval $[a, y]$ at time $t$, conditioned on the value of $\boldsymbol{\theta}(t)$, i.e., the cumulative distribution of the dual process $y(t)$ given $\boldsymbol{\theta}(t)$. We write it as follows

$$\tilde{\Phi}(y, t|\boldsymbol{\theta}; b) = \tilde{\mathbb{P}}(y(t) \leq y | \boldsymbol{\theta}(t) = \boldsymbol{\theta}; y(0) = b, \boldsymbol{\theta}(0)^{eq}), \tag{14.2.8}$$

where the tildes indicate that the observable concerns the dual process. We also use a semicolon to separate the events at time $t$ (the conditioning on $\boldsymbol{\theta}(t)$, on the left in Eq. (14.2.8)) from the initial condition (on the right). In the whole paper, a conditioning on $\boldsymbol{\theta}(0)^{eq}$ means that $\boldsymbol{\theta}(0)$ is drawn from $p_{eq}(\boldsymbol{\theta})$.

The main result of our paper [3] is that, based on the above assumptions, it is possible to establish the following duality relation at all time:

$$\boxed{E_b(x, \boldsymbol{\theta}, t) = \tilde{\Phi}(x, t|\boldsymbol{\theta}; b)}. \tag{14.2.9}$$

In simpler terms, this relation implies that the probability that the process $x(t)$ reaches the boundary $b$ before time $t$ when initiated with $\boldsymbol{\theta}(0) = \boldsymbol{\theta}$, is equal to the probability of finding the dual process $y(t)$ - where $\boldsymbol{\theta}(t) = \boldsymbol{\theta}$ - within the interval $[a, x]$ at time $t$, where $x = x(t = 0)$ and $y(t = 0) = b$. This is a new, explicit formulation of Siegmund duality, adapted for the study of many relevant physical models. Of course, the relation (14.2.9) has an equivalent for the exit probability at $x = a$, $E_a(x, \boldsymbol{\theta}, t)$, which can be immediately deduced by symmetry (one simply needs to revert the inequality in the definition of $\tilde{\Phi}$ and to choose $y(0) = a$). The relation (14.2.9) also remains valid in the limit where there is only one absorbing wall, for instance when $a \to -\infty$. Note that $E_b(x, \boldsymbol{\theta}, t)$ and $\tilde{\Phi}(x, t|\boldsymbol{\theta}; b)$ are both increasing functions of $x$ at fixed $t$ and increasing functions of $t$ at fixed $x$ (as long as $f$ does not depend explicitly on time).

The proof of Eq. (14.2.9) is provided in [3], and for simplicity, we do not include it in this thesis. The main idea is to write the Fokker-Planck equations associated with both $E_b(x, \boldsymbol{\theta}, t)$ and $\tilde{\Phi}(x, t|\boldsymbol{\theta}; b)$, and to show that these equations, along with their boundary and initial conditions, are identical. This is how we derived the duality for Brownian motion in Section 1.5.4. As additional examples, we provide this proof for the run-and-tumble particle subjected to an arbitrary force in Section 15.4, and for the resetting Brownian motion in Section 16.1.

A particular situation where the relation (14.2.9) can be useful is when the process $y(t)$ admits a stationary distribution, in which case one may consider the infinite time limit. If the system is ergodic, the stationary distribution is unique and independent of the initial condition. In this case (14.2.9) becomes at large times

$$E_b^{st}(x, \boldsymbol{\theta}) = \tilde{\Phi}^{st}(x|\boldsymbol{\theta}), \tag{14.2.10}$$

where the notation "*st*" means that the quantities are drawn from there stationary distribution ($E_b^{st}(x, \boldsymbol{\theta}) = E_b(x, \boldsymbol{\theta}, t \to \infty)$ and similarly for $\tilde{\Phi}$). In that case, one has $E_b^{st}(x, \boldsymbol{\theta}) + E_a^{st}(x, \boldsymbol{\theta}) = 1$.

One can also imagine situations where the initial value of $\boldsymbol{\theta}$ is unknown. In this case we assume that it is drawn from its equilibrium distribution $p_{eq}(\boldsymbol{\theta})$ and we define

$$E_b(x, t) = \int d\boldsymbol{\theta}\, p_{eq}(\boldsymbol{\theta}) E_b(x, \boldsymbol{\theta}, t)\,, \qquad (14.2.11)$$

as well as

$$\tilde{\Phi}(y, t|b) = \tilde{\mathbb{P}}(y(t) \leq y, t|y(0) = b, \boldsymbol{\theta}(0)^{eq}) = \int d\boldsymbol{\theta}\, p_{eq}(\boldsymbol{\theta}) \tilde{\Phi}(y, t|\boldsymbol{\theta}; b)\,. \qquad (14.2.12)$$

Averaging both sides of (14.2.9) over $p_{eq}(\boldsymbol{\theta})$ we obtain

$$E_b(x, t) = \tilde{\Phi}(x, t|b)\,. \qquad (14.2.13)$$

Finally, it is possible to derive the equivalent of the full Siegmund duality relation (14.1.1) for the Langevin dynamics (14.2.1) (see [3] for the complete derivation). The duality relation reads in this case

$$\boxed{\mathbb{P}(x(t) \geq y|x(0) = x, \boldsymbol{\theta}(0)^{eq}) = \tilde{\mathbb{P}}(y(t) \leq x|y(0) = y, \boldsymbol{\theta}(0)^{eq})}\,. \qquad (14.2.14)$$

This relation connects the full probability density of $x(t)$ (in the presence of absorbing walls) at any finite time with the probability density of $y(t)$ (in the presence of hard walls). The relation (14.2.13) can be recovered from Eq. (14.2.14) by taking $y = b$. It is worth mentioning that Eq. (14.2.14) still holds when there are no walls.

The duality relation (14.2.9) is illustrated schematically in Fig. 14.2, where typical trajectories contributing to the probabilities $E_b(x, \boldsymbol{\theta}, t)$ and $\tilde{\Phi}(x, t|\boldsymbol{\theta}; b)$ are shown. It can be intuitively understood as a form of time reversal symmetry (hence the minus sign that appears in front of the force for the dual, while the additional term $\partial_x \mathcal{T}$ compensates for the flux of probability generated by the gradient of diffusion coefficient). The detailed balance conditions (14.2.3)-(14.2.4) for the driving process $\boldsymbol{\theta}(t)$ play a crucial role in this symmetry.

### 14.2.2 Specialization to Active Particle Models

The most well-known active particle models can all be described as follows: $\boldsymbol{\theta}(t) = v(t)$ is a scalar which represents the random velocity, and the dynamics typically involve $f(x, v) = F(x) + \alpha(x)\, v(t)$, with $F(x)$ denoting an external force, and $\alpha(x)$ a (generally positive) function modulating the particle's intrinsic speed, while we choose the temperature $\mathcal{T}(x, v) = T(x)$ to be independent of $v$. The equation of motion is then

$$\dot{x}(t) = F(x(t)) + \alpha(x(t))\, v(t) + \sqrt{2T(x(t))}\, \xi(t)\,. \qquad (14.2.15)$$

This setting includes in particular (see also Table 14.1):

- the RTP, which corresponds to $g = 0$, $\mathcal{D} = 0$, and $\mathcal{W}(v'|v) = \gamma\, \delta(v + v')$ in Eq. (14.2.2), having for initial condition $v(0) = \pm v_0$, resulting in a constant velocity of $\pm v_0$ throughout.

- the AOUP, represented by $g(v) = -\frac{v}{\tau}$ and $\mathcal{D} = \frac{D}{\tau^2}$ (constant), with $\mathcal{W} = 0$.

- for a two-dimensional ABP projected into one dimension, it is simpler to choose $\boldsymbol{\theta}$ in (14.2.1) as the angle $\varphi$ between the particle's orientation and the $x$-axis. Here, the dynamics involve $f(x, \varphi) = F(x) + \alpha(x) \cos \varphi$ (we assume that the external force along the $x$-direction does not vary along the other space direction), with parameters $g = 0$, $\mathcal{D}$ a constant and $\mathcal{W} = 0$. However this can be easily rewritten under the form (14.2.15) by writing $v(t) = \cos \varphi$.

| **Model** | RTP | AOUP | ABP |
|---|---|---|---|
| $\boldsymbol{\theta}(t)$ | $v(t)$ | $v(t)$ | $\varphi(t)$ |
| Evolution of $\boldsymbol{\theta}(t)$ | $g=0,\ \mathcal{D}=0$ $\mathcal{W}(v'\vert v)=\gamma\delta(v+v')$ | $g(v)=-\frac{v}{\tau},$ $\mathcal{D}=\frac{D}{\tau^2},\ \mathcal{W}=0$ | $\dot{\varphi}(t)=2D\,\eta(t)$ |
| $f(x,\boldsymbol{\theta}(t))$ | $F(x)+\alpha(x)\,v(t)$ | $F(x)+\alpha(x)\,v(t)$ | $F(x)+\alpha(x)\,\cos\varphi(t)$ |

**Table 14.1:** This table illustrates how the dynamics in Eq. (14.2.1) can be adapted to describe different active particle models. The evolution of $\boldsymbol{\theta}(t)$ is given in Eq. (14.2.2).

For an active particle, the dual process of Eq. (14.2.15) is equivalent to

$$\boxed{\dot{y}(t)=\tilde{F}(y(t))+\alpha(y(t))\,\tilde{v}(t)+\sqrt{2T(y(t))}\,\xi(t)\quad,\quad \tilde{F}(x)=-F(x)+\partial_x T(x)}\,, \qquad (14.2.16)$$

where $\tilde{v}(t)$ has the same dynamics as $-v(t)$. This definition is more in line with the interpretation of $v$ as a velocity (changing the sign of $\alpha(x)$ would lead to a counter-inuitive situation where a positive velocity, i.e. positive value of $v(t)$, would push the particle in the $-$ direction). For all the examples considered in this thesis (RTP, AOUP and ABP), the equation of motion is invariant under the change $v\to -v$, so that $\tilde{v}(t)$ evolves as $v(t)$.

In this setting, Eq. (14.2.9) reads

$$\boxed{E_b(x,v,t)=\tilde{\Phi}(x,t\vert \tilde{v}(t)=-v;b)}\,. \qquad (14.2.17)$$

Here, it relates the exit probability of a particle with initial velocity $v$ to the cumulative distribution of the dual particle with velocity $-v$ at time $t$. Similarly, the results (14.2.10), (14.2.13) and (14.2.14) can be rewritten in terms of $v(t)$ and $\tilde{v}(t)$.

### 14.2.3 Discrete Time Random Walks

All the results presented above have analogues for discrete-time random walks $X_n$ with absorbing boundaries at $x=a^-$ and $x=b$, described by

$$X_n=X_{n-1}+W_n, \qquad (14.2.18)$$

where $W_n$ is a stationary stochastic process whose stationary distribution $p_{st}(w)$ satisfies the time-reversal property:

$$p_{st}(w_1)P(W_2=w_2,\ldots,W_T=w_T\mid W_1=w_1)=p_{st}(w_T)P(W_2=w_{T-1},\ldots,W_T=w_1\mid W_1=w_T). \qquad (14.2.19)$$

Here, $a^-$ should be interpreted as $a-\epsilon$ with $\epsilon\to 0$; that is, we regard $x=a^-$ as distinct from $x=a$ with $a^-<a$, but any negative jump from the position $x=a$ leads to $x<a^-$. This distinction is crucial for the proof presented in Ref. [3]. The increment $W_n$ may take either discrete or continuous values, but its evolution does not depend on the current position $X_n$.

The dual process (with hard walls at positions $a$ and $b$) is defined as

$$Y_n=Y_{n-1}-\tilde{W}_n, \qquad (14.2.20)$$

where $\tilde{W}_n$ follows the same stochastic dynamics as $W_n$ and is initialized according to its stationary distribution $p_{st}(w)$. This random walk can be viewed as the time-reversed counterpart of the original process $X_n$, but starting from an arbitrary initial position $Y_0$ and having hard walls at $a$ and $b$ rather than absorbing boundaries.

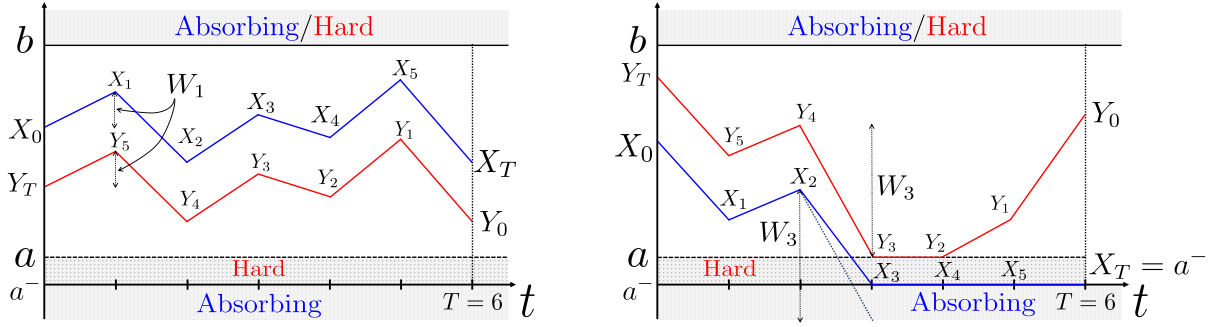

**Figure 14.3:** We show in blue two examples of trajectories of a process $X_n$ (defined in Eq. (14.2.18)), and in red we show the associated reversed dual trajectory $\hat{Y}_n = Y_{T-n}$, where $Y_n$ is the dual trajectory. **Left**: We illustrate the fact that when $X_n$ never reaches a wall, $\hat{Y}_n$ is parallel to $X_n$ and never crosses it. **Right**: $X_n$ reaches a wall at time $t_0 = 3$ and is absorbed at $x = a^-$. On the other hand, for $\hat{Y}_n$, $a$ and $b$ are hard walls and it can still move inside $[a, b]$. In both cases, the equivalence (14.2.23) holds.

As for the continuous case, we are interested in making a connection between the exit probability at $b$ for the process $X_n$,

$$E_b(x, w, T) = \mathbb{P}(X_T = b | X_0 = x, W_1 = w), \qquad (14.2.21)$$

and the conditional cumulative distribution of its dual $Y_n$, where $\tilde{W}_1$ is drawn from $p_{st}(w)$

$$\begin{aligned}\tilde{\Phi}(x, T|w; y) &= \tilde{\mathbb{P}}(Y_n \leq x | \tilde{W}_T = w, Y_0 = y, \tilde{W}_1^{eq}) \\ &\equiv \int dw \, p_{st}(w_1) \tilde{\mathbb{P}}(Y_T \leq x | \tilde{W}_T = w, Y_0 = y, \tilde{W}_1 = w_1).\end{aligned} \qquad (14.2.22)$$

Note that $E_b(x, w, T)$ is conditioned on the next step *after* time $n = 0$ (i.e. the jump that occurs between $X_0$ and $X_1$), while the cumulative is conditioned on the last step *before* time $T$ (i.e. between $Y_{T-1}$ and $Y_T$).

For a given time interval $[0, T]$ and a given $Y_0$, we define the dual trajectory of $X_n$ as the realization of $Y_n$ with for all $n$ $\tilde{W}_n = W_{T+1-n}$. Examples of trajectories along with their reversed dual trajectories $\hat{Y}_n = Y_{T-n}$ are shown in Fig. 14.3. What we have shown in [3] is a one to one mapping between the trajectories of $X_n$ and its dual trajectory (a realization of $Y_n$). For this purpose, in [3] we prove the equivalence

$$X_0 \geq Y_T \Leftrightarrow X_T \geq Y_0. \qquad (14.2.23)$$

Using this equivalence, we can show that if $W_1$ and $\tilde{W}_1$ are drawn from $p_{st}(w)$, we have

$$\mathbb{P}(X_T \geq y | X_0 = x, W_1 = w) = \tilde{\mathbb{P}}(Y_T \leq x | \tilde{W}_T = w, Y_0 = y, \tilde{W}_1^{st}). \qquad (14.2.24)$$

which is an instance of a Siegmund duality. Specializing (14.2.24) to $Y_0 = b$, we obtain the desired relation

$$\boxed{E_b(x, w, T) = \tilde{\Phi}(x, T|w; b)}. \qquad (14.2.25)$$

This construction can be generalized for continuous-time random walks [3]. In the left panel of Fig. 14.3, we also show that the equivalence (14.2.23) holds straightforwardly for trajectories that do not interact with the walls. However, when interactions with the walls occur, the situation becomes more subtle and the derivation more involved (see the right panel of Fig. 14.3) [3].

### 14.2.4 Stochastic Resetting

All the models considered here can be extended by adding Poissonian resetting of the position. During a time interval $[t, t + dt]$, the process $x(t)$ evolves according to (14.2.1) with probability $1 - \sum_{i=1}^{n} r_i dt$, or restarts it dynamics at position $X_r^i \in [a,b]$ with probability $r_i dt$, where $r_i$ ($i = 1, ..., n$) is the resetting rate associated with the resetting position $X_r^i$. In that case, the definition of the dual is a bit more subtle. Similarly to $x(t)$, the dual process $y(t)$ "resets" with rates $r_i$, $i = 1, ..., N$, but instead of resetting to a fixed position, the position $y(t^+)$ after the reset now depends on the position $y(t^-)$ just before the reset: if $y(t^-) \leq X_r^i$, then $y(t^+) = a$, but if $y(t^-) > X_r^i$, one has instead $y(t^+) = b$. Resetting can also be considered for discrete random walks. The details are given in Sec. 16. The construction of a Siegmund dual for stochastic resetting is another important new result of this thesis.

### 14.2.5 $N$ Independent Stochastic Processes

An interesting and non-trivial new result for extreme value statistics can be derived as a corollary of the duality relation (14.2.14). Consider $N$ independent processes $x_i(t)$ ($i \in \{1, 2, \ldots N\}$) following the Langevin dynamics (14.2.1) with absorbing walls located at $a$ and $b$, all starting at the same position $x_i(0) = x \in [a,b]$. Consider also $N$ independent realisations $y_i(t)$ of the dual process, which evolve according to (14.2.7) with hard walls at $a$ and $b$, and initial condition $y_i(0) = y \in [a,b]$ for all $i$. The driven equilibrium process $\boldsymbol{\theta}(t)$ is initialized in its equilibrium distribution in both cases. In [3], we show that there exists a relation between the cumulative distributions of the maximum (resp. minimum) of the $x_i$'s, and the minimum (resp. maximum) of the $y_i$'s,

$$\mathbb{P}\left(\max_{0 \leq i \leq N} x_i(t) \geq y \,\bigg|\, x_i(0) = x\right) = \mathbb{P}\left(\min_{0 \leq i \leq N} y_i(t) \leq x \,\bigg|\, y_i(0) = y\right) , \qquad (14.2.26)$$

and

$$\mathbb{P}\left(\min_{0 \leq i \leq N} x_i(t) \geq y \,\bigg|\, x_i(0) = x\right) = \mathbb{P}\left(\max_{0 \leq i \leq N} y_i(t) \leq x \,\bigg|\, y_i(0) = y\right) . \qquad (14.2.27)$$

When specializing to $y = b$, this relates the exit probability at time $t$ of the maximum (resp. minimum) of $x_i$ at $b$, to the cumulative distribution of the minimum (resp. maximum) of $N$ independent dual processes starting at $b$.

## 14.3 Discussion

Since the equations used to compute analytically the two quantities $E_b$ and $\tilde{\Phi}$ are quite similar (indeed in [3], we show that in the continuous case they satisfy the same differential equation with the same boundary and initial conditions), one could expect that the difficulty associated with both computations will be comparable in most cases. However, it is useful to know that they are connected by such a simple relation, so that one can always obtain the second quantity as a byproduct of the other.

If one wants to compute these quantities using numerical simulations however, there are cases where one quantity is clearly simpler to obtain than the other. In particular if one is interested in the large time limit, if the system is ergodic, the stationary density in the presence of hard walls can be obtained using a single simulation and averaging over time. In contrast, computing the exit probability requires a large number of simulations starting from every position $x$ and waiting for the particle to reach one of the walls every time, which requires much more computation time. This could also be useful in experimental settings, where it is often much simpler to obtain long time series of data from a single particle trajectory [395]. One could therefore imagine an experimental protocol to indirectly measure the exit probability of a particle via the measure of the distribution of positions of a "dual" particle. On the other hand, the probability density conditioned on the value of $\boldsymbol{\theta}$ can be challenging to compute numerically, while computing the exit probability for a given initial value of $\boldsymbol{\theta}$ is quite straightforward.

## Chapter 15

# Siegmund Duality for the One-Dimensional Run-and-Tumble Particle

In this chapter, we consider a run-and-tumble particle subjected to an arbitrary external potential under two types of boundary conditions, illustrated in Figure 14.1: absorbing walls and hard walls. Our goal is to highlight the surprising connection between these boundary conditions in the context of active particles. More precisely, we show that the exit probability of a run-and-tumble particle with absorbing boundaries is closely related to its cumulative distribution in the presence of hard walls.

We first derive in Sec. 15.1 the exit probability at infinite time for an RTP with an arbitrary external force. This computation was performed before in [222] but only for a free RTP (although with an additional Brownian noise). In Sec. 15.2, we then calculate the cumulative distribution of the position of an RTP when the walls are hard walls (drawing inspiration from [216, 370]). We clarify the connection that exists between the two quantities in Sec. 15.3. This relation is however not restricted to the stationary state, and in Sec. 15.4 we show that it holds at any time, given the right initial conditions. In Sec. 15.5, we derive the duality for the persistent random walk, which is the discrete-time version of the RTP.

Most of the results presented in this chapter are based on our joint work with Léo Touzo, published in Ref. [4].

## 15.1 Exit Probability of a Run-and-Tumble Particle

We consider an RTP subjected to a force $F(x)$, that may derive from a potential $V(x)$ through the relation $F(x) = -V'(x)$. The dynamics is as follows,

$$\dot{x}(t) = F(x) + v_0\,\sigma(t)\,, \tag{15.1.1}$$

where $v_0$ is the internal speed of the particle. The telegraphic noise $\sigma(t)$ switches value at rate $\gamma$ through

$$\sigma(t+dt) = \begin{cases} \sigma(t) & \text{, with probability } (1-\gamma\,dt) \\ -\sigma(t) & \text{, with probability } \gamma\,dt \end{cases}. \tag{15.1.2}$$

The transitions between these values are referred to as *tumbles*. The time duration $\tau$ between two consecutive tumbles follows an exponential distribution with a probability density function given by $p(\tau) = \gamma\,e^{-\gamma\tau}$. When the RTP is in the positive state $\sigma = +1$, we describe the particle as being in a "positive" (+) state, whereas when it is in the negative state $\sigma = -1$, we refer to it as being in a "negative" (−) state. Its initial position is $x(0) = x \in [a, b]$ and two absorbing walls are located at $x = a$ and $x = b$ (see Figure 14.1). In this section, we will assume that $|F(x)| < v_0$ on the interval $[a, b]$ such that the whole interval is accessible to the particle no matter its starting position[20].

We consider the probability for the RTP to exit at wall $b$, before or at time $t$, with a positive (resp. negative) initial velocity

$$E_b(x, \pm, t) = \mathbb{P}\left(x(t) = b \,|\, x(0) = x, \sigma(0) = \pm 1\right). \quad (15.1.3)$$

Recall that when the particle reaches $b$, it stays there forever – see Fig. 14.1. In this section we will be focusing on the infinite time limit

$$E_b(x, \pm) = \lim_{t \to +\infty} E_b(x, \pm, t), \quad (15.1.4)$$

i.e. we want to know the probability that after an infinite time, the particle has exited at $b$ and not at $a$. If we assume $\sigma(0) = +1$ or $\sigma(0) = -1$ with probability $1/2$, the exit probability of an RTP regardless of the initial speed is

$$E_b(x) = \frac{1}{2}\left(E_b(x, +) + E_b(x, -)\right). \quad (15.1.5)$$

We can write a pair of coupled first-order differential equations for $E_b(x, \pm)$. Let us evolve the particle during the time interval $[0, dt]$ and average over the possible trajectories. Suppose the RTP starts its motion at $x$, then for the positive state of the RTP, the only possible events are that the particle switches sign, with probability $\gamma\, dt$, and moves from $x$ to $x + [F(x) - v_0]\, dt$, or that it moves over a distance $[F(x) + v_0]\, dt$ while staying in the positive state, with probability $1 - \gamma dt$. This translates to

$$E_b(x, +, t+dt) = (1 - \gamma dt)\, E_b(x + [F(x) + v_0]\, dt, +, t) + \gamma dt\, E_b(x + [F(x) - v_0]\, dt, -, t). \quad (15.1.6)$$

After Taylor-expanding at order $dt$ (and repeating the operation for $E_b(x, -, t)$), we obtain

$$\partial_t E_b(x, +, t) = [F(x) + v_0]\, \partial_x E_b(x, +, t) + \gamma E_b(x, -, t) - \gamma E_b(x, +, t), \quad (15.1.7)$$
$$\partial_t E_b(x, -, t) = [F(x) - v_0]\, \partial_x E_b(x, -, t) + \gamma E_b(x, +, t) - \gamma E_b(x, -, t), \quad (15.1.8)$$

In the stationary state, this becomes

$$[F(x) + v_0]\, \partial_x E_b(x, +) + \gamma\, E_b(x, -) - \gamma\, E_b(x, +) = 0, \quad (15.1.9)$$
$$[F(x) - v_0]\, \partial_x E_b(x, -) + \gamma\, E_b(x, +) - \gamma\, E_b(x, -) = 0. \quad (15.1.10)$$

To solve the coupled equations (15.1.9) and (15.1.10), one needs two boundary conditions. When an RTP starts in the negative state at position $a$, it is absorbed at the wall $a$ and thus never reaches $b$ leading to $E_b(a, -) = 0$. On the other hand, if an RTP starts in the positive state at wall $b$ it will directly exit the interval such that $E_b(b, +) = 1$.

Writing equations (15.1.9)-(15.1.10) in terms of $E_b(x) = \frac{1}{2}\left(E_b(x, +) + E_b(x, -)\right)$ and $\Delta E_b(x) = \frac{1}{2}\left(E_b(x, +) - E_b(x, -)\right)$ we get

$$F(x)\, \partial_x E_b(x) + v_0\, \partial_x \Delta E_b(x) = 0, \quad (15.1.11)$$
$$F(x)\, \partial_x \Delta E_b(x) + v_0\, \partial_x E_b(x) - 2\gamma\, \Delta E_b(x) = 0. \quad (15.1.12)$$

---

[20] The case in which the force has turning points (i.e., points where $F(x) = \pm v_0$, creating regions where the external force exceeds the velocity of the RTP and making certain regions of space inaccessible to the particle) is slightly more subtle and is discussed in Ref. [4].

Using equation (15.1.11), we can replace $E_b(x)$ in equation (15.1.12) to get a first order equation for $\Delta E_b(x)$, which we solve. From this we deduce $E_b(x)$, using $E_b(a, -) = 0$ and $E_b(b, +) = 1$ to fix the integration constants. We obtain

$$E_b(x) = \frac{1}{Z} \left( 2\gamma v_0 \int_a^x dz \, \frac{\exp\left[-2\gamma \int_a^z du \, \frac{F(u)}{v_0^2 - F(u)^2}\right]}{v_0^2 - F(z)^2} + 1 \right), \quad (15.1.13)$$

$$Z = 2\gamma v_0 \int_a^b dz \, \frac{\exp\left[-2\gamma \int_a^z du \, \frac{F(u)}{v_0^2 - F(u)^2}\right]}{v_0^2 - F(z)^2} + \exp\left[-2\gamma \int_a^b du \, \frac{F(u)}{v_0^2 - F(u)^2}\right] + 1, \quad (15.1.14)$$

and using that $E_b(x, \pm) = E_b(x) \pm \Delta E_b(x)$, we finally get

$$\boxed{E_b(x, \pm) = \frac{1}{Z} \left( 2\gamma v_0 \int_a^x dz \, \frac{\exp\left[-2\gamma \int_a^z du \, \frac{F(u)}{v_0^2 - F(u)^2}\right]}{v_0^2 - F(z)^2} \pm \exp\left[-2\gamma \int_a^x du \, \frac{F(u)}{v_0^2 - F(u)^2}\right] + 1 \right).}$$
(15.1.15)

*Diffusive limit:* An RTP behaves as a diffusive particle when taking the limit $\gamma \to +\infty$, and $v_0 \to \infty$, while keeping the ratio $D_a = v_0^2/(2\gamma)$ fixed (where the subscript 'a' stands for active). In this diffusive limit, using Eq. (15.1.13), and if one assumes that the force derives from a potential $F(x) = -V'(x)$, we recover the well known result

$$E_b(x) = \frac{\int_a^x dz \, e^{\frac{V(z)}{D_a}}}{\int_a^b dz \, e^{\frac{V(z)}{D_a}}}. \quad (15.1.16)$$

*Free RTP:* When there is no force applied to the RTP, the particle is free and $F(x) = 0$. In this case, it is quite straightforward to compute the exit probabilities. From Eq. (15.1.13) and Eq. (15.1.15), one obtains

$$E_b(x) = \frac{\frac{1}{2} + \frac{\gamma}{v_0}(x-a)}{1 + \frac{\gamma}{v_0}(b-a)}, \quad E_b(x, +) = \frac{1 + \frac{\gamma}{v_0}(x-a)}{1 + \frac{\gamma}{v_0}(b-a)}, \quad \text{and} \quad E_b(x, -) = \frac{x-a}{\frac{v_0}{\gamma} + (b-a)}. \quad (15.1.17)$$

The result is linear, as in the Brownian case where the exit probability is simply $(x-a)/(b-a)$, which is also recovered taking the diffusive limit of RTPs in Eq. (15.1.17).

*Constant drift:* The exit probabilities $E_b(x, \pm)$ (15.1.15) can be computed exactly for a range of potentials. Let us start with the simplest case, namely a constant drift $F(x) = \alpha$ with $|\alpha| < v_0$. One finds

$$E_b(x, \pm) = \frac{1}{Z} \left[ \left(1 + \frac{v_0}{\alpha}\right) + \left(\pm 1 - \frac{v_0}{\alpha}\right) e^{-\frac{2\gamma\alpha}{v_0^2 - \alpha^2}(x-a)} \right], \quad (15.1.18)$$

$$Z = 1 + \frac{v_0}{\alpha} + \left(1 - \frac{v_0}{\alpha}\right) e^{-\frac{2\gamma\alpha}{v_0^2 - \alpha^2}(b-a)}. \quad (15.1.19)$$

*Harmonic potential:* It is also possible to obtain an explicit form for the exit probabilities of an RTP inside a harmonic potential $V(x) = \mu x^2/2$. The force is $F(x) = -\mu x$ and we impose $b < v_0/\mu$ and $a > -v_0/\mu$ such that $|F(x)| < v_0$ in $[a, b]$. The exit probabilities are given by (for

any $x$)

$$E_b(x, \pm) = \frac{1}{Z} \left\{ 1 \pm \left(\frac{v_0^2 - a^2\mu^2}{v_0^2 - \mu^2 x^2}\right)^{\frac{\gamma}{\mu}} + \frac{2\gamma}{v_0} \left(1 - \frac{a^2\mu^2}{v_0^2}\right)^{\frac{\gamma}{\mu}} \left[ x \, _2F_1\left(\frac{1}{2}, 1+\frac{\gamma}{\mu}, \frac{3}{2}, \frac{\mu^2 x^2}{v_0^2}\right) \right. \right.$$
$$\left. \left. - a \, _2F_1\left(\frac{1}{2}, 1+\frac{\gamma}{\mu}, \frac{3}{2}, \frac{a^2\mu^2}{v_0^2}\right) \right] \right\}, \quad (15.1.20)$$

where $_2F_1$ is the hypergeometric function. The expression of the normalisation constant is

$$Z = 1 + \left(\frac{v_0^2 - a^2\mu^2}{v_0^2 - b^2\mu^2}\right)^{\frac{\gamma}{\mu}} + \frac{2\gamma}{v_0}\left(1 - \frac{a^2\mu^2}{v_0^2}\right)^{\frac{\gamma}{\mu}} \left[ b \, _2F_1\left(\frac{1}{2}, 1+\frac{\gamma}{\mu}, \frac{3}{2}, \frac{b^2\mu^2}{v_0^2}\right) \right.$$
$$\left. - a \, _2F_1\left(\frac{1}{2}, 1+\frac{\gamma}{\mu}, \frac{3}{2}, \frac{a^2\mu^2}{v_0^2}\right) \right]. \quad (15.1.21)$$

One can check from Eq. (15.1.20) that $E_b(x, -)$ vanishes linearly as $x \to a^+$, and that $E_b(b^-, -) < 1$ (and similarly $E_b(a^+, +) > 0$, and $1 - E_b(x, +)$ vanishes linearly as $x \to b^-$).

In Figure 15.1, we compare our formulas to simulations and the agreement is perfect. For all these computations, we made sure that inside the interval $[a, b]$ the force verifies $|F(x)| < v_0$.

## 15.2 Distribution of the Position of a Run-and-Tumble Particle with Hard Walls

Let us now consider a seemingly completely different problem, namely the calculation of the stationary density of a run-and-tumble particle subjected to an external force $F(x)$, confined between two impenetrable walls at $x = a$ and $x = b$ (see Figure 14.1). We again assume $|F(x)| < v_0$ for any $x$ in the interval $[a, b]$. The solution without walls was found in [216] for an arbitrary $F(x)$, while the solution in the presence of walls but without the external force was presented in [370]. We have combined both methods to obtain the solution to our problem.

Let us denote $P(x, +)$ and $P(x, -)$ the densities of an RTP in the $+$ and $-$ states, which are normalised such that $\int_a^b dx \, [P(x, +) + P(x, -)] = 1$. The steady-state Fokker-Planck equations for these densities are given by

$$\partial_x \left[ (F(x) + v_0) P(x, +) \right] + \gamma P(x, +) - \gamma P(x, -) = 0, \quad (15.2.1)$$
$$\partial_x \left[ (F(x) - v_0) P(x, -) \right] - \gamma P(x, +) + \gamma P(x, -) = 0. \quad (15.2.2)$$

Introducing $P(x) = P(x, +) + P(x, -)$ and $Q(x) = P(x, +) - P(x, -)$ we get

$$\partial_x \left[ F(x) P(x) + v_0 Q(x) \right] = 0, \quad (15.2.3)$$
$$\partial_x \left[ F(x) Q(x) + v_0 P(x) \right] + 2\gamma Q(x) = 0, \quad (15.2.4)$$

as in [216]. The difference arises when considering the boundary conditions. Due to the persistent motion, the RTP may remain at either wall for a finite time. More precisely, the density $P(x, -)$ will have a finite mass $\kappa_a$ at $x = a$, and $P(x, +)$ will have a finite mass $\kappa_b$ at $x = b$. Since $\kappa_a$ and $\kappa_b$ are stationary, the total current $J(x) = J(x, +) + J(x, -)$, where $J(x, \pm) = (F(x) \pm v_0) P(x, \pm)$, should vanish at the boundaries. In addition the probability current of a $+$ (resp. $-$) particle at $x = a$ (resp. $x = b$) arises entirely from a $-$ (resp. $+$) particle stuck at the wall which switches sign. Therefore one can write $J(a,+) = -J(a,-) = \gamma \kappa_a$ and $J(b,+) = -J(b,-) = \gamma \kappa_b$ which translates to

$$[v_0 + F(a)] P(a,+) = [v_0 - F(a)] P(a,-) = \gamma \kappa_a, \tag{15.2.5}$$
$$[v_0 + F(b)] P(b,+) = [v_0 - F(b)] P(b,-) = \gamma \kappa_b, \tag{15.2.6}$$

and implies
$$F(a) P(a) + v_0 Q(a) = 0 \quad , \quad F(b) P(b) + v_0 Q(b) = 0. \tag{15.2.7}$$

Integrating (15.2.3) and taking this condition into account gives for all $x$

$$F(x) P(x) + v_0 Q(x) = 0, \tag{15.2.8}$$

which we can use to replace $Q(x)$ in (15.2.4). We thus obtain the equation

$$\partial_x[(v_0^2 - F(x)^2) P(x)] - 2\gamma F(x) P(x) = 0, \tag{15.2.9}$$

that is solved by

$$P(x) = \frac{1}{Z} \frac{2\gamma v_0}{v_0^2 - F(x)^2} \exp\left[2\gamma \int_a^x dz \frac{F(z)}{v_0^2 - F(z)^2}\right], \tag{15.2.10}$$

for any $x$ in $(a,b)$, with $Z$ a normalisation constant. This expression is the same as the density of an RTP without walls derived in [216]. The difference comes from the normalisation and the presence of delta functions at the walls. The expressions for $P(x, \pm)$ can be easily deduced from this result

$$P(x, \pm) = \frac{1}{2}(P(x) \pm Q(x)) = \frac{1}{2}\left(1 \mp \frac{F(x)}{v_0}\right) P(x) = \frac{\gamma}{Z} \frac{1}{v_0 \pm F(x)} \exp\left[2\gamma \int_a^x dz \frac{F(z)}{v_0^2 - F(z)^2}\right]. \tag{15.2.11}$$

Using (15.2.5)-(15.2.6) we deduce

$$\kappa_a = \frac{1}{Z} \quad , \quad \kappa_b = \frac{1}{Z} \exp\left[2\gamma \int_a^b dz \frac{F(z)}{v_0^2 - F(z)^2}\right]. \tag{15.2.12}$$

The constant $Z$ is then fixed by the normalisation condition $\int_{a^+}^{b^-} dx \, P(x) + \kappa_a + \kappa_b = 1$. Thus the full expression for $P(x)$ is

$$P(x) = \frac{1}{Z}\left(2\gamma v_0 \frac{\exp\left[2\gamma \int_a^x du \frac{F(u)}{v_0^2 - F(u)^2}\right]}{v_0^2 - F(x)^2} + \delta(x-a) + \exp\left[2\gamma \int_a^b du \frac{F(u)}{v_0^2 - F(u)^2}\right] \delta(x-b)\right),$$

$$Z = 2\gamma v_0 \int_a^b dz \frac{\exp\left[2\gamma \int_a^z du \frac{F(u)}{v_0^2 - F(u)^2}\right]}{v_0^2 - F(z)^2} + \exp\left[2\gamma \int_a^b du \frac{F(u)}{v_0^2 - F(u)^2}\right] + 1. \tag{15.2.13}$$

Let us now write the associated cumulative distribution. One has, for any $x$ in $[a, b]$,

$$\Phi(x) = \int_{a^-}^x dz \, P(z) = \frac{1}{Z}\left(2\gamma v_0 \int_a^x dz \frac{\exp\left[2\gamma \int_a^z du \frac{F(u)}{v_0^2 - F(u)^2}\right]}{v_0^2 - F(z)^2} + 1\right), \tag{15.2.14}$$

which is exactly the same result as (15.1.13) for the probability to exit at $x = b$ when starting at $x$, but with an opposite force $-F(x)$. From Eq. (15.2.11), we can express $P(x, \sigma)$ for $x \in [a, b]$ and $\sigma = \pm 1$ by adding Kronecker symbols $\delta_{\sigma,\pm}$ to take into account the finite probability masses of $-$ and $+$ at the walls. It gives

$$P(x,\sigma) = \frac{1}{Z}\left(\gamma \frac{\exp\left[2\gamma \int_a^x dy \frac{F(y)}{v_0^2-F(y)^2}\right]}{v_0 + \sigma F(x)} + \delta_{\sigma,-}\delta(x-a) + \delta_{\sigma,+}\exp\left[2\gamma \int_a^b dy \frac{F(y)}{v_0^2-F(y)^2}\right]\delta(x-b)\right). \tag{15.2.15}$$

Integrating over $x$ and dividing by $P(\sigma) = \frac{1}{2}$ (the probability for the particle to be in the state $\sigma$ in the stationary state), we obtain the cumulative of the positions of the dual process conditioned on the internal state $\sigma$ for $x \in (a,b)$,

$$\Phi(x|\sigma) = \frac{1}{P(\sigma)}\int_{a^-}^x dz\, P(z,\sigma) = \frac{2}{Z}\left(\gamma \int_a^x dz \frac{\exp\left[2\gamma \int_a^z du \frac{F(u)}{v_0^2-F(u)^2}\right]}{v_0 + \sigma F(z)} + \delta_{\sigma,-}\right)$$

$$= \frac{2}{Z}\left(\gamma v_0 \int_a^x dz \frac{\exp\left[2\gamma \int_a^z du \frac{F(u)}{v_0^2-F(u)^2}\right]}{v_0^2 - F(z)^2} - \sigma\gamma \int_a^x dz \frac{F(z)}{v_0^2-F(z)^2}\exp\left[2\gamma \int_a^z du \frac{F(u)}{v_0^2-F(u)^2}\right] + \delta_{\sigma,-}\right). \tag{15.2.16}$$

One can then simplify by noticing that the integrand of the second integral is the derivative of $1/(2\gamma)\exp\left[2\gamma \int_a^z du \frac{F(u)}{v_0^2-F(u)^2}\right]$. We thus finally obtain

$$\boxed{\Phi(x|\sigma=\pm) = \frac{1}{Z}\left(2\gamma v_0 \int_a^x dz \frac{\exp\left[2\gamma \int_a^z du \frac{F(u)}{v_0^2-F(u)^2}\right]}{v_0^2-F(z)^2} \mp \exp\left[2\gamma \int_a^x dz \frac{F(z)}{v_0^2-F(z)^2}\right] + 1\right).} \tag{15.2.17}$$

The weights of the delta peaks in Eq. (15.2.15) are recovered from $\Phi(a^+|\sigma) = \kappa_a\, \delta_{\sigma,-}$ and $1 - \Phi(b^-|\sigma) = \kappa_b\, \delta_{\sigma,+}$. In the next subsection we will relate $\Phi(x|\pm)$ to the exit probabilities $E_b(x,\pm)$.

## 15.3 Duality Relation for a Run-and-Tumble Particle in the Stationary State

The strong similarity between Eqs. (15.1.15) and (15.2.17) suggests the existence of a relation between the exit probability and the cumulative distribution of an RTP in the presence of hard walls. We will now clarify this connection.

Let us consider again the process $x(t)$ defined in Eq. (15.1.1). We define its dual process $y(t)$ through the dynamics

$$\dot{y}(t) = -F(y) + v_0\,\tilde{\sigma}(t), \tag{15.3.1}$$

where $\tilde{\sigma}(t)$ is a different realisation of the same telegraphic noise defined in Eq. (15.1.2). The process $y(t)$ describes the motion of a run-and-tumble particle subjected to the reversed force $-F(y)$. In addition, the process $y(t)$ has hard walls at $a$ and $b$. We denote by $\tilde{\Phi}(y)$ the cumulative distribution of the dual process $y(t)$. From Eq. (15.2.14), we observe that one can link the exit probability of an RTP (15.1.13) to the cumulative of its dual via the following relation

$$E_b(x) = \tilde{\Phi}(x). \tag{15.3.2}$$

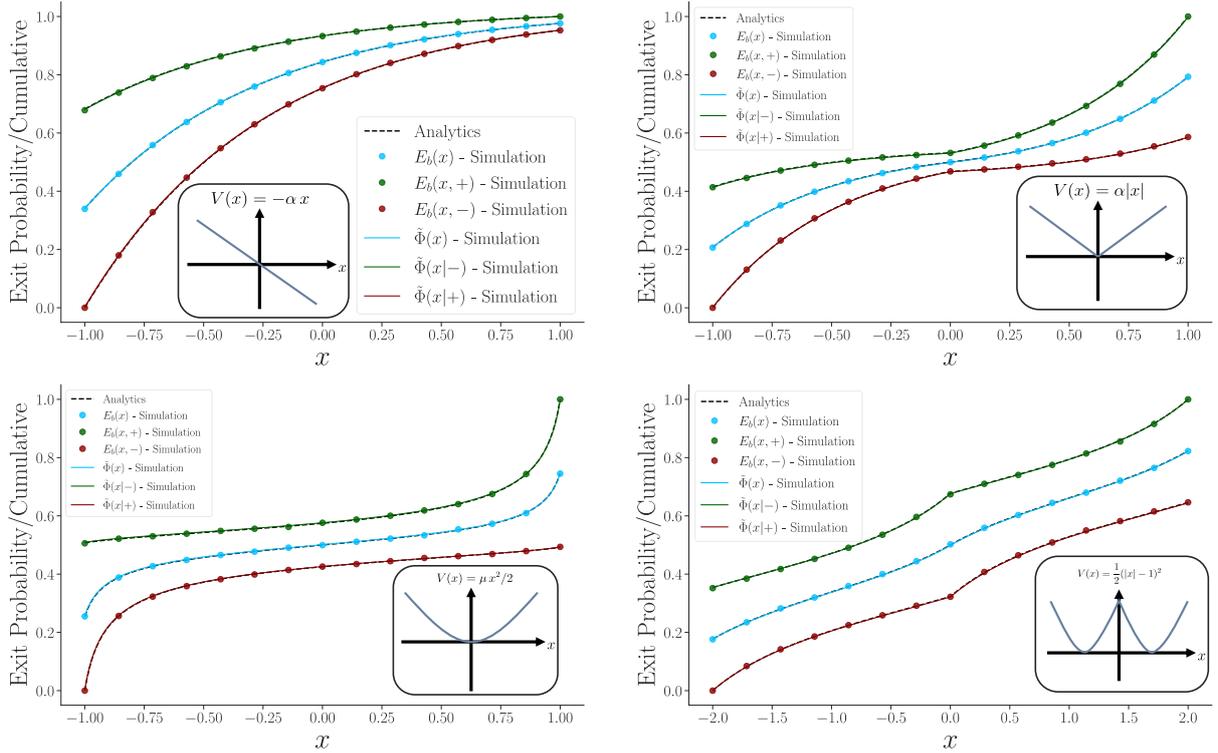

**Figure 15.1:** This figure illustrates the duality relations $E_b(x) = \tilde{\Phi}(x)$ (15.3.2) and $E_b(x, \pm) = \tilde{\Phi}(x|\mp)$ (15.3.3) for a run-and-tumble particle (RTP). We compute the exit probability (resp. the cumulative distribution) of an RTP in various external potentials $V(x)$ (resp. $-V(x)$), with absorbing walls (resp. hard walls) located at positions $x = a$ and $x = b$. The solid lines show numerical simulations of the cumulative $\tilde{\Phi}$ while the dots are obtained from simulations of the exit probability $E_b$. The data from $E_b$ and $\tilde{\Phi}$ overlap exactly. Explicit analytical solutions (dotted lines) are also plotted in dashed line. Throughout this analysis, we ensure that the condition $|F(x)| < v_0$ is satisfied. Top left: $V(x) = -\alpha x$, $\gamma = 1$, $\alpha = 0.5$, $v_0 = 1$, $a = -1$, $b = 1$. Top right: $V(x) = \alpha|x|$, $\gamma = 1$, $\alpha = 0.6$, $v_0 = 1$, $a = -1$, $b = 1$. Bottom left: $V(x) = \mu x^2/2$, $\gamma = 1$, $\mu = 1.9$, $v_0 = 2$, $a = -1$, $b = 1$. Bottom right: $V(x) = 1/2\,(|x| - 1)^2$, $\gamma = 1$, $v_0 = 2$, $a = -2$, $b = 2$.

One can also write a more general relation for the two states of the RTP relating Eq. (15.1.15) to Eq. (15.2.17),

$$\boxed{E_b(x, \pm) = \tilde{\Phi}(x|\mp)}. \tag{15.3.3}$$

This relation coincides with Eq. (14.2.17) for a run-and-tumble particle in the stationary state. Note that the derivation of the exit probability is simpler than the one of the cumulative, in particular because the boundary conditions for the density are more subtle. In Figure 15.1, we check numerically these relations for different forces $F(x)$ such that $|F(x)| < v_0$.

Let us stress an important consequence of this relation, which is specific to active particles (in the absence of diffusion). On the one hand, an RTP with initial position at $x = a^+$ can still go away from the wall and eventually exit at $x = b$ as long as its velocity is positive at early times, and thus $E_b(a^+, +)$ is generally non-zero (and similarly $E_b(b^-, -) < 1$). On the other hand, active particles tend to accumulate near hard walls, leading to delta peaks in the stationary density at $x = a$ and $x = b$ (where the RTP is in the $-$ and $+$ state respectively), such that $\Phi(a^+, -) > 0$ and $\Phi(b^-, +) < 1$. The identity (15.3.3) shows that these two phenomena are related, since $E_b(a^+, +) = \tilde{\Phi}(a^+, -)$ and $E_b(b^-, -) = \tilde{\Phi}(b^-, +)$ (see Fig. 15.1).

The identity (15.3.3) is actually valid not only in the stationary state, but at any time $t$, provided that the initial conditions are chosen correctly. This is what we show in the next section.

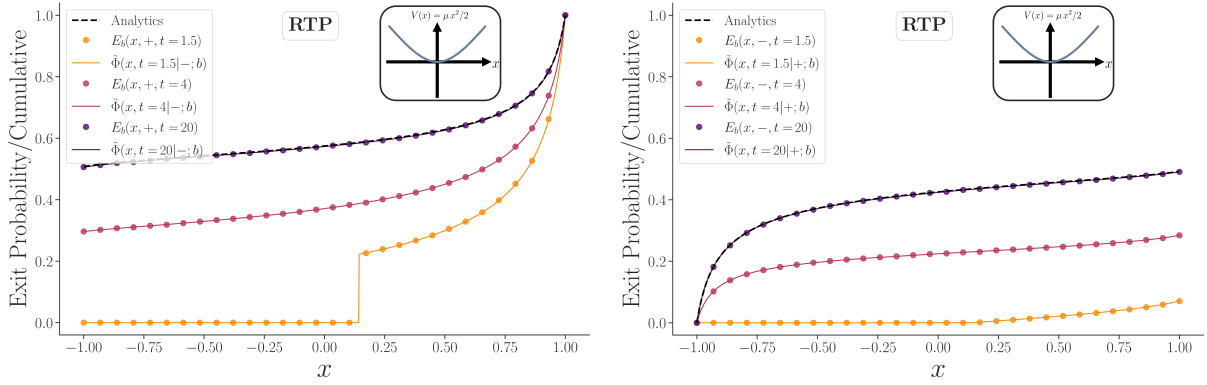

**Figure 15.2:** Illustration of the duality relation (15.4.14) at finite time $t$ for an RTP in a harmonic potential. **Right:** the dots represent the exit probability $E_b(x,+,t)$ in the presence of a potential $V(x) = \mu x^2/2$, with absorbing walls at $a = -1$ and $b = 1$, while the lines show the cumulative distribution of the dual $\tilde{\Phi}(x,t|-;b)$, with a potential $-V(x)$ and hard walls at $a$ and $b$. **Left:** Same plot with the $+$ and $-$ particles exchanged. All the results were obtained by averaging over $10^6$ simulated trajectories, with parameters $\mu = 1.9$, $v_0 = 2$ and $\gamma = 1$. The dashed black lines show the analytic predictions for the stationary state.

## 15.4 Duality Relation for a Run-and-Tumble Particle at Finite Time

In Section 15.3, we have established the duality relation (15.3.3) for RTPs in the stationary state, i.e, at infinite time. It is natural to ask whether such a duality is also valid at finite time. Explicitly obtaining the exit probability, or the cumulative distribution of an RTP at finite time is highly non trivial, especially in the presence of an external force. However, without computing them, we can show that they obey the same differential equation with the same boundary and initial conditions. Although this section is somewhat technical, we include it in this thesis to illustrate how Fokker-Planck equations can be used to derive the duality. This approach is the same as the one we used in [3] to obtain Eq. (14.2.9) for the general process (14.2.1).

Let us consider again the process $y(t)$ defined in (15.3.1) (still with hard walls at $a$ and $b$). In the following, we denote $\tilde{F}(x) = -F(x)$ the external force applied to $y(t)$, and we write with a tilde all the quantities related to $y(t)$. Previously we were only interested in the stationary distribution of this process, so that the initial condition did not matter. However, if we want to extend the duality to finite time, we need to specify the initial condition, for the position $y(0)$, but also for the state $\tilde{\sigma}(0)$. We will now show that the duality relation between $x(t)$ and $y(t)$ still holds at finite time if $y(0) = b$ and if $\tilde{\sigma}(0) = \sigma_0$ is drawn from its stationary distribution $\mathbb{P}(\sigma_0 = \pm 1) = 1/2$.

We begin by writing the forward Fokker-Planck equation satisfied by the density of the position of $y(t)$, $\tilde{P}(y,t|\tilde{\sigma}(t) = \pm 1; y(0) = b, \sigma_0)$, where the semicolon in the conditioning separates finite time conditions from initial conditions. We then deduce from it the differential equation verified by the cumulative distribution, defined as

$$\tilde{\Phi}(x,t|\tilde{\sigma}(t) = \pm 1; y(0) = b) = \frac{1}{2} \int_{a^-}^{x} dy \sum_{\tilde{\sigma}_0 = \pm 1} \tilde{P}(y,t|\tilde{\sigma}(t) = \pm 1; y(0) = b, \sigma_0). \tag{15.4.1}$$

For convenience, in the rest of this section we will denote $\tilde{P}_\pm(y,t) = \tilde{P}(y,t|\tilde{\sigma}(t) = \pm 1; y(0) = b, \sigma_0)$ and $\tilde{\Phi}_\pm(x,t|b) = \tilde{\Phi}(x,t|\tilde{\sigma}(t) = \pm 1; y(0) = b)$, and we will often omit the arguments. We start by writing the Fokker-Planck equations satisfied by the particle density

$$\partial_t \tilde{P}_+ = -\partial_x \left[(\tilde{F}(x) + v_0)\tilde{P}_+\right] - \gamma \tilde{P}_+ + \gamma \tilde{P}_-, \tag{15.4.2}$$

$$\partial_t \tilde{P}_- = -\partial_x \left[(\tilde{F}(x) - v_0)\tilde{P}_-\right] + \gamma \tilde{P}_+ - \gamma \tilde{P}_-. \tag{15.4.3}$$

Since we choose the initial distribution of $\sigma_0$ to be the stationary distribution $p_{st}(\tilde{\sigma} = \pm 1) = \frac{1}{2}$, the distribution of $\tilde{\sigma}(t)$ is independent of time. Using $\tilde{P}_\pm(x,t) = \partial_x \tilde{\Phi}_\pm(x,t)$ we write

$$\partial_x[-\partial_t \tilde{\Phi}_+ - (\tilde{F}(x) + v_0)\partial_x \tilde{\Phi}_+ - \gamma \tilde{\Phi}_+ + \gamma \tilde{\Phi}_-] = 0\,, \tag{15.4.4}$$

$$\partial_x[-\partial_t \tilde{\Phi}_- - (\tilde{F}(x) - v_0)\partial_x \tilde{\Phi}_- + \gamma \tilde{\Phi}_+ - \gamma \tilde{\Phi}_-] = 0\,. \tag{15.4.5}$$

We now want to integrate equations (15.4.4) and (15.4.5) over $x$. As we have already seen in Sec. 15.2, the boundary conditions need to be treated with care. Indeed the density $\tilde{P}_-$ (resp. $\tilde{P}_+$) has a finite mass at $x = a$ (resp. $x = b$), given by $\tilde{\Phi}_-(a^+, t|b) = \kappa_a(t)$ (resp $1 - \tilde{\Phi}_+(b^-, t|b) = \kappa_b(t)$). However, since there is no mass of $+$ particles at $x = a$ nor of $-$ particles at $x = b$, we have $\tilde{\Phi}_+(a^+, t|b) = 1 - \tilde{\Phi}_-(b^-, t|b) = 0$. The current of $+$ particles at $x = a^+$ is thus generated entirely by the $-$ particles at $x = a$ which switch sign. Thus the boundary condition at $x = a^+$ reads $(\tilde{F}(a^+) + v_0)\tilde{P}_+(a^+, t|b) = \gamma \kappa_a(t)$, or equivalently,

$$(\tilde{F}(a^+) + v_0)\partial_x \tilde{\Phi}_+(a^+, t|b) = \gamma \tilde{\Phi}_-(a^+, t|b)\,, \tag{15.4.6}$$

and similarly at $x = b^-$, exchanging $+$ and $-$ particles. Thus we have

$$0 = -\partial_t \tilde{\Phi}_+(a^+, t|b) - (\tilde{F}(a^+) + v_0)\partial_x \tilde{\Phi}_+(a^+, t|b) - \gamma \tilde{\Phi}_+(a^+, t|b) + \gamma \tilde{\Phi}_-(a^+, t|b)\,, \tag{15.4.7}$$

$$0 = -\partial_t \tilde{\Phi}_-(b^-, t|b) - (\tilde{F}(b^-) - v_0)\partial_x \tilde{\Phi}_-(b^-, t|b) + \gamma \tilde{\Phi}_+(b^-, t|b) - \gamma \tilde{\Phi}_-(b^-, t|b)\,. \tag{15.4.8}$$

Therefore there is no integration constant when integrating (15.4.4)-(15.4.5) and we can write, for any $x$ in $(a, b)$,

$$\partial_t \tilde{\Phi}_+ = -(\tilde{F}(x) + v_0)\partial_x \tilde{\Phi}_+ - \gamma \tilde{\Phi}_+ + \gamma \tilde{\Phi}_-\,, \tag{15.4.9}$$

$$\partial_t \tilde{\Phi}_- = -(\tilde{F}(x) - v_0)\partial_x \tilde{\Phi}_- + \gamma \tilde{\Phi}_+ - \gamma \tilde{\Phi}_-\,. \tag{15.4.10}$$

These are exactly the backward Fokker-Planck equations (15.1.7)-(15.1.8) for the exit probabilities of the process $x(t)$, $E_b(x, -, t)$ and $E_b(x, +, t)$, with an external force $F(x) = -\tilde{F}(x)$. In addition $\tilde{\Phi}_\pm(x, t|b)$ and $E_b(x, \pm, t)$ satisfy the same boundary conditions

$$E_b(a^+, -, t) = \tilde{\Phi}_+(a^+, t|b) = 0\,, \tag{15.4.11}$$

$$E_b(b^-, +, t) = \tilde{\Phi}_-(b^-, t|b) = 1\,, \tag{15.4.12}$$

as well as the same initial conditions

$$E_b(x, \pm, 0) = \tilde{\Phi}_\mp(x, 0|b) = \begin{cases} 0 \text{ if } x < b\,, \\ 1 \text{ if } x = b\,. \end{cases} \tag{15.4.13}$$

Therefore, denoting again $\tilde{\Phi}_\mp(x, t|b)$ as $\tilde{\Phi}(x, t|\mp; b)$, we can write the generalized version of (15.3.3)[21],

$$\boxed{E_b(x, \pm, t) = \tilde{\Phi}(x, t|\mp; b)}\,. \tag{15.4.14}$$

This relation (15.4.14) can be thought of as a sort of time reversal. For every trajectory of a particle starting at $x$ in the presence of a potential $V(x)$ and exiting at $b$ before or at time $t$, one can construct a trajectory of a dual particle (with hard walls) starting at $b$ with a reversed potential $-V(x)$ and located inside $[a, x]$ at time $t$.

The duality identity (15.4.14) is illustrated through numerical results in Fig.15.2, for an RTP in a harmonic potential. Of course the finite time version of (15.3.2) is also valid. Let us also note that the relation (15.4.14) (both at finite and infinite time) has an equivalent for the exit probability at $x = a$, $E_a(x, v, t)$, which can be immediately deduced by symmetry (one simply needs to replace $\tilde{\Phi}$ by $1 - \tilde{\Phi}$ and to choose $y(0) = a$).

---

[21] We admit here without proof the unicity of the solution of the PDEs with these initial and boundary conditions.

## 15.5 Duality for the Persistent Random Walk

In this section we show how Eq. (14.2.10) applies for a persistent random walk [270–272] – the discrete-time version of the RTP – on a one-dimensional lattice with $L+2$ sites ($i = 0, ..., L+1$). The motion follows the rule

$$x_n = x_{n-1} + \sigma_n, \tag{15.5.1}$$

where the steps $\sigma_n = \pm 1$ follow a Markov dynamics defined by

$$\sigma_n = \begin{cases} \sigma_{n-1} & \text{with probability } p \\ -\sigma_{n-1} & \text{with probability } 1-p \end{cases}. \tag{15.5.2}$$

The parameter $p \in [0,1]$ controls the "persistence" of the random walk. When $p = 1/2$, we retrieve the well-known symmetric random walk with uncorrelated steps. For $p > \frac{1}{2}$, the steps are correlated positively, leading to persistence in the motion of the walker, while for $p < \frac{1}{2}$ the steps are negatively correlated.

The first quantity of interest is the exit probability in the presence of absorbing boundary conditions at sites 0 and $L+1$ (i.e. if the particle reaches one of those two sites, it stays there forever),

$$E_i^\pm(n) = \mathbb{P}(\text{particle exits at } L+1 \text{ before or at time } n | \text{ particle starts at site } i \text{ \& first jump is } \pm 1). \tag{15.5.3}$$

Since the jump distribution is symmetric (i.e. we have $q(\sigma_{n+1} = \sigma_n|\sigma_n) = q(\sigma_{n+1} = -\sigma_n|-\sigma_n) = p$), the dual process of $x_n$, that we denote $y_n$, is naturally defined by the same recursive relation (see Eq. (14.2.20))

$$y_n = y_{n-1} + \sigma_n, \tag{15.5.4}$$

with $y_0 = L+1$. In this case, we consider hard walls at sites 1 and $L+1$, i.e. if the particle is on site 1 and jumps to the left it stays at 1, and similarly at the other end of the lattice (as explained in [3], the hard wall at $a$ for the dual is shifted by 1 compared to the absorbing wall of the original process in the case of lattice random walks). Here we are interested in the cumulative of the distribution of positions of $y_n$. Let us denote $p_i^\pm(n)$ the probability that the particle is on site $i$ at time $n$ and that the previous jump was $\pm 1$. The cumulative distribution reads

$$\tilde{\Phi}_i^\pm(n) = \sum_{j=1}^i P(j, n|\pm) = 2 \sum_{j=1}^i p_j^\pm(n) \tag{15.5.5}$$

where $P(i, n|\pm)$ is the probability that the particle is on site $i$ at time $n$ given that it has sign $\pm$ (we indeed have $P(\pm)P(i,n|\pm) = p_i^\pm(n)$, with $P(\pm) = 1/2$, the probability that $\sigma = \pm 1$). Then, according to (14.2.25), one has

$$\boxed{E_i^\pm(n) = \tilde{\Phi}_i^\mp(n)}. \tag{15.5.6}$$

We have computed these two quantities at different times using transfer matrices and we show the results in Figure 15.3. There is indeed a perfect overlap. The details of the numerical computations are presented in [3]. There we also compute both quantities in the stationary state. The stationary solution for the exit probability is (with $E_i^\pm = E_i^\pm(n \to +\infty)$ and similarly for $p_i^\pm$ below)

$$E_i^+ = \frac{1-p}{(1-p)L+p}(i-1) + \frac{1}{(1-p)L+p},$$
$$E_i^- = \frac{1-p}{(1-p)L+p}(i-1), \tag{15.5.7}$$

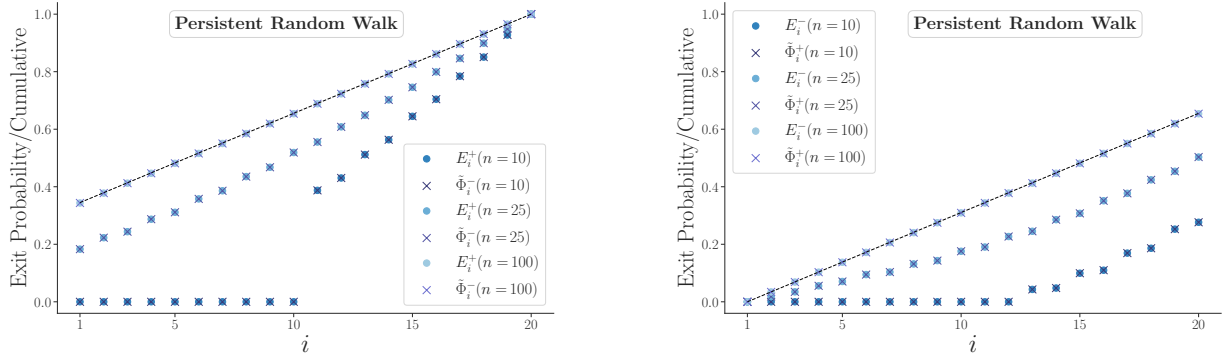

**Figure 15.3:** **Left**: For a persistent random walk, we compute numerically the exit probability $E_i^+(n)$ and the cumulative distribution with hard walls $\tilde{\Phi}_i^-(n)$ using transfer matrices, as described in [3], for $p = 0.9$ and $L = 20$. The dashed black line corresponds to the analytical result for the stationary state (15.5.7) and (15.5.8). **Right**: Same plot for $E_i^-(n)$ and $\tilde{\Phi}_i^+(n)$.

for $1 \leq i \leq L$. The stationary distribution of positions of the dual process $y_n$ is given by

$$p_i^+ = p_i^- = \frac{1}{2} \frac{1-p}{(1-p)L + p} \quad \text{for} \quad 2 \leq i \leq L,$$
$$p_1^- = p_{L+1}^+ = \frac{1}{2} \frac{1}{(1-p)L + p},$$
$$p_1^+ = p_{L+1}^- = 0, \tag{15.5.8}$$

and one can indeed verify that (15.5.6) holds in the stationary state.

# Chapter 16

# Duality for Stochastic Resetting

A particular class of stochastic processes has generated a lot of interest in recent years: stochastic processes under resetting [35,54,152], which we introduced in Chapter 3. The resetting procedure is the following: a random process $x(t)$ starts its dynamics at $x(0)$ and evolves, e.g. through the Langevin dynamics (14.2.1), but at random times, it is reset at a given position $X_r$ where it restarts its dynamics. A celebrated example is the resetting Brownian motion (rBm) where the dynamics is a simple Brownian motion, and the random times are exponentially distributed.

The duality presented in this chapter can be extended to processes undergoing resetting. The construction of the dual process is intuitive when considering non-instantaneous resetting [165,166,169]. Instead of teleporting the particle to $X_r$ at random times, one switches on a force for a finite duration $\Delta t$ that attracts the particle toward $X_r$, for example, $f(x) = -\mu(x - X_r)$. For the dual procedure, we expect this force to be reversed, i.e. $-f(x)$, and it will now drive the particle away from $X_r$. Taking the double limit $\mu \to \infty$ and $\Delta t \to 0$ gives the dual procedure of instantaneous resetting: the dual of a process $x(t)$ with instantaneous resetting to $X_r$ and absorbing walls at $a$ and $b$ is a process $y(t)$ which is reset at random times either to $a$ if $y(t) < X_r$, or to $b$ if $y(t) > X_r$, and where walls are now hard walls (of course $X_r \in [a,b]$). For a schematic description of the dual procedure, see Figure 16.1.

This duality transformation for the resetting events is actually independent of the boundary conditions. Indeed, a process $y(t)$ which resets at $X_r$ with fixed rate $r$, and hard walls located at $a$ and $b$, has a dual process $x(t)$ with absorbing walls at $a$ and $b$ which resets at $a$ if $x(t) < X_r$, or at $b$ if $x(t) > X_r$. Therefore, it is absorbed either at $a$ or $b$ after the first reset.

In this section we show that the main results of this paper still hold when processes are subjected to stochastic resetting if the dual is constructed as we explained above. We start by illustrating the duality with the resetting Brownian motion, for which explicit expressions can even be obtained in the stationary state. We then briefly explain how this duality can be extended to discrete time random walks.

## 16.1 Duality for the Resetting Brownian Motion

We consider a resetting Brownian motion in the presence of a force $F(x)$. With a fixed rate $r$, it restarts its dynamics at position $X_r$, such that it evolves through

$$x(t+dt) = \begin{cases} x(t) + \left(F(x(t)) + \sqrt{2T}\,\xi(t)\right) dt & \text{with probability } (1 - r\,dt) \\ X_r & \text{with probability } r\,dt \end{cases}, \quad (16.1.1)$$

where $T$ is a diffusion constant, and $\xi(t)$ is a Gaussian white noise with zero mean and correlation function given by $\langle \xi(t)\xi(t') \rangle = \delta(t - t')$. In the presence of two absorbing walls located at $a$ and $b$, we want to compute the exit probability at wall $b$ and time $t$ of the rBm. We can write the backward Fokker-Planck equation by considering the evolution of the process between times $0$ and $dt$. Starting from $x$, the exit probability at time $t + dt$ reads[22]

$$E_b^r(x, t+dt, X_r) = (1 - rdt)\, \mathbb{E}_\xi \left[ E_b^r\left( x + \left( F(x) + \sqrt{2T}\, \xi(t) \right) dt, t, X_r \right) \right] + rdt\, E_b^r(X_r, t, X_r), \tag{16.1.2}$$

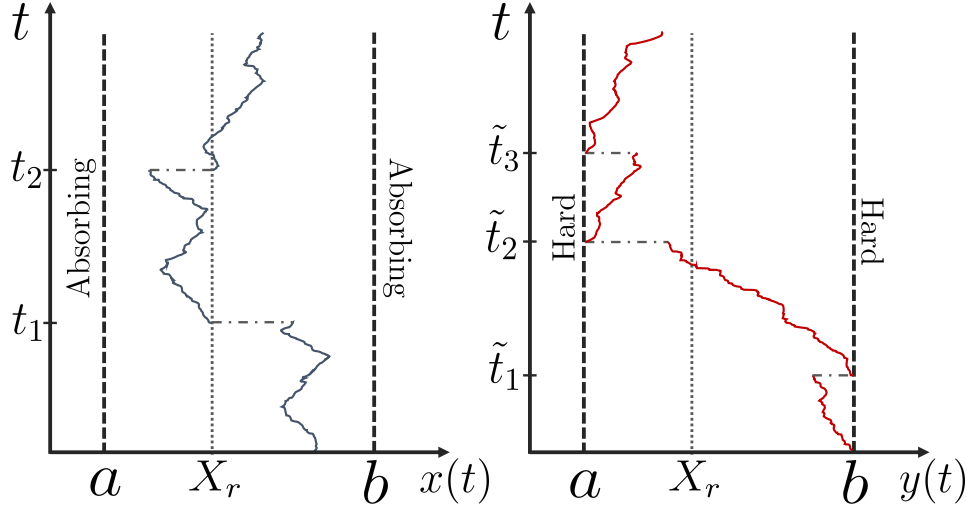

**Figure 16.1: Left**: Schematic of a trajectory of the process $x(t)$. At random times $t_i$ ($i \in \{1,2,3,...\}$), $x(t)$ restarts its dynamics at $X_r$. **Right:** On the other hand, at random times $\tilde{t}_i$ (with the same distribution as the $t_i$'s), the dual process of $x(t)$ that we denote $y(t)$, restarts its motion at $a$ if $y(t) < X_r$ just before the resetting, or at $b$ if $y(t) > X_r$.

The first term comes from the fact that with probability $1 - rdt$ there is no reset, and the process diffuses from $0$ to $dt$, and one has to average over the noise. When a reset happens, with probability $rdt$, the particle restarts its dynamics at $X_r$ at time $dt$. Expanding the first term on the right hand-side and averaging over $\xi(t)$ gives the backward Fokker-Planck equation corresponding to (16.1.1)

$$\partial_t E_b^r(x, t, X_r) = T\, \partial_x^2 E_b^r(x, t, X_r) + F(x)\, \partial_x E_b^r(x, t, X_r) - r E_b^r(x, t, X_r) + r E_b^r(X_r, t, X_r). \tag{16.1.3}$$

The boundary conditions are $E_b^r(a, t, X_r) = 0$, and $E_b^r(b, t, X_r) = 1$, while the initial condition reads $E_b^r(x, t=0, X_r) = 0$ for $x < b$, and $E_b^r(b, t=0, X_r) = 1$.

On the other hand, the dual of $x(t)$ is denoted $y(t)$ and evolves through

$$y(t+dt) = \begin{cases} y(t) + \left( \tilde{F}(y(t)) + \sqrt{2T}\, \xi(t) \right) dt & \text{with probability } (1 - r\, dt) \\ a & \text{with probability } r\, dt \text{ if } y(t) < X_r \\ b & \text{with probability } r\, dt \text{ if } y(t) > X_r \end{cases}, \tag{16.1.4}$$

and $y(0) = b$. We do not consider the case $y(t) = X_r$ in Eq. (16.1.4) as it has zero measure in continuous space. For discrete time random walks, one needs to be more careful (see Sec. 16.2). Let us write $\tilde{p}(y, t)$ the probability distribution of the position of the process (16.1.4). It obeys the following forward Fokker-Planck equation

$$\partial_t \tilde{p}(y, t) = T\, \partial_y^2 \tilde{p}(y, t) - \partial_y \left( \tilde{F}(y) \tilde{p}(y, t) \right) - r \tilde{p}(y, t)$$
$$+ r \left[ \int_a^{X_r} dz\, \tilde{p}(z, t) \right] \delta(y - a) + r \left[ \int_{X_r}^b dz\, \tilde{p}(z, t) \right] \delta(y - b). \tag{16.1.5}$$

---

[22] In this section we write the resetting rate as an exponent and the resetting position as an additional argument in the functions which depend on these parameters.

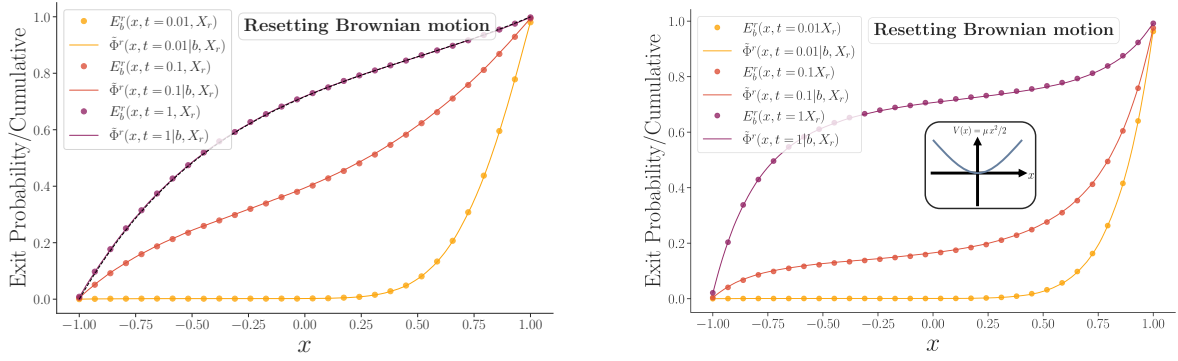

**Figure 16.2:** We verify the duality relation (14.2.13) for a Brownian motion subjected to Poissonian resetting. **Left**: Comparison of simulation results for the exit probability (dots) and cumulative distribution (lines) of the resetting Brownian motion without external potential, with $T = 4$, $r = 10$ and $X_r = 1.5$, at different times. The dashed black curve corresponds to the analytic prediction for the stationary state (16.1.8). **Right**: Same plot in the presence of a harmonic potential $V(x) = \frac{\mu}{2}x^2$ with $\mu = 20$ for $E_b$, and $-V(x)$ for $\tilde{\Phi}$. In both cases the overlap is perfect.

Now we want to relate the Fokker-Planck equation (16.1.3) on the exit probability to the one satisfied by the cumulative distribution of the dual process (the probability to find the dual particle in $[a, x]$ at time $t$), defined as

$$\tilde{\Phi}^r(x, t | b, X_r) = \int_{a^-}^x dy\, \tilde{p}(y, t)\,. \tag{16.1.6}$$

Integrating (16.1.5) between $a^-$ and $x$ leads to

$$\partial_t \tilde{\Phi}^r(x,t|b,X_r) = T\,\partial_x^2 \tilde{\Phi}^r(x,t|b,X_r) - \tilde{F}(x)\,\partial_x \tilde{\Phi}^r(x,t|b,X_r) - r\,\tilde{\Phi}^r(x,t|b,X_r) + r\,\tilde{\Phi}^r(X_r,t|b,X_r)\,, \tag{16.1.7}$$

which is exactly (16.1.3) if $\tilde{F}(x) = -F(x)$. Finally, we have $\tilde{\Phi}^r(a,t|b,X_r) = 0$, $\tilde{\Phi}^r(b,t|b,X_r) = 1$ and the initial conditions $\tilde{\Phi}^r(x,t=0|b,X_r) = 0$ for $x < b$ and $\tilde{\Phi}^r(b,t=0|b,X_r) = 1$. We thus once again obtain that the two quantities $E_b^r(x,t,X_r)$ and $\tilde{\Phi}^r(x,t|b,X_r)$ satisfy the same equations with the same boundary and initial conditions, and thus the identity (14.2.13) holds.

When $F(x) = 0$, it is possible to write explicitly the solution to (16.1.3)-(16.1.7) in the stationary state (see [167]). It reads, with $\alpha_r = \sqrt{r/T}$,

$$\boxed{E_b^r(x, t \to \infty, X_r) = \tilde{\Phi}^r(x, t \to \infty | b, X_r) = \frac{\sinh(\alpha_r(x - X_r)) + \sinh(\alpha_r(X_r - a))}{\sinh(\alpha_r(b - X_r)) + \sinh(\alpha_r(X_r - a))}}\,. \tag{16.1.8}$$

In Figure 16.2, we validate our statements with simulations.

In [3], we also consider a rBm with hard walls and resetting position $X_r$. We construct its dual which is a Brownian motion with absorbing walls which resets at $a$ if $x(t) < X_r$ and at $b$ if $x(t) > X_r$. Note that in this case, the particle is always absorbed at one of the two walls after the first resetting event.

## 16.2 Random Walks Subjected to Resetting

Resetting is not restricted to continuous-time stochastic processes. It can also be introduced for discrete time random walks [35, 155]. We consider the dynamics (14.2.18) and introduce an additional random variable $R_n$ such that at each time $n$, $R_n = 1$ with some probability $0 \leq r \leq 1$ and $R_n = 0$ otherwise, independently of the past history. Whenever $R_n = 1$, the process $X_n$ resets to some value $X_r \in [a, b]$. More precisely, $X_n$ now follows the Eq. (14.2.18) between times $n-1$ and $n$ if $R_n = 0$, but if $R_n = 1$, then $X_n = X_r$ independently of its previous value (unless it has already been absorbed at $X_n = a^-$ or $b$, in which case it remains there). As for continuous stochastic processes, one could allow for independent resettings at different points $X_r^i$ with different rates $r^i$, but we will restrict to $i = 1$ for simplicity.

We now define the dual process $Y_n$ as follows: let $\tilde{R}_n$ be a process with the same law as $R_n$. If $\tilde{R}_n = 0$, $Y_n$ is given by (14.2.20), but instead if $\tilde{R}_n = 1$, one has:

$$Y_n = \begin{cases} a & \text{if } Y_{n-1} \leq X_r, \\ b & \text{if } Y_{n-1} > X_r. \end{cases} \tag{16.2.1}$$

Note the asymmetry in the definition of the resetting events for $Y_n$. This is connected to the shift of the boundary condition from $a$ to $a^-$ in $X_n$. If one is interested in the exit probability at $a$, $E_a(x, n)$, then one should instead set $Y_n = a$ when $Y_{n-1} = X_r$. Again, this precision becomes irrelevant if the process is such that the event $Y_n = X_r$ has zero measure.

Finally, given a trajectory of the process $X_n$ and a value of $Y_0$, the dual trajectory is the realization of $Y_n$ with $\tilde{W}_n = W_{T+1-n}$ and $\tilde{R}_n = R_{T+1-n}$ for all $i$.

One can check that, with these definitions the duality (14.2.25) remains valid [3]. Since the $R_n$'s are independent, the hypothesis (14.2.19) is obviously still satisfied when considering the joint probability of $W_n$ and $R_n$.

## Chapter 17

# Numerical Results: from Active Particles to Fractional Brownian Motion

The duality relation (14.2.9), derived in [3], extends to a broad spectrum of continuous stochastic processes, including the most well-known models of active particles and random diffusivity models. In Chapter 15 and in Figure 15.2, we have already checked (14.2.9) for run-and-tumble particles at finite time for which $\boldsymbol{\theta}(t)$ takes discrete value. In this chapter, we consider several examples of frequently studied continuous processes, and show numerical evidence that this relation holds in these cases. For these examples, $\boldsymbol{\theta}(t)$ takes continuous values , and it is more convenient to verify Eq. (14.2.13) rather than (14.2.9), which we do. To this end, we performed direct simulations of the stochastic differential equation for each model and computed the two quantities of interest as follows:

- (i) For given initial values of $x$ and $\boldsymbol{\theta}$, we compute $E_b(x, \boldsymbol{\theta}, t)$ by simulating independently the trajectories of $N$ particles over a time interval $t$ and counting how many of them escape at $b$.

- (ii) For the cumulative distribution, we simply simulate independently the trajectories of $N$ particles over a time interval $t$, starting from $x = b$ and with the initial value of $\boldsymbol{\theta}$ drawn from $p_{eq}(\boldsymbol{\theta})$, and compute a histogram of positions at time $t$.

In both cases, we used $N \sim 10^5 - 10^7$ depending on the model[23].

The first three models considered in this section are well-known models of active particles, which can all be written under the form (14.2.15) (i.e. $\boldsymbol{\theta}$ plays the role of a driving velocity), and for which the dual process and duality relation take the forms (14.2.16) and (14.2.17) respectively. We then consider a diffusing diffusivity model, for which the parameter $\boldsymbol{\theta}$ now modifies the temperature of the process $x(t)$. We also consider two simple cases with a space-dependent temperature and a time-dependent external force.

Finally, we test the duality relation for fractional Brownian motion and for a process subjected to resetting noise. Although these processes do not satisfies the hypothesis under which we derived the results of Section 14.2.1, our numerical simulations suggest that Siegmund duality appears to hold in these cases as well.

---

[23] If we only wanted to compute the cumulative distribution in the stationary state, it would be more efficient to perform the simulation for a single particle and compute the histogram of positions over time once it has reached stationarity, as in [4], assuming the process is ergodic.

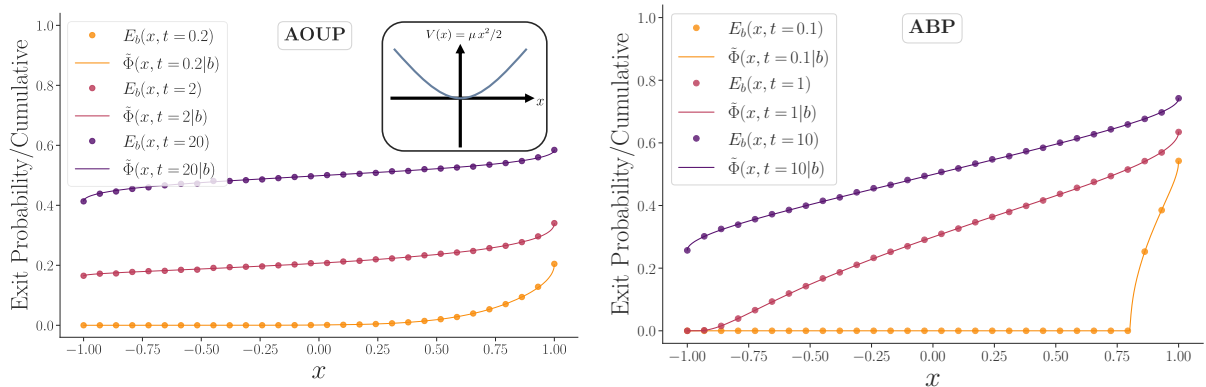

**Figure 17.1:** We validate the duality relation (14.2.13) numerically considering two models of active particles at finite times. The dots represent the exit probability, while the solid lines show the empirical cumulative distribution of the dual process, both being computed numerically through averages over many trajectories. **Left**: active Ornstein-Uhlenbeck particle (AOUP) with $D = 4$, $\tau = 1$, in the presence of a harmonic potential with $\mu = 1$ for $E_b$ and $-V(x)$ for $\tilde{\Phi}$. **Right**: active Brownian particle (ABP) with velocity $v_0 = 2$ and diffusion coefficient $D = 1$, in the absence of external potential. In each case, the exit probability overlaps the empirical cumulative of the dual process perfectly. The discontinuities observed in the yellow curve can be attributed to the fact that particles starting on the left side of these discontinuities do not have enough time to exit the interval, even if they remain in the positive state throughout the entire simulation.

## 17.1 Active Orstein-Uhlenbeck Particle

We first consider an active Orstein-Uhlenbeck particle [227, 228], for which the equation of motion is

$$\frac{dx}{dt} = F(x) + v(t) \quad , \quad \tau \frac{dv}{dt} = -v(t) + \sqrt{2D}\,\eta(t)\,, \tag{17.1.1}$$

where $F(x) = -V'(x)$ is an external force, $\tau$ is the persistence time, $D$ is a diffusion coefficient, and $\eta$ a Gaussian white noise. For the dual, the velocity $v$ should again be initialised in the equilibrium distribution, now given by

$$p_{eq}(v) = \sqrt{\frac{\tau}{2\pi D}}\,e^{-\frac{\tau v^2}{2D}}\,. \tag{17.1.2}$$

The central panel of Fig. 17.1 shows a perfect agreement between the two quantities, for an AOUP in a harmonic potential.

## 17.2 2d Active Brownian Particle Between Two Parallel Walls

Another common model of active particles is the active Brownian particle [50, 223]. It can only be defined in two dimensions or higher (in 1d it coincides with the RTP). Here we consider the 2d case in the particular situation where it can effectively be described in 1d. More precisely, we assume that the particle is confined between two parallel walls perpendicular to the $x$-direction, at $x = a$ and $x = b$, and that the external force in the $x$-direction is independent of the second coordinate $z$, so that the evolution along the $x$-axis is completely independent of the $z$ coordinate,

$$\frac{dx}{dt} = F_x(x) + v_0 \cos\varphi(t) \quad , \quad \frac{dz}{dt} = F_z(x,z) + v_0 \sin\varphi(t) \quad , \quad \frac{d\varphi}{dt} = \sqrt{2D}\,\eta(t), \tag{17.2.1}$$

where $F_x(x)$ and $F_y(x,z)$ are the projections of the external force, $v_0$ is a constant speed, $D$ is again a diffusion coefficient, and $\eta(t)$ a Gaussian white noise. Forgetting about the $z$ coordinate, this problem is very similar to the AOUP case. The equilibrium distribution of $\varphi(t)$ is simply the uniform distribution on $[0, 2\pi)$. The right panel of Figure 17.1 illustrates Eq. (14.2.13) in this case.

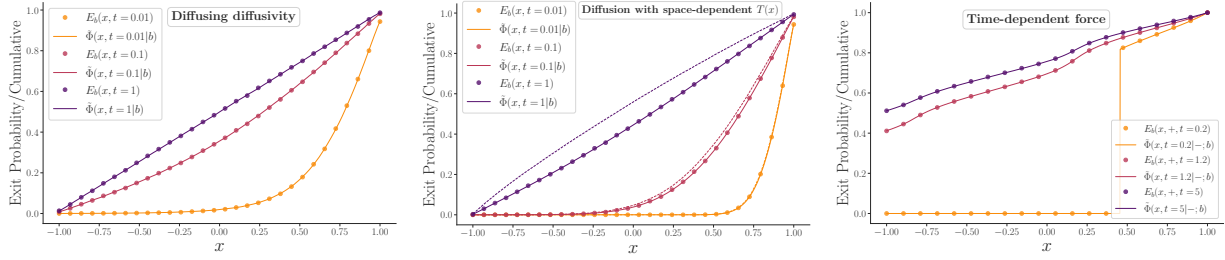

**Figure 17.2:** Here, we show numerically the validity of Eq. (14.2.13) and (14.2.17) at finite time in three different cases. Dots show the value of the exit probability while the solid lines are for the cumulative distribution for the dual process. **Left**: diffusing diffusivity model with diffusion coefficient $D = 4$ and relaxation time $\tau = 1$. **Center**: Brownian particle with space-dependent diffusion coefficient $T(x) = 1 + \frac{x}{2}$ (and no external potential). To compensate the variations in temperature, the dual process is subjected to a constant force $\tilde{f} = \partial_x T = \frac{1}{2}$. The dashed line shows the cumulative distribution in the absence of this force. **Right**: RTP initialized in the $+$ state, subjected to a time dependent force $f(t) = a \sin(2\pi \frac{t}{\tau})$. The dual process is in the presence of hard walls, with a force $\tilde{f}(t) = -a \sin(2\pi \frac{t_f - t}{\tau})$ where $t_f$ is the time at which we measure the exit probability. The cumulative distribution is conditioned on $\sigma(t) = -1$. Here $a = 1$, $\tau = 0.5$, $v_0 = 2$ and $\gamma = 1$.

## 17.3 Diffusing Diffusivity

We now consider a family of models which is not connected to active particles, but which has also attracted a lot of attention in recent years: diffusing diffusivity models [111, 117, 129, 130]. One example is a Brownian particle where the diffusion coefficient is itself the square norm of an Ornstein-Uhlenbeck process in arbitrary dimension $d$,

$$\frac{dx}{dt} = F(x) + \sqrt{2T(t)}\,\xi(t) \quad,\quad T(t) = \boldsymbol{\theta}^2(t) \quad,\quad \tau \frac{d\boldsymbol{\theta}}{dt} = -\boldsymbol{\theta}(t) + \sqrt{2D}\,\boldsymbol{\eta}(t)\,, \qquad (17.3.1)$$

where $F(x)$ is the external force, $D$ is a diffusion coefficient, and $\xi$ and $\eta$ are Gaussian white noises. For the simulations, we restrict ourselves to the simplest case $d = 1$, although the results are valid for any $d$. For $d = 1$, the equilibrium distribution of $\theta$ is the same as for $v$ in the AOUP case, i.e. (17.1.2). In the left panel of Fig. 17.2 we confirm the validity of the relation (14.2.13).

Note also that, according to the general definition (14.2.1) of $x(t)$, switching diffusion models are also included in this framework. Indeed, for the one-dimensional case ($d = 1$), by setting $f(x(t), \theta(t)) = 0$ and $\mathcal{T}(x(t), \theta(t)) = \theta(t)$, where $\theta(t)$ jumps discretely at random times and takes random values drawn from a given distribution, we recover the simple switching diffusion model introduced in Section 2.4.

## 17.4 Multiplicative Noise

In the central panel of Fig. 17.2 we illustrate the effect of a multiplicative noise on the duality relation with a simple example. We consider a Brownian particle, with no external force, but with a space-dependent diffusion coefficient $T(x) = 1 + \frac{x}{2}$. For the dual, this means that we need to add a constant force $\tilde{f} = \partial_x T = \frac{1}{2}$. In the infinite time limit, the exit probability is $E_b^{st}(x) = \frac{x-a}{b-a} = \frac{1}{2}(x + 1)$ (when $a = -b = 1$ as in Fig. 17.2), exactly like for a constant $T$. This is true for any $T(x)$ as long as there is no external force, since the stationary equation is $T(x)\partial_x^2 E_b^{st} = 0$, leading to a linear solution independent of $T(x)$. On the other hand, the equilibrium probability density with hard walls is the solution of $0 = -\tilde{f}(x)P_{\text{eq}} + \partial_x[T(x)P_{\text{eq}}]$. In the absence of $\tilde{f}$, the solution is $P_{\text{eq}}(x) = \frac{1}{T(x)}/\int_a^b \frac{1}{T(x)}$, i.e. the particle density is higher where the diffusion coefficient is smaller. For $\tilde{f} = \partial_x T$, the equation simplifies to $\partial_x P_{\text{eq}} = 0$, leading to a constant density compatible with (14.2.10).

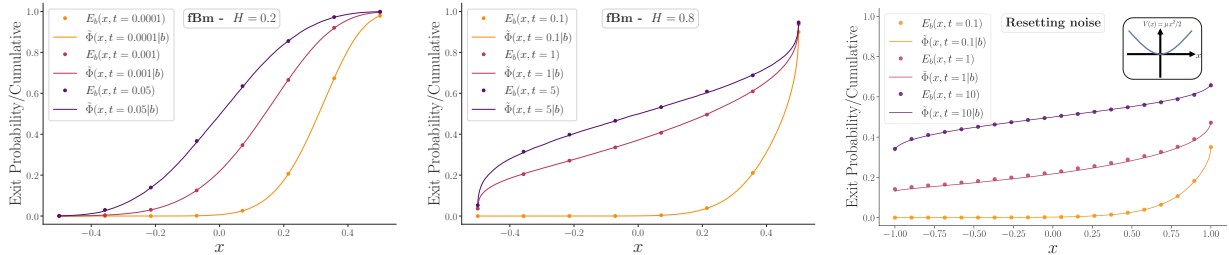

**Figure 17.3: Left and central panels**: We present simulation results indicating that the duality (14.2.13) seems to remain valid even for processes with non-Markovian noise (but stationary increments) such as the fractional Brownian motion. The dots show the value of the exit probability while the solid lines are the cumulative distributions. Both quantities are evaluated numerically by averaging over trajectories. We have tested the duality at different times for Hurst parameters $H = 0.2$ on the left plot, and $H = 0.8$ on the central one. **Right**: Same plot for a process subjected to a resetting noise in a harmonic potential $V(x) = \frac{\mu}{2}x^2$ with $\mu = 1$, $r = 1$ and $T = 4$. This model does not satisfy the detailed balance hypothesis, but the duality (14.2.13) seems to hold nonetheless.

## 17.5 Time-Dependent Force

We consider an RTP subjected to a force $f(t) = a\sin(2\pi\frac{t}{\tau})$. The duality relation (14.2.17) still applies, but now the dual process is subjected to a force $\tilde{f}(t) = -a\sin(2\pi\frac{t_f-t}{\tau})$ which depends on the time $t_f$ at which the exit probability is measured. In terms of numerical simulations, this means that if we want to obtain the exit probability at different times from the cumulative of the dual process, we need to run a different simulation of the dual for each time. Taking this into account, the left panel of Fig. 17.2 shows that (14.2.17) holds once again.

## 17.6 How General is this Duality?

In the derivation of the duality for continuous stochastic processes (see Section 14.2.1 and [3]), we have made the crucial assumptions that $x(t)$, whose dynamics is given by (14.2.1), is driven by an equilibrium Markov process $\boldsymbol{\theta}(t)$ that satisfies the detailed balance conditions (14.2.3)-(14.2.4). This way, it was possible to write the Fokker-Planck equations and derive the duality relation (14.2.9). However, for the discrete case in Sec. 14.2.3, we have used assumptions that are a bit weaker in that regard: the Markovianity and detailed balance assumptions are replaced by a simple time-reversal property (14.2.19). This may suggest that this duality relation is a bit more general in the case of continuous stochastic processes as well. In this section, we consider two processes that do not satisfy these assumptions, but for which the duality still seems to hold.

### 17.6.1 Fractional Brownian Motion

We first consider a fractional Brownian motion $x(t)$, i.e. a continuous-time centered Gaussian process with two-time correlations function [86]

$$\langle x(t_1)x(t_2)\rangle = t_1^{2H} + t_2^{2H} - |t_1 - t_2|^{2H}, \tag{17.6.1}$$

where $0 < H < 1$ the Hurst parameter. The fBm increments are generated by a fractional Gaussian noise (fGn) $\xi_f(t)$ which is a stationary non-Markovian process with correlations

$$\langle \xi_f(t_1)\xi_f(t_2)\rangle = 2H(2H-1)|t_1-t_2|^{2H-2}. \tag{17.6.2}$$

Due to the non-Markovian property of the fGn, it is not possible to write the standard Fokker-Planck equation as done in [3] for the proof of (14.2.13). The fBm breaks the assumptions we have made, but simulations show that the duality still holds at any time $t$. In Figure 17.3, we validate numerically the duality for $H = 0.2$, and $H = 0.8$.

### 17.6.2 Resetting Noise

The last example is a process $x(t)$ driven by a resetting noise as considered in [6]. To be more specific, $x(t)$ evolves in a harmonic trap of strength $\mu$ and is subjected to a resetting Brownian motion $y_r(t)$ such that it follows the Langevin dynamics

$$\frac{dx(t)}{dt} = -\mu\, x(t) + r\, y_r(t), \tag{17.6.3}$$

and the rBm evolves through

$$y_r(t+dt) = \begin{cases} 0 & \text{with probability } r\, dt \\ y_r(t) + \sqrt{2T}\,\xi(t)\, dt & \text{with probability } (1-r\, dt) \end{cases}, \tag{17.6.4}$$

where $\xi(t)$ is a Gaussian noise with zero mean and two times correlation function $\langle \xi(t)\xi(t')\rangle = \delta(t-t')$. The driven process $y_r(t)$ is Markovian but it violates detailed balance due to the resettings. Indeed, while it is possible for $y_r(t)$ to jump from $x \neq 0$ to the origin, the reversed event has zero probability. Thus, this process seems to violate our hypotheses more strongly than the fBm, since the driving process $\boldsymbol{\theta}(t) = y_r(t)$ is intrinsically irreversible. In the right panel of Figure 17.3, we have tested the duality relation (14.2.13) for this model. Although there seems to be a small discrepancy at intermediate times, the agreement is still quite surprising.

## 17.7 Conclusion

In this last part of the thesis, we have shed light on a relation between the exit probability from an interval of a stochastic process driven by a stationary noise, and the position distribution of its dual in the presence of hard walls. This duality holds in both continuous and discrete time, as shown in [3], and it establishes a link between two fundamental aspects of stochastic processes: first-passage properties and spatial behavior. While such duality relations between absorbing and reflecting boundary conditions have a long history in mathematics, to the best of our knowledge, it is the first time Siegmund duality is constructed in the context of physical models like those studied here, such as active particles, random diffusion models or stochastic resetting.

Although this framework includes most of the models studied by physicists, it does not include in theory the fractional Brownian motion or a process driven by a resetting noise that breaks the detailed balance assumption. However, our numerical simulations indicate that the duality is very likely to hold also in these cases, at least approximately. Additionally, a notable assumption in our framework is that the driven noise should have a stationary distribution. As a result, constructing the duality for processes like the random acceleration process, where the acceleration is a Gaussian white noise, remains an open challenge.

Another open question, of particular relevance for applications, is whether a similar connection can be formulated in higher dimensions. Such a duality in multi-dimension has been formulated in an abstract way in other contexts [376], but expressing this duality in a way which is useful to practical computations even for simple models seems much more challenging than in 1d. Finally, a recent work proposed the construction of the Siegmund dual of an $N$-particle system [396]. The study of $N$-particle systems is very relevant when considering active particles where collective motion emerges. In this respect, we have derived a new result for a system of $N$ independent particles with applications to extreme value statistics – see Section 14.2.5.

Understanding both the spatial behavior and first-passage properties of stochastic processes is crucial, particularly in the context of active particles. The duality presented here provides an elegant means of linking these aspects together. Computing both the exit probability and the cumulative distribution of the dual is challenging. However, the duality allows us to easily derive one from the other, offering a significant analytical advantage. Furthermore, measuring the splitting probability of a stochastic process via simulations or experiments requires many trajectories. In contrast, studying the dual process, if it is ergodic, would involve observing only one long trajectory, thereby simplifying the numerical evaluation of the splitting probability. We hope that this new connection will be useful to the large community of people working in these fields, and pave the way to new results, both analytical and numerical.

# Conclusion and Perspectives

In this thesis we have investigated a wide variety of stochastic processes, ranging from classical Brownian motion to active particle models, with a unifying focus on their non-equilibrium dynamics and first-passage properties. While Part I provided a review of the foundations of continuous-time stochastic processes, the remaining parts present original contributions. Below, we summarize the key insights from these chapters and outline several promising directions for future research inspired by our findings.

**Part II – Steady-State Distributions with Resetting Noise.** In the second part of this thesis, we introduced a method based on *Kesten variables* to derive an integral equation for the steady state of a harmonically confined stochastic process driven by colored noise with a renewal/resetting structure. This approach allows to handle both periodic and Poissonian resetting of the noise. We used it to obtain several exact results, including the non-equilibrium steady state of a generalized RTP, and in Part III, the exact NESS of a switching diffusion process in a harmonic trap.

The approach developed here based on Kesten variables is quite appealing and it would be nice to extend it in various directions. A first class of models where this approach could be useful are the "switching potential", studied for instance in Refs. [160, 169, 397]. In these models, a single Brownian particle is subjected to an external confining potential that is switched on and off stochastically, as in the recent experiments on resetting using optical tweezers [165, 166]. In fact, a similar approach has recently been applied to systems of multiple particles confined in switching harmonic traps [183].

One may also wonder whether this approach can be extended to two or several RTPs. A possible starting point could be the two RTPs model studied in [398]. It would also be interesting to use this framework to study the recently introduced active Dyson Brownian motion [206]. Similarly, it is natural to ask whether this approach can be extended to study models of active particles in two and higher dimensions, for which there exists very few analytical results. One could also study different resetting protocols acting on the noise term. Here we mainly focused on the periodic/sharp and Poissonian protocols but we could consider protocols which are more realistic from the experimental point of view [165]. In such protocols, the resetting is not instantaneous and it will be interesting to see how this feature would modify the results presented here. Also, one can wonder whether this Kesten variables approach can be used to obtain information about the large deviation form of the stationary distribution, a question that has recently attracted some attention [274, 399–401]. Finally, it would be interesting to extend this Kesten approach to resetting processes with non-Poissonian time distributions.

**Part III – Switching Diffusion and Large Deviations.** In this part, we investigated *switching diffusion*, where the diffusion coefficient $D(t)$ of a diffusing particle evolves as a stochastic process that randomly switches between values drawn from a distribution $W(D)$. This model, which belongs to the class of random diffusivity models, captures the behavior of particles in disordered or heterogeneous environments where diffusivity varies in time but leads to an average MSD growing linearly with time, i.e., $\langle x^2(t) \rangle \sim 2 \langle D \rangle t$, while the position distribution remains non-Gaussian. We derived exact expressions for all moments $\langle x^{2n}(t) \rangle$, and showed that, in the long-time limit, the cumulants $\langle x^{2n}(t)\rangle_c$ grow linearly in time and are proportional to the free cumulants associated with the distribution $W(D)$. This surprising connection to free cumulants, which has also recently observed in a model of random multiplicative growth [318], provides a new mathematical lens through which such processes could be understood. Furthermore, by using tools from large deviation theory, we computed the rate function and scaled cumulant-generating function of the PDF of the position, identifying dynamical transitions from Gaussian to non-Gaussian fluctuations induced by the switching dynamics.

We also extended our study to include a harmonic confining potential, for which we derived the exact stationary distribution of the position of the particle for arbitrary $W(D)$. Remarkably, the link to free cumulants persists in this setting in the limit $\beta \to +\infty$ where the stationary distribution becomes Gaussian with variance $\propto \langle D \rangle$.

Our work opens several natural extensions. A key question is the generalization to higher dimensions ($N > 1$) or, equivalently to $N$ particles subjected to simultaneous switching dynamics. This direction could build upon recent studies in the context of simultaneous resetting [182, 183]. Similar questions were recently studied for $N$ particles in a harmonic trap in the presence of switching stiffnesses [402] and switching centers [174, 403]. It would be interesting to investigate whether the connection to free cumulants still holds in such interacting or higher-dimensional settings.

While we have shown that the connection to free cumulants persists in the harmonic case, it remains an open question whether similar structures arise under other forms of confinement, such as a linear potential. As in both situations we considered – with and without harmonic confinement – the connection to free cumulants arises in the large-time or large-$\beta$ limit, when the stationary distribution converges to a Gaussian with variance proportional to the average diffusivity $\langle D \rangle$. It would therefore be interesting to study a case where this Gaussian behavior does not emerge, and investigate whether the connection to free cumulants still persists.

Finally, it would be interesting to study a more realistic model in which the random switching times are correlated with the random diffusion coefficient, as is sometimes done to fit experimental data [133].

**Part IV – First-Passage Times for Run-and-Tumble Particles within a Potential.**
In Part IV, we focused on the first-passage properties of run-and-tumble particles, in particular the mean first-passage time for an RTP moving in an arbitrary external potential. We developed a general method to compute this quantity exactly. Depending on the nature of the turning points in the dynamics, we identified four distinct dynamical phases, each associated with a specific analytical expression for the MFPT. For arbitrary force profiles involving multiple turning points, the full MFPT can be obtained by appropriately gluing together the solutions from each phase. We further demonstrated that for a wide class of confining potentials, the MFPT can be optimized with respect to the tumbling rate, revealing conditions under which the search process is most efficient. Finally, we outlined potential applications of this approach, including the calculation of Kramers' escape rate for RTPs.

The present work can be extended in several interesting directions. For instance, although run times were traditionally believed to follow an exponential distribution (see e.g., [262]), more recent measurements suggest that a power-law distribution may be more accurate in some cases [264]. In this paper, we considered run times that are exponentially distributed, which is commonly assumed in RTP models and is considered a reasonable approximation based on experimental data [263]. However, it would be interesting to compute the MFPT for other distributions of run times. Another interesting generalization would be to consider a space-dependent tumbling rate $\gamma(x)$ [231, 404]. Following our derivations, it should be possible to write an explicit expression for this specific case.

It would also be natural to consider the effect of adding a white noise in addition to the telegraphic process. It would be particularly interesting to investigate how this additional noise modifies the mean first-passage time and whether it introduces qualitatively new behaviors com-pared to the pure run-and-tumble case. On another note, studying the first-passage properties of a system of $N$ run-and-tumble particles, potentially with interactions as in [206], would be an interesting direction for future research.

Another interesting future direction would be to study the MFPT of an RTP in higher dimensions [401, 405]. For example, in $d = 2$, it would be interesting to consider an anisotropic harmonic potential $V(x,y) = \mu_1 x^2/2 + \mu_2 y^2/2$. Here we have studied the limiting case $\mu_2 \to \infty$ where the problem becomes effectively one-dimensional. We expect our results to be valid (up to some renormalization of time scales) when $\mu_2$ is finite but $\mu_2 \gg \mu_1$, while the results for the isotropic case (when $\mu_2 = O(\mu_1)$) may differ significantly and it would be interesting to explore this, since this is also relevant for experiments in optical traps.

Finally, in this work, we have provided a detailed analysis of the first moment ($n = 1$), and a natural extension would be to compute higher-order moments ($n > 1$), which could reveal additional features of the underlying first-passage distribution.

**Part V – Siegmund Duality.** In the last part of this thesis, we have constructed the Siegmund dual for all the stochastic processes studied throughout this thesis – from simple Brownian motion to more complex active particle models. A striking and useful feature of this duality is that it provides a direct correspondence between the first-passage properties of the original process and the spatial distribution of its dual. We have shown how this relationship can be beneficial both analytically, by allowing one quantity to be inferred from the other, and numerically, as simulating the dual process can, in certain cases, provide a more efficient way to computing difficult observables. While our results are currently limited to one-dimensional systems, this framework already proves to be highly versatile. However, for broader applications – particularly in higher-dimensional systems – a natural and important next step would be to extend this duality to multiple dimensions.

A natural and promising direction to generalize the results obtained on Siegmund duality is to extend the framework to higher dimensions. However, doing so is not straightforward. One key difficulty lies in defining the appropriate geometry of confinement, that is, the multi-dimensional equivalent of absorbing or hard walls – especially in systems where the different components of the position of the particle may interact, for example, through coupling forces.

Nevertheless, a simple yet insightful approach would be to consider a multi-dimensional process where each coordinate evolves independently and satisfies a one-dimensional Siegmund duality. In this case, the particle moves inside a hypercube bounded by absorbing walls – e.g., a square in two dimensions, a cube in three. One could then compute the exit probability through a specific face of the hypercube. The dual process would evolve in the same geometric domain, but with hard walls, and the quantity of interest becomes the probability that the dual particle lies within a region determined by the Siegmund duality applied to the coordinate corresponding to the exit face. While simplified, this construction could serve as a first step toward understanding Siegmund duality in higher dimensions and might offer valuable insight.

Finally, we have obtained a result using on the duality for the extreme value statistics of $N$ independent particles. It would be interesting to investigate whether this approach can be extended to systems of particles with interactions.

**Concluding Remarks.** This thesis has explored a range of non-equilibrium stochastic processes, revealing exact results through new analytical approaches. From resetting dynamics to active particles and switching diffusions, we have demonstrated that meaningful analytical progress is possible even in complex settings by leveraging tools such as Kesten variables, large deviation theory, and Siegmund duality. Beyond the specific models studied, the methods developed here offer a foundation for future work, particularly in higher-dimensional or many-particle systems. We hope these contributions offer insights that contribute to a deeper understanding of stochastic processes far from equilibrium.

# Appendix

# A Overdamped Langevin Equation

We study here the role of inertia in Eq. (1.2.4) at different timescales. We introduce the characteristic timescale $\tau = m/\gamma$, where $m$ is the mass of the particle and $\gamma$ is the drag coefficient, which represents the viscous resistance exerted by the fluid on the particle. This timescale characterizes how quickly the particle's velocity decays due to viscous damping from the surrounding medium.

To analyze the particle's motion, we integrate Eq. (1.2.6) to obtain the time evolution of its position. Without any loss of generality, we set $\mathbf{x}(0) = 0$. The position is then given by

$$\mathbf{x}(t) = \tau \left[1 - \exp\left(-\frac{t}{\tau}\right)\right] \dot{\mathbf{x}}(0) + \frac{1}{\gamma} \int_0^t \left[1 - \exp\left(-\frac{t-t'}{\tau}\right)\right] \boldsymbol{\eta}(t') dt', \tag{A.1}$$

where we performed an integration by parts to obtain the integral on the right-hand side. We now compute the variance

$$\langle \mathbf{x}^2(t) \rangle = \tau^2 \left[1 - \exp\left(-\frac{t}{\tau}\right)\right]^2 \langle \dot{\mathbf{x}}(0)^2 \rangle \tag{A.2}$$

$$+ \frac{1}{\gamma^2} \int_0^t \int_0^t \left[1 - \exp\left(-\frac{t-t_1}{\tau}\right)\right] \left[1 - \exp\left(-\frac{t-t_2}{\tau}\right)\right] \langle \boldsymbol{\eta}(t_1)\boldsymbol{\eta}(t_2) \rangle dt_1 dt_2. \tag{A.3}$$

Since the initial velocity is uncorrelated with the thermal white noise, we use $\langle \boldsymbol{\eta}(t)\dot{\mathbf{x}}(0) \rangle = 0$ which simplifies the variance to

$$\langle \mathbf{x}^2(t) \rangle = \tau^2 \left[1 - \exp\left(-\frac{t}{\tau}\right)\right]^2 \langle \dot{\mathbf{x}}(0)^2 \rangle + \frac{6D}{\gamma^2} \left\{ t - 2\tau \left[1 - \exp\left(-\frac{t}{\tau}\right)\right] + \frac{\tau}{2} \left[1 - \exp\left(-2\frac{t}{\tau}\right)\right] \right\} \tag{A.4}$$

Taking the limits at short and long timescales, we obtain

$$\langle \mathbf{x}^2(t) \rangle \underset{t \ll \tau}{\approx} \dot{\mathbf{x}}(0)^2 t^2 \quad , \quad \langle \mathbf{x}^2(t) \rangle \underset{t \gg \tau}{\approx} \frac{6D}{\gamma^2} t. \tag{A.5}$$

This result tells us that at short times ($t \ll \tau$), the displacement is *ballistic*: the particle moves in a straight line at a constant speed. However, at long times ($t \gg \tau$), the displacement becomes *diffusive*: the initial velocity is damped, and the motion is dominated by random fluctuations from the noise. In this regime, inertia is no longer observable and can be neglected. We plot Eq. (A.4) and its asymptotics given in Eq. (A.5) in Fig. 17.4.

# B Derivation of White Noise

## B.1 Velocity of a Brownian motion and white noise

By definition, the increments of a Brownian motion $x(t)$ over an infinitesimally small time interval $dt$ are normally distributed as $\mathcal{N}(0, 2Ddt)$ where $D$ is the diffusion coefficient. A natural way to define an approximate velocity of the Brownian motion over the interval $dt$ is

$$\eta_{dt}(t) = \frac{x(t+dt) - x(t)}{dt}, \tag{B.1}$$

However, computing its mean squared value using Eq. (1.3.8), we find

$$\langle \eta_{dt}(t)^2 \rangle = \frac{\langle (x(t+dt) - x(t))^2 \rangle}{dt^2} = \frac{2Ddt}{dt^2} = \frac{2D}{dt}. \tag{B.2}$$

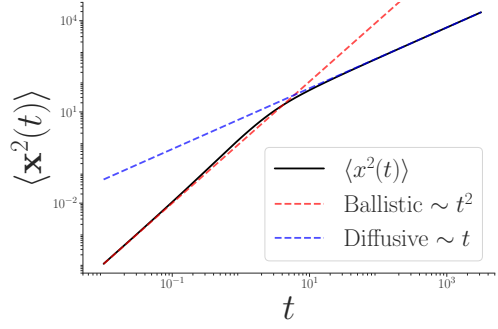

**Figure 17.4:** The solid line shows $\langle \mathbf{x}^2(t)\rangle$ given in Eq. (A.4), while the dotted lines represent its short- and long-time asymptotics. For times $t \ll \tau$, the motion is ballistic, whereas for $t \gg \tau$, the motion is diffusive. The parameters are $\langle \dot{\mathbf{x}}(0)^2\rangle = 1$, $D = 1$, $\gamma = 1$, and $\tau = 1$.

Since this expression diverges as $dt \to 0$, it follows that the instantaneous velocity $\eta(t)$ does not have a well-defined finite variance at any given time. In other words, the typical magnitude of $\eta_{dt}(t)$ diverges as

$$\eta_{dt}(t) \sim \sqrt{\frac{2D}{dt}}\,. \tag{B.3}$$

Thus, the notion of velocity as a well-defined function must be replaced by white noise, which is not a classical function but rather a generalized function (a distribution). Let us also mention that, from Eq. (B.1), since the displacements are distributed as $x(t+dt) - x(t) \stackrel{d}{\sim} \mathcal{N}(0, 2Ddt)$, it follows straightforwardly that $\langle \eta(t)\rangle = 0$. Now, since $\eta(t)$ is a Gaussian process, it is fully determined by its mean and its correlation function, which we derive below.

## B.2 Two-time correlation function of a white noise

The two-time correlation function of a white noise can be derived from the result given in Eq. (1.3.8). To see this, we consider the two-time correlation of the approximate velocity $\eta_{dt}(t)$, given by

$$\langle \eta_{dt}(t)\eta_{dt}(t')\rangle = \left\langle \frac{x(t+dt) - x(t)}{dt}\frac{x(t'+dt) - x(t')}{dt}\right\rangle . \tag{B.4}$$

Using Eq. (1.3.8), it follows that

$$\langle (x(t+dt) - x(t))(x(t'+dt) - x(t'))\rangle \tag{B.5}$$
$$= 2D\left[\min(t+dt, t'+dt) - \min(t+dt, t') - \min(t, t'+dt) + \min(t, t')\right]. \tag{B.6}$$

For $|t - t'| > dt$, we find $\langle \eta_{dt}(t)\eta_{dt}(t')\rangle = 0$, which is consistent with the fact that the increments are uncorrelated. However, for $|t - t'| \leq dt$, one can show that

$$\langle (x(t+dt) - x(t))(x(t'+dt) - x(t'))\rangle = 2D\left(dt - |t - t'|\right). \tag{B.7}$$

Thus, we obtain

$$f(t - t') = \langle \eta_{dt}(t)\eta_{dt}(t')\rangle = \begin{cases} \frac{2D}{dt^2}\left(dt - |t - t'|\right) &,\quad |t - t'| \leq dt, \\ 0 &,\quad |t - t'| > dt\,. \end{cases} \tag{B.8}$$

To show that this expression converges to a Dirac delta function in the limit $dt \to 0$, we integrate it against a test function $g(t)$. We compute

$$\int_{-\infty}^{+\infty} d\tau \, g(\tau) f(\tau) = \int_{-dt}^{+dt} d\tau \, g(\tau) \left[ \frac{2D}{dt^2} \left( dt - |\tau| \right) \right]. \tag{B.9}$$

Applying the change of variable $u = \tau/dt$, we obtain

$$2D \int_{-1}^{1} du \, g(u \, dt) \left( 1 - |u| \right). \tag{B.10}$$

Taking the limit $dt \to 0$, we find

$$\lim_{dt \to 0} 2D \int_{-1}^{1} du \, g(u \, dt) \left( 1 - |u| \right) = 2Dg(0) \int_{-1}^{1} du \, (1 - |u|) = 2D \, g(0). \tag{B.11}$$

which completes the proof. Therefore, based solely on the definition of the free Brownian motion, we have shown that

$$\langle \eta(t_1) \eta(t_2) \rangle = 2D \delta(t_1 - t_2). \tag{B.12}$$

## C  Discretization of the Langevin Equation

Here, we introduce both Itō and Stratonovich conventions for stochastic calculus. The two conventions lead to different results, particularly when studying multiplicative noise, i.e., when $\eta(t)$ is multiplied by a function of $x$, as in $\dot{x}(t) = 2D(x)\eta(t)$. However, throughout this thesis, we will exclusively consider and use the Itō convention, as it simplifies calculations by treating white noise as independent of the position, i.e., $\langle x(t)\eta(t) \rangle = 0$.

### C.1  A paradox

Consider the Langevin equation of a free Brownian motion (1.3.1). For an initial condition $x(0) = 0$, is trivially solved by

$$x(t) = \int_0^t \eta(t') dt'. \tag{C.1}$$

Using that $\langle \eta(t_1)\eta(t_2) \rangle = 2D\delta(t_1 - t_2)$, we obtain the correlation function which was predicted by Einstein (see Section 1.1),

$$\langle x^2(t) \rangle = \int_0^t \int_0^t \langle \eta(t_1)\eta(t_2) \rangle dt_1 dt_2 = 2D \int_0^t dt_1 = 2Dt. \tag{C.2}$$

One can also compute the derivative of the variance of $x(t)$ by differentiating directly

$$\frac{d}{dt} \langle x^2(t) \rangle = 2D. \tag{C.3}$$

However, an alternative approach would be to write

$$\frac{d}{dt} \langle x^2(t) \rangle = \langle 2x(t) \frac{dx}{dt} \rangle = 2 \langle x(t) \eta(t) \rangle = 0, \tag{C.4}$$

where in the last equality, we have used the fact that the position $x(t)$ is uncorrelated with the noise $\eta(t)$.

This result appears to contradict the previous computation. As we will see below, this apparent paradox arises due to the discretization, which must be carefully defined.

## C.2 Itō discretization

The discretization of the Langevin equation (1.3.1) is given by

$$\frac{x_{t+dt} - x_t}{dt} = \eta_{dt}, \tag{C.5}$$

where, for better readability, we write $x(t) \equiv x_t$. In the Itō convention, within a time interval $[t, t+dt]$, the integrands are evaluated at the beginning of each interval. A significant feature of the Itô integral is that it assumes the position at an instant is independent of the noise within the interval $[t, t+dt]$. The associated discretization of the correlations is given by

$$\langle f(x(t))\rangle \to \langle f(x_t)\rangle, \quad \langle f(x(t))\dot{x}(t)\rangle \to \langle f(x_t)\eta_{dt}\rangle, \quad \text{and} \quad \langle f(x(t))\dot{x}(t)\rangle \to \left\langle f(x_t)\frac{x_{t+dt} - x_t}{dt}\right\rangle. \tag{C.6}$$

which leads to Itô's chain rule

$$\frac{d}{dt}\langle f(x(t))\rangle = \left\langle f'(x(t))\frac{dx}{dt}\right\rangle + D\langle f''(x(t))\rangle. \tag{C.7}$$

*Proof.* We start from

$$\frac{d}{dt}\langle f(x(t))\rangle = \lim_{dt\to 0}\frac{\langle f(x(t)+dt)\rangle - \langle f(x(t))\rangle}{dt} = \lim_{dt\to 0}\left\langle \frac{f(x_{t+dt}) - f(x_t)}{dt}\right\rangle \equiv \left\langle \frac{f(x_{t+dt}) - f(x_t)}{dt}\right\rangle, \tag{C.8}$$

where we dropped the limit in the last equality for notation simplicity. Let us expand $f$ in a Taylor series around $x_t$ in powers of $x_{t+dt} - x_t$,

$$f(x_{t+dt}) = f(x_t) + f'(x_t)(x_{t+dt} - x_t) + \frac{f''(x_t)}{2}(x_{t+dt} - x_t)^2 + \cdots. \tag{C.9}$$

Substituting this into Eq. (C.8) and using the discretized version of the Langevin equation, we obtain:

$$\frac{d}{dt}\langle f(x(t))\rangle = \left\langle f'(x_t)\frac{x_{t+dt} - x_t}{dt}\right\rangle + \left\langle \frac{f''(x_t)}{2}\frac{(x_{t+dt} - x_t)^2}{dt}\right\rangle + \cdots \tag{C.10}$$

$$= \left\langle f'(x_t)\frac{x_{t+dt} - x_t}{dt}\right\rangle + \left\langle \frac{f''(x_t)}{2}\eta_{dt}^2 dt\right\rangle + \cdots. \tag{C.11}$$

Since $x_t$ and $\eta_{dt}$ are independent,

$$\frac{d}{dt}\langle f(x(t))\rangle = \left\langle f'(x_t)\frac{x_{t+dt} - x_t}{dt}\right\rangle + \left\langle \frac{f''(x_t)}{2}\right\rangle\langle \eta_{dt}^2\rangle dt, \tag{C.12}$$

and using that $\eta_{dt} \sim \sqrt{2D/dt}$ (see Eq. (B.3)),

$$= \left\langle f'(x_t)\frac{x_{t+dt} - x_t}{dt}\right\rangle + D\langle f''(x_t)\rangle, \tag{C.13}$$

thus, by applying Itô's rules for the discretization of the correlations, we recover the announced chain rule.

## C.3 Stratonovich discretization

The Stratonovich interpretation evaluates the integrand at the midpoint of each interval. This convention preserves the standard chain rule of calculus; however, in this case, we have $\langle x(t)\eta(t)\rangle \neq 0$. The Stratonovich convention proposes a different discretization of the correlations

$$\langle f(x(t))\rangle \rightarrow \left\langle \frac{f(x_{t+dt})+f(x_t)}{2}\right\rangle, \tag{C.14}$$

$$\langle f(x(t))\eta(t)\rangle \rightarrow \left\langle \frac{f(x_{t+dt})+f(x_t)}{2}\eta_t\right\rangle, \tag{C.15}$$

$$\langle f(x(t))\dot{x}(t)\rangle \rightarrow \left\langle \frac{f(x_{t+dt})+f(x_t)}{2}\frac{x_{t+dt}-x_t}{dt}\right\rangle, \tag{C.16}$$

which leads to the usual chain rule:

$$\frac{d}{dt}\langle f(x(t))\rangle = \left\langle f'(x(t))\frac{dx}{dt}\right\rangle. \tag{C.17}$$

*Proof.* We again have

$$\frac{d}{dt}\langle f(x(t))\rangle = \left\langle \frac{f(x_{t+dt})-f(x_t)}{dt}\right\rangle = \left\langle f'(x_t)\frac{x_{t+dt}-x_t}{dt}\right\rangle + \left\langle \frac{f''(x_t)}{2}\frac{(x_{t+dt}-x_t)^2}{dt}\right\rangle + \cdots. \tag{C.18}$$

On the other hand, using Stratonovich rules, we have

$$\left\langle f'(x(t))\frac{dx}{dt}\right\rangle = \left\langle \frac{f'(x_{t+dt})+f'(x_t)}{2}\frac{x_{t+dt}-x_t}{dt}\right\rangle. \tag{C.19}$$

Using a Taylor expansion of $f'$ near $x_t$ in powers of $(x_{t+dt}-x_t)$, namely

$$f'(x_{t+dt}) = f'(x_t) + (x_{t+dt}-x_t)f''(x_t) + \dots, \tag{C.20}$$

we recover

$$\left\langle f'(x(t))\frac{dx}{dt}\right\rangle = \left\langle f'(x_t)\frac{x_{t+dt}-x_t}{dt}\right\rangle + \left\langle \frac{f''(x_t)}{2}\frac{(x_{t+dt}-x_t)^2}{dt}\right\rangle + \dots, \tag{C.21}$$

which is the right-hand side of Eq. (C.18), hence proving the announced chain rule.

## C.4 Fixing the paradox of the introduction

First, using Itō's chain rule with $f(x(t)) = x(t)^2$, we have

$$\frac{d}{dt}\langle x^2(t)\rangle = 2\left\langle x(t)\frac{dx}{dt}\right\rangle + 2D, \tag{C.22}$$

and using that $\frac{dx}{dt} = \eta(t)$, we obtain

$$\frac{d}{dt}\langle x^2(t)\rangle = 2\langle x(t)\eta(t)\rangle + 2D = 2D. \tag{C.23}$$

Performing the same calculation using the Stratonovich convention leads to

$$\frac{d}{dt}\langle x^2(t)\rangle = 2\left\langle x(t)\frac{dx}{dt}\right\rangle = 2\left\langle \frac{x_{t+dt}+x_t}{2}\eta_t\right\rangle = \langle x_{t+dt}\eta_t\rangle + \langle x_t\eta_t\rangle. \tag{C.24}$$

According to the Langevin equation, we have $x_{t+dt} = x_t + \eta_t dt$, with $\langle x_t\eta_t\rangle = 0$, such that

$$\frac{d}{dt}\langle x^2(t)\rangle = 2\langle x_t\eta_t\rangle + dt\langle \eta_t^2\rangle = 2D. \tag{C.25}$$

Thus, using both conventions, we recover the result from the introduction. The apparent paradox lay in the discretization.

## C.5 Simulation of Brownian Motion

To numerically study the statistical properties of Brownian motion, one typically samples multiple trajectories of the particle and computes the average of the observables of interest. The simplest and most efficient way to sample Brownian motion is to generate increments from a normal distribution

$$x(t + \Delta t) = x(t) + \mathcal{N}(0, 2D\Delta t), \tag{C.26}$$

where $\Delta t > 0$.

However, in a more general setting, such as for a Brownian motion evolving in a potential $V(x)$ (see the next section), or even beyond Brownian motion, one often needs to sample trajectories at very small time increments, i.e., $x(t+dt)$, where $dt \ll 1$. For this reason, one typically discretizes the Langevin equation as follows:

$$x(t + dt) = x(t) - V'(x)dt + \sqrt{2Ddt}\, \xi, \tag{C.27}$$

where $\xi \stackrel{d}{\sim} \mathcal{N}(0, 1)$. For a diffusion coefficient independent of the position, this discretization is valid in both Itō and Stratonovich conventions. In Fig. 1.2 and in the left panel of Fig. 1.4, we show trajectories of Brownian motions.

# D  Derivation of Forward Fokker-Planck Equation by Averaging over Paths

We propose here an alternative and more intuitive derivation of the forward Fokker-Planck equation. Consider a Brownian particle evolving through Eq. (1.4.1) located at $x'$ at time $t$ and at $x$ at time $t + dt$. We aim to compute $p(x, t+dt)$ by averaging over all possible paths from $x'$ to $x$ within a small time step $dt$

$$p(x, t + dt) = \langle \delta(x - x(t+dt)) \rangle_{\xi, x'} = \left\langle \delta\left(x - \left[x' + F(x')dt + \sqrt{2Ddt}\,\xi\right]\right) \right\rangle_{\xi, x'}, \tag{D.1}$$

where $\langle \cdot \rangle_{\xi, x'}$ denotes an average over both the white noise and the position $x'$. We have

$$p(x, t + dt) = \int_{-\infty}^{+\infty} dx' \int_{-\infty}^{+\infty} d\xi\, \frac{e^{-\frac{\xi^2}{2}}}{\sqrt{2\pi}}\, p(x', t)\, \delta\left(x - x' - F(x')dt - \sqrt{2Ddt}\,\xi\right). \tag{D.2}$$

Using the property $\delta(ax) = \delta(x)/|a|$, the Dirac delta function simplifies the integral over $\xi$, leading to

$$p(x, t + dt) = \int_{-\infty}^{+\infty} dx'\, \frac{e^{-\frac{(x-x'-F(x')dt)^2}{4Ddt}}}{\sqrt{4\pi Ddt}}\, p(x', t). \tag{D.3}$$

Since $(x' - x) = O(\sqrt{dt})$, we expand both $F(x')$ and $p(x', t)$ in powers of $(x' - x)$

$$F(x') \approx F(x) + (x' - x)\frac{\partial F(x)}{\partial x}, \tag{D.4}$$

$$p(x', t) \approx p(x, t) + (x' - x)\frac{\partial p(x, t)}{\partial x} + \frac{(x' - x)^2}{2}\frac{\partial^2 p(x, t)}{\partial x^2}. \tag{D.5}$$

The exponential term in Eq. (D.3) simplifies as follows

$$e^{-\frac{(x-x'-F(x')dt)^2}{4Ddt}} \approx e^{-\frac{(x-x')^2}{4Ddt}} e^{-\frac{(x'-x)}{2D}F(x')} \approx e^{-\frac{(x-x')^2}{4Ddt}}\left(1 - \frac{(x'-x)}{2D}F(x')\right) \tag{D.6}$$

$$\approx e^{-\frac{(x-x')^2}{4Ddt}}\left(1 - \frac{(x'-x)}{2D}F(x) - \frac{(x'-x)^2}{2D}\frac{\partial F(x)}{\partial x}\right). \tag{D.7}$$

Substituting these expansions into Eq. (D.3), we obtain

$$p(x, t+dt) = \int_{-\infty}^{+\infty} dx' \frac{e^{-\frac{(x-x')^2}{4Ddt}}}{\sqrt{4\pi Ddt}} \left(1 - \frac{(x'-x)}{2D}F(x) - \frac{(x'-x)^2}{2D}\frac{\partial F(x)}{\partial x}\right) \quad \text{(D.8)}$$

$$\times \left(p(x,t) + (x'-x)\frac{\partial p(x,t)}{\partial x} + \frac{(x'-x)^2}{2}\frac{\partial^2 p(x,t)}{\partial x^2}\right). \quad \text{(D.9)}$$

Using the integrals

$$\int_{-\infty}^{+\infty} dx' \, (x'-x) \frac{e^{-\frac{(x-x')^2}{4Ddt}}}{\sqrt{4\pi Ddt}} = 0, \quad \text{(D.10)}$$

$$\int_{-\infty}^{+\infty} dx' \, (x'-x)^2 \frac{e^{-\frac{(x-x')^2}{4Ddt}}}{\sqrt{4\pi Ddt}} = 2Ddt, \quad \text{(D.11)}$$

we finally obtain

$$p(x, t+dt) = p(x,t) - dt \left(\frac{\partial F(x)}{\partial x} p(x,t) + F(x)\frac{\partial p(x,t)}{\partial x}\right) + dt \, D \frac{\partial^2 p(x,t)}{\partial x^2}. \quad \text{(D.12)}$$

Taking the limit $dt \to 0$, we recover the forward Fokker-Planck equation

$$\frac{\partial p(x,t)}{\partial t} = -\frac{\partial}{\partial x}(F(x)p(x,t)) + D\frac{\partial^2 p(x,t)}{\partial x^2}. \quad \text{(D.13)}$$

# E  Numerical Method for Switching Diffusion

In this section, we explain the algorithm used to compute numerically the scaled cumulant generating function $\Psi(q)$ and the large deviation function $I(y)$. To obtain high numerical accuracy, we implemented the code in `Julia`, utilizing the `BigFloat()` type for arbitrary precision arithmetic.

## E.1  Numerical evaluation of $\Psi(q)$

Recall that the scaled cumulant generating function (SCGF) is defined as $\Psi(q) = \lim_{t\to\infty} \ln \hat{p}_r(q,t)/t$. We will directly compute numerically $\hat{p}_r(q,t)$. As shown in Eq. (9.1.7), $\hat{p}_r(q,t)$ satisfies an integral equation, which we solve numerically. The equation is given by

$$\hat{p}_r(q,t) = e^{-rt}\hat{G}_0(q,t) + \int_0^t d\tau \, r \, e^{-r\tau} \hat{G}_0(q,\tau) \hat{p}_r(q, t-\tau), \quad \hat{G}_0(q,t) = \int_0^{D_{\max}} dD \, W(D) \, e^{Dq^2 t}, \quad \text{(E.1)}$$

$$\hat{p}_r(q,t) = \langle e^{qx} \rangle = \int_{-\infty}^{+\infty} dx \, e^{qx} \, p_r(x,t), \quad \hat{p}_r(0,t) = 1, \quad \text{(E.2)}$$

where $\hat{p}_r(0,t) = 1$ is just the normalization. Here, $\hat{G}_0(q,t)$ represents the bilateral Laplace transform of the Brownian motion propagator with diffusion constant $D$, averaged over the distribution $W(D)$. Interestingly, this integral equation resembles that of resetting Brownian motion. Indeed, the cumulative distribution $Q_r(M,t)$ of the maximum of a rBM starting at the origin up to time $t$ obeys the same integral equation (with the identification $q \to M$), but with a different function $\hat{G}_0(q,t)$. We solve this equation numerically using a recursive method, discretizing time into small intervals $\Delta t$, as explained in Section IV of [406].

## E.2 Numerical evaluation of $I(y)$

The rate function is defined as $I(y = x/t) = \lim_{t\to\infty} -\ln p_r(x,t)/t$. To compare with our analytical prediction, we will compute the Fourier transform of the distribution of the position of the particle $\hat{p}_r(k,t)$ defined in Eq. (9.1.12), and then invert it. We choose to work in Fourier space because it is easier to numerically invert than the bilateral Laplace transform. We first compute numerically $\hat{p}_r(k,t)$ as explained in the previous section by replacing $q^2 \to -k^2$. The inverse Fourier transform is approximated as

$$\begin{aligned} p_r(x,t) &= \frac{1}{2\pi} \int_{-\infty}^{+\infty} dk\, e^{-ikx} \hat{p}_r(k,t) = \frac{1}{\pi} \int_0^{+\infty} dk\, \cos(kx)\, \hat{p}_r(k,t) \\ &\approx \frac{1}{\pi} \int_0^{k_{\text{Max}}} dk\, \cos(kx)\, \hat{p}_r(k,t)\,, \end{aligned} \qquad (\text{E.3})$$

where the second equality we have used the fact that $p_r(x,t)$ is symmetric and real. The last approximation comes from the fact that we cannot integrate up to $+\infty$ as we evaluate $\hat{p}_r(k,t)$ numerically. Therefore, we need to specify an upper bound for the integral which we call $k_{\text{Max}}$. We choose the value $k_{\text{Max}}$ to achieve the desired precision in our evaluation. As we have a prediction for the tail of $I(y)$ (see Eq. (10) of the letter), we can estimate the precision required to compute $p_r(x = yt, t)$ for a given $x$ as

$$p_r(x = yt, t) \underset{t\to\infty}{\approx} \exp\left(-\frac{y}{4D_{\text{max}}}\right)\,. \qquad (\text{E.4})$$

The value of $k_{\text{Max}}$ is then chosen such that $\hat{p}_r(k_{\text{Max}}, t)$ is of the same order as the right hand side of Eq. (E.4) in order to estimate the integral $\int_0^{k_{\text{Max}}} dk\, \cos(kx)\, \hat{p}_r(k,t)$ with the required precision. To probe the large deviations, one needs to go at high values of $t$ and $x$ (typically, $x$ is of the order of $10^3$ to $10^4$). Therefore, the cosine in the integral has a really small period and the integrand highly oscillates. To numerically compute the integral in Eq. (E.3), we employ Filon's method [407], which is effective for oscillatory integrals.

# F Proof of the Validity of the Series Expansion of $\hat{p}(k)$

We recall that for a switching diffusion particle evolving in a harmonic potential, based on an analytical guess, we found that the Fourier transform of the non-equilibrium steady-state distribution is given by

$$\hat{p}(k) = \sum_{n=0}^{+\infty} \frac{(ik)^{2n}}{(2\mu)^n} \frac{\Gamma\left(\frac{\beta}{2}\right)}{\Gamma\left(\frac{\beta}{2}+n\right)} \frac{B_n\left(1!\frac{\beta}{2}\frac{\langle D\rangle}{1}, \ldots, n!\frac{\beta}{2}\frac{\langle D^n\rangle}{n}\right)}{n!}\,. \qquad (\text{F.1})$$

In this Appendix, we demonstrate that our proposed series expansion, Eq. (8.3.2), indeed satisfies the integral equation Eq. (10.1.12) for $\hat{p}(k)$, which we recall below

$$\hat{p}(k) = \int_0^{D_{\text{max}}} dD\, W(D) \int_0^1 dU\, \beta U^{\beta-1}\, e^{-\frac{k^2 \alpha(U)}{2}}\, \hat{p}(kU)\,, \quad \alpha(U) = \frac{D}{\mu}(1-U^2)\,. \qquad (\text{F.2})$$

This can be verified by substituting the series expansion into the right-hand side of the integral equation. The right-hand side indeed reads

$$\int_0^{D_{\text{max}}} dD\, W(D) \int_0^1 dU\, \beta U^{\beta-1}\, e^{-\frac{k^2 \alpha(U)}{2}}\, \hat{p}(kU) \qquad (\text{F.3})$$

$$= \sum_{i=0}^{+\infty} \int_0^{D_{\text{max}}} dD\, W(D) \int_0^1 dU\, \beta U^{\beta-1+2i} e^{-\frac{k^2 D(1-U^2)}{2\mu}} \left(\frac{-k^2}{2\mu}\right)^i \frac{\Gamma\left(\frac{\beta}{2}\right)}{\Gamma\left(\frac{\beta}{2}+i\right)} \frac{B_i\left(1!\frac{\beta}{2}\frac{\langle D\rangle}{1}, \ldots, i!\frac{\beta}{2}\frac{\langle D^i\rangle}{i}\right)}{i!}\,.$$

The integration over $U$ can be performed as follows

$$\int_0^1 dU\, \beta U^{\beta-1+2i} e^{-\frac{k^2 D(1-U^2)}{2\mu}} = \frac{\frac{\beta}{2}}{\frac{\beta}{2}+i}\, {}_1F_1\left(1, \frac{\beta}{2}+1+i, -\frac{Dk^2}{2\mu}\right), \tag{F.4}$$

where ${}_1F_1$ is the confluent hypergeometric function. By definition, we have

$$ {}_1F_1\left(1, \frac{\beta}{2}+1+i, -\frac{Dk^2}{2\mu}\right) = \sum_{j=0}^{+\infty} \frac{\Gamma\left(\frac{\beta}{2}+1+i\right)}{\Gamma\left(\frac{\beta}{2}+1+i+j\right)} \left(\frac{-Dk^2}{2\mu}\right)^j, \tag{F.5}$$

such that the right-hand side of Eq. (10.1.12) now reads

$$\sum_{i=0}^{+\infty}\sum_{j=0}^{+\infty} \left(\frac{-k^2}{2\mu}\right)^{i+j} \langle D^j \rangle \frac{\frac{\beta}{2}}{\frac{\beta}{2}+i} \frac{\Gamma\left(\frac{\beta}{2}+1+i\right)}{\Gamma\left(\frac{\beta}{2}+1+i+j\right)} \frac{\Gamma\left(\frac{\beta}{2}\right)}{\Gamma\left(\frac{\beta}{2}+i\right)} \frac{B_i\left(1!\frac{\beta}{2}\frac{\langle D\rangle}{1},\ldots,i!\frac{\beta}{2}\frac{\langle D^i\rangle}{i}\right)}{i!}. \tag{F.6}$$

It can be simplified as

$$\sum_{i=0}^{+\infty}\sum_{j=0}^{+\infty} \left(\frac{-k^2}{2\mu}\right)^{i+j} \langle D^j \rangle \frac{\frac{\beta}{2}}{\frac{\beta}{2}+i+j} \frac{\Gamma\left(\frac{\beta}{2}\right)}{\Gamma\left(\frac{\beta}{2}+i+j\right)} \frac{B_i\left(1!\frac{\beta}{2}\frac{\langle D\rangle}{1},\ldots,i!\frac{\beta}{2}\frac{\langle D^i\rangle}{i}\right)}{i!}. \tag{F.7}$$

We now want to identify the term of order $k^{2n}$ and compare to Eq. (F.1). It is given by the case $i+j=n$, i.e.,

$$\left(\frac{-k^2}{2\mu}\right)^n \frac{\Gamma\left(\frac{\beta}{2}\right)}{\Gamma\left(\frac{\beta}{2}+n\right)} \frac{\frac{\beta}{2}}{\frac{\beta}{2}+n} \sum_{i=0}^{n} \langle D^{n-i}\rangle \frac{B_i\left(1!\frac{\beta}{2}\frac{\langle D\rangle}{1},\ldots,i!\frac{\beta}{2}\frac{\langle D^i\rangle}{i}\right)}{i!}. \tag{F.8}$$

We can simplify by extracting the $n^{\text{th}}$ term of the sum as follows

$$\left(\frac{-k^2}{2\mu}\right)^n \frac{\Gamma\left(\frac{\beta}{2}\right)}{\Gamma\left(\frac{\beta}{2}+n\right)} \frac{1}{\frac{\beta}{2}+n} \left[\frac{\beta}{2}\frac{B_n\left(1!\frac{\beta}{2}\frac{\langle D\rangle}{1},\ldots,n!\frac{\beta}{2}\frac{\langle D^n\rangle}{n}\right)}{n!} + \sum_{i=0}^{n-1}\frac{\beta}{2}\langle D^{n-i}\rangle \frac{B_i\left(1!\frac{\beta}{2}\frac{\langle D\rangle}{1},\ldots,i!\frac{\beta}{2}\frac{\langle D^i\rangle}{i}\right)}{i!}\right]. \tag{F.9}$$

One can use the recursive definition of the complete Bell polynomial, i.e.,

$$B_{n+1}(x_1, x_2, \ldots, x_{n+1}) = \sum_{k=0}^{n} \binom{n}{k} B_k(x_1, x_2, \ldots, x_k) x_{n-k+1}, \tag{F.10}$$

to show that

$$\sum_{i=0}^{n-1} \frac{\beta}{2}\langle D^{n-i}\rangle \frac{B_i\left(1!\frac{\beta}{2}\frac{\langle D\rangle}{1},\ldots,i!\frac{\beta}{2}\frac{\langle D^i\rangle}{i}\right)}{i!} = n\frac{B_n\left(1!\frac{\beta}{2}\frac{\langle D\rangle}{1},\ldots,n!\frac{\beta}{2}\frac{\langle D^n\rangle}{n}\right)}{n!}. \tag{F.11}$$

In the end, the right-hand side is given by

$$\sum_{n=0}^{+\infty} \left(\frac{-k^2}{2\mu}\right)^n \frac{\Gamma\left(\frac{\beta}{2}\right)}{\Gamma\left(\frac{\beta}{2}+n\right)} \frac{B_n\left(1!\frac{\beta}{2}\frac{\langle D\rangle}{1},\ldots,n!\frac{\beta}{2}\frac{\langle D^n\rangle}{n}\right)}{n!} = \hat{p}(k), \tag{F.12}$$

which concludes the proof.

# G  Derivation of the Small $\beta$ Limit of the Cumulants

We have shown in Section 10.1.3 that in the small $\beta$ limit, the Fourier transform of the NESS of the switching diffusion inside a harmonic trap is given by

$$\hat{p}(k) \underset{\beta \to 0}{\approx} \int_0^{D_{\max}} dD\, W(D) e^{\frac{-k^2 D}{2\mu}}. \tag{G.1}$$

To derive the limiting behavior of the cumulants in the limit $\beta \to 0$, we recall that the cumulant generating function is defined as

$$\log \hat{p}(k) = \sum_{n=1}^{\infty} \frac{(ik)^{2n}}{(2n)!} \langle x^{2n} \rangle_c. \tag{G.2}$$

From Eq. (G.1), we have

$$\log \hat{p}(k) \underset{\beta \to 0}{\approx} \log\left(1 + \sum_{n=1}^{+\infty} \frac{\left(\frac{-k^2}{2\mu}\right)}{n!} \langle D^n \rangle\right) \tag{G.3}$$

$$= \sum_{n=1}^{+\infty} (-1)^{n-1} \frac{(n-1)!}{n!} \left[\sum_{m=1}^{+\infty} \left(\frac{-k^2}{2\mu}\right)^m \frac{1}{m!} \langle D^m \rangle\right]^n \tag{G.4}$$

Using the following property of Bell polynomials

$$\left[\sum_{m=1}^{+\infty} \langle D^m \rangle z^m\right]^n = n! \sum_{m=n}^{+\infty} B_{m,n}(a_1, \ldots, a_{m-n+1}) \frac{z^m}{m!} \quad , \quad a_i = i! \langle D^i \rangle, \tag{G.5}$$

which specialized in our case to

$$\left[\sum_{m=1}^{+\infty} \left(\frac{-k^2}{2\mu}\right)^m \frac{1}{m!} \langle D^m \rangle\right]^n = n! \sum_{m=n}^{+\infty} \left(\frac{-k^2}{2\mu}\right)^m \frac{1}{m!} B_{m,n}(\langle D \rangle, \ldots, \langle D^{m-n+1} \rangle), \tag{G.6}$$

we obtain

$$\log \hat{p}(k) \underset{\beta \to 0}{\approx} \sum_{n=1}^{+\infty} \sum_{m=n}^{+\infty} (-1)^{n-1} (n-1)! \left(\frac{-k^2}{2\mu}\right)^m \frac{1}{m!} B_{m,n}(\langle D \rangle, \ldots, \langle D^{m-n+1} \rangle). \tag{G.7}$$

$$= \sum_{m=1}^{+\infty} \left(\frac{-k^2}{2\mu}\right)^m \frac{1}{m!} \sum_{n=1}^{m} (-1)^{n-1} (n-1)! B_{m,n}(\langle D \rangle, \ldots, \langle D^{m-n+1} \rangle). \tag{G.8}$$

The relation between cumulants and moments given in Eq. (8.1.6) yields

$$\log \hat{p}(k) \underset{\beta \to 0}{\approx} \sum_{m=1}^{+\infty} \left(\frac{-k^2}{2\mu}\right)^m \frac{1}{m!} \langle D^m \rangle_c. \tag{G.9}$$

Comparing with Eq. (G.2), we finally find

$$\langle x^{2n} \rangle_c = \frac{(2n-1)!!}{\mu^n} \langle D^n \rangle_c, \tag{G.10}$$

which is the anounced result in the first line of Eq. (10.1.36).

**Remark.** A more direct derivation consists in comparing the two expansions:

$$\log\left(\langle e^{sD} \rangle\right) = \log\left(\int_0^{D_{\max}} dD\, W(D) e^{sD}\right) = \sum_{n=1}^{+\infty} \frac{s^n}{n!} \langle D^n \rangle_c, \tag{G.11}$$

and

$$\log [\hat{p}(k)] \underset{\beta \to 0}{\approx} \log \left( \int_0^{D_{\max}} dD \, W(D) \, e^{-\frac{k^2 D}{2\mu}} \right) = \sum_{n=1}^{+\infty} \frac{(ik)^{2n}}{(2n)!} \langle x^{2n}(t) \rangle_c \,. \tag{G.12}$$

Taking $s = -k^2/(2\mu)$ leads to

$$\langle x^{2n}(t) \rangle_c = \frac{(2n)!}{2^n n!} \frac{1}{\mu^n} \langle D^n \rangle_c = \frac{(2n-1)!!}{\mu^n} \langle D^n \rangle_c \,. \tag{G.13}$$

# H Differential Equations for the MFPT $\tau_\gamma$, and Relations for $\tau_\gamma^\pm$

In this Section, we derive the differential equation for $\tau_\gamma$ as well as the expression of $\tau_\gamma^\pm$ with respect to $\tau_\gamma$ and its first derivative. As recalled in Section 11.2, and derived in [355], $\tau_\gamma^\pm(x_0)$ obey the following coupled differential equations

$$[f(x_0) + v_0] \, \partial_{x_0} \tau_\gamma^+(x_0) - \gamma \tau_\gamma^+(x_0) + \gamma \tau_\gamma^-(x_0) = -1 \,, \tag{H.1}$$

$$[f(x_0) - v_0] \, \partial_{x_0} \tau_\gamma^-(x_0) + \gamma \tau_\gamma^+(x_0) - \gamma \tau_\gamma^-(x_0) = -1 \,. \tag{H.2}$$

The first goal is to derive an ordinary differential equation (ODE) for the MFPT $\tau_\gamma(x_0) = 1/2 \left( \tau_\gamma^+(x_0) + \tau_\gamma^-(x_0) \right)$. Equations (H.1) and (H.2) can be also written in terms of operators $\mathcal{L}_\pm$ acting on $\tau_\gamma^\pm$ as

$$\mathcal{L}_+ \tau_\gamma^+(x_0) = \left[ (f(x_0) + v_0) \, \partial_{x_0} - \gamma \right] \tau_\gamma^+(x_0) = -1 - \gamma \tau_\gamma^-(x_0) \,, \tag{H.3}$$

$$\mathcal{L}_- \tau_\gamma^-(x_0) = \left[ (f(x_0) - v_0) \, \partial_{x_0} - \gamma \right] \tau_\gamma^-(x_0) = -1 - \gamma \tau_\gamma^-(x_0) \,. \tag{H.4}$$

Applying $\mathcal{L}_-$ on the left hand side (LHS) of Eq. (H.3) and using Eq. (H.4) gives a second order differential equation on $\tau_\gamma^+$. A similar procedure where one applies $\mathcal{L}_+$ on the LHS of Eq. (H.4) instead gives the equation for $\tau_\gamma^-$. It gives

$$\left[ v_0^2 - f(x_0)^2 \right] \partial_{x_0}^2 \tau_\pm(x_0) + \left[ 2\gamma f(x_0) - f(x_0) f'(x_0) \pm v_0 f'(x_0) \right] \partial_{x_0} \tau_\pm(x_0) = -2\gamma \,. \tag{H.5}$$

Next, by summing up these two equations for $\tau_\gamma^+$ and $\tau_\gamma^-$, one finds

$$2 \left[ v_0^2 - f(x_0)^2 \right] \partial_{x_0}^2 \tau_\gamma(x_0) + 2 \left[ 2\gamma f(x_0) - f(x_0) f'(x_0) \right] \partial_{x_0} \tau_\gamma(x_0) + v_0 f'(x_0) \partial_{x_0} \left( \tau_\gamma^+(x_0) - \tau_\gamma^-(x_0) \right) = -4\gamma \,. \tag{H.6}$$

We now want to express the difference $\left( \tau_\gamma^+(x_0) - \tau_\gamma^-(x_0) \right)$ in terms of the MFPT $\tau_\gamma(x_0)$. To this end, one can sum the two equations (H.1) and (H.2). We obtain

$$\partial_{x_0} \left( \tau_\gamma^+(x_0) - \tau_\gamma^-(x_0) \right) = -\frac{2}{v_0} \left( 1 + f(x_0) \partial_{x_0} \tau_\gamma(x_0) \right) \,. \tag{H.7}$$

One can then insert Eq. (H.7) in Eq. (H.6) and simplify to obtain the ODE on $\tau_\gamma(x_0)$

$$\left[ f(x_0)^2 - v_0^2 \right] \partial_{x_0}^2 \tau_\gamma(x_0) + 2 f(x_0) \left[ f'(x_0) - \gamma \right] \partial_{x_0} \tau_\gamma(x_0) = 2\gamma - f'(x_0) \,, \tag{H.8}$$

which is Eq. (11.2.10) given in text.

In addition, from the knowledge of $\tau_\gamma(x_0)$, it is possible to obtain the full expressions of $\tau_\gamma^-(x_0)$ and $\tau_\gamma^+(x_0)$. By definition we have

$$\tau_\gamma(x_0) = \frac{1}{2} \left( \tau_\gamma^+(x_0) + \tau_\gamma^-(x_0) \right) \,, \tag{H.9}$$

so that
$$\tau_\gamma^+(x_0) = 2\tau_\gamma(x_0) - \tau_\gamma^-(x_0)\,. \tag{H.10}$$

Substituing this relation in Eq. (H.7) gives
$$\partial_{x_0}\left(2\tau_\gamma(x_0) - 2\tau_\gamma^-(x_0)\right) = -\frac{2}{v_0}\left(1 + f(x_0)\partial_{x_0}\tau_\gamma(x_0)\right), \tag{H.11}$$

which can also be written as
$$\partial_{x_0}\tau_\gamma^-(x_0) = \left(1 + \frac{f(x_0)}{v_0}\right)\partial_{x_0}\tau_\gamma(x_0) + \frac{1}{v_0}\,. \tag{H.12}$$

Substituting Eqs. (H.10) and (H.12) in Eq. (H.2) yields an expression for $\tau_\gamma^-(x_0)$ in terms of $\tau_\gamma(x_0)$ and $\partial_{x_0}\tau_\gamma(x_0)$. And using again Eq. (H.10) one finally obtains

$$\tau_\gamma^-(x_0) = \tfrac{1}{2\gamma}\tfrac{f(x_0)}{v_0} - \tfrac{v_0}{2\gamma}\left(1 - \tfrac{f(x_0)^2}{v_0^2}\right)\partial_{x_0}\tau_\gamma(x_0) + \tau_\gamma(x_0) \tag{H.13}$$

$$\tau_\gamma^+(x_0) = -\tfrac{1}{2\gamma}\tfrac{f(x_0)}{v_0} + \tfrac{v_0}{2\gamma}\left(1 - \tfrac{f(x_0)^2}{v_0^2}\right)\partial_{x_0}\tau_\gamma(x_0) + \tau_\gamma(x_0)\,, \tag{H.14}$$

as given in Section 11.2.

# I Higher Order Moments of the Distribution of the First-Passage Time

In this Appendix, we derive a recursive relation for the moments of the distribution of the first-passage time $F_{\text{fp}}(x_0, t) = -\partial_t Q(x_0, t)$ where we recall $Q(x_0, t) = \frac{1}{2}(Q^+(x_0, t) + Q^-(x_0, t))$. We thus define $\langle T_\pm^n \rangle = -\int_0^{+\infty} dt\, t^n\, \partial_t Q^\pm(x_0, t)$ such that $\langle T^n \rangle = \frac{1}{2}(\langle T_+^n \rangle + \langle T_-^n \rangle)$. Here, we assume all moments $\langle T^n \rangle$ to be well-defined. We also have the identity $\langle T_\pm \rangle \equiv T_\gamma^\pm(x_0)$. Following from the derivation of equations (11.2.7) and (11.2.8), it is easy to generalised the two coupled equations to the $n^{\text{th}}$ moments yielding

$$(f(x_0) + v_0)\,\partial_{x_0}\langle T_+^n \rangle - \gamma\langle T_+^n \rangle + \gamma\langle T_-^n \rangle = -n\,\langle T_+^{n-1} \rangle\,, \tag{I.1}$$
$$(f(x_0) - v_0)\,\partial_{x_0}\langle T_-^n \rangle + \gamma\langle T_+^n \rangle - \gamma\langle T_-^n \rangle = -n\langle T_-^{n-1} \rangle\,. \tag{I.2}$$

It is possible to re-write the system in terms of the differential operator $\mathcal{L}_\pm$

$$\mathcal{L}_+\langle T_+^n \rangle = \left[(f(x_0) + v_0)\,\partial_{x_0} - \gamma\right]\langle T_+^n \rangle = -n\,\langle T_+^{n-1} \rangle - \gamma\langle T_-^n \rangle\,, \tag{I.3}$$
$$\mathcal{L}_-\langle T_-^n \rangle = \left[(f(x_0) - v_0)\,\partial_{x_0} - \gamma\right]\langle T_-^n \rangle = -n\langle T_-^{n-1} \rangle - \gamma\langle T_+^n \rangle\,. \tag{I.4}$$

Applying $\mathcal{L}_-$ on eq. (I.3) and $\mathcal{L}_+$ on eq. (I.4) gives the second order differential equations

$$\left(f(x_0)^2 - v_0^2\right)\partial_{x_0}^2\langle T_+^n \rangle + \left[(f(x_0) - v_0)f'(x_0) - 2\gamma f(x_0)\right]\partial_{x_0}\langle T_+^n \rangle$$
$$= -n\left(f(x_0) - v_0\right)\partial_{x_0}\langle T_+^{n-1} \rangle + 2n\gamma\langle T^{n-1} \rangle\,, \tag{I.5}$$
$$\left(f(x_0)^2 - v_0^2\right)\partial_{x_0}^2\langle T_-^n \rangle + \left[(f(x_0) + v_0)f'(x_0) - 2\gamma f(x_0)\right]\partial_{x_0}\langle T_-^n \rangle$$
$$= -n\left(f(x_0) + v_0\right)\partial_{x_0}\langle T_-^{n-1} \rangle + 2n\gamma\langle T^{n-1} \rangle\,. \tag{I.6}$$

Summing the two second order equations gives

$$2\left(f(x_0)^2 - v_0^2\right)\partial_{x_0}^2\langle T^n \rangle + 2f(x_0)\left[f'(x_0) - 2\gamma\right]\partial_{x_0}\langle T^n \rangle - v_0 f'(x_0)\partial_{x_0}(\langle T_+^n \rangle - \langle T_-^n \rangle)$$
$$= -2nf(x_0)\partial_{x_0}\langle T^{n-1} \rangle + 4n\gamma\langle T^{n-1} \rangle + nv_0\,\partial_{x_0}(\langle T_+^{n-1} \rangle - \langle T_-^{n-1} \rangle)\,. \tag{I.7}$$

Similarly, by summing up Eqs. (I.1) and (I.2) gives us the following useful relation

$$v_0 \, \partial_{x_0}(\langle T_+^n \rangle - \langle T_-^n \rangle) = -2n\langle T^{n-1} \rangle - 2f(x_0)\partial_{x_0}\langle T^n \rangle \,. \tag{I.8}$$

Plugging this inside (I.7) leads to the final result announced in the main text in equation (11.2.13) for $n \geq 2$

$$\left(v_0^2 - f(x_0)^2\right) \partial_{x_0}^2 \langle T^n \rangle + 2f(x_0)\left[\gamma - f'(x_0)\right]\partial_{x_0}\langle T^n \rangle$$
$$= n\left(f'(x_0) - 2\gamma\right)\langle T^{n-1} \rangle + 2nf(x_0)\partial_{x_0}\langle T^{n-1} \rangle + n^2 \langle T^{n-2} \rangle \,. \tag{I.9}$$

In principle, Eq. (I.9) is of first order and could be solved formally. However, to express the explicit solution we need two conditions that may depend on the value of $n$. Note that in the diffusive limit, i.e., when $v_0 \to \infty$ and $\gamma \to \infty$ with $D = \frac{v_0^2}{2\gamma}$ fixed we retrieve the recursive relation [77]

$$D \, \partial_{x_0}^2 \langle T^n \rangle + f(x_0)\partial_{x_0}\langle T^n \rangle = -n\langle T^{n-1} \rangle \,, \tag{I.10}$$

which is sometimes called in the literature "Pontryagin equation" [77].

# J  Second Condition on the MFPT in Phase I and III

## J.1  Second Condition for Phase I

We consider here a situation where $|f(x)| < v_0$ and the RTP moves in the presence of a Reflecting Wall at $x = L$. What happens if the particle starts exactly at the wall $x(0) = L$? The constraint that we need to impose is that if the net velocity of the particle $f(L) \pm v_0$ is positive, it has to stay at the wall, i.e., $x(dt) = L$ (since this is a hard wall). Hence we need to distinguish two cases:

- *Case 1*: $f(L) + v_0 < 0$: here the velocity of the particle is negtive in both states and it moves towards the origin – just as if there were no wall at $x = L$, i.e.,

$$x(dt) = \begin{cases} L + [f(L) + v_0] \, dt & \text{, with probability } 1 - \gamma \, dt \text{ and } \sigma(t + dt) = +1 \\ L + [f(L) - v_0] \, dt & \text{, with probability } \gamma \, dt \text{ and } \sigma(t + dt) = -1 \,. \end{cases} \tag{J.1}$$

In this case, the dynamical evolution of $Q^\pm(L, t)$ is not affected by the wall, as given in Eqs. (11.2.3)-(11.2.4).

- *Case 2*: $f(L) + v_0 > 0$. The particle has positive speed in the '+' state and tries to cross the reflecting wall, while it is negative in the '-' state and moves normally towards the origin. Thus the dynamics reads

$$x(dt) = \begin{cases} L & \text{, with probability } 1 - \gamma \, dt \text{ and } \sigma(t + dt) = +1 \\ L + [f(L) - v_0] \, dt & \text{, with probability } \gamma \, dt \text{ and } \sigma(t + dt) = -1 \end{cases} \tag{J.2}$$

As a consequence the backward Fokker-Planck equations are modified at the wall. It leads to

$$Q^+(L, t + dt) = (1 - \gamma \, dt)Q^+(L, t) + \gamma \, dt \, Q^-(L + [f(L) - v_0] \, dt, t) \,, \tag{J.3}$$

where we recall that $Q^\pm$ are the survival probabilities in the states $\sigma = \pm 1$. We therefore have the following differential equation

$$\partial_t Q^+(L, t) = -\gamma \, Q^+(L, t) + \gamma \, Q^-(L, t) \,. \tag{J.4}$$

If we compare it to the Fokker-Planck equation for the survival probability of an RTP – Eq. (11.2.3) in the main text – we see that we have to impose

$$\theta(f(L) + v_0) \, \partial_{x_0} Q^+(x_0, t)\Big|_{x_0 = L} = 0 \,. \tag{J.5}$$

We conclude that when $f(L) > -v_0$, Eq. (J.5) applies and we can impose

$$\partial_{x_0} Q^+(x_0, t)\Big|_{x_0 = L} = 0 \,. \tag{J.6}$$

In other words, as we have

$$\tau_\gamma^+(x_0) = -\int_0^{+\infty} dt \, t \, \partial_t Q^+(x_0, t) \,, \tag{J.7}$$

it is equivalent to impose

$$\partial_{x_0} \tau_\gamma^+(x_0)\Big|_{x_0 = L} = 0 \,. \tag{J.8}$$

## J.2   Second Condition for Phase III

To obtain a second boundary condition in the Phase III, let us analyse the case where the RTP starts from the stable fixed point at $x_0 = x_-^s$, with $f(x_-^s) = -v_0$ in the initial state '+'. In the time interval $[0, dt]$, the position of the particle evolves according to

$$x(dt) = \begin{cases} x_-^s & \text{, w. proba. } 1 - \gamma \, dt, \; \sigma(dt) = +1 \\ x_-^s - 2v_0 \, dt & \text{, w. proba. } \gamma \, dt, \; \sigma(dt) = -1 \end{cases}, \tag{J.9}$$

where the first line corresponds to the case where $\sigma$ remains unchanged (i.e., positive) while the second line applies if $\sigma$ has flipped from '+' to '-'. One can then write a backward equation for $Q^+(x_0, t + dt)$ by decomposing the time interval into $[0, dt]$ and $[dt, t + dt]$. Using the evolution during the time interval $[0, dt]$ in (J.9), one clearly has

$$Q^+(x_-^s, t + dt) = Q^+(x_-^s, t)(1 - \gamma dt) + Q^-(x_-^s - 2v_0 dt, t)\gamma dt \,. \tag{J.10}$$

Expanding for small $dt$ one finds

$$\partial_t Q^+(x_-^s, t) = -\gamma \, Q^+(x_-^s, t) + \gamma \, Q^-(x_-^s, t) \,. \tag{J.11}$$

Therefore, to ensure the compatibility between Eq. (11.2.3) and Eq. (J.11), we need to impose the second condition

$$\lim_{x_0 \to x_-^s} \left[(f(x_0) + v_0) \, \partial_{x_0} Q^+(x_0, t)\right] = 0 \,. \tag{J.12}$$

As before, taking one time derivative of (J.12), multiplying by $t$ and integrating it over $t$, one obtains

$$\lim_{x_0 \to x_-^s} \left[(f(x_0) + v_0) \, \partial_{x_0} \tau_\gamma^+(x_0)\right] = 0 \,. \tag{J.13}$$

Let us now show that condition (J.13) on $\tau_\gamma^+$ also applies to $\tau_\gamma$. Using the same procedure that was used to derive Eq. (H.12), it is easy to show that we also have

$$\partial_{x_0} \tau_\gamma^+ = \left(1 - \frac{f(x_0)}{v_0}\right) \partial_{x_0} \tau_\gamma - \frac{1}{v_0} \,. \tag{J.14}$$

As by definition $f(x_-^s) = -v_0$, the condition (J.13) also imposes

$$\lim_{x_0 \to x_-^s} (f(x_0) + v_0) \, \partial_{x_0} \tau_\gamma(x_0) = 0 \,. \tag{J.15}$$

Therefore, for a force with one unique stable negative fixed point $f(x_-^s) = -v_0$, we conclude that $\tau_\gamma(x_0)$ is the unique solution of the differential equation (11.2.10) satisfying two following conditions

$$\tau_\gamma^-(x_0 = 0) = 0 \,, \tag{J.16}$$

$$\lim_{x_0 \to x_-^s} \left[ (f(x_0) + v_0) \, \partial_{x_0} \tau_\gamma(x_0) \right] = 0 \,. \tag{J.17}$$

| Mathis GUÉNEAU | 16 juin 2025 |

# Sujet : Dynamiques hors d'équilibres et temps de premier passage des processus stochastiques : du mouvement brownien aux particules actives

**Résumé** : Dans cette thèse, nous développons des méthodes analytiques pour étudier des processus stochastiques hors d'équilibre, dont le bruit est coloré, c'est-à-dire un bruit avec des corrélations temporelles. Ces processus non-markoviens sont plus difficiles à analyser que ceux avec un bruit blanc, comme le mouvement brownien. Nous nous concentrons principalement sur les systèmes de particules actives, en particulier la particule "run-and-tumble" soumise à une force arbitraire. Nous obtenons des expressions exactes pour son temps moyen de premier passage et sa probabilité de sortie d'un intervalle, en utilisant l'équation de Fokker-Planck. En particulier, nous montrons que le temps moyen de premier passage peut être optimisé en fonction de la fréquence des "tumbles". Nous étudions aussi des modèles de réinitialisation stochastique et de diffusion avec changement d'état. Pour ces derniers – qui sont des exemples de "diffusions Browniennes mais non gaussiennes" – nous utilisons une approche par renouvellement et la théorie des grandes déviations pour obtenir des résultats exacts sur plusieurs observables : la distribution de position de la particule, ses moments, mais aussi ses cumulants, qui permettent de caractériser les fluctuations non gaussiennes. De façon surprenante, nous mettons en évidence un lien inattendu entre ce modèle et les cumulants libres. Nous étudions également ces modèles dans un potentiel harmonique, en utilisant les variables de Kesten. Cette méthode nous permet d'écrire une équation intégrale pour la distribution stationnaire, que nous résolvons dans certains cas. Enfin, nous généralisons la dualité de Siegmund – un concept peu connu dans la littérature physique – aux particules actives, aux modèles de diffusion aléatoire, à la réinitialisation stochastique, et aux marches aléatoires en temps continu. Cette dualité établit un lien direct entre les observables de premier passage et les propriétés spatiales d'un processus dual, que nous construisons explicitement.

**Mots clés** : processus stochastiques; temps de premier passage; matière active; dualité de Siegmund; grandes déviations; cumulants libres.

# Subject : Non-Equilibrium Dynamics and First-Passage Properties of Stochastic Processes: From Brownian Motion to Active Particles

**Abstract**: In this thesis, we develop analytical methods to study out-of-equilibrium stochastic processes driven by colored noise, i.e., noise with temporal correlations. These non-Markovian processes pose significant analytical challenges compared to processes driven by white noise, such as Brownian motion. A primary focus is on active particle systems, specifically the run-and-tumble particle subjected to an arbitrary force. We derive exact expressions for its mean first-passage time (MFPT) and exit probability from an interval using the backward Fokker-Planck equation. Remarkably, we find that the MFPT can be optimized as a function of the tumbling rate. Additionally, we investigate stochastic resetting and switching diffusion models. For switching diffusion models which are examples of "Brownian yet non-Gaussian diffusions", we use a renewal approach and large deviation theory to derive exact results for various observables. These include the distribution of the position of the particle and its moments, but also its cumulants which are key observables to characterize non-Gaussian fluctuations. Notably, we uncover an unexpected connection between this model and free cumulants. We also examine these models in the presence of a harmonic potential by using Kesten variables. This approach enables us to write an integral equation for the steady-state distribution, which we solve in specific cases. Furthermore, we extend Siegmund duality – a concept that is not widely known in the physics literature – to active particles, random diffusion models, stochastic resetting, and continuous-time random walks. This duality establishes a direct relation between first passage observables and the spatial properties of a dual process, which we explicitly construct.

**Keywords**: stochastic processes; first-passage properties; active matter; Siegmund duality; large deviations; free cumulants.